\documentclass[twocolumn]{aastex63}
\accepted{in ApJ 18.01.2020}
\shortauthors{N. Wittenburg, P. Kroupa \& B. Famaey}
\shorttitle{The formation of exponential disk galaxies in MOND}

\begin{document}

\title{The formation of exponential disk galaxies in MOND}

\correspondingauthor{Nils Wittenburg}
\email{nwittenbrug@astro.uni-bonn.de}

\author{Nils Wittenburg}
\affiliation{University of Bonn, Helmholtz-Institut f{\"u}r Strahlen- und Kernphysik, Nussallee 14-16, D-53115 Bonn, Germany}

\author{Pavel Kroupa}
\affiliation{University of Bonn, Helmholtz-Institut f{\"u}r Strahlen- und Kernphysik, Nussallee 14-16, D-53115 Bonn, Germany}
\affiliation{Charles University in Prague, Faculty of Mathematics and Physics, Astronomical Institute, V Hole\v{s}ovi\v{c}k\'{a}ch 2, CZ-180 00 Praha 8, Czech Republic}

\author{Benoit Famaey}
\affiliation{Universit\'{e} de Strasbourg, Observatoire Astronomique de Strasbourg, CNRS UMR 7550, 11 rue de l'Universit\'{e}, F-67000 Strasbourg, France}

\begin{abstract}

The formation and evolution of galaxies is highly dependent on the dynamics of stars and gas, which is governed by the underlying law of gravity. To investigate how the formation and evolution of galaxies takes place in Milgromian gravity (MOND), we present full hydrodynamical simulations with the Phantom of Ramses (POR) code. These are the first-ever galaxy formation simulations done in MOND with detailed hydrodynamics, including star formation, stellar feedback, radiative transfer and supernovae. These models start from simplified initial conditions, in the form of isolated, rotating gas spheres in the early Universe. These collapse and form late-type galaxies obeying several scaling relations, which was not a priori expected. The formed galaxies have a compact bulge and a disk with exponentially decreasing surface mass density profiles and scale lengths consistent with observed galaxies, and vertical stellar mass distributions with distinct exponential profiles (thin and thick disk). This work thus shows for the first time that disk galaxies with exponential profiles in both gas and stars are a generic outcome of collapsing gas clouds in MOND. These models have a slight lack of stellar angular momentum because of their somewhat compact stellar bulge, which is connected to the simple initial conditions and the negligible later gas accretion. We also analyse how the addition of more complex baryonic physics changes the main resulting properties of the models and find this to be negligibly so in the Milgromian framework.

\end{abstract}


\section{Introduction}
Galaxies, groups of galaxies and clusters of galaxies are the largest gravitationally bound baryonic structures of our Universe. Yet, the formation and evolution of galaxies are still among the most challenging phenomena in astrophysics, partly because it is, even today, not possible to perform high resolution simulations that contain all dynamical and hydrodynamical effects from galaxy-scales down to the scale of a single star.\\
\indent
On the other hand, concerning their general properties, galaxies seem to be simple objects. The vast majority of galaxies with stellar mass above $\approx10^{10}M_\odot$ are late-type galaxies at all redshifts (e.g., \citealp{Delgado-Serrano2010}, \citealp{Tamburri2014}) with pure ellipticals representing less than 5 per cent (e.g., \citealp{Delgado-Serrano2010}). \cite{Disney2008} analysed hundreds of galaxies with respect to six of their properties, which describe the galaxy completely. Correlation matrices for all parameters were analysed with a principal component analysis. Surprisingly, they found that all parameters are strongly correlated and that they are determined by only one of them, which they could not determine with their method, thus only one principal component is evident from observations. Furthermore, the mass-discrepancy-acceleration relation/radial-acceleration relation (MDAR/RAR) (\citealp{McGaugh2004,Lelli2017}), which generalises several other galaxy-scaling relations such as the mass-asymptotic-speed relation (MASR aka Baryonic-Tully-Fisher relation, BTFR)\footnote{To prevent confusion about the velocity that has to be used here, we choose to use the name MASR.} (\citealp{Sanders1990,McGaugh2000}), the "dichotomy" between HSB and LSB galaxies (\citealp{McGaugh&deBlock1997,Tully&Verheijen1997}), the central density relation (\citealp{Lelli2013,Lelli2016c}), the "baryon-halo conspiracy" \citep{vanAlba&Sancisi1986}, Renzo's rule  \citep{Sancisi2004}, the Faber-Jackson relation \citep{Faber&Jackson1976} and the $\sigma_{\star}-V_{\mathrm{HI}}$ relation \citep{Serra2016}, while having little to none intrinsic scatter, strengthen the argument that galaxies follow a tight and seemingly simple law. This is further emphasized by late-type (i.e. disk) galaxies lying on a star-forming main sequence with a small dispersion, over a broad range of redshifts \citep{Speagle2014}.\\
\indent
Especially late-type galaxies show ordered structures, regardless of e.g. the total mass or morphology, as the majority of them have an exponentially decreasing stellar surface mass density profile. This phenomenon has been known for more than 40 years \citep{Freeman1970}, but its origin is still not understood. For example \cite{Herpich2017} show that these profiles might appear because of the dynamics within the disc, but there are also explanations that use the initial conditions of the formation of the galaxy for their reasoning (\citealp{Fall&Efstathiou1980,DalcatonSpergel&Summers1997,Dutton2009}).\\
\indent
The order and simplicity of galaxies, amongst other observations on galactic scales \citep{Kroupa2015}, is difficult to explain in the standard model of cosmology, as the evolution of galaxies is, in that framework, based on many subsequent mergers that happen stochastically. Fossils left over from this merging history orbit the galaxy as satellite galaxies, but the observed very significant phase-space correlation of satellite galaxies around the MW (\citealp{Kroupaetal2005,Pawlowski2012,Pawlowski2019}), Andromeda (\citealp{Metz2007}; \citealp{Ibata2013}) and Cen A \citep{Muller2018} is inconsistent with the observed satellites being such fossils \citep{Pawlowski2018}. The mutual alignment of the MW and Andromeda satellite population have not found an explanation in standard-dark-matter models, but appear to arise naturally in Milgromian dynamics \citep{Bilek2018,Banik2018}. Moreover, given the non-detection of dark-matter-particles despite a significant effort to find them \citep{Bertone2018} and astronomical arguments suggesting dark matter not to exist \citep{Kroupa2010,Kroupa2012,Kroupa2015}, it is necessary to also begin to investigate if non-dark-matter-based models lead to objects which resemble observed galaxies. This is of course an immense task, which we merely start taking on here, with rather simplified initial configurations allowing us to investigate some very generic trends. As we will show, some interesting regularities will nevertheless appear.\\
\indent
On galaxy-scales Milgromian dynamics (MOND) \citep{Milgrom1983,Famaey&McGaugh2012,Scholarpedia2014Moti}, accounts for the observations regarding the shape of the rotation curves of galaxies and galaxy-scaling relations such as the MASR (BTFR) (\citealp{Sanders1990,McGaugh2000,McGaugh2012}). As the baryonic content is the only source for the gravitational potential, MOND is also capable of predicting the rotation curve of a galaxy solely from its baryonic content and furthermore, due to its predictive nature, it lead to the discovery of new scaling relations, e.g. the MDAR \citep{Sanders1990,McGaugh2004,McGaugh2005,McGaugh2012,McGaugh2014}, which is mathematically the same as the RAR (see also Eq. 2 of \citealp{Milgrom1983} as a first prediction of the MDAR/RAR).\\
\indent
For completeness we note that \cite{Dicintio2016,Ludlow2017,Keller2017,Navarro2017} argue that these relations can be reproduced within the standard dark-matter based LCDM models, although they were not predicted and despite certain questions remaining open about their scatter (see, e.g. \citealp{Dicintio2016,Lelli2017,Desmond2017a,Desmond2017b}).\\
\indent
Although the MOND framework is analytically successful on galaxy-scales, it has hardly been tested in simulations. Part of the reason is its non-linearity and therefore the lack of simulation codes that are able to calculate the MOND potential. Especially, self consistent simulations with star formation and a full hydrodynamical treatment of gas were impossible to perform until the recent development of the Phantom of Ramses (POR) \citep{Lueghausen2015} and RAyMOND \citep{Candlish2015} codes, which are customized versions of the publicly available simulation code Ramses \citep{Teyssier2002}.\\
\indent
Previous applications of POR to state-of-the art problems are available in the very-high-resolution simulations of Antennae-like galaxies \citep{Renaud2016}, and (without hydro) in the simulation of the Sagittarius satellite galaxy and its stream by \cite{Guillaume2017} as well as the computations of streams from globular clusters \citep{Thomas2018}, and the computation of the Andromeda-Milky-Way encounter by \cite{Bilek2018}. Prior to this, Milgromian simulations with gas \citep{Tiret2008,Combes2014} were carried out by using sticky particles and a MOND Poisson solver \citep{Brada1999,Tiret2007}. These publications were concerned with the evolution of pre-existing disc galaxies and/or their satellites, but none were concerned with the formation of galaxy discs.\\
\indent
This work presents the first fully self consistent hydrodynamical simulations of galaxy formation done in the Milgromian framework. It is however worth insisting here that the MOND paradigm, as it stands today, is essentially mute on cosmology, so that the paradigm must necessarily be incomplete and needs to be embedded into a self-consistent cosmological model. Nevertheless, although details will of course depend on the specific parent theory, we can study some general consequences of a MOND-like force in cosmology. Firstly, because of the stronger force, dynamical measures of the cosmic mass density will be overestimated, just as in galaxies. Also, MOND forms structures more rapidly than Newtonian gravity with the same source perturbation \citep{Llinares2008}. In the early Universe, perturbations would (in principle) not grow because the baryons are coupled to the photon fluid.  Since the mass density is lower in MOND, matter domination would occur later than in $\Lambda$CDM. Consequently, MOND structure formation initially would lag behind $\Lambda$CDM. However, as the influence of the photon field would then decline and perturbations begin to enter the MOND regime, structure formation would rapidly accelerate. Large galaxies could form as early as $z \approx 10$ from the collapse of gas clouds which we model hereafter. Note that this means that there would be fewer mergers than in $\Lambda$CDM at intermediate redshifts, and in those rare mergers and much more numerous encounters between spirals, tidal dwarf galaxies would be formed and survive more easily than in $\Lambda$CDM \citep{Sanders1998}. This may indeed plausibly be the source of satellite galaxies and their observed phase-space correlation \citep{Banik2018,Bilek2018,Okazaki2000,Javanmardi2020}\\
\indent
At this point we want to focus on one important difference between simulations done in $\Lambda$CDM and MOND. Simulations done in the $\Lambda$CDM framework depend strongly on the feedback processes that are implemented in the code that is used, which are assumed by the community to be very uncertain. E.g. supernovae, AGN activity and ram-pressure introduce baryonic outflows from the galaxy that lead to a lower value of the baryon-to-DM fraction within the galaxy compared to the cosmic mean value. Therefore, the implementation of these processes and their fine-tuning are very important for the shape and properties of the galaxy. However, in MOND this should not be the case, because there is no large gravitating dark-matter-halo with its own dynamical history and no reason to eject material to obtain the right baryon fraction. We will show that in MOND it is indeed not necessary to introduce complex baryonic physics, as the general properties of the galaxies will not be significantly different. Furthermore, in $\Lambda$CDM simulations, angular momentum can be exchanged between the baryons and the dark halo. Again, this is not the case in simulations done in MOND.\\
\indent
A short introduction into the MOND framework and the numerical methods that were used is given in Section 2, afterwards the models are described in Sec. 3. Sec. 4 contains the detailed analysis of the formed galaxies. In Sec. 5 the computations are compared with observed scaling relations (MDAR and MASR) and in Sec. 6 the results are discussed and further work will be outlined that should be done in the future.
\section{Theory \& numerical methods}
\subsection{Modified Newtonian dynamics (MOND)}
Since galaxies show systematic and tightly correlated dynamical deviations from Newtonian behaviour, here we assume this to be the result of effective gravitation deviating from the classical Newtonian law. This is plausible, because the law of universal gravitation was empirically motivated only by Solar System objects, both in the case of Newton \citep{Newton1687} and Einstein \citep{Einstein1916}. We follow the empirical finding that galaxies follow scale-invariant dynamics (SID) in the low-acceleration limit. Scale invariance is the symmetry under which the gravitational and dynamical (actual) acceleration transform in the same way under scaling of lengths and time by a factor $\lambda$. Thus, if the actual acceleration $g$ (which scales as $\lambda^{-1}$) is a function of the Newtonian gravitational acceleration, $g_\mathrm{N}$ (which scales as $\lambda^{-2}$), their relation has to be of the form shown in Eq. \ref{eq:1}. As SI only leads to proportionality in Eq. \ref{eq:1}, $a_0$ needs to be normalized such that equality holds (\citealp{Milgrom1983}; \citealp{Milgrom2009}; \citealp{Wu+Kroupa2015}). Thus, when scale-invariance applies the true gravitational acceleration, $g$, is given by
\begin{eqnarray}\label{eq:1}
g=\sqrt{a_0g_\mathrm{N}},
\end{eqnarray}
where $a_0\approx10^{-10}\mathrm{ms^{-2}}\approx3.7\mathrm{pcMyr^{-2}}$ is Milgrom's constant and $g_\mathrm{N}$ is the Newtonian acceleration.\\
\indent
Eq.\ref{eq:1} is the basic equation underlying the MOND paradigm and it is only valid in the low acceleration deep-MOND limit $g\ll a_0\approx10^{-10}$ms$^{-2}$. To also encapsulate the symmetry breaking in the Newtonian regime, Milgrom's law is formulated as follows:
\begin{eqnarray}\label{eq:Milgrom}
\mathbf{g}=\nu\left(\frac{g_\mathrm{N}}{a_0}\right)\mathbf{g_\mathrm{N}},
\end{eqnarray}
with $\nu(y)$, $y=g_\mathrm{N}/a_0$, being the transition function, which is defined by its limits:
\begin{eqnarray}
\nu(y)\rightarrow1\ \mathrm{for}\ y\gg1\ \mathrm{and}\ \nu(y)\rightarrow y^{-1/2}\ \mathrm{for}\ y\ll1.
\end{eqnarray}
Eq. \ref{eq:1} correctly predicts galaxy scaling relations, e.g. the MASR (BTFR), and led to the discovery of new scaling relations like the MDAR \citep{Sanders1990,McGaugh2004,McGaugh2005,McGaugh2012}. Additionally it is the best fitting model to the aforementioned RAR, which can be considered to be the most general galaxy-scaling relation to date. MOND can be formulated as a classical gravitation theory with a fully developed Lagrangian formalism (\citealp{Bekenstein&Milgrom1984,Famaey&McGaugh2012}) and it may be related to the quantum physics of the vacuum (\citealp{Milgrom1999,Smolin2017}).
\subsection{Simulation code}
RAMSES is a publicly available hydrodynamical N-body code based on the Adaptive Mesh Refinement technique (AMR), hence it is a grid based code with a tree-like data structure so that the grid can be recursively refined on a cell-by-cell basis \citep{Teyssier2002}. Simulations start with the coarsest grid and the cells in this grid will split up into $2^{\mathrm{dim}}$ child cells (octs) if certain conditions are fulfilled, e.g. if the density or particle number threshold for the cell is reached. This also works vice versa, so that octs can be destroyed if e.g. the density becomes too low. Therefore the resolution increases in areas of interest.\\
\indent
The basic steps for grid based codes to update the positions and velocities for the particles are the following \citep{Teyssier2002}:
\begin{enumerate}
\item
The mass density $\rho$ on the mesh is calculated using a ``Cloud-In-Cell'' (CIC) interpolation scheme.
\item
The potential $\phi$ on the mesh is calculated by solving the Poisson equation (see the description of POR below for the MOND case).
\item
The acceleration on the mesh is calculated using a standard finite-difference approximation of the gradient.
\item
The particles' accelerations are computed using an inverse CIC interpolation scheme.
\item
The particles' velocities are updated according to their accelerations.
\item
The particles' positions are updated according to their velocities.
\end{enumerate}
To be more precise, the Poisson equation is solved in RAMSES by minimizing the residual $\triangle\phi-4\pi G\rho$ using the Gauss-Seidel method and the particles' positions and velocities are updated using the Leap-Frog scheme with adaptive time steps.\\
\indent
It is important to note that RAMSES uses a "one-way interface" scheme, i.e. the finer level solutions are updated using the coarser level solutions and not vice versa, which would be a "two-way interface" scheme. But this basic scheme is not exactly true for all simulations done in this work, as most of the simulations were computed with MOND (see Section 2.2.3).\\
\indent
As mentioned before not only the trajectories and accelerations of particles can be computed with RAMSES, but it also includes a hydrodynamical solver. Therefore star formation and cooling/heating processes of a given gas content are available as well as dynamical phenomena like shocks. In the code the Euler equations in their conservative form are solved using a second-order Godunov scheme with a Riemann solver (see \citealp{Teyssier2002} and references therein).

\subsubsection{Star formation and sink particles}
Star formation is very important for the simulations done in this work, because all particles are born from the initial gas content.\\
\indent
Star formation in general occurs due to the fragmentation of a gas cloud and the formation of dense clumps. If these clumps are heavier than the Jeans mass, they should collapse under self-gravity and form stars. However this mechanism is not implemented in codes that are made for cosmological and galactic scales due to several reasons. The resolution of the finest grid would have to be smaller than the size of a star, which would increase the computation time drastically. Further, if the resolution is on that level, all physical processes that act on this scale would have to be implemented, e.g. a correct IMF \citep{Kroupa2013,Jerabkova2018}, the evolution of single stars and most importantly star-star interactions and binary stars (e.g. \citealp{Oh2016}), altogether amounting to a prohibitive computational cost.\\
\indent
In RAMSES, star formation is implemented as follows. Stellar particles are only created if two criteria are met. First, the gas mass-volume-density, $\rho$, needs to exceed a user defined threshold and second, a dimensionless integer number, $n_\star$, drawn from a Poisson distribution has to be unequal to zero. The distribution has a mean value of $\rho_{\mathrm{SFR}}d^3_{x}dt/M_\star$ and depends on the local star formation rate density, $\rho_{\mathrm{SFR}}$, the length of the time step, $dt$, the 1D size of the cell, $d_x$, in which the density threshold is exceeded and $M_\star$, which is the mass scale of newly born stars (either user defined or given by $M_\star=\rho_\mathrm{thres}d_{x}^3$). The SFR follows the Schmidt law, $\rho_{\mathrm{SFR}}=\rho/t_{\star}\propto\rho^{3/2}$, where $t_{\star}=t_\mathrm{ff}(\rho/\rho_\mathrm{thres})^{-1/2}$ is the star formation time scale, which is proportional to the local free fall time, $t_\mathrm{ff}=\sqrt{3\pi/32G\rho}$. The code checks at every time step if any cell exceeds the density threshold, then $n_\star$ is drawn for the respective cell and if it is non-zero, a stellar particle is created by converting the amount of gas, $n_\star M_\star$, into the mass of the stellar particle. In our work the mass of these particles exceed $10^4\mathrm{M_\odot}$ due to our resolution, so they may be viewed as clusters of stars rather than single stars.\\
\indent
All particles produced this way are disconnected from the hydrodynamical evolution and only interact with the gas through gravity and feedback, which starts after a certain delay or directly depending on the user's choice (here a delay of $10\mathrm{Myr}$ is applied).\\
\indent
Also a different kind of particle can be produced within the simulation, namely sink particles (sinks), which are meant to stabilize the simulation.\\
\indent
In galaxy-scale simulations it is nearly impossible to resolve single stars as mentioned before, so following the gravitational collapse of a gas cloud has to be artificially stopped at some point due to the maximum resolution. But not resolving the Jeans length and mass can lead to artificial fragmentation of the gas in collapsing regions \citep{Truelove1998}. This can be avoided using different techniques. One is to introduce a barotropic equation of state, so that the gas heats up if a certain density threshold is met. The downside of this implementation is that the gas formations stay extended and are more vulnerable to disruption (e.g through shocks or tidal stripping). Also because of this A. Bleuler and R. Teyssier invented and implemented a new sink particle scheme into RAMSES. The idea is that collapsing gas clouds are artificially stopped if a certain density is exceeded and the gas is condensed into a point mass, which is decoupled from the hydrodynamical evolution as mentioned above. So the user can choose the density threshold according to a maximum resolution or certain physical scale (for further details about the implementation and tests of the sink particle scheme see \citealt{Teyssier2014sink}).
\subsubsection{Cooling/Heating and radiative transfer}
As explained before, RAMSES uses a second-order Godunov scheme to solve the Euler equations in their conservative form. There is also a radiative-transfer (RT) addition to RAMSES called RAMSES-RT \citep{Rosdahl2013rt}, which uses a first-order Godunov solver, which will only briefly be explained at the end of this section.\\
\indent
In the code itself the function, which contains radiative cooling and heating, is computed separately, therefore the thermochemistry scheme can be changed without changing the whole hydrodynamical solver. It depends on the gas density, temperature, ionization states of the gas (e.g. Helium, Hydrogen) and metallicity, but in simulations without the addition of the RT scheme, collisional ionization equilibrium (CIE) is assumed. That means that the ionization states can be calculated with the temperature and the density alone, so they do not need to be tracked in the code. Cooling and heating of the gas is therefore computed by using tables that are included in the code, which describe the cooling and heating rates due to several physical processes. The heating term includes photoionization processes and the radiative cooling processes are: collisional excitation, collisional ionization, recombination, bremsstrahlung and Compton scattering. The \citet{Sutherland&Dopita1993} cooling model is used in RAMSES with look-up tables in the temperature and metallicity plane and also the tables from \citet{Courty+Alimi2004} are used for the different cooling/heating processes.\\
\indent
In fact there is no rigorous treatment of RT by using this scheme, because the ionization states are computed by assuming CIE so that the cooling and heating rates can be computed with the temperature and density as described above. In the end the temperature is updated and with it the total energy density by also taking the rest of the Euler equations into account.\\
\indent
As mentioned at the beginning of this section there is an option called RT which can be enabled for simulations. By doing so the simulation is run with a different hydrodynamical solver. With this addition photon fluxes are introduced and the code keeps track of the ionization states, because they are now correctly computed by using photons, collisions and non-equilibrium thermochemistry. So radiative transfer between sources like stellar particles is implemented.\\
\indent
Using the RT option increases the memory requirement massively, because the \textit{conservative} state vector $\mathcal{U} = (\rho,\rho\bf{u},E,\rho Z )$, which stores the hydrodynamical properties of each cell (gas density $\rho$, momentum density $\rho \bf{u}$, total energy density $E$ and metal mass density $\rho Z$) becomes $\mathcal{U}=(\rho,\rho\bf{u},E,\rho Z,\rho x_{\mathrm{H_{II}}},\rho x_{\mathrm{He_{II}}},\rho x_{\mathrm{He_{III}}},N_i,\bf{F}_i)$, where new variables, connected to the photons, are introduced (ionization fraction densities of Hydrogen and Helium $\rho x_{\mathrm{H_{II}}},\rho x_{\mathrm{He_{II}}},\rho x_{\mathrm{He_{III}}}$ with e.g. $x_{\mathrm{H_{II}}}=n_{\mathrm{H_{II}}}/n_\mathrm{H}$, photon number density $N_i$ and photon Fluxes $\bf{F}_i$), and this roughly increases the memory requirement by a factor of 3.5.
\subsubsection{Phantom of Ramses (POR)}
Every computation done in this work used a customized version of RAMSES, called Phantom of Ramses (POR). POR \citep{Lueghausen2015} is the only publicly available code, which is capable of running full hydrodynamical simulations with a MONDian Poisson solver. Part of the reason why this is the only publicly available code is that the generalization of the Poisson equation in MONDian theories is non-linear, which makes it very challenging or impossible to implement into an N-body code. However there is a formulation of MOND, which can be implemented without changing the Poisson solver of the code. It is called the quasi-linear formulation of MOND (QUMOND, \citealp{Milgrom2010}), which is derived from an action and obeys the standard conservation laws, with the generalised Poisson equation,
\begin{eqnarray}
\label{eq:8}
&\triangle \Phi(\mathbf{x}) = 4\pi G\rho_\mathrm{b}(\mathbf{x})+\nabla \cdot [\tilde{\nu}(|\nabla\phi|/a_0)\nabla\phi(\mathbf{x})],\\
\label{eq:9}
&\triangle \Phi(\mathbf{x}) = 4\pi G(\rho_\mathrm{b}(\mathbf{x})+\rho_\mathrm{ph}(\mathbf{x})).
\end{eqnarray}
Here $\rho_\mathrm{b}(\mathbf{x})$ is the baryonic density, $\phi(\mathbf{x})$ the Newtonian potential, which fulfills the standard Poisson equation $\triangle\phi(\mathbf{x})=4\pi G\rho_\mathrm{b}(\mathbf{x})$, $\Phi(\mathbf{x})$ is the total gravitational potential and $\tilde{\nu}$ is the transition function between the Newtonian and the MOND regime, which is $\nu$ from Eq. \ref{eq:Milgrom} minus 1 with the limits:  $\tilde{\nu}(y)\rightarrow 0$ if $y\gg1$ (Newtonian regime) and $\tilde{\nu}(y)\rightarrow y^{-1/2}$ if $y\ll1$ (MOND regime) with $y=g_N/a_0$. Several functions that fulfill this criterion have been used in the literature (see \citealp{Lueghausen2015}), but here
\begin{eqnarray}\label{eq:nu}
\tilde{\nu}(y)=-\frac{1}{2}+\frac{1}{2}\sqrt{1+\frac{4}{y}}
\end{eqnarray}
is used.\footnote{Note that Eq. \ref{eq:nu} corresponds to the `simple' interpolating function, see, e.g., Sect. 6.2 of \citet{Famaey&McGaugh2012}.}\\
\indent
Note that the second term on the right hand side of Eq.\ref{eq:8} was condensed into
\begin{eqnarray}\label{eq:10}
\rho_\mathrm{ph}(\mathbf{x})=\frac{\nabla \cdot [\tilde{\nu}(|\nabla\phi|/a_0)\nabla\phi(\mathbf{x})]}{4\pi G},
\end{eqnarray}
so the generalised Poisson equation also visualizes its quasi-linearity. Eq.\ref{eq:9} shows that the total gravitational potential in MOND depends on the baryonic density and an additional term, which is called $\rho_\mathrm{ph}$, also has the units of matter density and depends on the Newtonian potential alone. Therefore the total gravitational potential can be written as $\Phi=\phi+\Phi_\mathrm{ph}$, where it is divided into a Newtonian and a MONDian part. $\rho_\mathrm{ph}$ is called "phantom dark matter" (PDM) density, but it is not a real matter distribution. It is a mathematical formulation to compare more easily MOND with Newtonian dynamics. It is also exactly the density that would be interpreted as dark matter in the standard cosmology framework.\\
\indent
As mentioned before, the generalised Poisson equation is not linear (through the computation of $\rho_\mathrm{ph}$), so the Poisson solver in POR is different from the one in RAMSES. However the QUMOND formulation makes the implementation much easier, because the already existing standard Poisson solver can be used and no new solver needs to be implemented. Therefore, to solve the complete Poisson equation for QUMOND, three steps were implemented in the code:
\begin{enumerate}
\item
After the smoothed matter density distribution is calculated in a given cell from gas and particles, the standard Poisson equation,
\begin{eqnarray}\label{eq:11}
\triangle\phi(\mathbf{x})=4\pi G \rho_\mathrm{b}(\mathbf{x}),
\end{eqnarray}
is solved to compute the Newtonian potential $\phi$ and its gradient.
\item
The PDM density is calculated using Eq.\ref{eq:10} (for the detailed scheme see \citealp{Lueghausen2015}).
\item
With the matter density and the PDM density, the whole Poisson equation in QUMOND (Eq.\ref{eq:9}) is solved to compute the total gravitational potential. The gradient of this potential at location $\mathbf{x}$ then yields the acceleration, $\mathbf{a}$, at $\mathbf{x}$, $\mathbf{a}(\mathbf{x})=-\mathbf{\nabla}\Phi(\mathbf{x})$.
\end{enumerate}
We stress here that only the calculation of the potential is slightly changed, such that the total gravitational potential in MOND is calculated. Therefore the hydrodynamical solver is unchanged, but uses the total gravitational potential from the MONDian gravity solver. 
\section{The Models}
The computations start from pure gas clouds, setting up identical models, except for the initial rotation velocity, radius and mass. This was done to see whether it is possible to reproduce roughly the Hubble Sequence \citep{Hubble1926} by changing the initial rotation velocity from 0 to a certain value. According to \citet{Disney2008} galaxies are simple objects and their properties only depend on a single parameter, so if MOND contains the right description of gravitational dynamics\footnote{Right in the sense of reproducing the observed properties of galaxies.}, it should be possible to reproduce the Hubble Sequence in simulations, that is if all members of the sequence can form in isolation from one gas cloud without further accretion of gas.\\
\indent
The initial distribution of gas is a uniform sphere with the same mass density for all simulations, which leads to a surface mass density distribution, $\Sigma_\mathrm{init}(r_\mathrm{cyl})$ that only depends on the initial spherical radius, $r_\mathrm{init}$, and on the cylindrical radius, $r_\mathrm{cyl}$,
\begin{eqnarray}
\Sigma_\mathrm{init}(r_\mathrm{cyl})=2\rho_\mathrm{init}\sqrt{r_\mathrm{init}^2-r_\mathrm{cyl}^2},
\end{eqnarray}
with $\rho_\mathrm{init}$ being the constant initial mass density. Hence the models are initially morphologically unrelated to late type galaxies, which are rotationally-supported exponential disks. So if a disk galaxy forms, it forms because of the dynamics and not because of the initial distribution of gas. 
\subsection{Initial conditions}
The initial conditions used may be viewed to be a first rough approximation of an early-Universe gas cloud, which is gravitationally unstable. Later work will investigate initially turbulent gas clouds. We emphasize that the initial conditions used here are not based on standard cosmology, as it would be unlogical and unphysical to use initial conditions from the LCDM model in a model of galaxy formation without dark matter. In particular, since possible parent theories of MOND might be inherently non-local, using e.g. the high-z power spectrum in a MOND context is practically impossible, and we need to use some initial conditions, which are likely to not be very unphysical (i.e. pure gas clouds). Ultimately, by studying which initial conditions lead to which types of galaxy, we should be able to constrain MONDian cosmological theory.\\
\indent
Here it is assumed that neutral gas clouds form by a redshift of about $15$ to $50$ \citep{Barkana&Loeb2001}. The calculations thus begin with a neutral cloud with the following initial properties (see also Table \ref{tab:initconditions}). The angular momentum of the gas cloud is expected to result from its internal motions as it cools and begins to collapse and from tidal torquing. In the past, the angular momentum of galaxies has been associated with tidal torquing \citep{Efstathiou1979,Wesson1985,Voglis1994,Catelan1996}. In the models explored here a very simple initial law, which significantly differs from that of galaxies, is assumed in order to ascertain that the properties of the final galaxies are not assured through the specific initial conditions.\\
\indent
The models start as a constant-density sphere of gas, where the initial rotation velocity has a radial dependence,
\begin{eqnarray}
\mathbf{v}_\mathrm{i}(\mathbf{r})=\eta \left( \mathbf{r} \times \mathbf{1}_z \right),
\end{eqnarray}
with $\eta$ being an angular velocity parameter, $\mathbf{r}$ the radius vector, $\mathbf{v}_\mathrm{i}$ the initial rotation velocity vector and $\mathbf{1}_z$ the unit vector in the direction of the z-axis. This is motivated by the first Larson relation \citep{Larson1981}, which relates the velocity dispersion of a molecular gas cloud proportionally to the radius. Only three parameters are available to change the model: $\eta$ for the initial rotation velocity distribution, the initial 3D radius, $r_\mathrm{init}$, and mass, $M_\mathrm{init}$, of the sphere. Important to note here is that the initial velocity and density distribution are vastly different from those of observed disk galaxies.\\
\indent
\begin{table*}[h]
\caption{Initial conditions of all models. The first column shows the name and number of the model, the second one the initial mass, the third the initial radius of the sphere, column 4 shows the initial velocity parameter $\eta$, column 5 which options are added/changed (sn=supernova, sink=sink particles, rt=radiative transfer, $Z$=metallicity, $\mathrm{res_{max}}$=maximum resolution) and column 6 gives information about which Poisson solver is used. Note that $\eta$ for model M1const has the units $\mathrm{kms}^{-1}$, because it starts with a constant rotation velocity throughout the whole sphere of gas.}
\centering
\begin{tabular}[c]{l l l l l l}
\hline
Model & $M_\mathrm{init}$ & $r_\mathrm{init}$ & $\eta$ & Additions & Poisson \\
name/no. & $[10^9M_\odot]$ & $[\mathrm{kpc}]$ & $[\mathrm{kms}^{-1}\mathrm{kpc}^{-1}]$ & & solver \\
\hline
M1/1 & $6.4$ & $20$ & $1.44$ & - & MOND \\

M1sn/2 & $\shortparallel$ & $\shortparallel$ & $\shortparallel$ & sink,rt,sn & $\shortparallel$ \\

M1N/3 & $\shortparallel$ & $\shortparallel$ & $\shortparallel$ & - & Newton \\

M1const/4 & $\shortparallel$ & $\shortparallel$ & $6.56\mathrm{kpc}$ & - & MOND\\
M1Zpoor/5 & $\shortparallel$ & $\shortparallel$ & $1.44$ & $Z=10^{-4}z_\odot$ & $\shortparallel$\\
M1Zpoorsn/6 & $\shortparallel$ & $\shortparallel$ & $\shortparallel$ & $Z=10^{-4}z_\odot$ & $\shortparallel$\\
&&&&sink,rt,sn&\\
M1l11/7 & $\shortparallel$ & $\shortparallel$ & $\shortparallel$ & $\mathrm{res_{max}}$ & $\shortparallel$\\
&&&& $=468.75\mathrm{pc}$ &\\
M1l13/8 & $\shortparallel$ & $\shortparallel$ & $\shortparallel$ & $\mathrm{res_{max}}$ & $\shortparallel$\\
&&&& $=117.19\mathrm{pc}$ &\\

M2/9 & $21.6$ & $30$ & $0.39$ & - & $\shortparallel$ \\

M2sn/10 & $\shortparallel$ & $\shortparallel$ & $\shortparallel$ & sink,rt,sn & $\shortparallel$ \\

M2N/11 & $\shortparallel$ & $\shortparallel$ & $\shortparallel$ & - & Newton \\

M3/12 & $\shortparallel$ & $\shortparallel$ & $0.96$ & - & MOND \\

M3sn/13 & $\shortparallel$ & $\shortparallel$ & $\shortparallel$ & sink,rt,sn & $\shortparallel$ \\

M4/14 & $100.0$ & $50$ & $0.58$ & - & $\shortparallel$ \\

M4sn/15 & $\shortparallel$ & $\shortparallel$ & $\shortparallel$ & sink,rt,sn & $\shortparallel$ \\
\hline
\end{tabular}
\label{tab:initconditions}
\end{table*}
The starting temperature of the gas is $T=10^4\mathrm{K}$ and the density threshold for star formation to take place is set to $\rho_\mathrm{star}=0.1\mathrm{Hcm}^{-3}$ with a star formation efficiency, $\mathrm{sfe}$, of $5\%$ \citep{Dubois+Teyssier2008}. Also all models evolve in isolation and there is no UV-background radiation. The initial density is equal for all simulations with a value of $\rho_\mathrm{init}=7.83\times10^{-3}\mathrm{Hcm}^{-3}$. To mimic isolation the density of the intergalactic medium outside the gas sphere is set to $\rho_\mathrm{IGM}=10^{-5}\times\rho_\mathrm{init}$ such that after the collapse of the gas cloud, further accretion from the ambient intergalactic medium onto the formed galaxy is not significant. The minimum mass of a stellar particle is $M_{\star}\approx3\times10^{4}\mathrm{M}_\odot$.The initial metallicity is solar, except for M1Zpoor and M1Zpoorsn, and constant throughout the simulations. The maximum resolution for most simulations is $234.375\mathrm{pc}$ (for M1l11 and M1l13 see Tab. \ref{tab:initconditions}), which is the length of the smallest grid-cell and the size of the coarsest grid-cell (the simulation box itself) is set to $960\mathrm{kpc}$. Milgrom's constant is set to $a_0=1.12\times10^{-10}\mathrm{ms^{-2}}$ within the code.
Supernovae and sink particles will be discussed later, when such models are explained (Sec. 4.1).\\
\indent
15 different models were calculated for this work with the following properties (see also Tab. \ref{tab:initconditions}):
\begin{itemize}
\item
M1 assumes simple cooling/heating and no additional or more complex baryonic physics, such as e.g. supernovae. It starts with $M_\mathrm{init}=6.4\times10^9\mathrm{M}_\odot$, $r_\mathrm{init}=20\mathrm{kpc}$ and $\eta=1.44\mathrm{kms}^{-1}\mathrm{kpc}^{-1}$ and is simulated with the MONDian Poisson solver.
\item
M1sn has the same initial conditions as M1 with the addition of supernovae, sink particles and explicit radiative transfer.
\item
M1N also is initially identical to M1, but it is simulated with the Newtonian Poisson solver.
\item
M1const is set up with a different (constant) initial rotation law with $\eta=0.1(\mathrm{kpc}/r)\times65.6\mathrm{kms}^{-1}$ and is otherwise identical to M1.
\item
M1Zpoor and M1Zpoorsn are M1 and M1sn with lower metallicity $Z=10^{-4}\times Z_\odot$.
\item
M1l13 and M1l11 are M1 with a higher and smaller maximum refinement level respectively and therefore different maximum resolution, $\mathrm{res_{max}}=117.1875\mathrm{pc}$ and $\mathrm{res_{max}}=468.75\mathrm{pc}$.
\item
M2: $M_\mathrm{init}=21.6\times10^9\mathrm{M}_\odot$, $r_\mathrm{init}=30\mathrm{kpc}$, $\eta=0.39\mathrm{kms}^{-1}\mathrm{kpc}^{-1}$
\item
M3: $M_\mathrm{init}=21.6\times10^9\mathrm{M}_\odot$, $r_\mathrm{init}=30\mathrm{kpc}$, $\eta=0.96\mathrm{kms}^{-1}\mathrm{kpc}^{-1}$
\item
M4: $M_\mathrm{init}=100\times10^9\mathrm{M}_\odot$, $r_\mathrm{init}=50\mathrm{kpc}$, $\eta=0.58\mathrm{kms}^{-1}\mathrm{kpc}^{-1}$
\item
The name convention for M2, M3 and M4 is identical to M1, so sn means with sinks, supernovae and radiative transfer and N means with the Newtonian Poisson solver. Only $M_\mathrm{init}$, $r_\mathrm{init}$ and $\eta$ are different.
\end{itemize}
\subsection{Galaxy models}
The initial collapse time is similar for all models, except for the Newtonian ones (this will be further discussed in Sec. 4.7), $\approx0.5\mathrm{Gyr}$. 
So after approximately $0.5\mathrm{Gyr}$ the spheres collapse and form a rotating dense, thin disk with less-dense gas surrounding it. All figures and videos of the simulations are made by projecting the relevant property of the galaxy onto the respective plane (edge-on/face-on view), i.e. a figure consists of equally sized cells ($0.46875\times0.46875\mathrm{kpc^2}$) and for every figure-cell the mass weighted average of the stellar surface mass density and the gas density, along the axis that is not shown, is computed.\\
\begin{figure*}[h]
\begin{minipage}[t]{0.49\linewidth}
\includegraphics[width=1.0\linewidth]{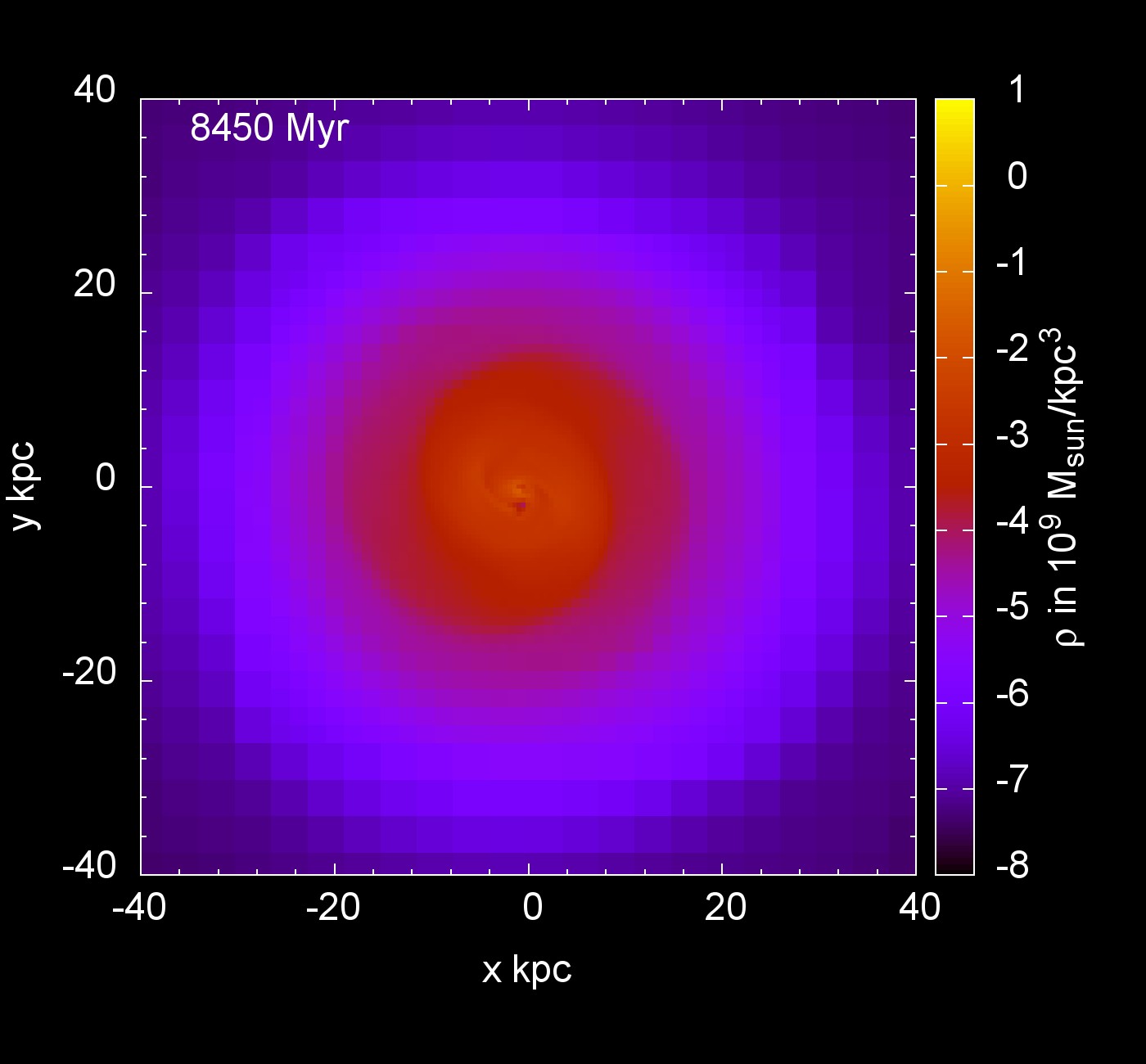}
\end{minipage}
\begin{minipage}[t]{0.49\linewidth}
\includegraphics[width=1.0\linewidth]{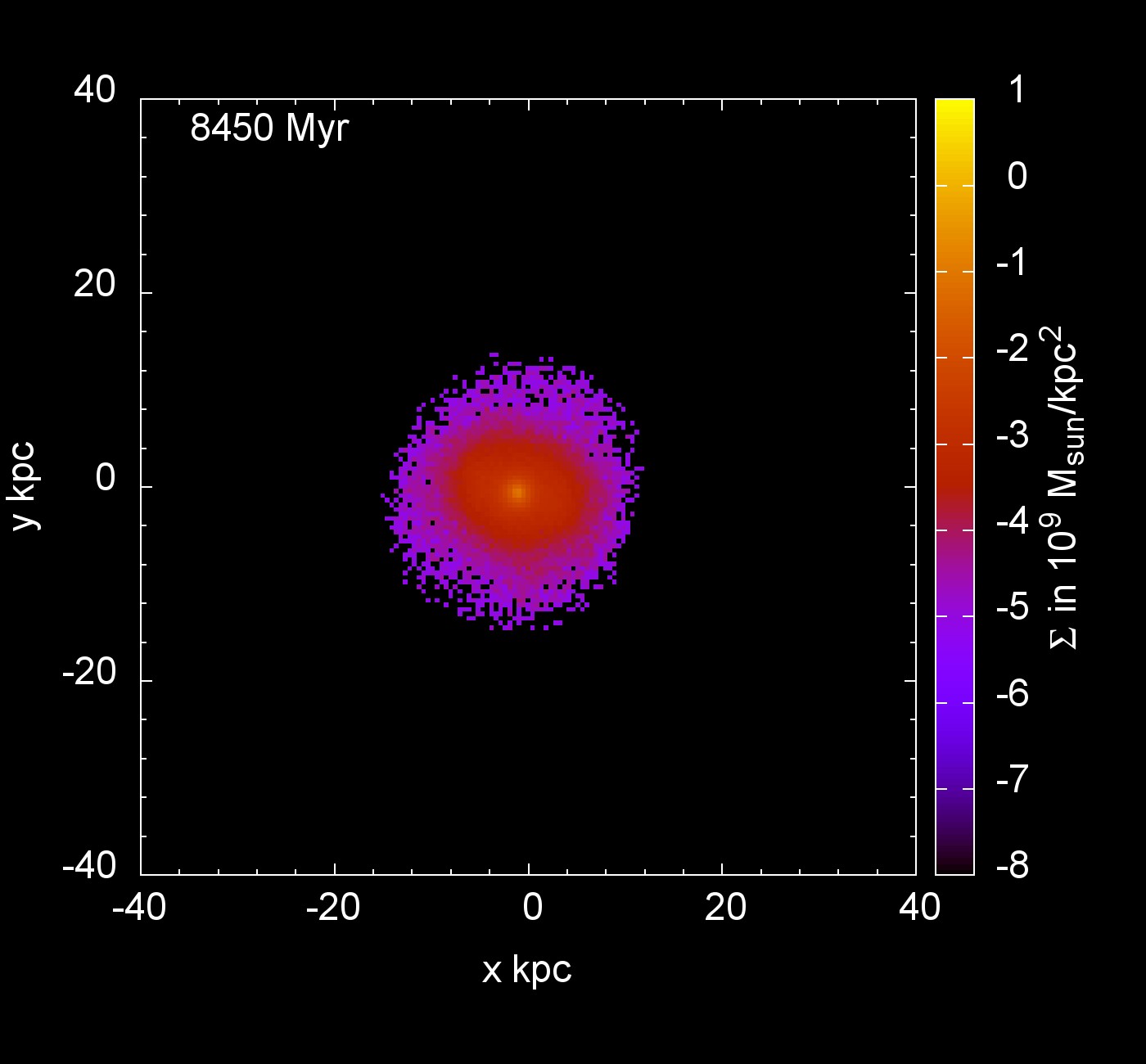}
\end{minipage}
\begin{minipage}[t]{0.49\linewidth}
\includegraphics[width=1.0\linewidth]{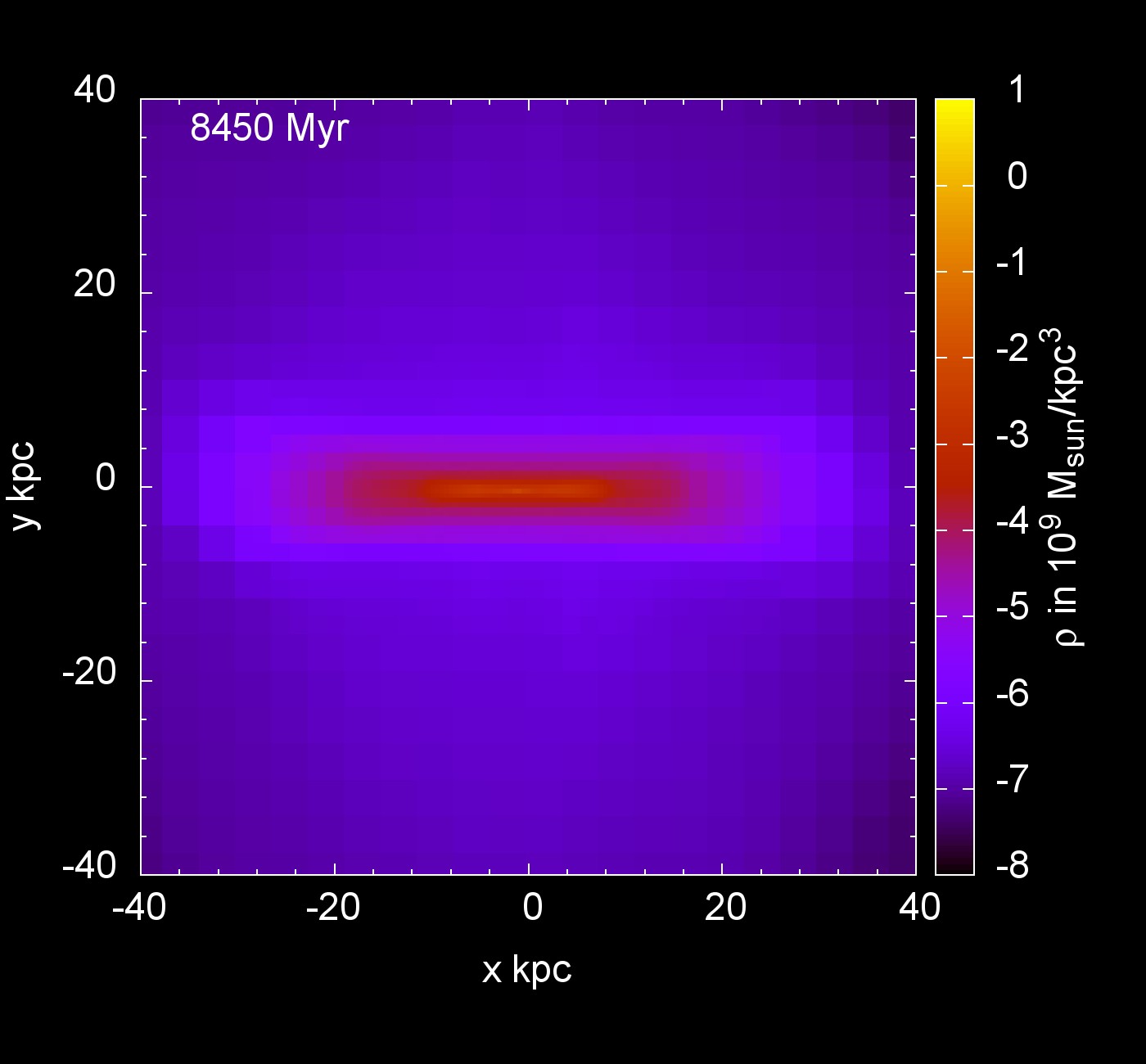}
\end{minipage}
\begin{minipage}[t]{0.49\linewidth}
\includegraphics[width=1.0\linewidth]{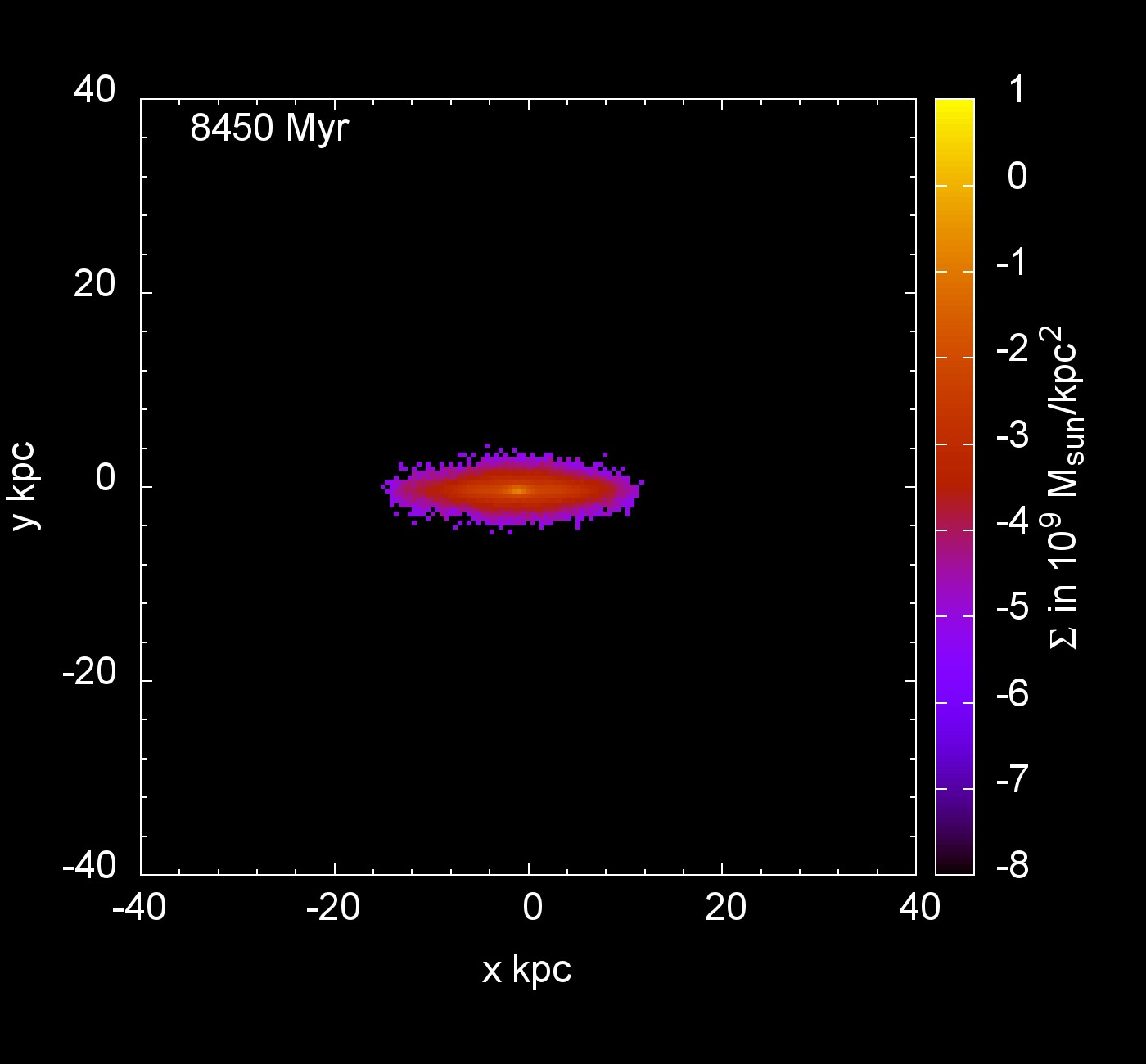}
\end{minipage}
\caption{M1sn after $8.45\mathrm{Gyr}$ shown face-on (top panels) and edge-on (bottom panels). \textit{Left panels:} Gas-density, $\rho$, of M1sn colour-coded as indicated by the scale on the right of the plot. Also, a void-like structure can be seen near the centre, which is a 'recent' supernova explosion (top left panel). \textit{Right panels:} Stellar surface mass density, $\Sigma$, plotted with the same colour-scheme as $\rho$. For movies of the formation of these galaxies see Youtube entry:"Formation of disk galaxies in MOND by Nils Wittenburg".}
\label{fig:M1snexample}
\end{figure*}
\indent
During the collapse stellar particles form a thin, dense disk with a stellar halo surrounding the centre and a less dense component in the outskirts. Very shortly after the formation of the disk-like structure, a spiral pattern arises, which is especially visible in the gaseous component (face-on).\\
\indent
Until the end of the simulation ($10\mathrm{Gyr}$) the appearance of the galaxy does not change much, but it spreads radially, especially the less dense gaseous component does so. A spiral pattern in the face-on view of the gas remains, while the stellar component shows a very dense core (face-on) and a thin, dense disk surrounded by a flattened stellar halo (edge-on), see for example Fig. \ref{fig:M1snexample}.\\
\indent
The left panel of Fig. \ref{fig:rccomparisonr2010Gyr} shows the rotation curve of M1 (red crosses) and the 'pristine' MOND rotation curve (Eq. \ref{eq:Milgrom}, purple boxes) after $10\mathrm{Gyr}$. The rotation curve is calculated by averaging the cylindrically radial accelerations of the stellar particles in radial bins of $500\mathrm{pc}$ and then computing the circular velocity according to $v_\mathrm{c}=\sqrt{ar}$, with $a$ being the acceleration and $r$ the radius of the radial bin. The uncertainties are calculated from the scatter around the average value of the acceleration in the bins and then propagated by Gaussian error propagation.\\
\indent
The 'pristine' rotation curve is calculated by using Eq. \ref{eq:Milgrom}, assuming circular motion and calculating the Newtonian circular velocity beforehand. In general the relation between $g$ and $g_N$ is equivalent for QUMOND and Milgrom's law up to a curl field correction \citep{Brada1995}, so every difference between the two rotation curves shows directly the effect of this curl field. Therefore, comparing the rotation curve of the simulation (QUMOND) with the 'pristine' rotation curve (approximation that stems from Milgrom's law) not only shows the effect of the curl field, but also how significant this curl field correction is.\\
\indent
As is evident from Fig. \ref{fig:rccomparisonr2010Gyr} the rotation curve rises steeply in the centre ($r<1\mathrm{kpc}$), decreases afterwards for approximately $2\mathrm{kpc}$, this being the Newtonian regime, and then becomes flat for the majority of the galaxy as indicated by the fit. To calculate the asymptotically flat rotation velocity of the galaxy, $v_\mathrm{rot,flat}$, a constant fit function,
\begin{eqnarray}\label{eq:flatvc}
c(r)=v_\mathrm{rot,flat},
\end{eqnarray}
is used here, which results in $v_\mathrm{rot,flat}=103.03\pm0.19\mathrm{kms^{-1}}$ for M1. The fit is performed for $3<r/\mathrm{kpc}<20$ (all fits with Eq. \ref{eq:flatvc} are calculated from the beginning of the flat part of the rotation curve until the end of the stellar disk, which changes depending on the model).\\
\indent
The differences between the 'pristine' curve and the measured curve are most prominent in two regions of the galaxy. First around $3\mathrm{kpc}$ away from the centre, where the rotation velocity decreases significantly compared to the initial peak, but is still outside the flat part of the curve and the second region is the end of the stellar disk, where the measured rotation velocity increases again slightly in contrast to the behaviour of the 'pristine' curve.\\
\indent
The left panel of Fig. \ref{fig:ddcomparisonr2010Gyr} shows the surface mass density distribution, $\Sigma$, of stars, gas and the sum of both after 10$\mathrm{Gyr}$. $\Sigma$ is calculated by adding the mass of every star or gas-cell within a radial bin and dividing the total mass of the bin by its area. The uncertainties are Poisson uncertainties. Both the stellar and gaseous surface mass distributions are fitted by a simple exponential function to verify whether exponential disks form during the simulation. The function
\begin{eqnarray}\label{eq:sigma}
\Sigma_\mathrm{exp}(r)=n\times \exp(-r/r_\mathrm{e})
\end{eqnarray}
is used, with $n$ being the normalization and $r_\mathrm{e}$ the radial exponential scale length. For $\Sigma_\mathrm{stars}$ only the stellar disk is used for the fit, meaning that the central part ($r<2\mathrm{kpc}$) as well as radial bins with less than 10 stellar particles are neglected. $\Sigma_\mathrm{gas}$ is fitted outside of the stellar disk to trace the outer mass distribution of the galaxy. Indeed, both components show a radial exponential profile, so the total surface mass density is essentially the sum of two exponential profiles. The inner part of the galaxy is dominated by the stellar particles, therefore $\Sigma_\mathrm{tot}$ decreases according to $\Sigma_\mathrm{stars}$ until it becomes smaller than the surface mass density of the gas. At that point the decrease becomes shallower and $\Sigma_\mathrm{tot}$ follows $\Sigma_\mathrm{gas}$. This is also reflected in the different exponential scale lengths of the stellar and gaseous distribution: $r_\mathrm{e,stars}=1.61\pm0.04\mathrm{kpc}$ and $r_\mathrm{e,gas}=5.75\pm0.06\mathrm{kpc}$. The exponential scale lengths of every model are discussed in more detail in Fig. \ref{fig:overview} with additional properties of the models, while the numerical values of all fit parameters are shown in the Appendix. The more extended gas disks result from the star-formation condition not being fulfilled at larger radii such that the gas is not consumed there, while in the inner regions (the region occupied by the stellar particles) star formation has consumed the vast majority of the gas.\\
\indent
It should be noted here that observed galaxies show exponential stellar surface mass density profiles, so it is a major result that this very simple simulation without complex baryonic physics is showing a similar behaviour. What is more, it is also noteworthy that all galaxy models form an exponential disk shortly after their collapse (roughly $0.5 - 1\mathrm{Gyr}$ after the computation begins). Additionally we want to stress that the occurrence of exponential radial profiles in star-forming rotationally-supported galaxies is still an unsolved problem. There are several different approaches attempting to account for the observed exponential surface density profiles, such as: Inclusion of scattering of stars in idealised models by \cite{Struck&Elmegreen2017} generates thick and warm-to-hot stellar disks, while a phenomenological model of cloud disruption launching gas to large distances suggests the settling of the gas into an exponential disk which may form stars \citep{Struck&Elmegreen2018}. Although potentially promising, such models of baryonic processes remain, to some degree, ad hoc.\\
\indent
After the discussion of M1, the differences between models computed with simple and more complex baryonic physics will be shown in the following.
\section{Results}
In this section the resulting galaxies are analysed.
\subsection{Impact of different physical processes}
In order to understand the role of including more complex baryonic physics in galaxy formation and evolution within the MONDian framework, identical simulations were carried out with different physical processes involved.\\
\indent
The "simple" model (M1) was calculated with star formation and simple cooling/heating, while for the "complex" model (M1sn) star formation, radiative transfer, sink particle formation and supernova explosions are enabled.\\
\indent
The density threshold for the sink particle formation was set to $2.77\mathrm{Hcm}^{-3}$, such that the Jeans-length is resolved by at least 4 grid cells. The parameters for the implementation of the supernovae were taken from \citet{Dubois+Teyssier2008} and \citet{Teyssier2013}. The supernova scheme in RAMSES injects the kinetic part of the energy as spherical blast waves with the size of galactic superbubbles (here $r_{SN}=150\mathrm{pc}$). If the energy would be injected only as thermal energy, then most of it would be radiated away very quickly and the effect of the supernovae on the kinematics of the stars and the gas would be minor.\\
\indent
After the collapse, both show a dense disk in the gaseous and stellar component with a low density environment as well as a dense stellar core. Differences can be seen in the extent of the low density environment in the gaseous component and a spiral pattern is evident in the gas-disk of M1 seen face-on, while M1sn does not show this pattern shortly after the collapse. The differences between both models most likely stem from supernova explosions, as they make the gas more turbulent, so that early on no spiral pattern can form in model M1sn. The gas is also spread wider, so low density gas can be seen at larger radii.\\
\indent
After $10\mathrm{Gyr}$ both models are still morphologically quite similar. Both show a spiral pattern in the gaseous component face-on and the stellar components look nearly identical. The only major difference is the vertical extent of the gas, because supernovae push the gas in every direction in M1sn, while for M1 the gas stays within the disk.\\
\indent
\begin{figure*}[h]
\begin{minipage}[t]{0.49\linewidth}
\includegraphics[width=1.0\linewidth]{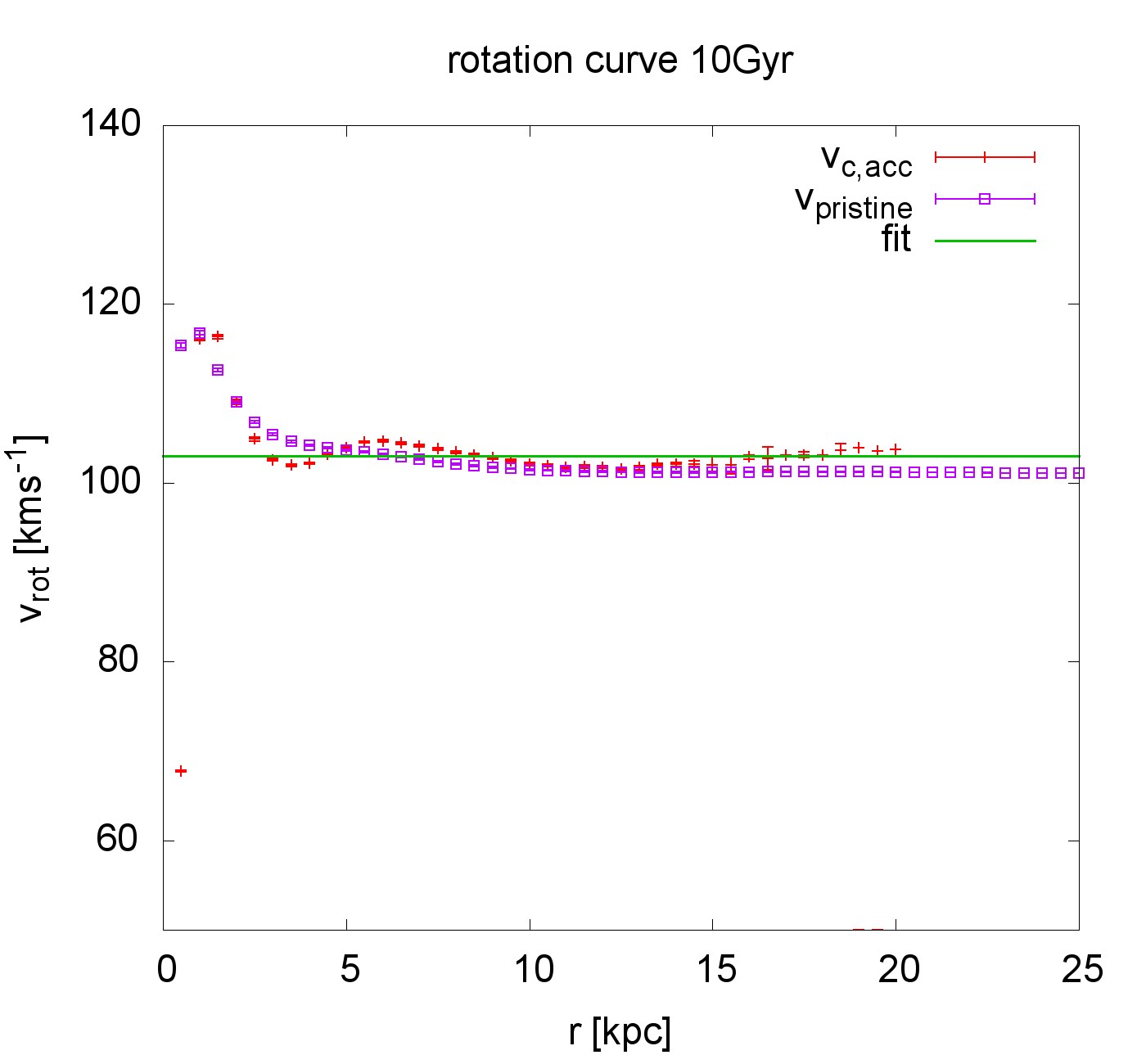}
\end{minipage}
\begin{minipage}[t]{0.49\linewidth}
\includegraphics[width=1.0\linewidth]{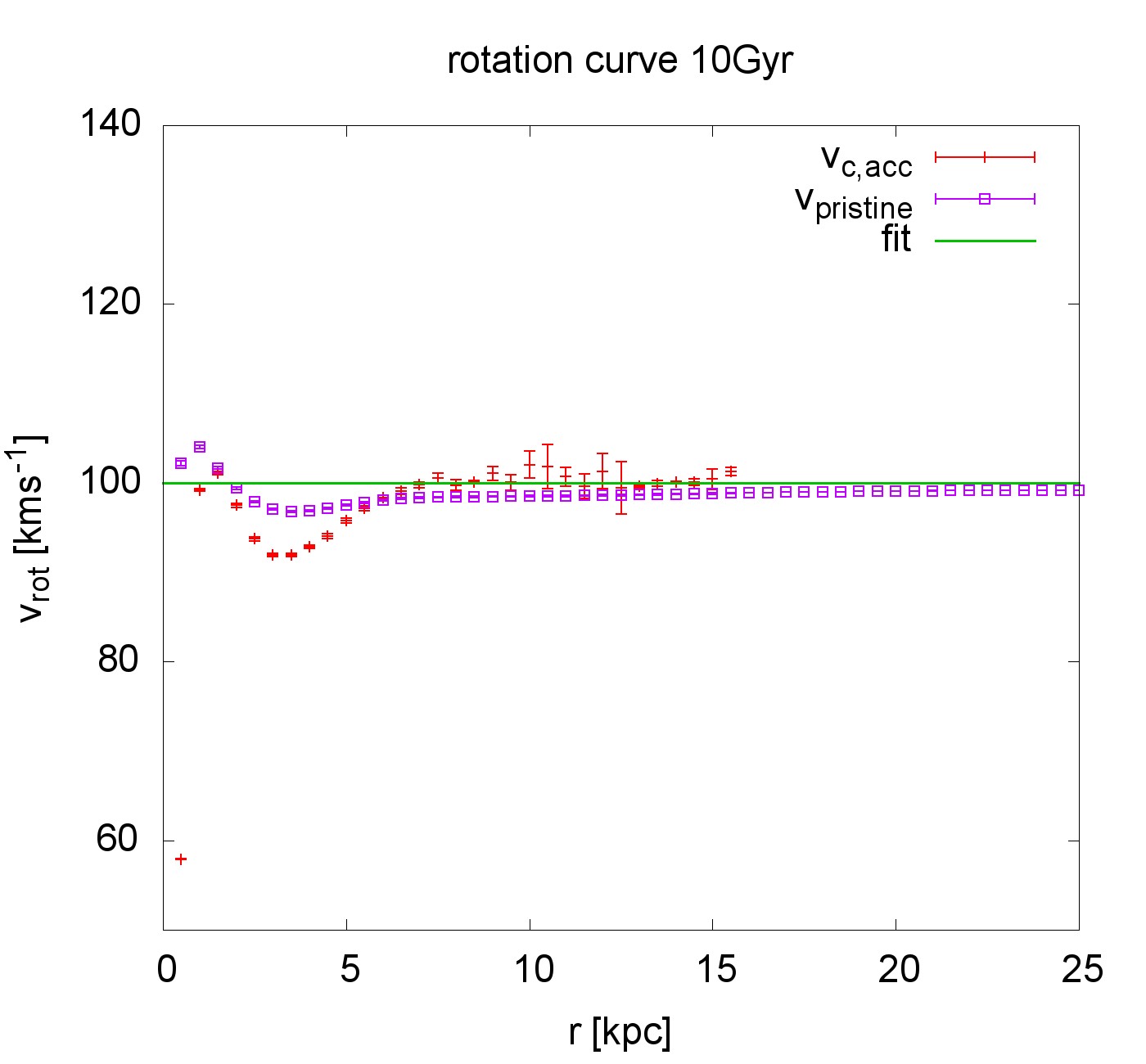}
\end{minipage}
\caption{Rotation curves of both models with $r_\mathrm{init}=20\mathrm{kpc}$ shown in red, $v_\mathrm{c,acc}$, and the 'pristine' rotation curve of the respective model in purple. $v_\mathrm{c,acc}$ shows the rotation velocity calculated from the mean radial acceleration in the respective radial bin.\textit{Left panel}: M1, \textit{Right panel}: M1sn. See text for further detail on the computation and analysis.}
\label{fig:rccomparisonr2010Gyr}
\end{figure*}
\begin{figure*}[h]
\begin{minipage}[t]{0.49\linewidth}
\includegraphics[width=1.0\linewidth]{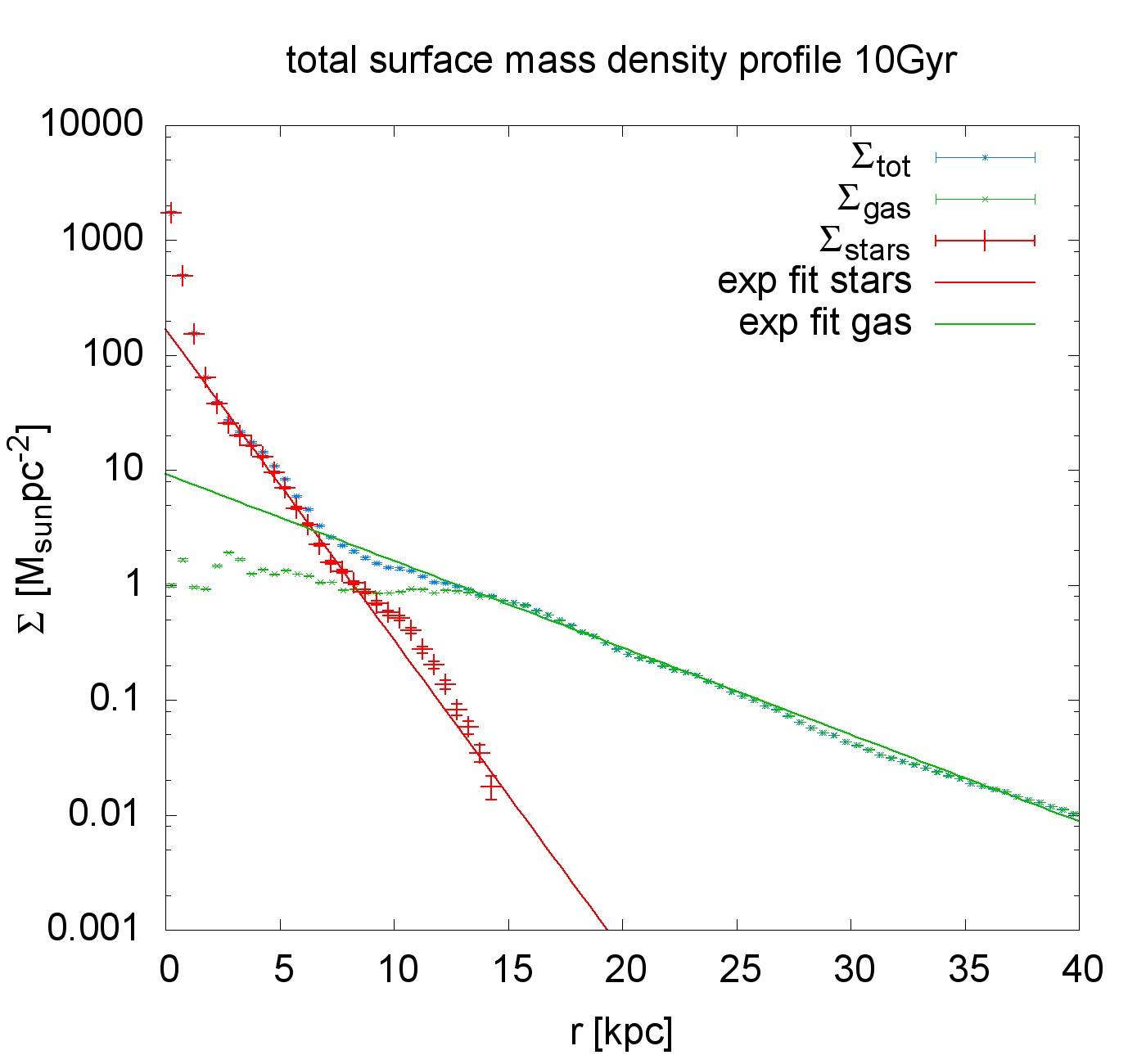}
\end{minipage}
\begin{minipage}[t]{0.49\linewidth}
\includegraphics[width=1.0\linewidth]{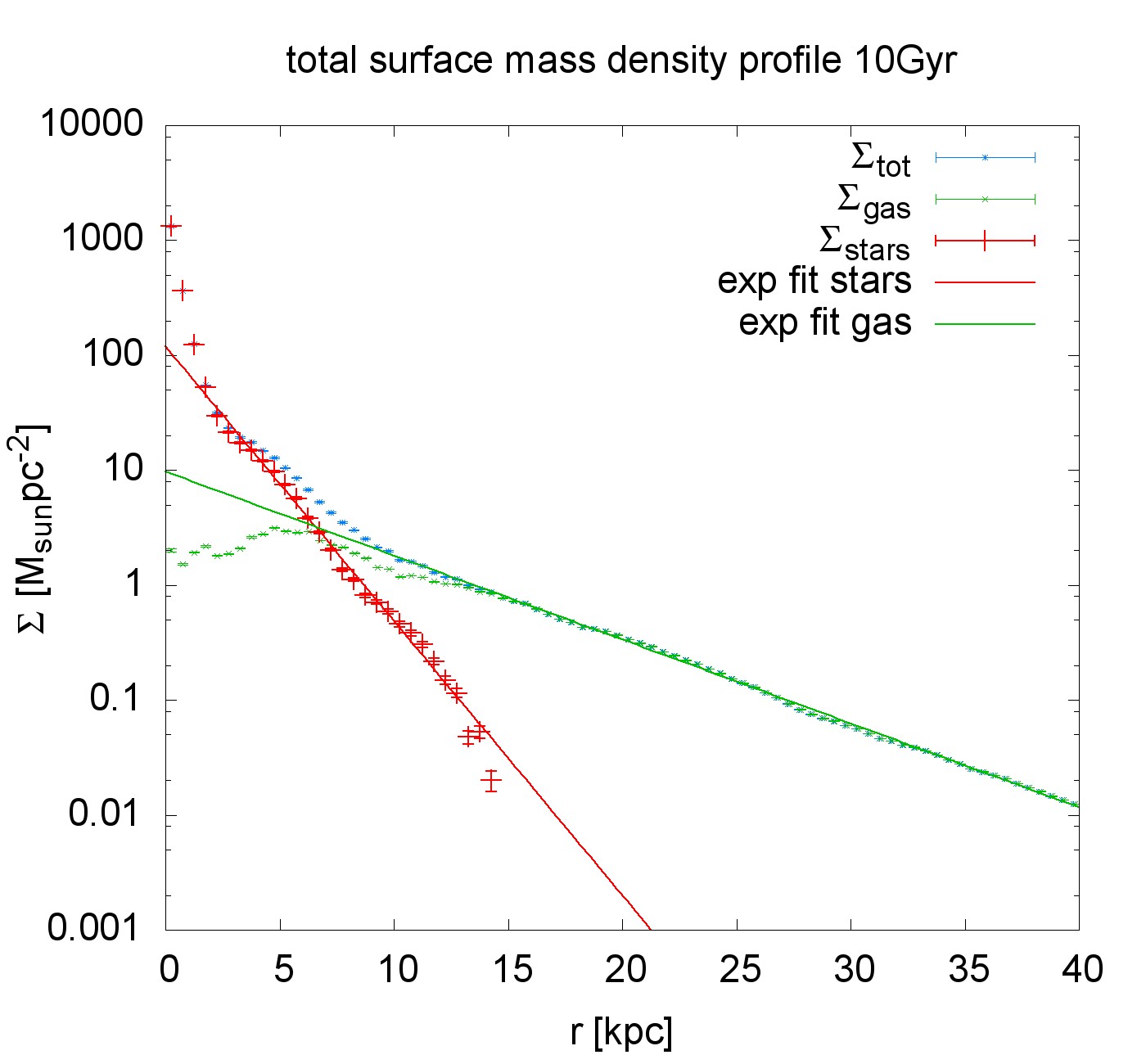}
\end{minipage}
\caption{Surface mass density profiles of both models with $r_\mathrm{init}=20\mathrm{kpc}$. \textit{Left panel}: M1, \textit{Right panel}: M1sn. The red points correspond to the stellar surface mass density, the green ones to the surface mass density distribution of the gas and the blue ones show the total baryonic surface mass density distribution. The lines show the respective exponential fit to data.}
\label{fig:ddcomparisonr2010Gyr}
\end{figure*}
Figs. \ref{fig:rccomparisonr2010Gyr} - \ref{fig:ddcomparisonr2010Gyr} show the comparison of the analysis of the two simulations.
The inner non-flat part of the rotation curves depicts the major difference between the two models, which can be explained by the effects of supernova explosions on the gas in M1sn which heat the gas thereby reducing the central deep collapse of the baryons. At the radius near $4\mathrm{kpc}$ the surface mass density increases for model M1sn in contrast to the slow decrease that can be seen for model M1. Most of the supernovae take place in the centre of the model, so the gas-density is decreased significantly in this region and therefore the surface mass density rises again to a certain degree outside of the centre. This also leads to a difference in the gravitational potential and so to a difference in the rotational velocity. Also, this is the region, where both models deviate most from their 'pristine' curve. In general the transition-region between the flat part of the rotation curve and the decreasing flank after the peak is mostly influenced by the curl field correction, which will be underlined further in the following sections.\\
\indent
Apart from the different distributions of gas in the centre, the surface mass density distributions are quantitatively in agreement. Both models show a steep decrease at the centre for the first few $\mathrm{kpc}$, after which $\Sigma_{\mathrm{stars}}$ exponentially decreases. In the outer parts of the galaxy $\Sigma_{\mathrm{gas}}$ becomes greater than $\Sigma_{\mathrm{stars}}$ and it decreases also exponentially. So both components for both models show exponentially decreasing profiles except in the centre. What is more, the exponential scale lengths are also slightly, but not substantially different, because of the redistribution of gas due to supernovae, leading to $r_\mathrm{e,stars,M1}=1.61\pm0.04\mathrm{kpc}$, $r_\mathrm{e,stars,M1sn}=1.82\pm0.06\mathrm{kpc}$ and $r_\mathrm{e,gas,M1}=5.37\pm0.10\mathrm{kpc}$, $r_\mathrm{e,gas,M1sn}=6.28\pm0.09\mathrm{kpc}$.\\
\indent
Although the supernovae have an obvious impact on the simulation, the values of the rotation velocity in the flat part of the two rotation curves are very comparable, $v_\mathrm{rot,flat,M1}=103.03\pm0.19\mathrm{kms^{-1}}$, $v_\mathrm{rot,flat,M1sn}=99.98\pm0.09\mathrm{kms^{-1}}$, the fits (Eq.\ref{eq:flatvc}) being obtained over $3<r/\mathrm{kpc}<20$ and $6<r/\mathrm{kpc}<15.5$ respectively.\\
\indent
All in all both models are very similar, despite the differences that a simulation with and without supernova explosions has, e.g. supernovae lead to a larger extent of the gaseous component, a higher gas fraction and therefore lower mass in stellar particles.\\
\indent
The next step is to see whether different initial values of the rotation velocity parameter, $\eta$, or a different initial rotation law changes the properties and the morphology of a model.
\subsection{Comparison between different initial rotation laws}
\begin{figure*}[h]
\includegraphics[width=1.0\linewidth]{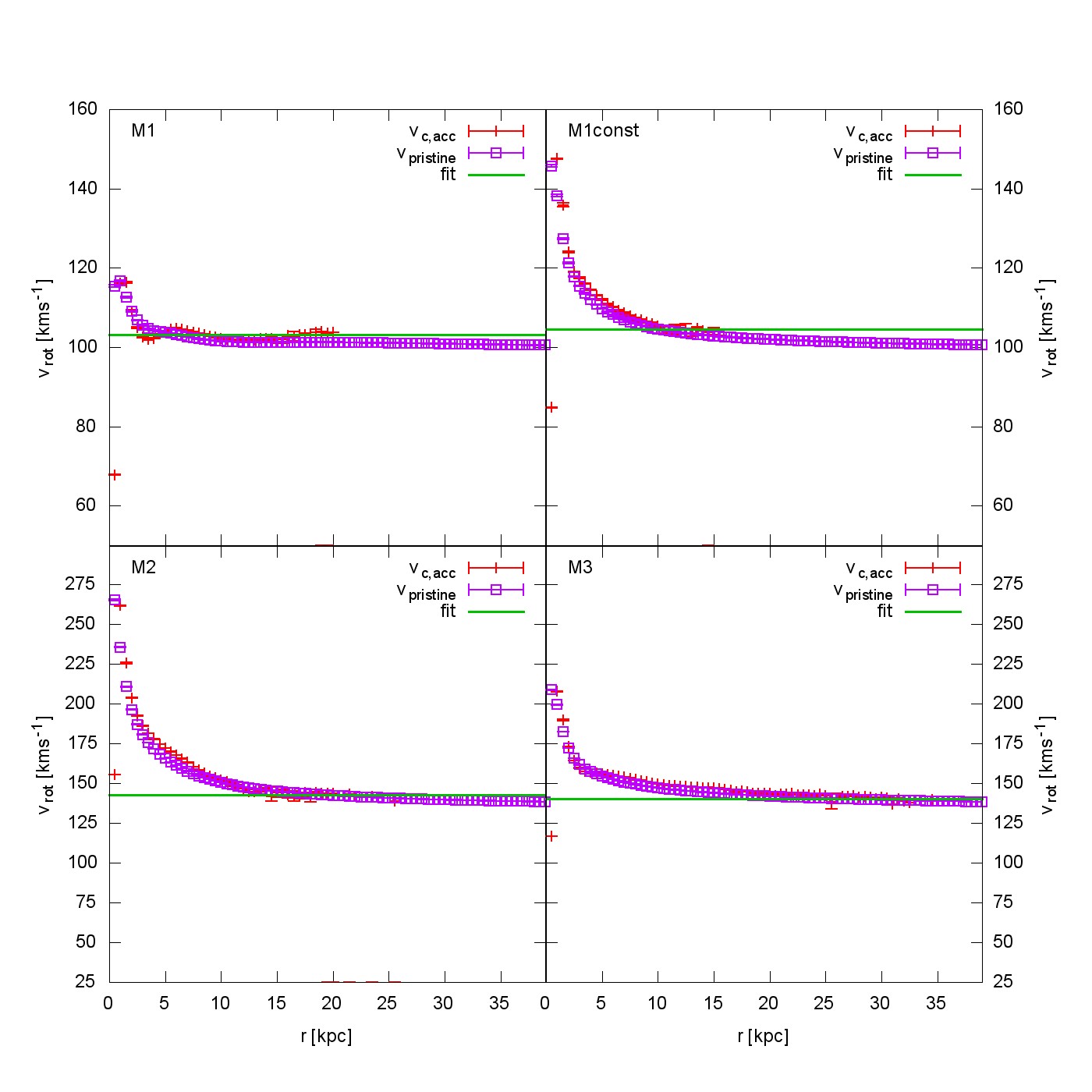}
\caption{Comparison of the rotation curves between models with different rotation laws (top) and different values of $\eta$ (bottom). From top left to bottom right: M1, M1const, M2 and M3. See text for further detail.}
\label{fig:rccomparisonr3010Gyr}
\end{figure*}

\begin{figure*}[h]
\includegraphics[width=1.0\linewidth]{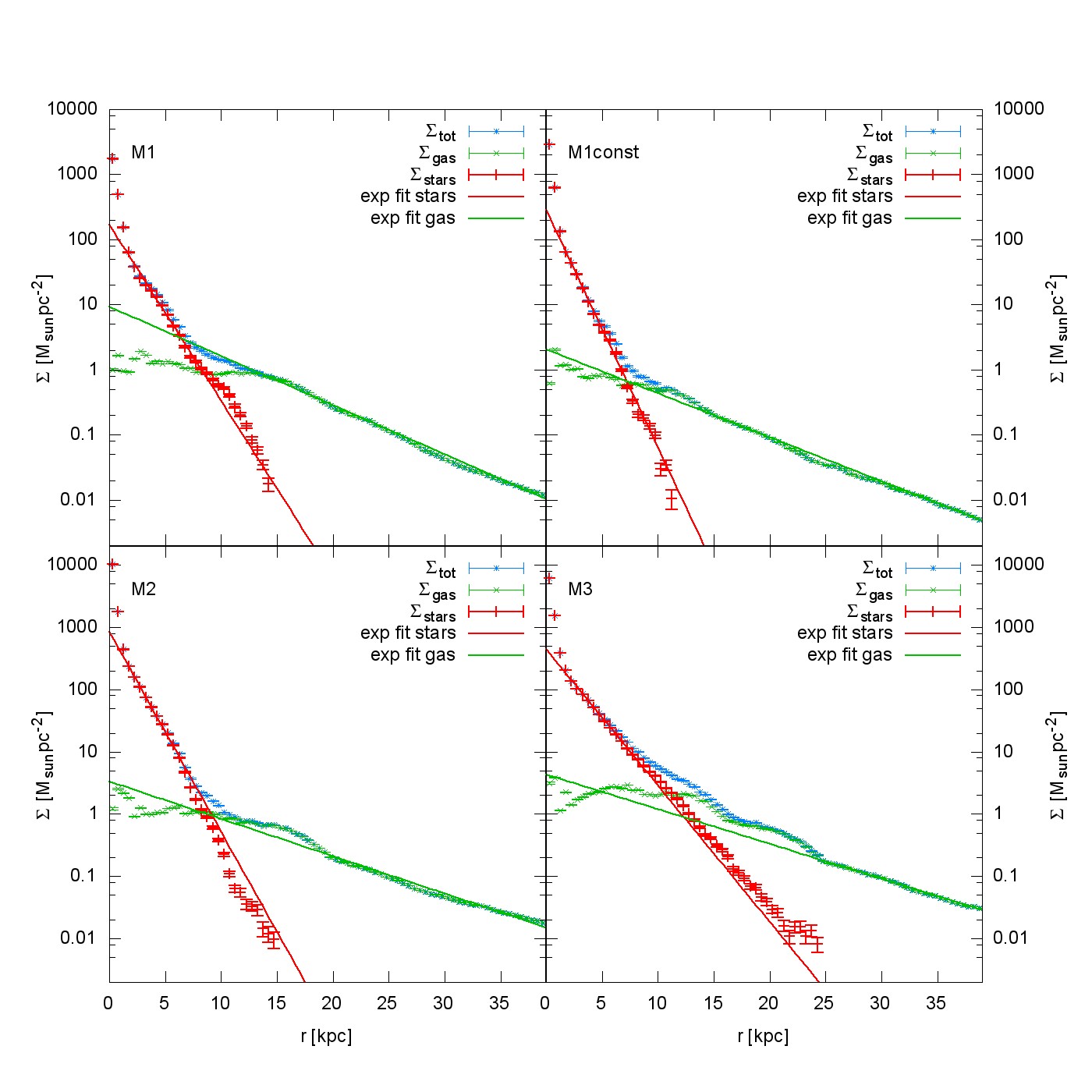}
\caption{Comparison of the surface mass density distributions between models with different rotation laws (top) and different values of $\eta$ (bottom). From top left to bottom right: M1, M1const, M2 and M3. See text for further detail.}
\label{fig:ddcomparisonr3010Gyr}
\end{figure*}

In this section the impact of different initial rotation laws and $\eta$ will be examined.\\
\indent
The comparison is shown in Fig. \ref{fig:rccomparisonr3010Gyr} \& \ref{fig:ddcomparisonr3010Gyr}. Four models are compared (M1, M1const, M2, M3) and the differences between M1 and M1const and between M2 and M3 are discussed. All initial parameters can be seen in Table \ref{tab:initconditions}. M2 and M3 assume a more extended and more massive gas cloud than M1, whereby M3 rotates faster than M2. M1const has a constant rotation velocity throughout the cloud.\\
\indent
Differences can be seen already after the collapse of the gaseous sphere. The collapse of M3 is not as smooth as the one for M2, because three dense clumps form prior to the formation of the whole galaxy, due to the higher rotation velocity than for model M2. These clumps encounter each other and merge asymmetrically, so an asymmetric, rotating gas distribution forms around the dense disk. Also, the whole galaxy is more extended (especially the stellar component), which is shown clearly in the rotation curve and the surface mass density distribution (Fig.\ref{fig:rccomparisonr3010Gyr} and Fig.\ref{fig:ddcomparisonr3010Gyr} respectively).\\
\indent
Despite these initial differences, the evolution of both models is very similar. Both form a thin, rotating, dense disk in both components.\\
\indent
M1const is initially different to all other models, because the initial rotation velocity distribution is constant, $v_\mathrm{i}=6.6\mathrm{kms}^{-1}$. Therefore the collapse of the sphere at the beginning of the simulation is more violent, because especially the infall of gas from the outermost radii is not slowed down by a higher rotation velocity. So, compared to M1, M1const is overall smaller, which is in agreement with the findings from the comparison between M2 and M3.\\
\indent
Not only the initial collapse is different depending on the initial rotation velocity, but also the radial extent of star formation differs, as the rotation curve of M3 reaches radii that are $10\mathrm{kpc}$ larger than the one for M2, while the stellar disk of M1 reaches out approximately $5\mathrm{kpc}$ further than the disk of M1const. M1const and M2 have higher peak-rotation-velocities compared to M1 and M3 respectively and the effect of the curl field correction is most prominent in the region directly after the peak and before the flat part, although the rotation curve of M3 is nearly identical to its 'pristine' curve. The flat rotation velocities are very comparable between M1 and M1const and between M2 and M3, but not identical as the mass within the stellar disk varies slightly due to different extents and gas densities, $v_\mathrm{rot,flat,M2}=142.41\pm0.46\mathrm{kms^{-1}}$,  $v_\mathrm{rot,flat,M3}=140.35\pm0.21\mathrm{kms^{-1}}$, $v_\mathrm{rot,flat,M1}=103.03\pm0.19\mathrm{kms^{-1}}$ and $v_\mathrm{rot,flat,M1const}=104.50\pm0.14\mathrm{kms^{-1}}$. The corresponding ranges for the fits (Eq. \ref{eq:flatvc}) are $3<r/\mathrm{kpc}<20$ (M1), $10<r/\mathrm{kpc}<15.25$ (M1const), $10<r/\mathrm{kpc}<18.25$ (M2) and $20<r/\mathrm{kpc}<26.75$ (M3)\\
\indent
Similar to the models shown in the last section, the surface mass density profiles for the stellar and the gaseous component are exponentially decreasing. Also, the difference in the extent of the stellar component is visible via the slope of the fit of this component, while the fit of $\Sigma_{gas}$ for every model shows that the surface mass density distribution of the gas varies as well, which is shown by the exponential scale lengths in Table \ref{tab:rexp}.\\
\indent
\begin{table}[h]
\caption{exponential scale lengths of M1, M1const, M2 and M3 for the stellar and gaseous surface mass distributions.}
\centering
\begin{tabular}[c]{l c c}
\hline
Model name & $r_\mathrm{e,stars}$[$\mathrm{kpc}$] & $r_\mathrm{e,gas}$[$\mathrm{kpc}$]\\
\hline
M1 & $1.61\pm0.04$ & $5.75\pm0.06$ \\
M1const & $1.19\pm0.01$ & $6.44\pm0.07$ \\
M2 & $1.35\pm0.01$ & $7.24\pm0.16$ \\
M3  & $1.98\pm0.02$ & $7.79\pm0.10$ \\
\hline
\end{tabular}
\label{tab:rexp}
\end{table}
Hence, changing the initial rotation velocity or the rotation law to a constant one does not alter the fact that a flat rotation curve and an exponentially decreasing disk is obtained. However, the initial collapse is different, which changes the morphology (especially the radial extent/scale-length) and accordingly the density distribution. This suggests that low surface brightness galaxies result from collapsing gas clouds with high specific angular momentum. As a side remark, note that, initially, we have constant temperature and density, so constant pressure. Different initial temperatures, and/or different profiles might also lead to slightly different collapse patterns.
\subsection{Comparison between Newtonian and MONDian models}
Due to the fact that exponential stellar surface mass density profiles occur in all the here calculated MONDian galaxy models and for the purpose of testing the correct behaviour of the code and the role of the gravitational law, M1 and M2 were simulated again, while only using the Newtonian Poisson solver, but still without dark matter.\\
\begin{figure*}[h]
\includegraphics[width=1.0\linewidth]{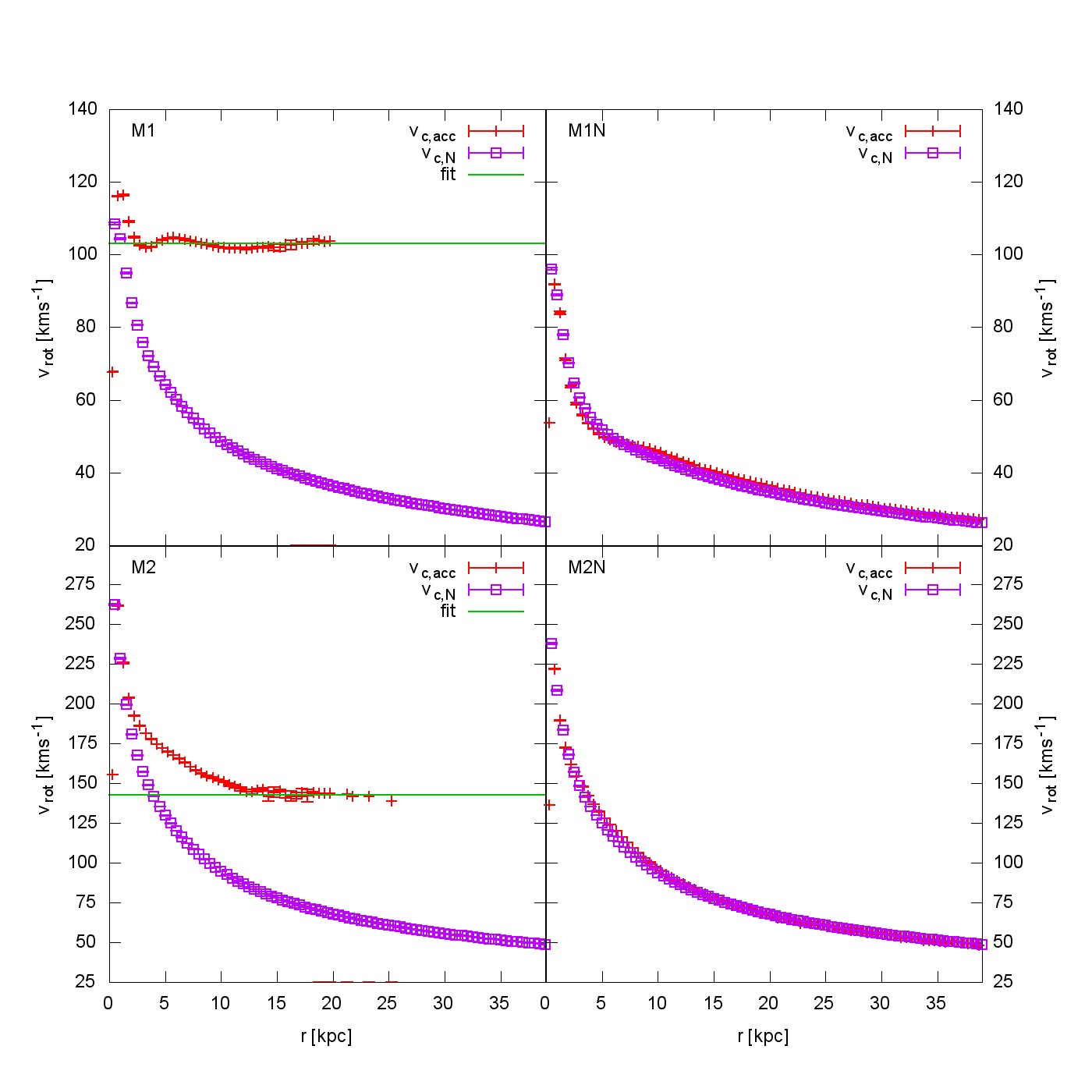}
\caption{Comparison between rotation curves of MONDian and Newtonian simulations for two different initial conditions. From top left to bottom right: M1, M1N, M2 and M2N. See text for further detail.}
\label{fig:rccomparisonNewtonMond10Gyr}
\end{figure*}
\begin{figure*}[h]
\includegraphics[width=1.0\linewidth]{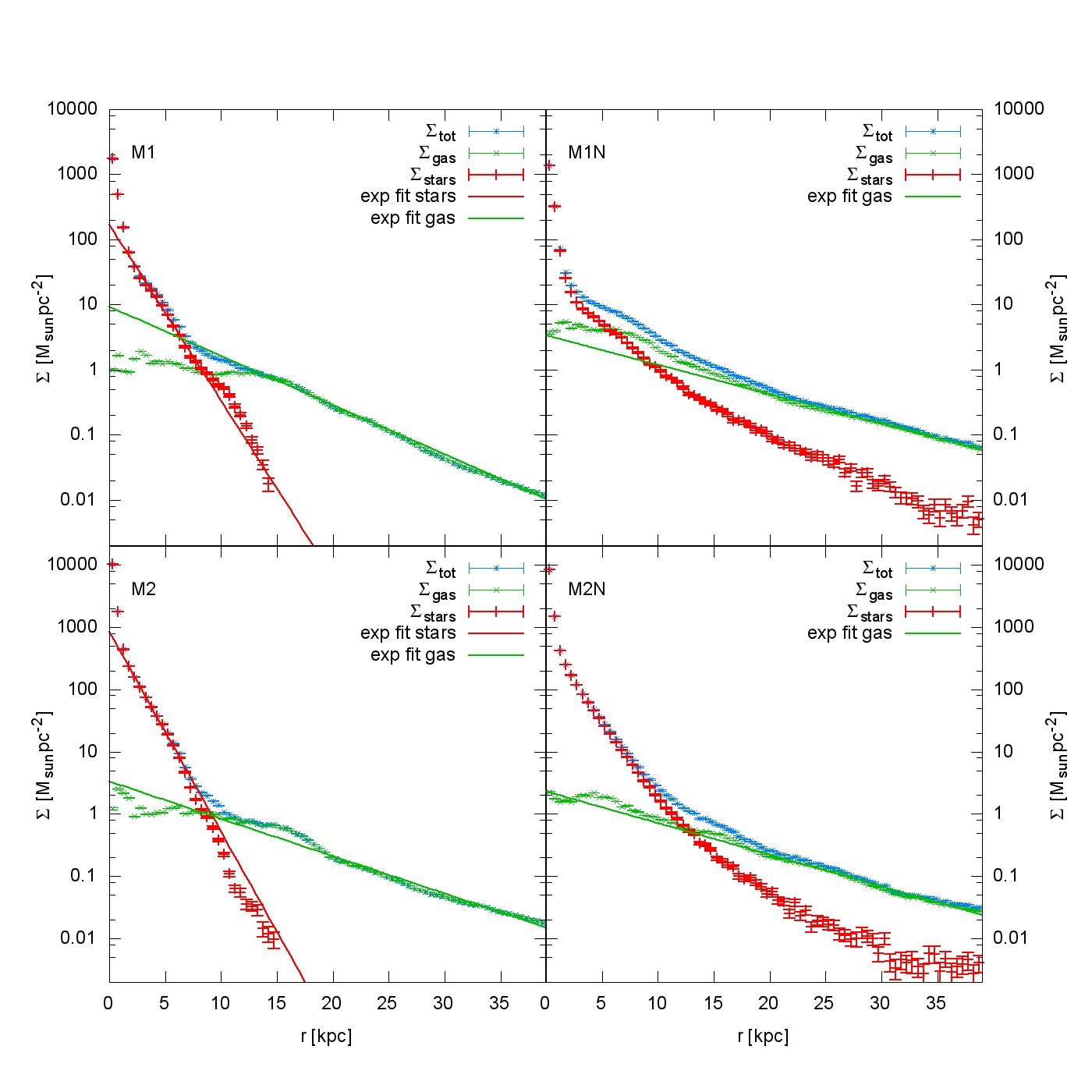}
\caption{Surface mass density profiles of MONDian and Newtonian simulations for two different initial conditions. From top left to bottom right: M1, M1N, M2 and M2N. See text for further detail.}
\label{fig:ddcomparisonNewtonMond10Gyr}
\end{figure*}
\indent
Fig. \ref{fig:ddcomparisonNewtonMond10Gyr} and Fig. \ref{fig:rccomparisonNewtonMond10Gyr} show, respectively, the surface mass density profiles and the rotation curves of the equivalent Newtonian and MONDian models. First of all, the rotation curve of the Newtonian models show no flatness but a Keplerian decrease, which is expected and demonstrates that the code works properly in this regard. On the other hand, the majority of the rotation curves of the MONDian models are flat, as seen before. Also, the measured rotation curves of the Newtonian models are very close to the theoretical spherical-equivalent curve, $v_{c,N}$ (purple boxes), which are calculated by using the mass inside the respective radius, $M(<r)$. Comparing $v_{c,N}$ with the MONDian curves reveals that the centre of M2 lies within the Newtonian acceleration regime of our interpolating function, because the curves overlap in this region. Contrary to that, the accelerations in M1 do not reach values well above $10^{-10}\mathrm{ms^{-2}}$, as the curves never meet. This overlapping region depends on the depth of the potential and extends further in radius the more massive a model is. Of course, the gap between the MONDian rotation curve and the theoretical Newtonian one increases with radius, because the rotation curves become asymptotically flat for M1 and M2, while the theoretical Newtonian curves show a Keplerian decrease.\\
\indent
Qualitatively the comparison of the surface mass densities shows an encouraging result, as the stellar surface mass density profile of the Newtonian models do not show an exponential profile, but are rather curved. However, $\Sigma_\mathrm{gas}$ shows an exponentially decreasing profile similar to the simulations done with MOND, but with a slower radial falloff, therefore leading to different exponential scale lengths, $r_\mathrm{e,gas,M1}=5.37\pm0.10\mathrm{kpc}$, $r_\mathrm{e,gas,M1N}=9.62\pm0.11\mathrm{kpc}$ and $r_\mathrm{e,gas,M2}=7.24\pm0.16\mathrm{kpc}$, $r_\mathrm{e,gas,M2N}=8.55\pm0.13\mathrm{kpc}$.\\
\indent
Although the forces for the two gravitation laws are different, when using the same mass content, it is still noteworthy that $\Sigma_\mathrm{stars}$ is decreasing exponentially for all models simulated with the MONDian Poisson solver, while the Newtonian models do not show this behaviour. This indicates that the natural appearance of exponentially decreasing stellar surface mass density profiles seems to be a feature of the MONDian framework and not a feature of the initial conditions.
\subsection{Initial size/mass limit}
As seen in Fig. \ref{fig:rccomparisonr3010Gyr} \& \ref{fig:ddcomparisonr3010Gyr}, the slope, $\eta$, of the rotation law used here changes the size of the computed galaxy noticeably in contrast to the subtle differences that occur by using more complex baryonic physics. Furthermore the models are not only sensitive to the slope of the initial rotation law, but also to $r_\mathrm{init}$ and $M_\mathrm{init}$. The models M4 and M4sn show significantly different stellar surface mass density profiles and formation histories compared to all other simulations.\\
\indent
Again, the initial collapse is very important for the evolution of these models. M4 and M4sn have nearly five times more mass than even the intermediately massive models, but they are collapsing into a comparably sized region. Therefore the collapse itself is more violent and dense and more stars are formed that have an elongated orbit. Also, many more stars are outside of the dense disk, which leads to a more populated and more extended lower density stellar component. Furthermore, like for M3, several dense clumps form very shortly before the overall collapse happens, but in addition to that in M4 one dense spiral-arm-like structure drifts away from the collapse and small satellites form out of it, which collide with the galaxy at about $6\mathrm{Gyr}$ after the start of the simulation. Supernovae seem to suppress the formation of these elongated dense clouds and also weaken the collapse overall, but still a small satellite forms during the collapse and merges very shortly afterwards with the main galaxy in M4sn. Considering the stability criterion for disks in MOND \citep{Banik2018a}
\begin{eqnarray}\label{MONDstability}
c_\mathrm{s} &\geq& \frac{\pi\  G\ \nu\ \Sigma\ \left(1+\frac{K_0}{2}\right)}{\Omega_r} \\
\qquad && \qquad\qquad \updownarrow \nonumber \\
1 &\leq& \frac{c_\mathrm{s}\ \Omega_r}{\pi\  G\ \nu\ \Sigma\ \left(1+\frac{K_0}{2}\right)}\equiv Q_\mathrm{MOND}, \nonumber 
\end{eqnarray}
where $c_\mathrm{s}$ is the sound speed of the gas, $\nu\propto \sqrt{a_0/|\bf{g_\mathrm{N}}|}$ the transition function, $\Sigma$ the surface mass density, $K_0=d\ln[\nu(y)]/d\ln(y)$ at $y=g_\mathrm{N0}/a_0$, where the subscript $_0$ for $g_\mathrm{N0}$ and $K_0$ indicates that these are unperturbed values (see \citealp{Banik2018a}) and $\Omega_r^2=-3g_r/r-\partial g_r/\partial r$, the formation of clumps for the more massive models, in contrast to M1 for example, becomes clearer. Eq. \ref{MONDstability} can be approximated using the analytical proportionalities of $\nu,\Sigma\ \mathrm{and}\ \Omega_\mathrm{r}$, as the rest is constant. With $\nu\propto r/\sqrt{M}$, $\Sigma\propto M/r^2$ and $\Omega_\mathrm{r}\propto v_\mathrm{rot}/r\propto \sqrt[4]{M}/r$, $Q_\mathrm{MOND}$ scales as $1/\sqrt[4]{M}$, which shows that there is a mass limit above which multiple fragments are likely to form for a given sound speed, $c_\mathrm{s}$. In reality this is more complicated due to the explicit form of  $\nu,\ \Sigma\ \mathrm{and}\ \Omega_\mathrm{r}$, which for example can then lead to central instabilities and the formation of a bulge, while the outer disk is stable.\\
\indent
\begin{figure*}[h]
\begin{minipage}[t]{0.49\linewidth}
\includegraphics[width=1.0\linewidth]{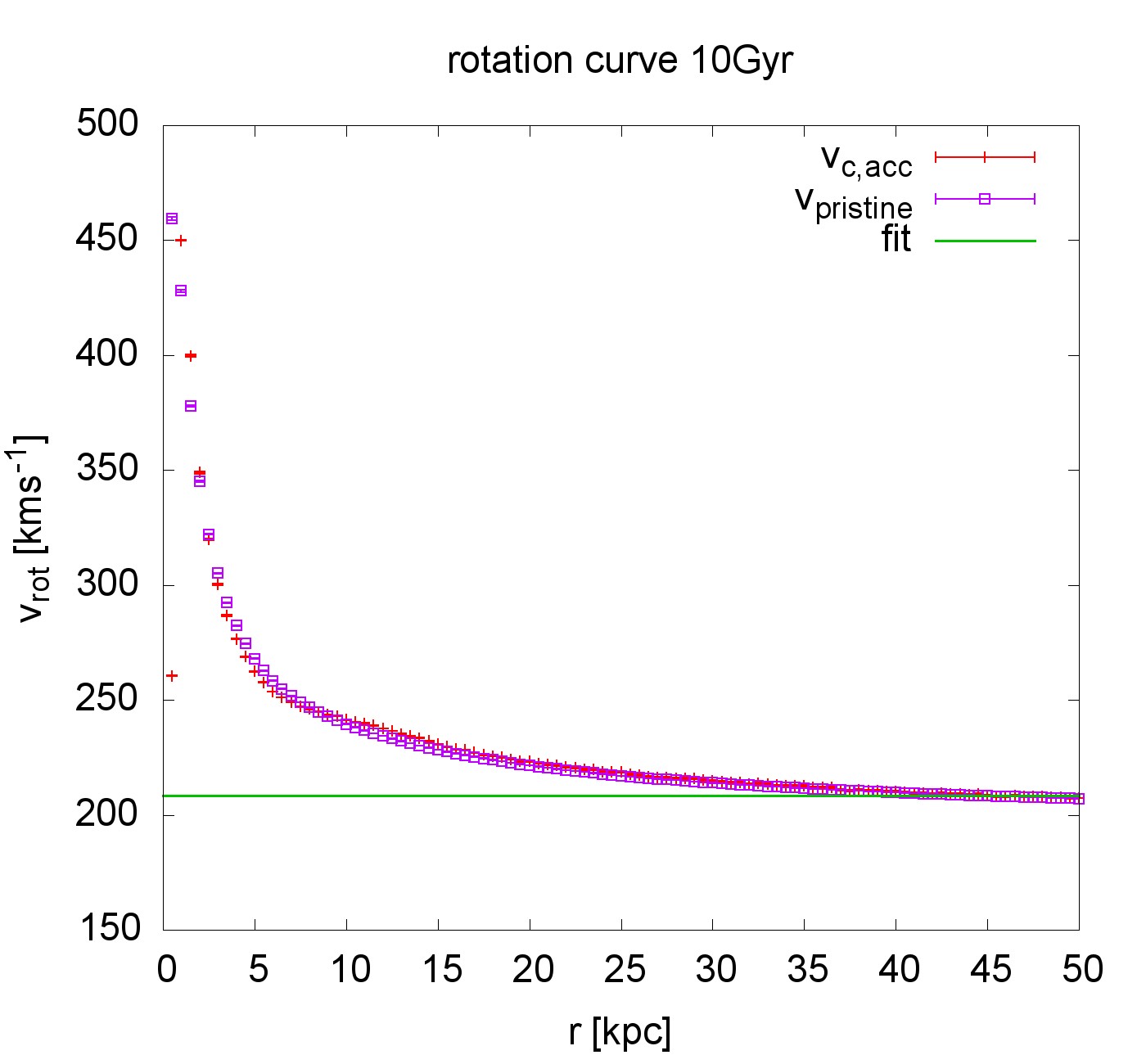}
\end{minipage}
\begin{minipage}[t]{0.49\linewidth}
\includegraphics[width=1.0\linewidth]{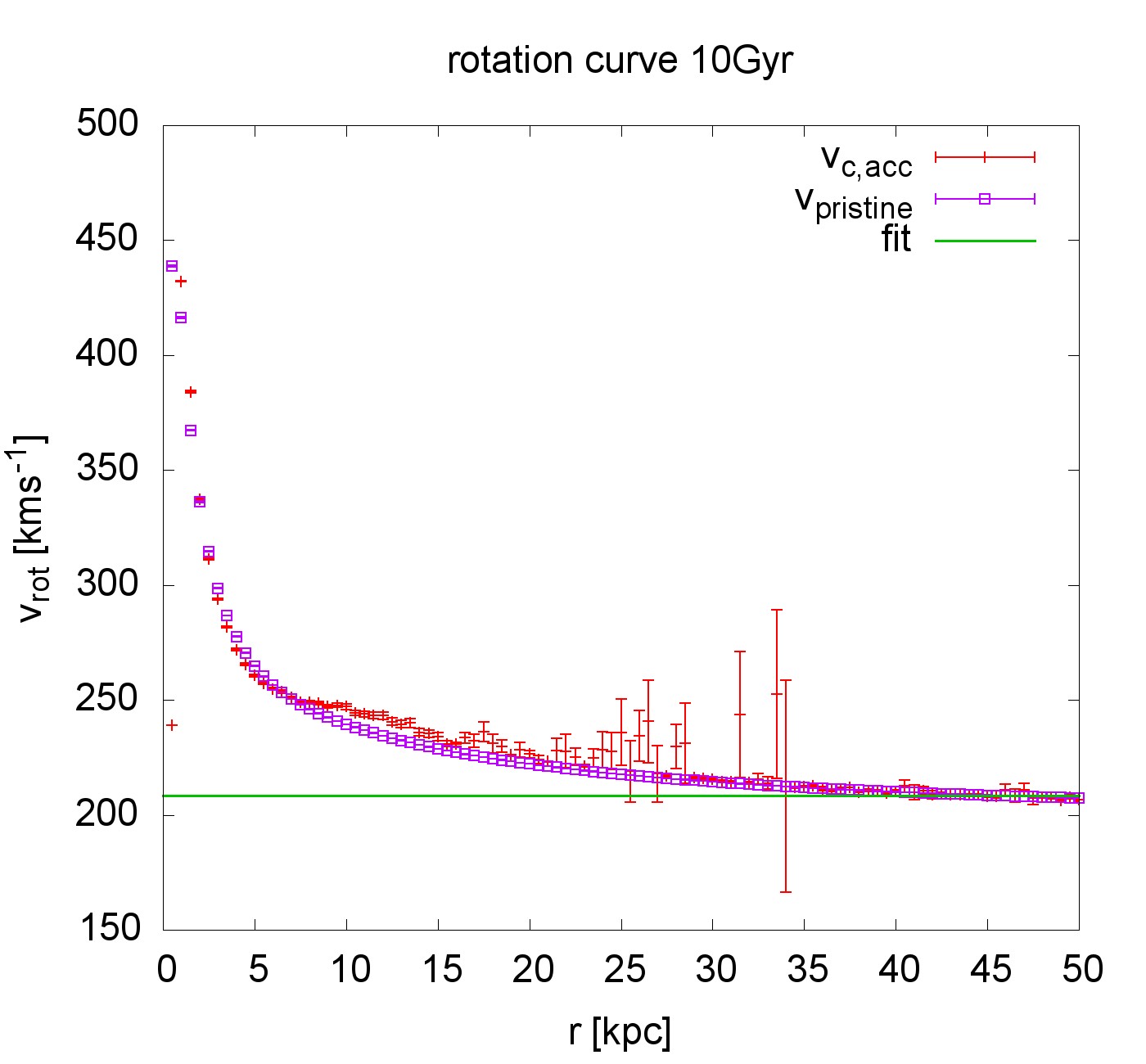}
\end{minipage}
\caption{Rotation curves of both models with $r_\mathrm{init}=50\mathrm{kpc}$. \textit{Left panel}: M4, \textit{Right panel}: M4sn. See text for further detail.}
\label{fig:rccomparisonr5010Gyr}
\end{figure*}
\begin{figure*}[h]
\begin{minipage}[t]{0.49\linewidth}
\includegraphics[width=1.0\linewidth]{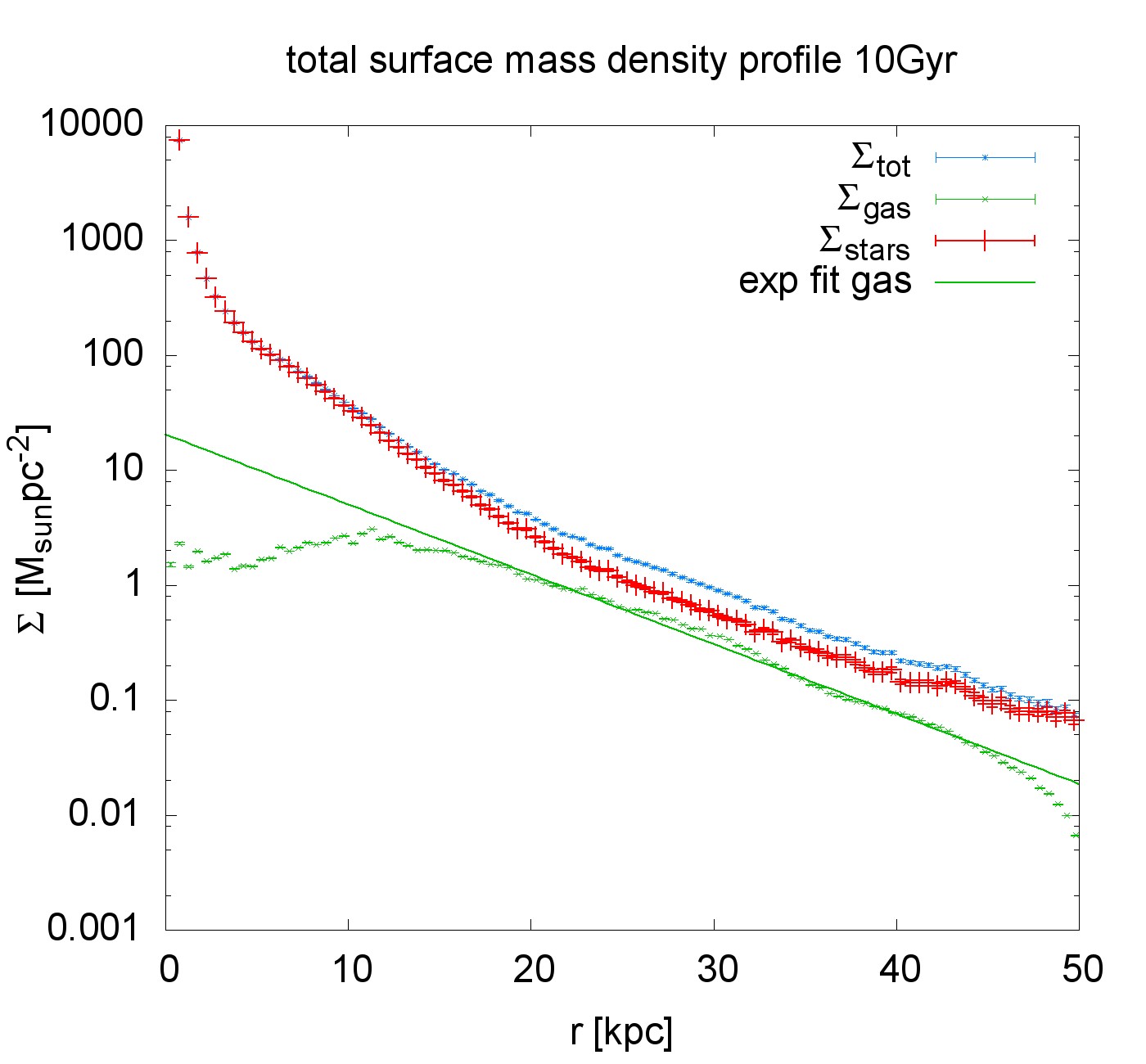}
\end{minipage}
\begin{minipage}[t]{0.49\linewidth}
\includegraphics[width=1.0\linewidth]{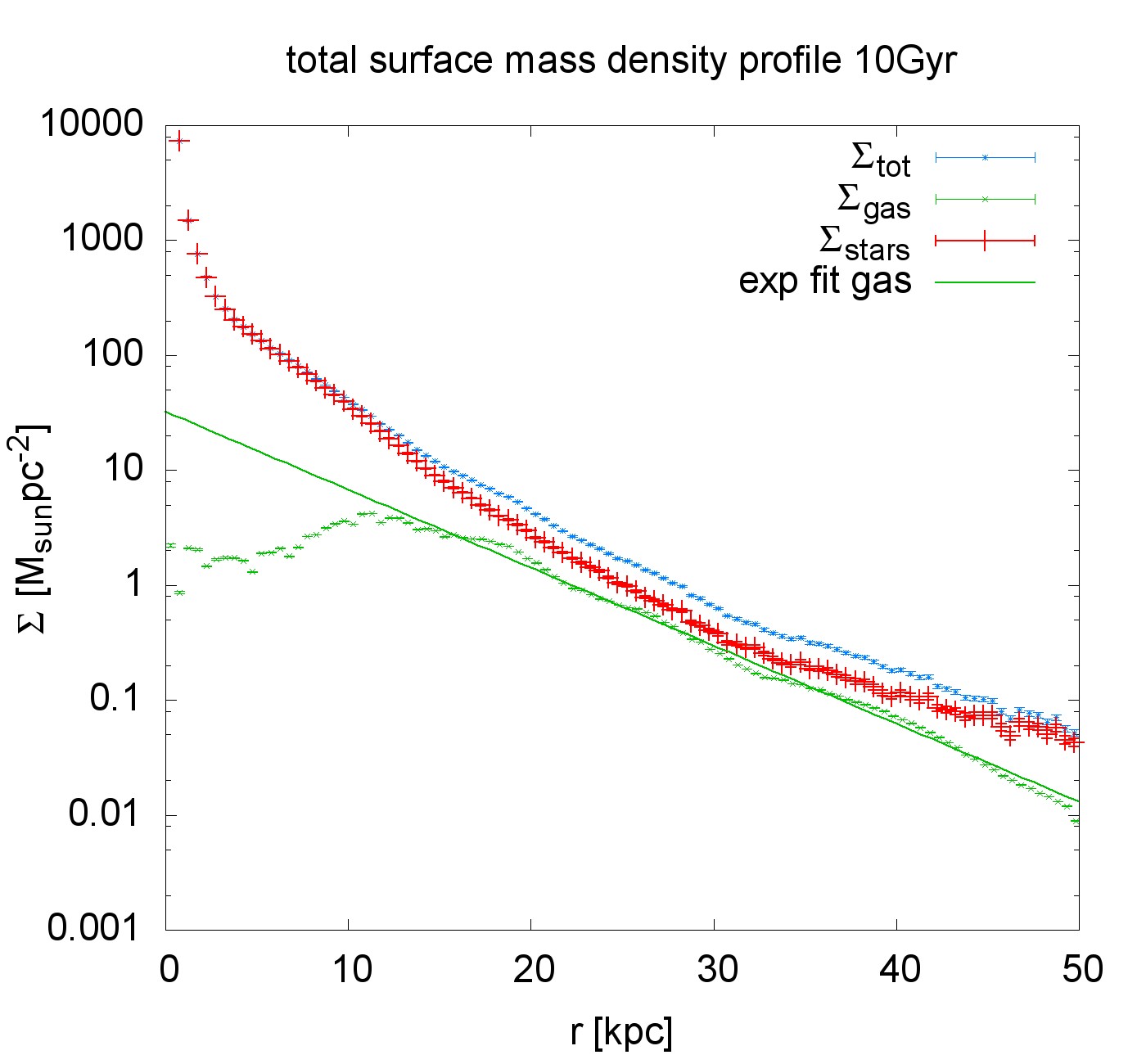}
\end{minipage}
\caption{Stellar surface mass density profiles of both models with $r_\mathrm{init}=50\mathrm{kpc}$. \textit{Left panel}: M4, \textit{Right panel}: M4sn. The stellar surface mass density distribution does not follow a single exponential profile as the formation of these models is dominated by mergers of smaller satellites during and shortly after the initial collapse. The surface density profile may be approximated by more than one exponential profile (not attempted here though). See text for further detail.}
\label{fig:ddcomparisonr5010Gyr}
\end{figure*}
The comparison between M4 and M4sn is shown in Figs. \ref{fig:rccomparisonr5010Gyr} \& \ref{fig:ddcomparisonr5010Gyr}. Although the initial collapse in M4 and M4sn gives rise to the formation of small satellites and a very extended stellar density profile, it does not effect the rotation curve. As for every other model simulated with the MONDian Poisson solver, the rotation curve flattens in the outer part of the galaxies. The rotation curve of M4 is nearly identical to the 'pristine' curve except in the very centre, while deviations are also apparent in the transition region for M4sn.\\
\indent
The major difference of M4/M4sn to all other models is the surface mass density distribution. The decrease for  $\Sigma_\mathrm{stars}$ is shallower than for the less massive models, because the low density stellar component is very populated and extended to a degree where $\Sigma_\mathrm{stars}$ nearly becomes flat at about $50\mathrm{kpc}$. Furthermore $\Sigma_\mathrm{stars}$ is higher than $\Sigma_\mathrm{gas}$ within a radius of $50\mathrm{kpc}$, which can be explained by the low gas fractions of the two models, $f_\mathrm{g,M4}=0.04$ and $f_\mathrm{g,M4sn}=0.05$. The central galaxy (not the satellites) forms partially from mergers of the smaller satellites during and shortly after the initial collapse and the resulting stellar surface mass density profile ceases to be a simple exponential profile. This suggests that galaxies, which do not have a single exponential stellar surface mass density profile, may have been partially formed by baryonic mergers.\\
\indent
Increasing the initial mass further will lead to more satellites and an even more violent collapse, so there seems to be a limit for the formation of isolated galaxies from a single rotating gas cloud with insignificant later accretion of additional gas. Moreover, the formation of dense clumps prior to the collapse can be enhanced by increasing the initial mass and also by increasing $\eta$. If both are done, it is possible to not only produce satellites, but groups of galaxies, which will be shown in a forthcoming contribution (Wittenburg et al. in prep.).\\
\indent
To test whether our models that show an exponentially decreasing stellar surface mass density profile are compatible with observations, we compared our models with the size-mass relation \citep{Lange2015},
\begin{eqnarray}\label{size-mass}
r_\mathrm{eff}=0.13(M_\mathrm{s})^{0.14}\left(1+\frac{M_\mathrm{s}}{14.03\times 10^{10}\mathrm{M_\odot}} \right)^{0.77},
\end{eqnarray}
where $r_\mathrm{eff}$ is the effective radius with $r_\mathrm{eff}=1.678r_\mathrm{e}$ for exponential disks (where $r_e$ is the scale-length) and $M_\mathrm{s}$ the stellar mass.\\
\begin{figure}[h]
\includegraphics[width=1.0\linewidth]{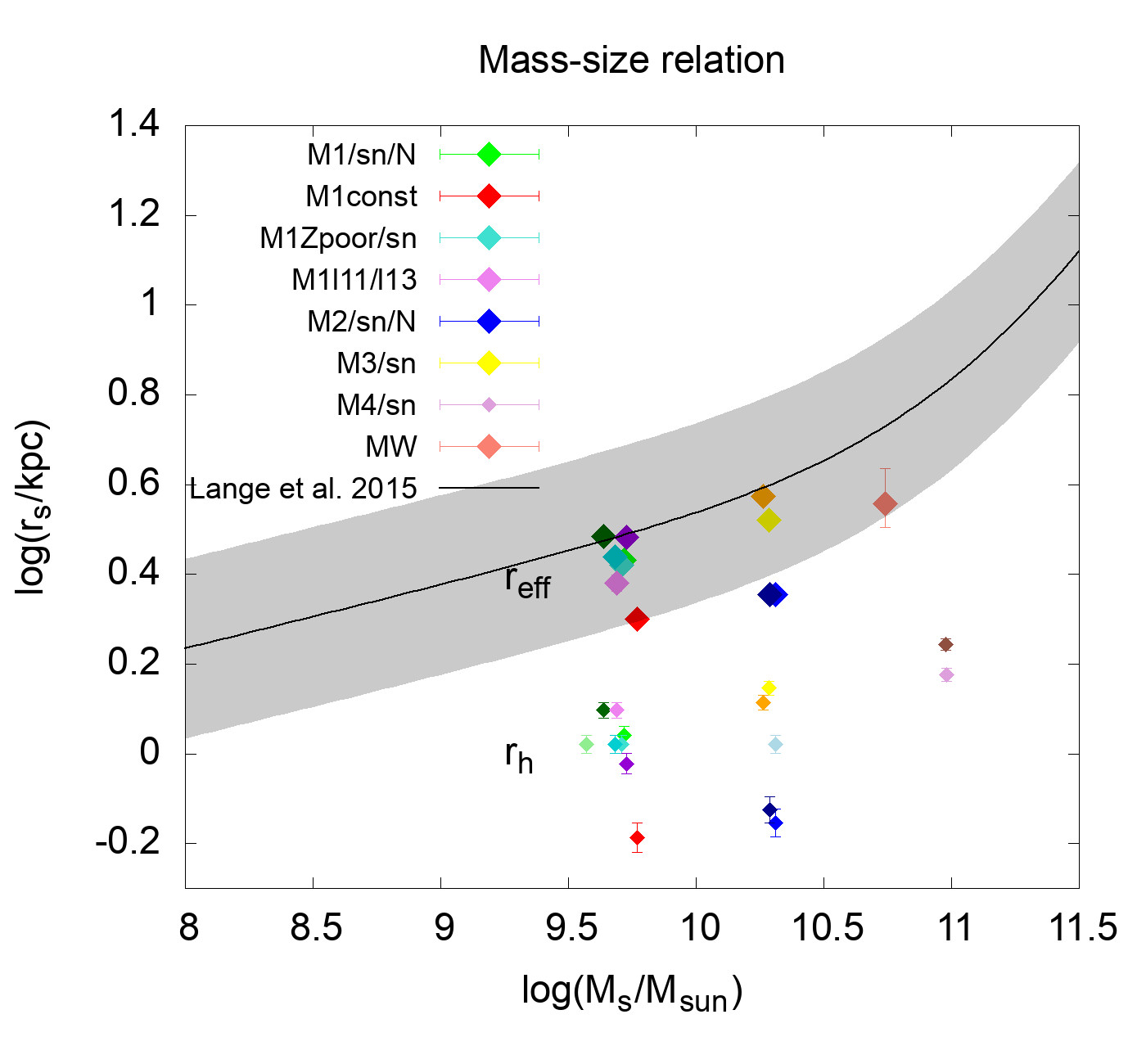}
\caption{The $r_\mathrm{eff}-M_\mathrm{s}$ relation (Eq. \ref{size-mass}) compared with our models and the Milky Way. Our models are coloured according to the key with darker colour indicating sn and lighter colour Newtonian models (e.g. M1=green, M1sn=dark-green and M1N=light-green). The data for the Milky Way is taken from \citealp{BanickZhao2018} (mass) and \citealp{Mor2018} (exp. scale length). The upper distribution of points shows the relation between stellar mass and effective radius, where the latter is based on the fit to the exponential stellar disks, $r_\mathrm{eff}=1.678r_\mathrm{e}$. The lower distribution shows the same for the stellar half-mass radius, $r_\mathrm{h}$. Missing points correspond to non-existing exponential stellar surface mass density profiles, e.g. M4 (plum).}
\label{fig:RvM}
\end{figure}
\indent
Fig. \ref{fig:RvM} depicts the $r_\mathrm{eff}-M_\mathrm{s}$ relation from \cite{Lange2015} (black line) with the uncertainty shown as the grey area, our models are colour-coded as indicated by the key (the sn models are shown with the respective darker colour and the Newtonian with lighter colour, e.g. M1=green, M1sn=dark-green and M1N=light-green) and the Milky Way is depicted with a mass estimate from \cite{BanickZhao2018} and the scale length measurement from \cite{Mor2018}.\\
\indent
Indeed, most of our models follow the observed relation within the uncertainties, when using the effective radius, which is based on the exponential fit to the stellar disk. When using the stellar half-mass radius, $r_\mathrm{h}$, the distribution looks similar, but it is offset to too low radii. There are two important results that emerge from these different distributions. First, the surface density profiles of the stellar disks of our models are compatible with observed ones. Second, because $r_\mathrm{eff}$ is defined as the half-mass radius of a purely exponential disk, the discrepancy between $r_\mathrm{eff}$ and $r_\mathrm{h}$ should be much smaller if at all evident in disc-dominated galaxies. Because $r_\mathrm{eff}>r_\mathrm{h}$ here, our models have a somewhat massive central region, which may stem from the formation process, as the galaxies do not grow, but form in practically one monolithic collapse. These compact bulges may also be related to the observed red nuggets \citep{delaRosa2016}. This feature of our models will also be important in Section 4.6, where we show that they slightly lack stellar angular momentum.\\
\indent
Important to note here is that it is a priori not clear that MONDian models should follow the size-mass relation, but this appears to be the natural result in MOND.
\subsection{Star formation history and gas-depletion time}
Although the structure and morphology of the present models is very comparable to real galaxy discs (see also the comparison between observational scaling relations and the data of the models in Sec. 5), the simplicity of the initial conditions, i.e. a collapsing gas-sphere with insignificant further accretion of gas, gives rise to a different star formation history compared to observations of real galaxies.\\
\indent
\begin{figure*}[h]
\includegraphics[width=1.0\linewidth]{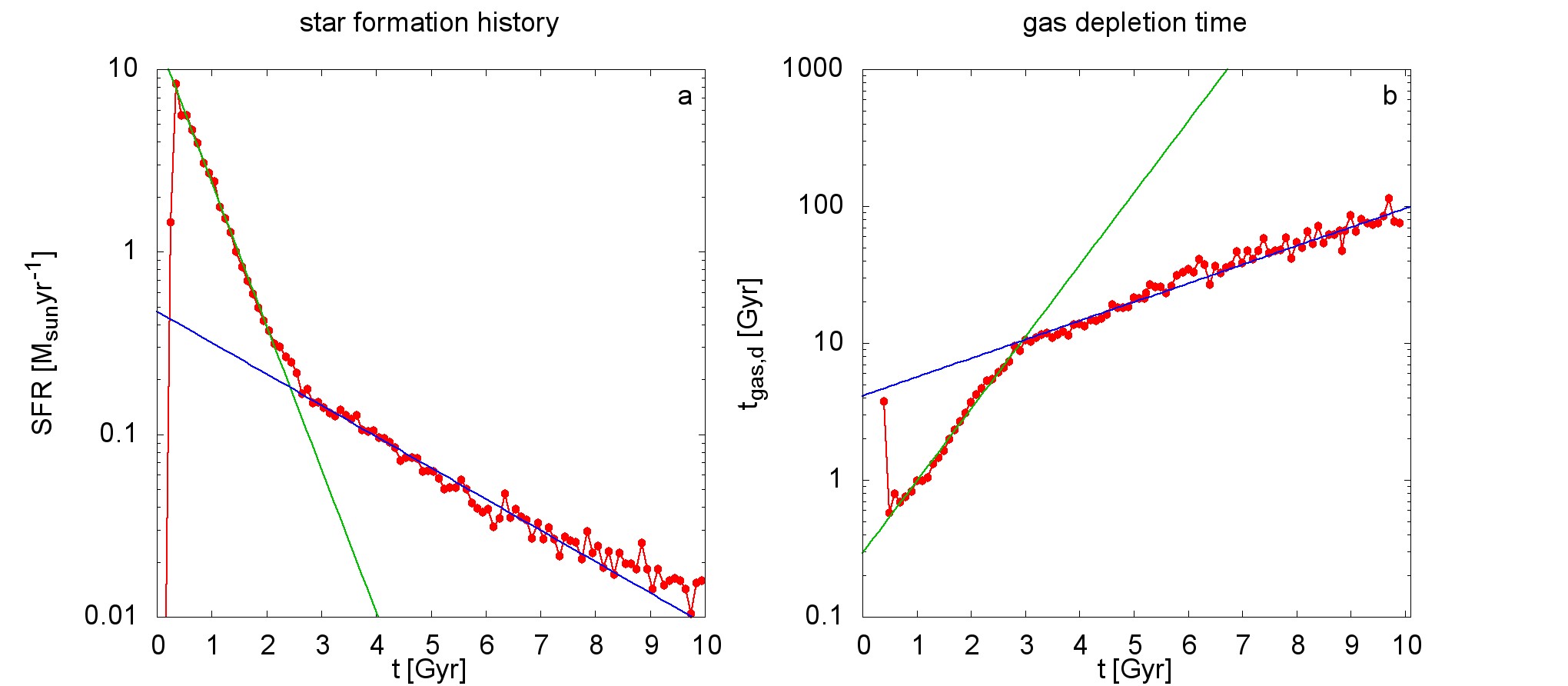}
\caption{The star formation history (SFH) and the evolution of the gas-depletion time for M1. The red dots show the model data, the green line shows the exponential fit for the first part of the decaying star formation rate (SFR) / increasing gas-depletion time directly after the initial collapse. The blue line shows the exponential fit for the shallower part after most of the gas was converted into stellar particles during and shortly after the collapse.}
\label{fig:SFHM1}
\end{figure*}
\begin{figure*}[h]
\includegraphics[width=1.0\linewidth]{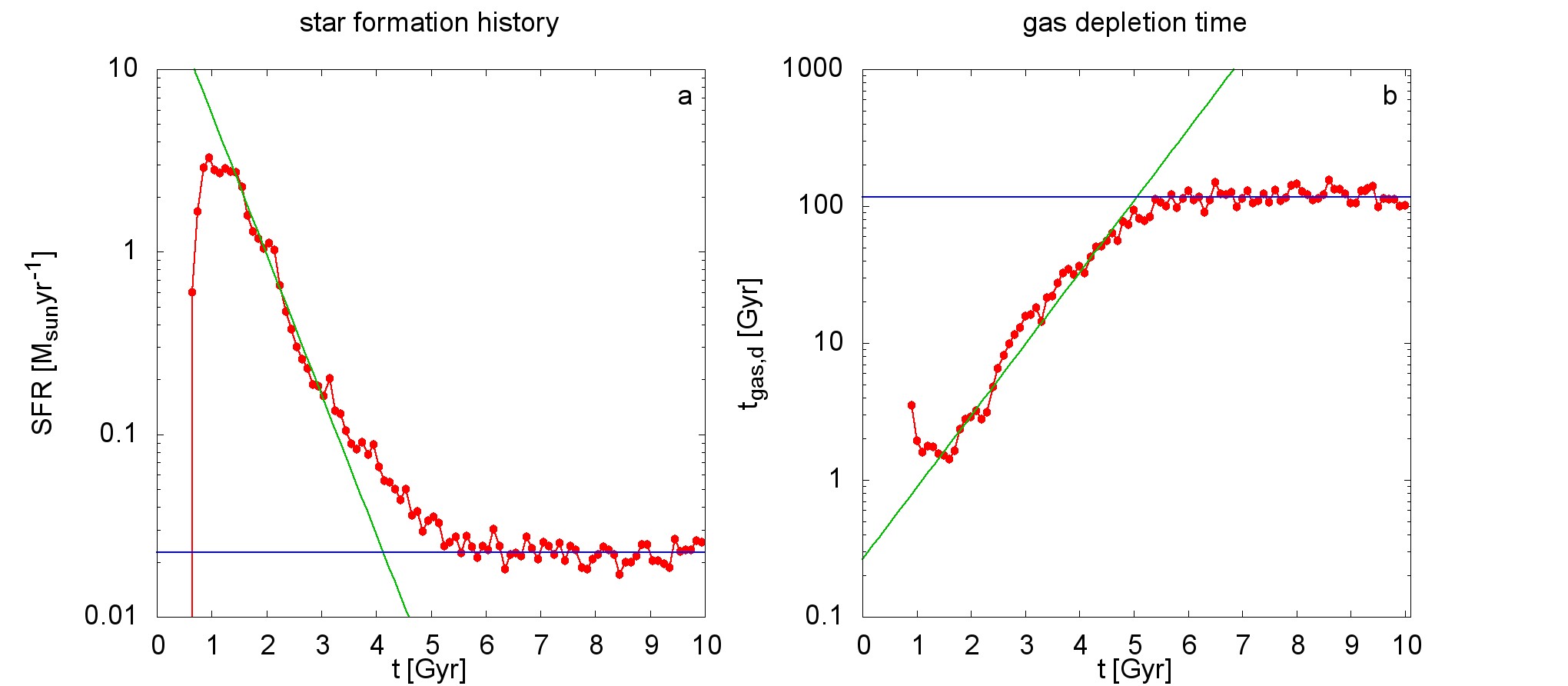}
\caption{As Fig. \ref{fig:SFHM1} but for M1N.}
\label{fig:SFHM1N}
\end{figure*}
The left panel of Fig. \ref{fig:SFHM1} and Fig. \ref{fig:SFHM1N} shows the star formation history of M1 and M1N respectively. The star formation rate (SFR) is calculated by separating all stellar particles in bins of $100\mathrm{Myr}$ according to their age, summing up the stellar mass in every time-bin and dividing this by the length of the time-bin.\\
\indent
The SFR increases sharply at the time of the collapse of the sphere until the maximum is reached (the halo forms during this time) and then it decreases exponentially as shown by the fit. The function
\begin{eqnarray}\label{SFHfit}
h(t)=e\times \exp(-t/t_\mathrm{e}),
\end{eqnarray}
with $e$ and $f$ being the fit-parameters, is used to fit the first (green line, 1) and second (blue line, 2) part of the decline for M1, and for M1N the second part is fitted by a constant, $c$. This results in $t_\mathrm{e,M1,1}=0.55\pm0.02\mathrm{Gyr}$,  $e_\mathrm{M1,1}=14.67\pm0.56\mathrm{M_\odot yr^{-1}}$, $t_\mathrm{e,M1,2}=2.53\pm0.06\mathrm{Gyr}$, $e_\mathrm{M1,2}=0.47\pm0.02\mathrm{M_\odot yr^{-1}}$ for M1 and in $t_\mathrm{e,M1N,1}=0.57\pm0.02\mathrm{Gyr}$,  $e_\mathrm{M1N,1}=33.46\pm4.10\mathrm{M_\odot yr^{-1}}$, $c_\mathrm{M1N}=0.02\pm4\times10^{-4}\mathrm{M_\odot yr^{-1}}$ for M1N.\\
\indent
The right panel of Fig. \ref{fig:SFHM1} and Fig. \ref{fig:SFHM1N} shows the gas-depletion time of M1 and M1N at every output-time.
\begin{eqnarray}\label{gdt}
t_\mathrm{gas,d}(t)=\frac{M_\mathrm{gas}(t)}{\mathrm{SFR}(t)}.
\end{eqnarray}
$t_\mathrm{gas,d}(t)$ corresponds to the time it would take to convert the remaining gas-mass, $M_\mathrm{gas}$, at time $t$ with the SFR at that time into stellar particles.\\
\indent
The evolution of the gas-depletion time is very similar to the SFH (just inverted), so the initial collapse corresponds to the minimum at the beginning of the curve, after that $t_\mathrm{gas,d}$ rises exponentially until most of the gas has been consumed, then the increase becomes shallower or, in the case of M1N, $t_\mathrm{gas,d}$ becomes constant. The exponential parts (directly after the collapse: green line, shallower/constant part: blue line) are again fitted by a simple exponential function, $g(t)$,
\begin{eqnarray}\label{gdtfit}
g(t)=i\times \exp(t/j),
\end{eqnarray}
where i and j are the fit parameters. The constant part of M1N is also again fitted by a constant, $c$. The results are $j_\mathrm{M1,1}=0.85\pm0.02\mathrm{Gyr}$,  $i_\mathrm{M1,1}=0.36\pm0.02\mathrm{Gyr}$, $j_\mathrm{M1,2}=3.17\pm0.08\mathrm{Gyr}$, $i_\mathrm{M1,2}=4.24\pm0.21\mathrm{Gyr}$ for M1 and in $j_\mathrm{M1N,1}=0.84\pm0.02\mathrm{Gyr}$,  $i_\mathrm{M1N,1}=0.32\pm0.04\mathrm{Gyr}$, $c_\mathrm{M1N}=117.92\pm2.27\mathrm{Gyr}$ for M1N. Again, indices 1 and 2 correspond to the steeper (green) and shallower/constant (blue) part respectively (the SFH and the evolution of $t_\mathrm{gas,d}$ for every model can be seen in the Appendix of the arXiv version of this paper).\\
\indent
Observed galaxies show a mildly decreasing SFR with cosmic time \citep{Speagle2014}, such that the present-day gas depletion times are independent of galaxy mass and about $2.8\mathrm{Gyr}$ \citep{JPFA2009}. Galaxies therefore need to accrete gas accordingly. This is different for the models computed here, because there is no reservoir of gas surrounding the model galaxies such that accretion of gas is insignificant and so the gas-density in the galaxy model decreases continuously. Therefore the SFR decreases and the gas-depletion time increases. In the case of M1N the gas spreads further, due to the weaker potential compared to MOND, so the peak of the SFH is smaller but also more extended in time. After approximately $5\mathrm{Gyr}$ the SFR and gas-depletion time become constant in contrast to the behaviour of M1, because the gas-density is higher for M1N compared to M1 due to the initial collapse not consuming as much gas. This also leads to different gas-fractions as can be seen in Fig. \ref{fig:overview}.\\
\begin{figure}[h]
\includegraphics[width=1.0\linewidth]{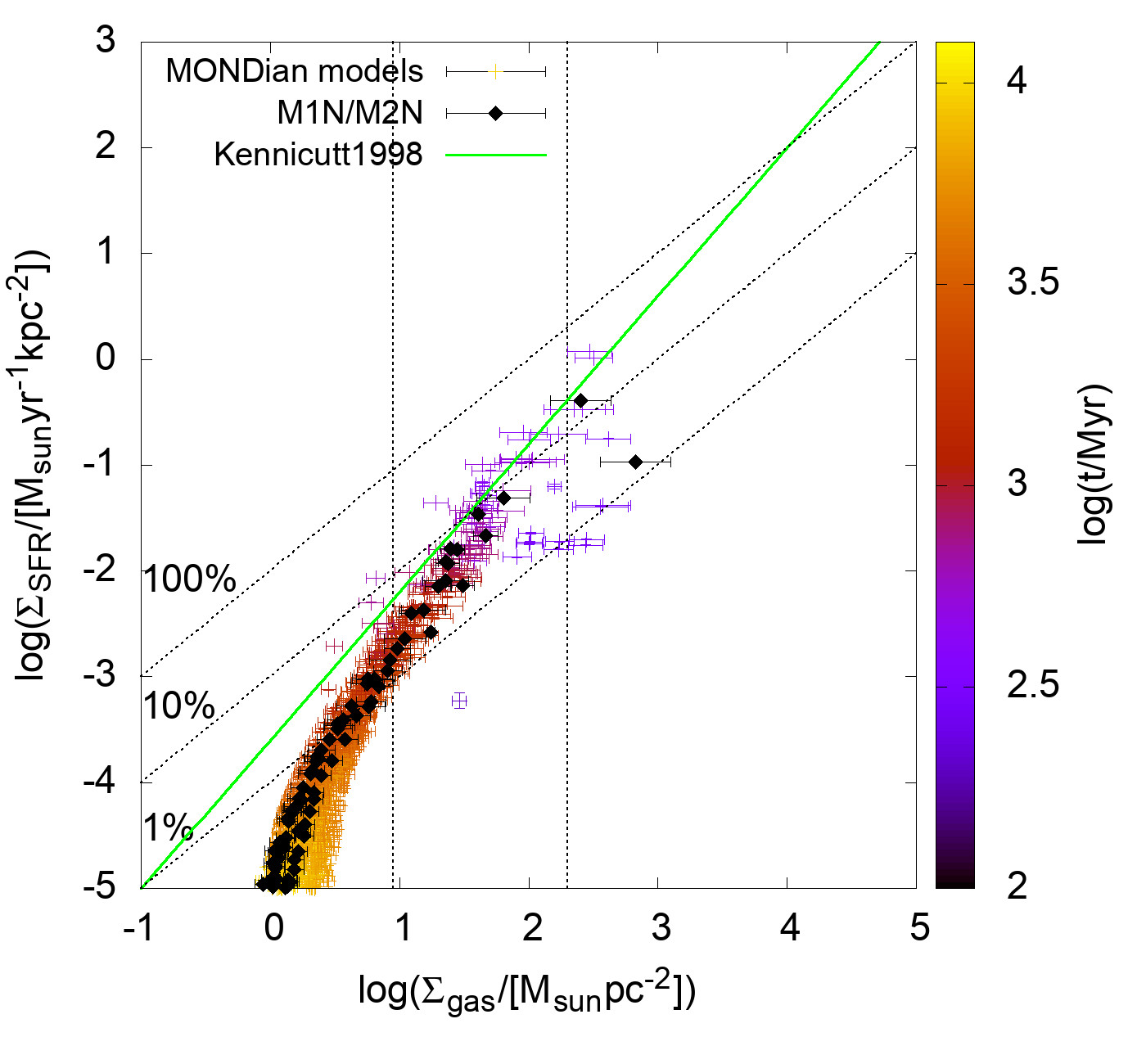}
\caption{The KS-diagram for all 15 models shown as a time-evolution, while the age is indicated by colour. The Newtonian models are emphasized as black diamonds, though they show the same trend as all other models. The green line corresponds to the best fit to observational data by \cite{Kennicutt1998}, while the vertical dotted lines correspond to the dividing lines between different regimes according to \cite{Bigiel2008} (see text). The diagonal dotted lines show star formation efficiencies of 1\%, 10\% and 100\% respectively, meaning that 1\%, 10\% and 100\% of the remaining gas mass is consumed within $10^8\mathrm{yr}$.}
\label{fig:KS}
\end{figure}
\indent
Fig. \ref{fig:KS} shows the Kennicutt-Schmidt-diagram \citep{Kennicutt1998} for all simulations. We adopted the layout of Fig. 15 of  \citealp{Bigiel2008}, as they further investigated the relation between the star-formation-rate-density, $\Sigma_\mathrm{SFR}$, and the gas-surface-mass-density, $\Sigma_\mathrm{gas}$, and to compare their results with our data. $\Sigma_\mathrm{gas}$ is calculated by averaging the gas-surface-mass-density over the stellar disk and $\Sigma_\mathrm{SFR}$ shows the SFR from e.g. Fig. \ref{fig:SFHM1} divided by the area of the stellar disk.\\
\indent
The diagonal dotted lines in Fig. \ref{fig:KS} correspond to star-formation-efficiencies of 1\%, 10\% and 100\% as indicated in the plot, meaning that these lines show the star-formation-rate needed to turn 1\%, 10\% or 100\% of the remaining gas-mass into stars within $10^{8} \mathrm{yr}$. The two vertical black and dotted lines show the limits \cite{Bigiel2008} established to divide three distinctly different regimes in the $\Sigma_\mathrm{SFR}$-$\Sigma_\mathrm{gas}$-plane. The left vertical line corresponds to the saturation of HI at a surface density of $\Sigma_\mathrm{gas}\approx9 \mathrm{M_\odot pc^{-2}}$, while the second line at $\Sigma_\mathrm{gas}\approx200\mathrm{M_\odot pc^{-2}}$ is the transition from "normal" galactic SFRs to starburst galaxies (see \cite{Bigiel2008}). The green line shows the best fit, equation (4), of \cite{Kennicutt1998},
\begin{eqnarray}
\label{Kennicutteq4}
&&\Sigma_\mathrm{SFR} \left(\mathrm{M_\odot yr^{-1} kpc^{-2}}\right) \\
&&=(2.5\pm 0.7)\times 10^{-4}\left(\frac{\Sigma_\mathrm{gas}}{1 \mathrm{M_\odot pc^{-2}}}\right)^{1.4\pm 0.15}. \nonumber
\end{eqnarray}
The black diamonds are the two Newtonian models, M1N and M2N, while the coloured crosses depict all MONDian simulations. The evolution of each model is plotted in steps of $100\mathrm{Myr}$, while the age of every point is shown by its colour. Similar to fig. 15 of \cite{Bigiel2008}, our simulations show a shallower increase of $\Sigma_\mathrm{SFR}$ in the central regime compared to the points with $\Sigma_\mathrm{gas}\leq9\mathrm{M_\odot pc^{-2}}$. Due to the fact that very few of our models reach gas-surface-mass-densities well within the starburst regime, we cannot make any reasonable statement for MONDian models in this regime. However, we can also confirm that the fit from \cite{Kennicutt1998} is offset to higher SFRs, because the author used starburst and non-starburst galaxies in one fit, as \cite{Bigiel2008} stated.\\
\indent
In general, our models are qualitatively similar to the observational data from \cite{Bigiel2008}, showing two distinct regimes with a transition close to the HI saturation line. On the other hand, the rapid evolution of our models is also evident, as the slope in the intermediate region, $\Sigma_\mathrm{gas}\geq9\mathrm{M_\odot pc^{-2}}$, is steeper than observed. Important to note is that also the Newtonian models show a similar, but slightly shallower, behaviour as the MONDian models, which suggests that the critical law here is the description of star-formation itself, rather than gravity. Additionally, we stress that we show the time-evolution of our models, while \cite{Bigiel2008} plot the SFR of nearby galaxies.\\
\indent
To investigate the evolution of the SFRs of our models from another angle and to test whether they lie near the observed main sequence of galaxies \citep{Speagle2014}, we also plotted the star formation rate, SFR, versus stellar mass, $M\mathrm{_{s}}$, dependency.
We use the function $\mathrm{SFR_{ms}}$,\footnote{Note that $\log\equiv\log_{10}$ throughout.}
\begin{eqnarray}
\log(\mathrm{SFR_{ms}}(M_\mathrm{s})) &=& \log\left(\frac{a}{1\mathrm{M_\odot yr^{-1}}}\right) \\
&+& b\times\log\left(\frac{M_\mathrm{s}}{\mathrm{M_\odot}}\right), \nonumber
\end{eqnarray}
with $a$ and $b$ being the fit-parameters, to fit our data.\\
\begin{figure*}[h]
\begin{minipage}[t]{0.49\linewidth}
\includegraphics[width=1.0\linewidth]{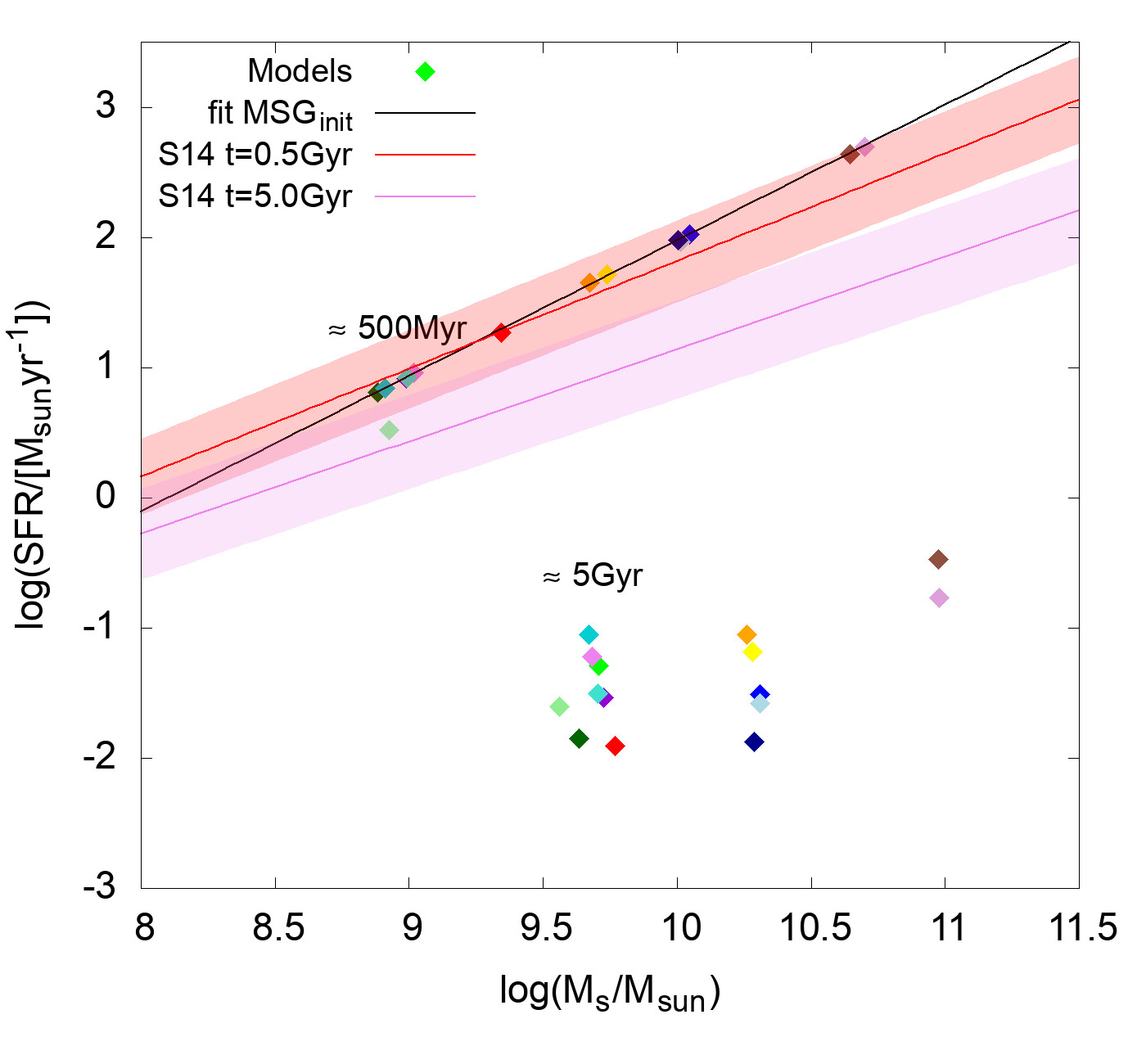}
\end{minipage}
\begin{minipage}[t]{0.49\linewidth}
\includegraphics[width=1.0\linewidth]{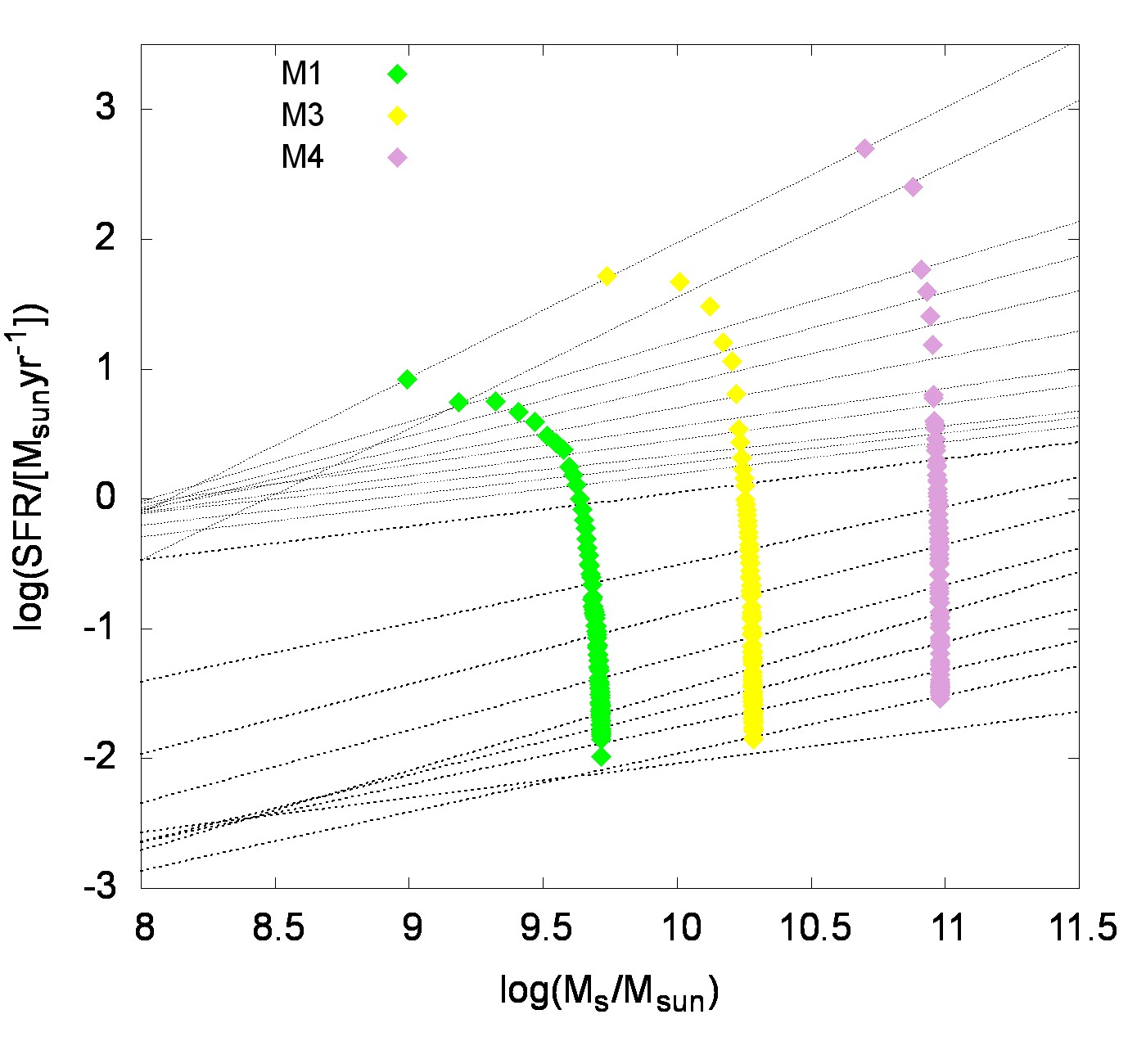}
\end{minipage}
\caption{\textit{Left panel}: The SFR-$M_\mathrm{s}$ plane with all models at their respective peak-SFR at the top and the distribution of the models $5\mathrm{Gyr}$ afterwards at the bottom. The black line shows the linear fit to the peak-SFR data, resulting in  $a=3.66\pm0.35\times 10^{-9}\mathrm{M_\odot yr^{-1}}$ and $b=1.041\pm0.004$. The colours are identical to Fig. \ref{fig:RvM} with the addition of M4=plum, M4sn=dark-plum and the Newtonian models indicated by light colours (e.g. M2=blue and M2N=light-blue). The red and violet areas show the observed main sequence of galaxies of \cite{Speagle2014} at $t=0.5\mathrm{Gyr}$ and $t=5.0\mathrm{Gyr}$, respectively (Eq. \ref{eq:Speagle}). \textit{Right panel}: The evolution of the linear fit to all models shown as the dashed lines at various times (see text) and the complete time-evolution of M1 (green), M3 (yellow) and M4 (plum).}
\label{fig:MSG}
\end{figure*}
\indent
The left panel of Fig. \ref{fig:MSG} shows our models directly at their peak-SFR (upper part of the plot) and after $\approx5\mathrm{Gyr}$ compared to the observed main sequence of galaxies of \cite{Speagle2014} at $t=0.5\mathrm{Gyr}$ (red area) and $t=5.0\mathrm{Gyr}$ (violet area). We only fitted the upper part resulting in $a=3.66\pm0.35\times 10^{-9}\mathrm{M_\odot yr^{-1}}$ and $b=1.041\pm0.004$, while the distribution of the models after $\approx5\mathrm{Gyr}$ suggests that the scatter of the correlation increases with time. At $t=0.5\mathrm{Gyr}$ the fit to our models suggests a steeper slope than Eq. \ref{eq:Speagle}, but the models lie within the uncertainties of the observed relation. After $t=5.0\mathrm{Gyr}$ the picture is different, as the scatter is strongly increased and the SFR of our models is offset by approximately two orders of magnitude compared to the observed relation. The right panel of Fig. \ref{fig:MSG} shows the evolution of M1, M3 and M4 with time in the SFR-$\mathrm{M_s}$ plane and the evolution of the linear fit to all models at various times. The thin lines are the fits from the peak SFR at $\approx500\mathrm{Myr}$ to $\approx1.5\mathrm{Gyr}$ (the first thick, dotted line) in $100\mathrm{Myr}$ steps. From that time onwards, the stepsize is increased to $1\mathrm{Gyr}$ and the last line shows the fit at $9.5\mathrm{Gyr}$. The slopes, $b$, of the fits range from $1-0.25$, while the uncertainty from the fit increases with time up to $85\%$ of the fitted value. We note that the observed main sequence of galaxies follows 
\begin{eqnarray}\label{eq:Speagle}
&&\log(\mathrm{SFR}(M_\mathrm{s},t)) = \\
&&(0.84\pm 0.02-0.026\pm 0.003 t) \log M_\mathrm{s} \nonumber \\
&&-(6.51\pm 0.24-0.11\pm 0.03 t), \nonumber
\end{eqnarray}
with t being the age of the universe \citep{Speagle2014}, such that the model result appears to be promising in terms of possibly helping to understand the origin and evolution of the main sequence. Clearly much more work is required, in particular to address the issue of continued gas accretion.\\
\indent
Different models react differently to the depletion of gas, due to the insignificant accretion, as indicated by the left panel of Fig. \ref{fig:MSG}. For the same $\eta$ and initial mass, there is a difference between models of approximately one order of magnitude in SFR after $\approx5\mathrm{Gyr}$ (points at the left end). For example, the highest SFR is evident in M1l13, which is shown by the dark-turquoise diamond. This is expected, as the density threshold for a star-forming event to take place needs to be met in a smaller volume compared to all other simulations. The comparison with the \cite{Speagle2014} relations underlines the fact that the SFR decreases more rapidly than observed, which we would expect due to the lack of further accretion. The most prominent effect of the absence of accretion can be seen in the right panel of Fig. \ref{fig:MSG}. The evolutionary track of M4 in the SFR-$\mathrm{M_s}$ plane becomes nearly vertical $\approx300\mathrm{Myr}$ after its peak SFR, while the track of M1 shows that behaviour more than $\approx1\mathrm{Gyr}$ later. This is a direct consequence of the main sequence of galaxies, because the more massive a galaxy is, the more gas it has to accrete to sustain its SFR.
\subsection{Stellar angular momentum and disk vs halo populations, thin and thick disks}
After analysing the SFH and showing that there might be a mass and initial cloud angular momentum limit to the single collapsing gas-sphere approach, the question remains how the angular momentum distribution of the stars is shaped by the simple initial conditions used here. We note that a series of simulations of the same model (M1) with different resolution (M1l11 and M1l13) yield the same total angular momentum in the stellar particles within 5\%.\\
\indent
\begin{figure*}[h]
\begin{minipage}[t]{0.49\linewidth}
\includegraphics[width=1.0\linewidth]{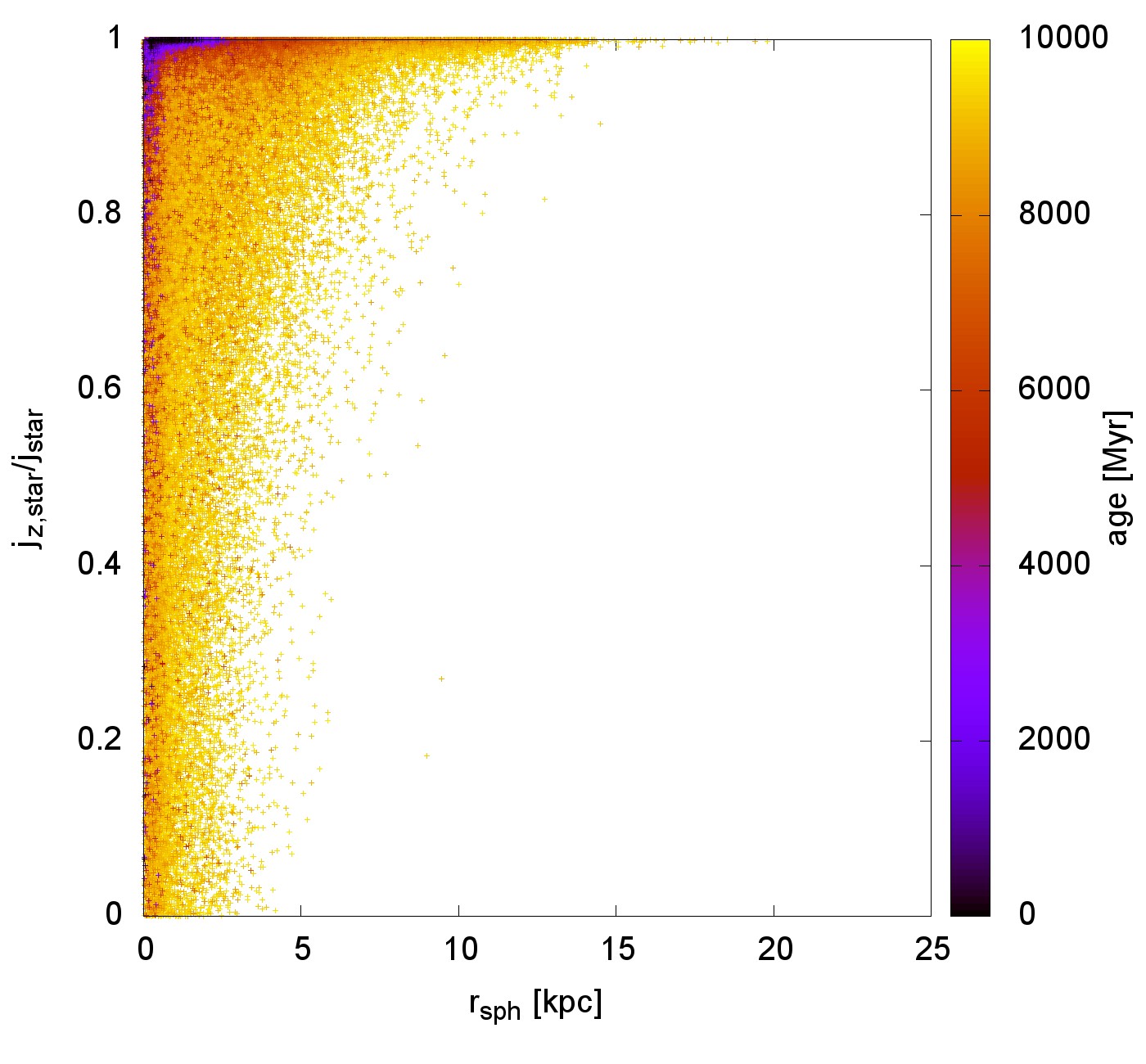}
\end{minipage}
\begin{minipage}[t]{0.49\linewidth}
\includegraphics[width=1.0\linewidth]{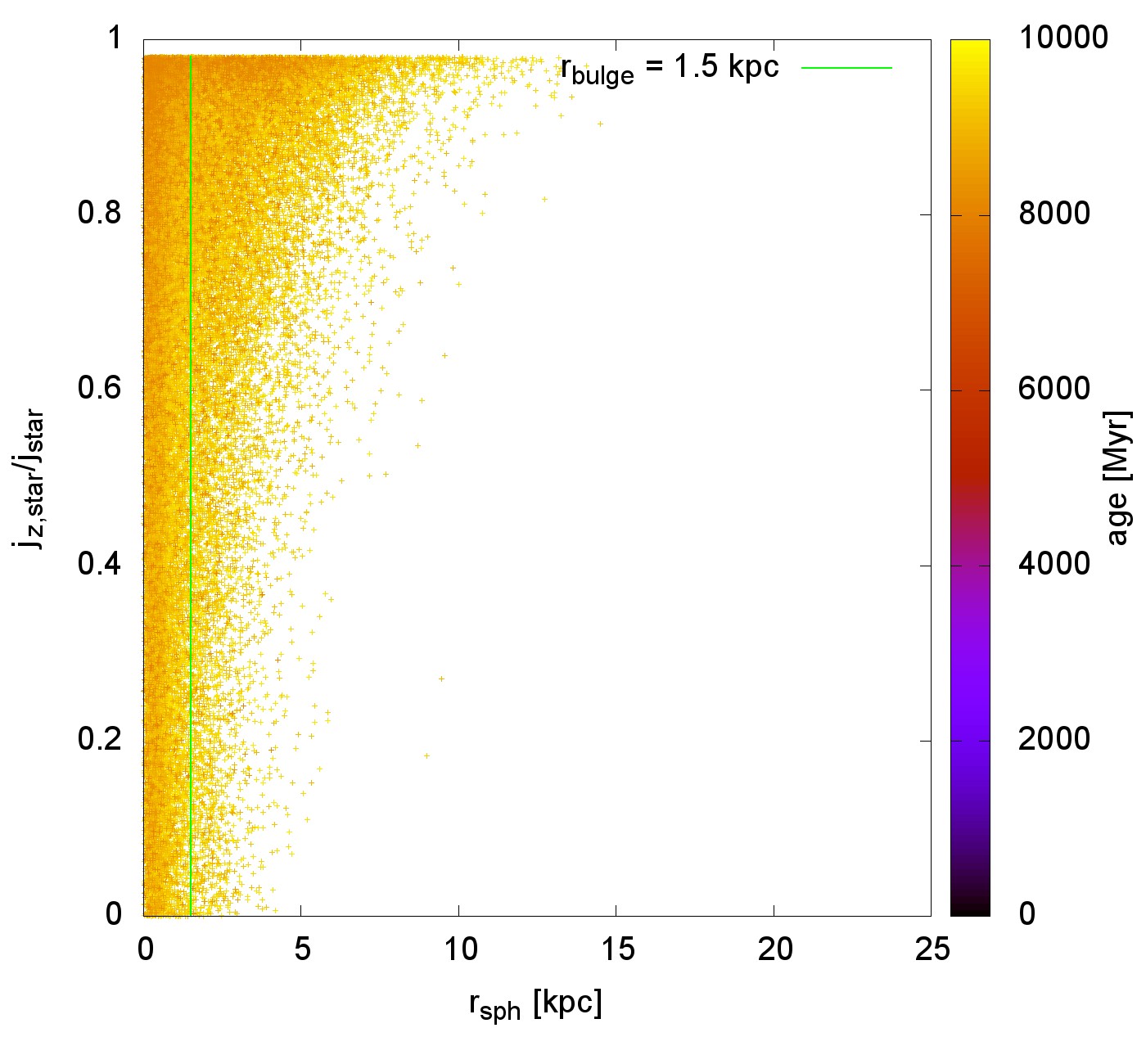}
\end{minipage}
\caption{Radial distribution of the ratio of the specific stellar angular momentum and its z-component for M1 after $10\mathrm{Gyr}$ with $r_{sph}$ being the spherical radius, $j_{star}$ the specific angular momentum and $j_{z,star}$ its z-component. The age of every star is shown via colour from $10\mathrm{Gyr}$ (bright yellow) to a few $\mathrm{Myr}$ (black). \textit{Left panel:} All stellar particles. \textit{Right panel:} Only bulge/halo stars that are older than $8\mathrm{Gyr}$ (see text for details).} 
\label{fig:jz/jM1}
\end{figure*}
Fig. \ref{fig:jz/jM1} shows the ratio of the specific stellar angular momentum, $j_\mathrm{star}$, and its z-component, $j_\mathrm{z,star}$, for all stars (left panel) and only for old, $> 8\mathrm{Gyr}$, halo/bulge stars (right panel) against the spherical radius, $r_{sph}$, for M1 (the plots for every other model can be seen in the Appendix of the arXiv version of this paper). The age of every stellar particle is shown via colour from $\approx10\mathrm{Gyr}$ (bright yellow) to a few $\mathrm{Myr}$ (black) as indicated by the coloured age scale. In order to obtain the contribution of population II particles (older than $8\mathrm{Gyr}$) to the mass of the bulge and halo, the disk particles were sorted out. These are assumed to have $j_{z,star}/j_{star}\geqq0.98$, and an age smaller than $8\mathrm{Gyr}$. The green line in the right panel corresponds to the dividing line between halo and bulge, which is defined here as the radius at which the smallest ratio of densities in neighbouring radial bins occurs, i.e. $n_\mathrm{star}(i)/n_\mathrm{star}(i-1)$, where $i$ is the number of the radial bin, is computed for all bins and the radius of bin $i$ with the smallest ratio within a reasonable distance to the centre, $r_\mathrm{sph}<10\mathrm{kpc}$, is chosen to be the radius of the bulge, $r_\mathrm{bulge}$.\\
\indent
The left panel of Fig. \ref{fig:jz/jM1} reveals that the youngest stars are concentrated in the centre of the galaxy and most of them move within the disk as the contribution of $j_\mathrm{z,star}$ to $j_\mathrm{star}$ is greater or equal to $0.98$. The rest of the young stars can be found within the bulge, but not within the halo. This is very different to observed late-type galaxies, because especially the most massive ones show a higher star formation rate in the disk compared to the centre. On the other hand, the discrepancy between the models and real galaxies is expected as the models computed here do not accrete gas during their evolution, hence the centre is the only region, where star formation continues after the galaxy has formed.\\
\indent
The right panel of Fig. \ref{fig:jz/jM1} shows the radial distribution of $j_\mathrm{z,star}/j_\mathrm{star}$ for all stars older than $8\mathrm{Gyr}$. The bulge and the halo are separated by the green line, although the bulge is also visible by eye, as it is the region where a significant amount of stars is not rotating around the z-axis. Comparing the left panel with the right one displays that there are many stars (especially older ones) that do not move within the disk, which already hints that there is a slight lack of stellar angular momentum. This is further emphasized by Tab. \ref{tab:bulge/halo}, where the masses and mass-fractions of the halo and bulge for all models are shown. The result is that more than $50\%$ of the total stellar mass is not in the disk, which will be put into perspective in Fig. \ref{fig:angularmomentum}.\\
\indent
\begin{table*}[h]
\caption{Stellar masses and mass fractions of the bulge and the halo and also the total stellar and gas mass of every model after $10\mathrm{Gyr}$. Note that the mass fraction of the bulge decreases the weaker the initial collapse is.}
\centering
\begin{tabular}[c]{l c c c c c c}
\hline
Model name & $\frac{M_\mathrm{tot,star}}{\mathrm{10^9M_\odot}}$ &
$\frac{M_\mathrm{tot,gas}}{\mathrm{10^9M_\odot}}$ & $\frac{M_\mathrm{b,star}}{\mathrm{10^9M_\odot}}$ & $\frac{M_\mathrm{b,star}}{M_\mathrm{tot,star}}$ & $\frac{M_\mathrm{h,star}}{\mathrm{10^9M_\odot}}$ & $\frac{M_\mathrm{h,star}}{M_\mathrm{tot,star}}$\\
\hline
M1 & $5.21$ & $1.19$ & $2.00$ & $0.38$ & $0.98$ & $0.19$\\
M1sn & $4.33$ & $1.69$ & $1.75$ & $0.40$ & $1.07$ & $0.25$ \\
M1N & $3.72$ & $2.66$ & $1.34$ & $0.36$ & $0.75$ & $0.20$ \\
M1const & $5.88$ & $0.52$ & $3.19$ & $0.54$ & $0.60$ & $0.10$\\
M1Zpoor & $5.12$ & $1.28$ & $2.11$ & $0.41$ & $0.86$ & $0.17$\\
M1Zpoorsn & $4.83$ & $1.44$ & $1.85$ & $0.38$ & $0.89$ & $0.18$\\
M1l11 & $4.90$ & $1.50$ & $2.27$ & $0.46$ & $0.68$ & $0.14$\\
M1l13 & $5.36$ & $1.04$ & $2.53$ & $0.47$ & $1.16$ & $0.22$\\
\hline
M2 & $20.48$ & $1.12$ & $10.75$ & $0.52$ & $2.03$ & $0.10$\\
M2sn & $19.47$ & $1.82$ & $10.25$ & $0.53$ & $2.26$ & $0.12$\\
M2N & $20.39$ & $1.19$ & $9.79$ & $0.48$ & $3.55$ & $0.17$\\
M3 & $19.23$ & $2.37$ & $7.47$ & $0.39$ & $3.44$ & $0.18$\\
M3sn & $18.38$ & $3.03$ & $7.27$ & $0.40$ & $3.35$ & $0.18$\\
\hline
M4 & $95.62$ & $4.07$ & $36.49$ & $0.38$ & $20.78$ & $0.22$\\
M4sn & $94.90$ & $4.91$ & $35.51$ & $0.37$ & $21.73$ & $0.23$\\
\hline
\end{tabular}
\label{tab:bulge/halo}
\end{table*}
\begin{figure}[h]
\includegraphics[width=1.0\linewidth]{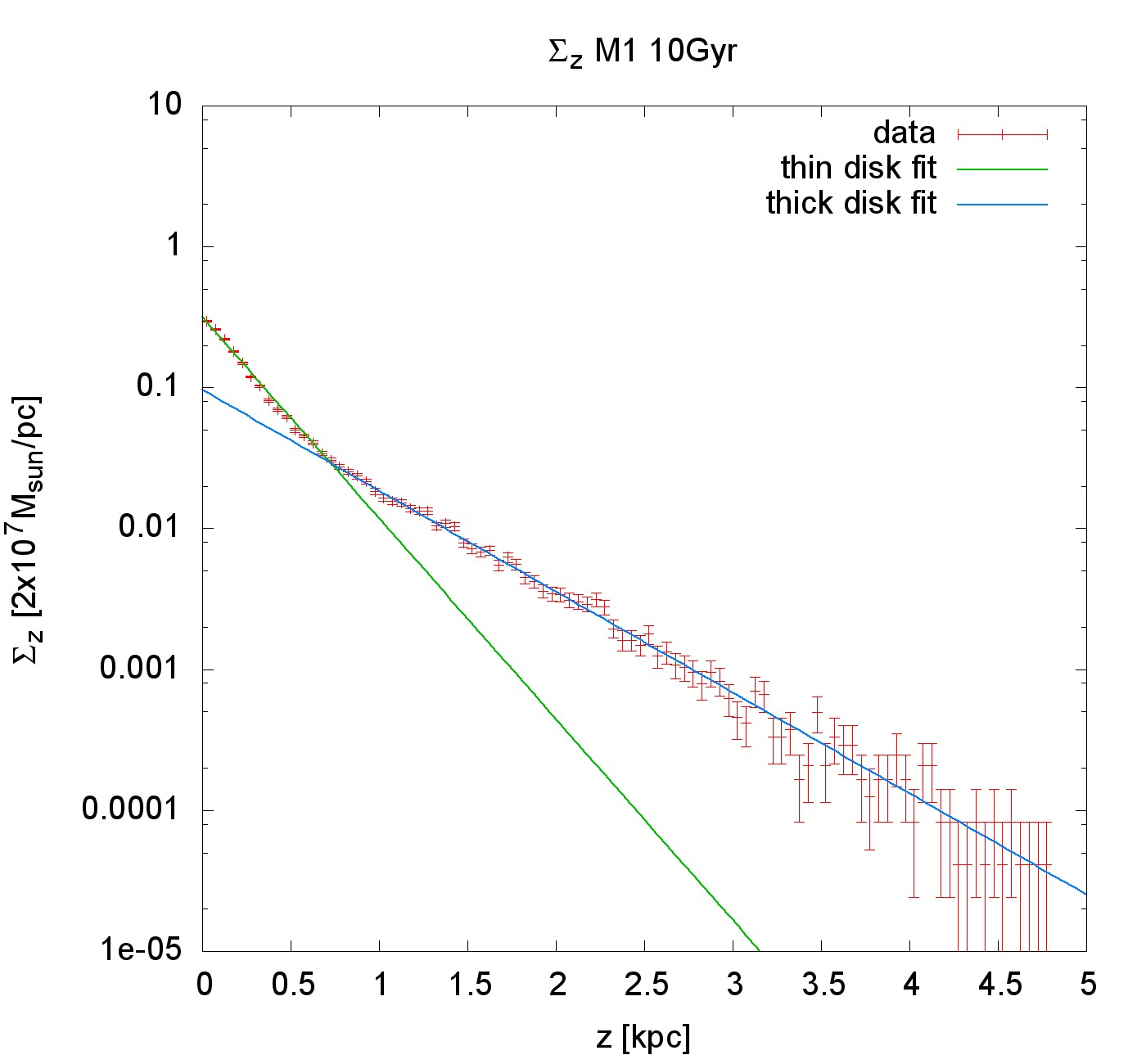}
\caption{Mass distribution along the rotation axis (z-axis), $\Sigma_z$. The green line corresponds to the exponential fit for the thin disk and the blue line to the thick disk, fitted in the range $0\ -\ 1\mathrm{kpc}$ and $1\ -\ 3\mathrm{kpc}$ respectively (see text for details).}
\label{fig:sigmazM1}
\end{figure}
Fig. \ref{fig:sigmazM1} shows the distribution of mass along the z-axis, $\Sigma_\mathrm{z}$, versus the vertical distance (height not thickness), $z$, for M1. In contrast to the distribution of mass outside of the disk, $\Sigma_\mathrm{z}$ is comparable to observations as the distribution can be fitted by two simple exponential functions, which would correspond to the profile of a thin and thick disk. For example, also the Milky Way shows a vertical stellar mass distribution with a thin and a thick disk. The fits are obtained with a similar function as for the surface mass profiles,
\begin{eqnarray}\label{eq:sigmaz}
\Sigma_\mathrm{z}(z)=n_\mathrm{z}\times\exp(-z/z_\mathrm{e}),
\end{eqnarray}
resulting in $n_\mathrm{z,thin,M1}/2\times10^7\mathrm{M_\odot}/\mathrm{pc}=0.316\pm0.009$, $z_\mathrm{e,thin,M1}=0.305\pm0.007\mathrm{kpc}$ and $n_\mathrm{z,thick,M1}/2\times10^7\mathrm{M_\odot}/\mathrm{pc}=0.096\pm0.004$, $z_\mathrm{e,thick,M1}=0.606\pm0.011\mathrm{kpc}$.\\
\indent
There are a few differences between the models computed here. Models based on M1 show a very smooth transition between thick disk and halo, where the halo can only be recognized in Fig. \ref{fig:sigmazM1} by the increasing scatter around the blue line of the thick disk fit. All other MONDian models (i.e. based on M2, M3 and M4) show a shallower halo profile. All Newtonian models do not show a thick disk, but decrease according to a power law beyond $z\approx1\mathrm{kpc}$,
\begin{eqnarray}\label{eq:powerlaw}
\Sigma_\mathrm{z}(z)=az^{-b},
\end{eqnarray}
where $a$ and $b$ are the fit parameters for the Newtonian models for the power law part (all vertical profiles can be seen in the Appendix of the arXiv version of this paper).\\
\indent
\begin{figure}[h]
\includegraphics[width=1.0\linewidth]{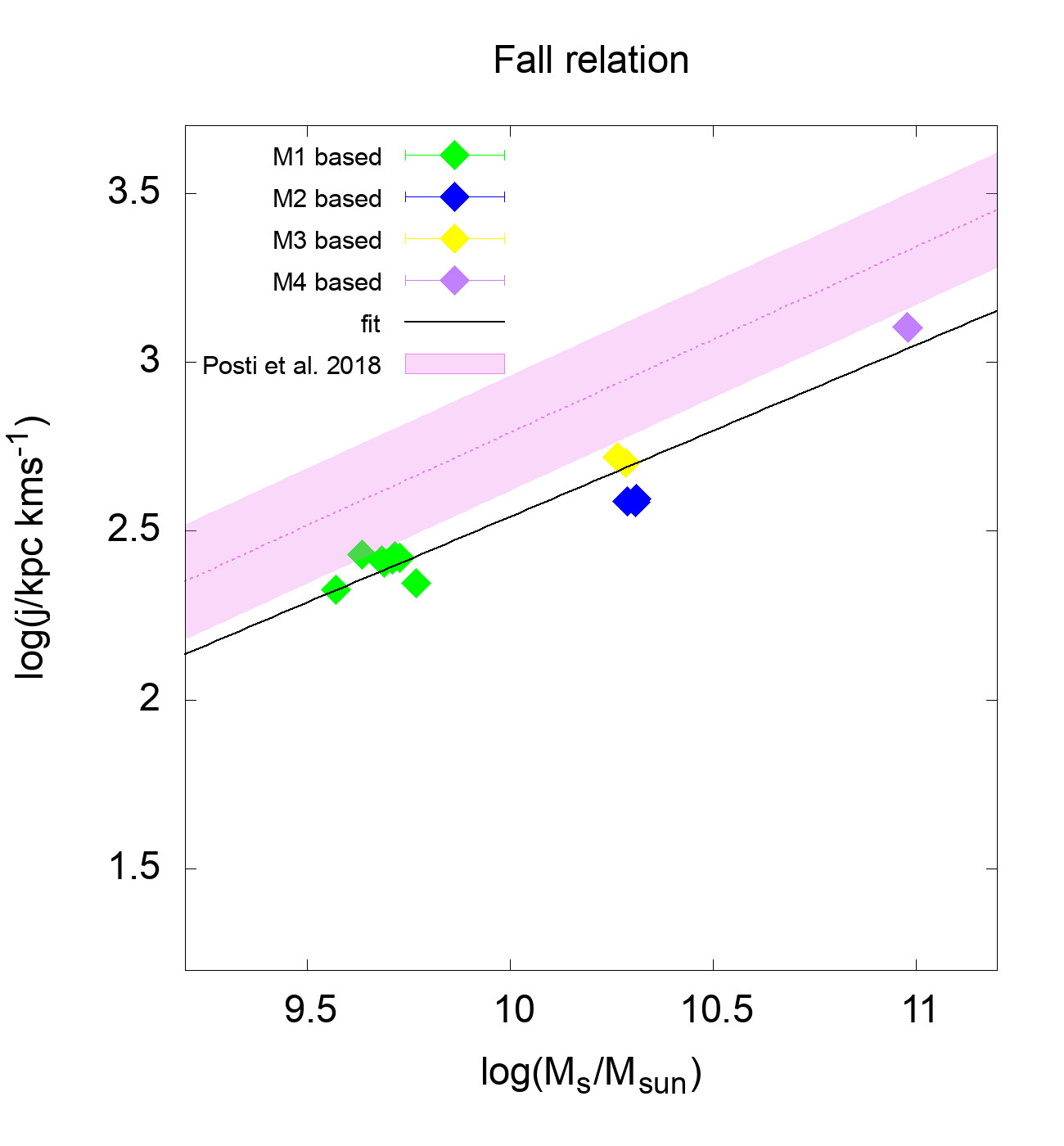}
\caption{Relation between the specific stellar angular momentum, $j_\mathrm{star}$, of all models with their respective stellar mass, $M_\mathrm{star}$ (Fall relation, computed here as for observations). The magenta area with the dashed line shows the observed data from \cite{Posti2018}, the symbols for the different models are shown in the legend and the black line (Eq. \ref{eq:Fallfit}) shows the best fit to the models. See text for further detail.}
\label{fig:angularmomentum}
\end{figure}
Fig. \ref{fig:angularmomentum} depicts the Fall relation, which compares the total specific stellar angular momentum of a galaxy, $j_\mathrm{star}$, with its stellar mass, $M_\mathrm{star}$. Note that, for this plot, we do not compute the true specific stellar angular momentum from individual stellar particles, but rather estimate it exactly as for observations \citep{Posti2018}. The magenta area with the dashed line shows the best fit of the observational data from \cite{Posti2018}, while the black line shows the best fit for the data of the models. The same functions as used by \cite{Posti2018} are here applied to calculate $j_\mathrm{star}$ and fit the data,
\begin{eqnarray}\label{eq:jstar}
j_\mathrm{star}(<R)&=&\frac{\int_{0}^{R}dR'R'^2\Sigma_\mathrm{star}(R')V_\mathrm{star,rot}(R')}{\int_{0}^{R}dR'R'\Sigma_\mathrm{star}(R')},\\
\label{eq:Fallfit}
\log j_\mathrm{star}&=&\alpha[\log(M_\mathrm{star}/M_\odot)-11]+\beta.
\end{eqnarray}
Eq. \ref{eq:jstar} and \ref{eq:Fallfit} are Eq. 1 and 4 of \cite{Posti2018}, where $\Sigma_\mathrm{star}$ is the stellar surface mass density, $V_\mathrm{star,rot}$ the stellar rotation curve of the galaxy and $\alpha$, $\beta$ are the fit parameters. \cite{Posti2018} find their best fit with $\alpha=0.55\pm0.02$ and $\beta=3.34\pm0.03$, while the best fit for the models here is $\alpha=0.51\pm0.04$ and $\beta=3.05\pm0.04$. Fig. \ref{fig:angularmomentum} together with these results shows that the simulations done in this work follow the Fall relation closely ($\alpha$ is compatible with observations), but they are slightly offset to lower values of $j_\mathrm{star}$. So the slope $\alpha$ of the specific stellar angular momentum is very similar to observed galaxies, but $j_\mathrm{star}$ is slightly too small. A large fraction of the initial angular momentum of the original cloud remains in the gas component. This is connected to the simplistic initial conditions, especially the major collapse at the beginning and the absence of accretion, resulting in mass-concentrated cores of the models (see again Fig. \ref{fig:RvM} and the corresponding explanation) and hence $j_\mathrm{star}$ is offset from the observed value. Such compact spheroidal components may have been observed as red nuggets \citep{delaRosa2016}. This may also be partially related to the detailed subgrid prescriptions used to form stars, that are fully applicable in the $\Lambda$CDM framework which they were calibrated on, but that are not necessarily applicable in MOND. Nevertheless, that the model and observed values of $\alpha$ are so similar is unexpected and suggests again that the formation of galaxies in Milgromian gravitation may contain non-trivial aspects of reality.
\subsection{Overview of all modelled galaxies}
\begin{figure*}[h]
\includegraphics[width=1.0\linewidth]{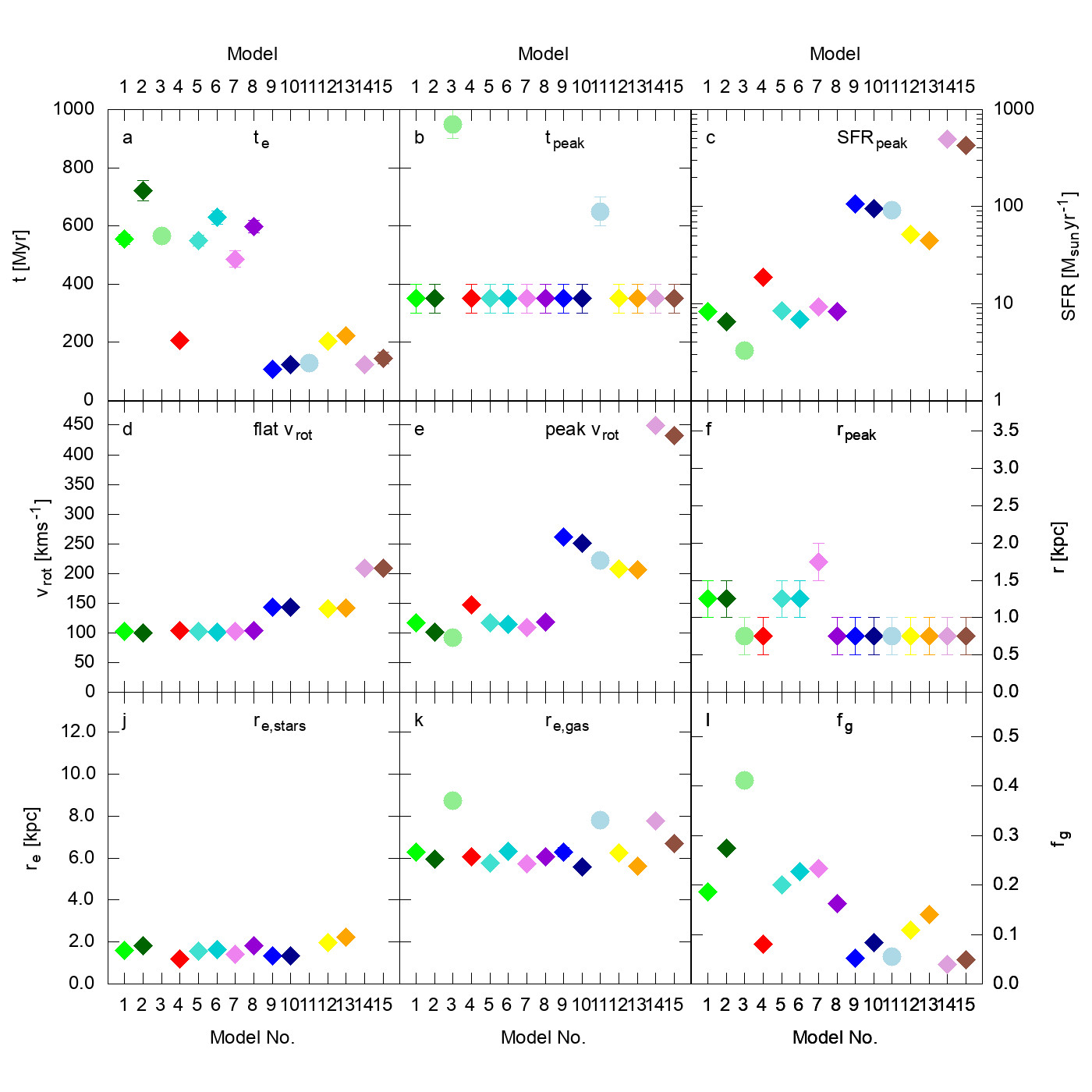}
\caption{Properties of all galaxy models at $10\mathrm{Gyr}$. The colour for every point is the same as in Fig. \ref{fig:RvM} (See Tab. \ref{tab:initconditions} for the number and description of the models), also the Newtonian models are shown as circles instead of diamonds. Panels a-c are linked to the SFH, a: Exponential star-formation decay time (see Eq. \ref{SFHfit}), b: Time of the highest SFR, c: Value of the highest SFR. Panels d-f show properties of the rotation curves, d: flat rotation velocity (flat part of the rotation curve, Eq. \ref{eq:flatvc}), e: highest rotation velocity, f: distance from the centre at which the highest $v_\mathrm{rot}$ occurs. Panels j and k show the exponential scale lengths for the stellar and gas profiles respectively (see Eq. \ref{eq:sigma}) and panel l shows the gas fraction of every model (Eq. \ref{gas-fraction}). Missing points imply that the model does not have the respective property, e.g. Newtonian models do not have an asymptotic rotation velocity.}
\label{fig:overview}
\end{figure*}
\noindent
Fig. \ref{fig:overview} shows an overview of the analysis of all models at the final time ($10\mathrm{Gyr}$ after the start of the simulation). We also include additional simulations with lower metallicity, $Z=10^{-4}\times Z_\odot$, and higher and lower resolution ($117.1875 \mathrm{pc}$ and $468.75\mathrm{pc}$) to confirm that these parameters have only a minor effect on our results. The panels a to c present properties of the star formation history, d to e of the rotation curve, j and k of the surface mass density profiles and l shows the gas fraction, $f_\mathrm{g}$,
\begin{eqnarray}\label{gas-fraction}
f_\mathrm{g}=\frac{M_\mathrm{gas}}{M_\mathrm{tot}},
\end{eqnarray}
with $M_\mathrm{gas}$ being the gas-mass and $M_\mathrm{tot}$ the total baryonic mass.\\
\indent
Panel a shows the exponential star-formation decay time of every model, which is decreasing with increasing $M_\mathrm{init}$, but increasing when using a steeper slope (larger $\eta$) of the initial rotation law. This emphasizes the explanations made earlier about how violent or strong the collapse is. As the initial mass increases, not only the peak height of the star formation rate increases but also the flanks of the peak become steeper and therefore the exponential decay time decreases. On the other hand, if the initial rotation velocity is increased, the collapse is weakened and a high star formation rate is evident over a longer period of time, so the peak broadens and the exponential decay time increases. Panel b shows that for all MONDian models the collapse takes place at the same time, $t=350\mathrm{Myr}$, regardless of $M_\mathrm{init}$ or $\eta$, while the two Newtonian models collapse later (M1N: $t=950\mathrm{Myr}$, M2N: $t=650\mathrm{Myr}$). Panel c shows the aforementioned trend that higher initial mass corresponds to a higher peak SFR, but also that supernovae and higher initial rotation velocities lead to lower peak SFRs and therefore weaken the initial collapse.\\
\indent
Panel e is related to the findings about the star formation histories, as it also shows that the peak rotation velocity increases with initial mass, but decreases with the addition of supernovae and steeper slopes, $\eta$, of the rotation law used. This comes about because $v_\mathrm{rot}$ is obtained from the acceleration at the respective radii and is therefore closely linked to the gravitational potential, which depends on the mass distribution. The mass distribution is shaped by the initial collapse and rotation velocity as seen before. Panel f shows that the highest rotation velocity is reached close to the centre of the models and d shows that regardless of the complexity of the baryonic physics or $\eta$, the flat rotation velocity is nearly equal for the same initial mass, which is an alternative way of describing the MASR (BTFR).\\
\indent
The exponential scale lengths shown in panel j and k for stars and gas respectively demonstrate that changes of the initial mass or rotation velocity and the addition of supernovae affect the stellar and gaseous distribution differently. While the stellar surface mass density distribution steepens with increasing mass and becomes shallower with the addition of supernovae and higher $\eta$, the steepest gas distributions, i.e. with the smallest radial exponential scale lengths, for models with the same $M_\mathrm{init}$ are evident for the ones with supernovae enabled, except for M1Zpoorsn. Supernovae seem to push considerable amounts of gas outside of the stellar disk such that $\Sigma_\mathrm{gas}$ is larger directly beyond the stellar disk for the supernova-models compared to the simple ones. However, outside of the gaseous disk, both the sn and the simple-feedback models show comparable surface mass densities, so the slope of the supernova-models is steeper and therefore $r_\mathrm{e,gas}$ is smaller. Moreover, a higher initial rotation velocity at the same $M_\mathrm{init}$ does not change the slope of the gas distribution significantly for the supernova-models (no. 10 and 13), but it increases the exponential scale length for model M3 (no. 12) compared to M2 (no. 9). The shallowest surface mass density profiles for the gaseous component are produced by the Newtonian models, but this is expected, because the gravitational force is weaker in these models, so overall the mass distribution is more extended.\\
\indent
Panel l depicts the gas fraction of every model and this again shows the expected behaviour already seen in panel c and e. The gas fraction decreases with higher initial mass, because the collapse is more violent and more gas is turned into stars and it increases again with higher initial rotation velocity and the addition of supernovae, as these weaken the initial collapse. Real galaxies are well known to have an increasing gas fraction with decreasing baryonic mass, so this aspect is qualitatively reproduced by the present models.\\
\indent
We also want to emphasize the differences between the Newtonian and MONDian models, as the panels closely related to the dynamics of the system (d,e,j,k) show a significant difference between the two. In the Newtonian models there is no flat rotation curve and the peak-rotation-velocity is smaller than that of the corresponding MOND model, the stellar surface-mass-density-profile is not described by an exponential profile, while the profile for the gas is significantly shallower. Additionally, star formation starts later and the evolution of the SFR is also shallower compared to the MONDian models.\\
\indent
Although M1const is simulated with a different (constant) rotation law, it underlines the aforementioned explanations about the initial collapse, because the major difference between the evolution of M1 and M1const is the more violent collapse for M1const. As can be seen in Fig. \ref{fig:overview} M1const shows the expected behaviour compared to M1 in every panel that is related to the initial collapse.\\
\indent
At this point it is important to reiterate what has been shown for the non-Newtonian models, before proceeding to the next section.
\begin{itemize}
\item
All stellar surface mass density profiles decrease exponentially within the respective stellar disk (except M4/M4sn).
\item
Changing the initial rotation velocity does not alter the resulting galaxy regarding the formation of a dense disk and an exponentially decreasing disk.
\item
Adding more complex baryonic physics also does not change the fact that an exponentially decreasing disk and the same shape for the star formation history come about. Though, star formation is slightly suppressed, because supernovae push gas out of dense regions.\\
\indent
More complex baryonic physics does therefore not yield significantly different results for the same combination of $M_\mathrm{init}$, $\eta$ and $r_\mathrm{init}$.
\item
All vertical mass density profiles show a thin and thick disk, while distinguishable disk, halo and bulge components within the stellar distribution are evident.
\item
The models slightly lack stellar angular momentum, which is linked to the initial conditions and the absence of accretion and thus the formation of a massive population II halo and bulge. This may also be related to the detailed prescriptions used to form stars, which are fully applicable in the standard cosmological framework they were calibrated on, but may not necessarily be applicable in MOND. There is substantial observational evidence that galactic discs can grow around bulges (i.e. early-formed elliptical galaxies or red nuggets) through accretion \citep{Mancini2019,delaRosa2016}.
\end{itemize}
These simulations with very simple initial conditions seem to match major properties of real galaxies. Galaxies with a stellar halo and nearly completely flat rotation curves form and exponentially decreasing disks also seem to appear naturally within MOND.
\section{The MASR and MDAR/RAR}
After analysing the differences between simple and more complex baryonic physics, different initial rotation laws and initial masses, we finally study as a consistency check, whether the models comply with the MASR and MDAR/RAR, which arise from analytical MOND formulations. Of course, it is well known that these two relations should always be obeyed in MOND as they are natural laws in this paradigm. But it is re-assuring to see that our numerical models do comply with them, since they were initialised as pure gas clouds. The MDAR/RAR combines several scaling relations including the MASR (BTFR) and has, for real galaxies, very little to no intrinsic scatter \citep{Lelli2017}. It relates the observed acceleration, $g_\mathrm{obs}$, which is obtained from the rotation curve, with the baryonic acceleration, $g_\mathrm{bar}$, which is the gravitational acceleration coming from the baryonic potential using the Newtonian gravitational law.
\begin{figure*}[h]
\includegraphics[width=1.0\linewidth]{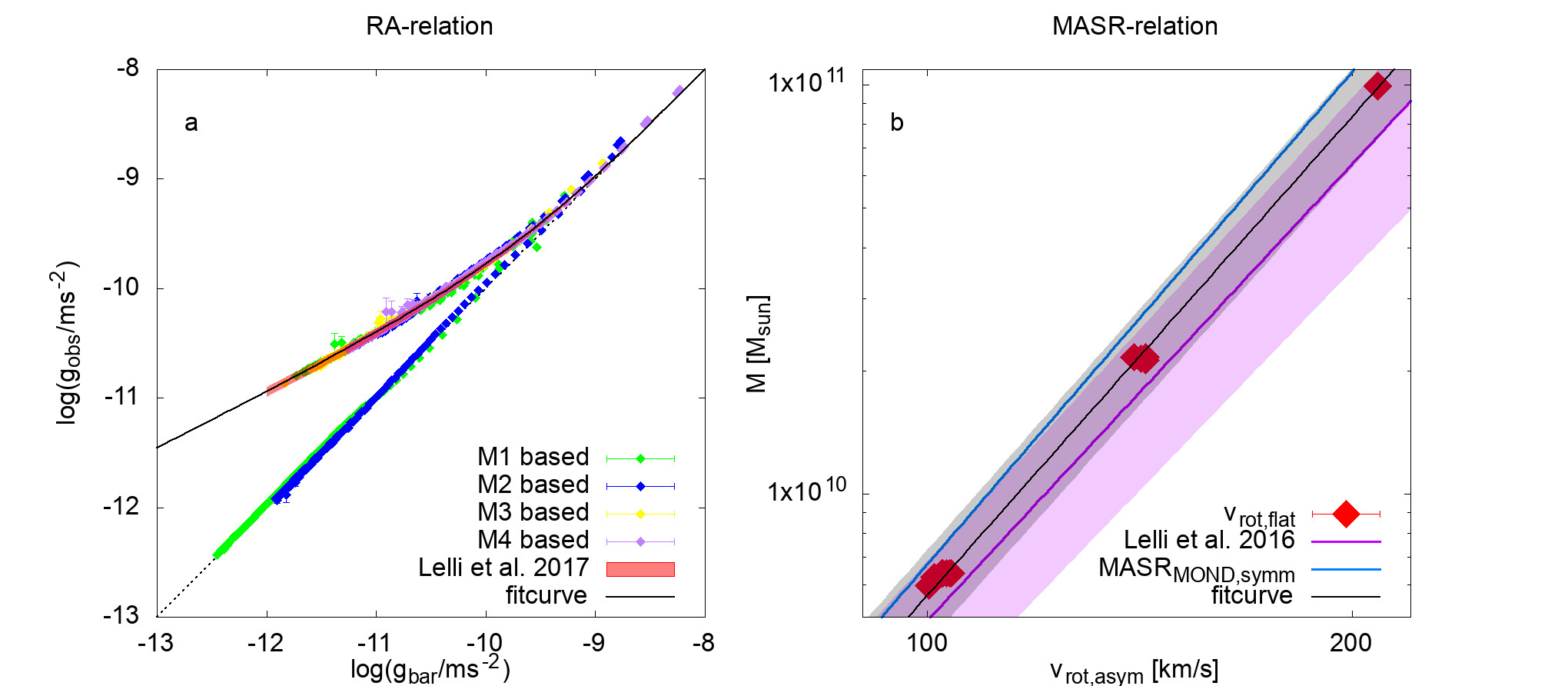}
\caption{Comparison of the simulations with observed galaxy-scaling relations. Panel a shows the radial-acceleration relation of the data obtained directly from the accelerations in the simulations and the observational data from \cite{Lelli2017} (red area), the best fit to the data from all models (black solid line) and the purely Newtonian line, where $g_\mathrm{obs}=g_\mathrm{bar}$ (black dotted line). The colours of the models are shown in the legend. Panel b shows the MASR (BTFR) of the models (red diamonds), the best fit (black solid line) with uncertainties (grey area), the MASR (BTFR) in MOND assuming spherical symmetry (blue line) and the observational data from \cite{Lelli2016a} (magenta area).}
\label{fig:RAR_BTFR}
\end{figure*}
To compare the simulations with observations, the same fit function was used and the best fit with uncertainties of the observational data is plotted together with the data from all models,
\begin{eqnarray}\label{eq:7}
g_\mathrm{obs}={\cal F} (g_\mathrm{bar})=\frac{g_\mathrm{bar}}{1-e^{-\sqrt{g_\mathrm{bar}/g_\dagger}}},
\end{eqnarray}
Eq. \ref{eq:7} here being Eq. 11 in \cite{Lelli2017}, where only one free parameter, $g_\dagger$, is available to fit the data. This parameter is in MOND equivalent to $a_0$ and ${\cal F}/g_\mathrm{bar}$ to the interpolation function $\nu$. The fit function could be precisely transformed into an interpolating function of the theory, which we could have implemented into PoR to ensure perfect agreement between the models and data. However, the simple interpolating function, which we chose to use in this work, is a good approximation to the Lelli et al. fit function \citep{Famaey&McGaugh2012}: we demonstrate this here once again.\\
\indent
The observational fit is shown as the red area in Fig. \ref{fig:RAR_BTFR} with $g_\dagger=1.20\pm0.02\times 10^{-10}\mathrm{ms}^{-2}=3.87\pm0.06\mathrm{pc}/\mathrm{Myr}^2$ and a systematic error of $0.24\times10^{-10}\mathrm{ms}^{-2}=0.77\mathrm{pc}/\mathrm{Myr}^2$. The best fit to the simulation data is the black line in panel a with $g_\dagger=1.217\pm0.006\times10^{-10}\mathrm{ms}^{-2}=3.92\pm0.02\mathrm{pc}/\mathrm{Myr}^2$, which is well within the uncertainties of the observational data. Note however that we do not have galaxies which are deeply in low-acceleration regime in their central parts, where the difference between the MDAR/RAR and QUMOND would be expected to be more important. The dashed black line shows $g_\mathrm{bar}=g_\mathrm{obs}$, which corresponds to a purely Newtonian gravitational acceleration even at the lowest accelerations. As can be seen, all Newtonian models lie on this line, which again tests the correct behaviour of the code.\\
\indent
Panel b shows the MASR (BTFR), where $v_\mathrm{rot,flat}$ is the flat rotation velocity of the models (red points), the blue line shows the theoretical MASR (BTFR) in MOND within the spherical approximation ($v_\mathrm{rot}=\left(GMa_0\right)^{1/4}$) with $a_0=1.12\times10^{-10}\mathrm{ms}^{-2}$, which is well known to deviate somewhat from the true MOND predicted MASR (BTFR), as we confirm here. This is simply due to the fact that a flattened mass distribution spins slightly faster than the equivalent spherical mass distribution \citep{StacyM2011reply} and that we use the flat rotation velocity and not the true asymptotic velocity. The black line shows the best fit to the model data with uncertainties shown as the grey area and the magenta area shows the observational data from \cite{Lelli2016a}. Again, the same fit function was used as in \cite{Lelli2016a} (the baryonic mass, $M_\mathrm{bar}$, is in units of $M_\odot$, $v_\mathrm{rot,flat}$ in units of $\mathrm{kms^{-1}}$ and $A$ is in units of $\mathrm{M_\odot(skm^{-1})^{\gamma}}$),
\begin{eqnarray}
\log_{10}(M_\mathrm{bar})=\gamma\times\log_{10}(v_\mathrm{rot,flat})+\log_{10}(A),
\end{eqnarray}
where $v_\mathrm{rot,flat}$ is the rotation velocity of the flat part of the rotation curve, $M_\mathrm{bar}$ the corresponding baryonic mass, $\log_{10}(A)$ the normalization and $\gamma$ the slope of the relation. Lelli et al. find $\gamma=3.71\pm0.08$ and $\log_{10}(A)=2.27\pm0.18$ and the best fit to the simulation data is $\gamma=3.97\pm0.04$ and $\log_{10}(A)=1.78\pm0.08$. The best fit including uncertainties lies within the observational constraints, while the theoretical MASR (BTFR) in MOND is also covered by the fit.\\
\indent
After analyzing both relations separately, we will focus on the difference between the inputted value of $a_0$, $a_0=1.12\times10^{-10}\mathrm{ms^{-2}}$ and the one from the RAR, $a_0=1.217\pm0.006\times10^{-10}\mathrm{ms}^{-2}$. There are several reasons to why this discrepancy may occur: first, the two interpolation functions used (the fit for the RAR and the one implemented in POR) can differ up to 5\% and also, we do not expect an exact MDAR/RAR in QUMOND, meaning that we also do not expect the best-fit value for $a_0$ to be exactly the same as the inputted one. Future work performed at significantly higher numerical resolution will allow us to revisit this problem.
\section{Discussion and future work}
In this work we carried out the first full hydrodynamical simulations in a Milgromian framework to study the formation and evolution of single galaxies. Despite using very simple initial conditions and a computationally-constrained resolution, our results are remarkably close to observations.
\begin{itemize}
\item
The work shows that late-type galaxies with exponentially decreasing surface mass density profiles, thin and thick disks, bulge and population II halos form naturally in the MOND framework.
\item
We compared simulations with simple and more complex baryonic physics to analyse if the additions lead to major differences between the models. Surprisingly, the properties of simulations with and without more complex baryonic physics do not change significantly. This result underlines the hypothesis that the evolution of galaxies seems to follow a strict and simple law, where complex but realistic feedback processes play a very minor role \citep{Kroupa2015}.
\item
We set up simulations with gas spheres having different initial rotation laws, which resulted in comparable flat rotation velocities but different spatial extents and therefore different density distributions.
\item
To check whether the occurrence of exponentially decreasing stellar surface mass density distributions is connected to our initial conditions, we computed  several models again with the Newtonian Poisson solver. Indeed only the Milgromian models show this feature, so the different gravitational theory and not the initial set up leads to an exponentially decreasing stellar surface mass density profile.
\item
We also showed that the formation of a single galaxy in isolation has its limits with our method. Above a certain mass and rotation velocity scale, the initial collapse produces satellites in addition to the main galaxy and, as mentioned in Sec. 4.4, this can also lead to the formation of a group of galaxies (Wittenburg et al. in prep.). Additionally, during the collapse of such models, a few proto-galaxies form, which merge due to gas-dissipative processes. The resulting surface mass density profiles deviate somewhat from the single exponential form (Eq. \ref{eq:sigma}).
\item
The specific stellar angular momentum is slightly too small for the models as a whole, which is connected to insignificant later accretion and the simple initial conditions and may also be connected to the detailed prescriptions used to form stars as mentioned in Sec. 4.7. The models presented here assume an extremely small gas density beyond the primordial cloud such that accretion onto the collapsed and evolving galaxy models is negligible. There is substantial observational evidence that galactic discs can grow around bulges (i.e. early-formed elliptical galaxies or red nuggets) through accretion \citep{Mancini2019,delaRosa2016}.
\end{itemize}
Overall our simulations suggest that major properties of galaxies, e.g. exponentially decreasing stellar surface mass density profiles, are rather a result of the dynamics produced by the gravitation potential, than the complexity of the baryonic processes in the Milgromian framework. It will be interesting to investigate whether the disk-internal matter and angular momentum redistribution mechanism suggested by \cite{Herpich2017} is at work here or a completely different one. Important to note here is that these findings are somewhat surprising, as the effectively stronger gravity in MOND could have led to compact, pressure supported systems that are well within the Newtonian regime due to a stronger collapse. We also began running simulations with initial fluctuation of the velocity- and density-distribution of up to 20\% of the unperturbed value. The first impression is that the properties of the resulting galaxies are not changed significantly.\\
\indent
The recent observation of the gas kinematics of galaxies at a redshift of $\mathrm{z}=6.8$ revealed that these young objects are rotation-dominated and not as turbulent as suggested before \citep{Smit2018}. This supports our findings that baryonic processes like supernova explosions have a minor effect on the dynamics and morphology of a galaxy as a whole (even for the earliest population of galaxies).\\
\indent
We want to stress that all simulations done in the MOND framework show a tight MDAR/RAR, which compares with the observed MDAR/RAR and discs compatible with the size-mass relation. Hence full hydrodynamical simulations in MOND produce objects that show dynamics comparable to real galaxies. The models however neither match the observed present-day SFHs nor the observed gas consumption timescales and have systematically slightly too small specific stellar angular momentum, but they do reproduce the slope of the Fall-relation of observed galaxies and show that more massive galaxies have a smaller gas fraction, while also reproducing qualitatively the same picture in the KS-diagram as real galaxies and initially match the main sequence of star forming galaxies.\\
\indent
Although the simulations done in this work are surprisingly close to real galaxies in many aspects, we only made the first steps in analysing the formation and evolution of galaxies in the MOND framework. Therefore more simulations have to be done with higher resolution, initially turbulent gas clouds, gas accretion, an external field and more baryonic processes (e.g. UV background radiation) in order to compare more observational results with our simulations. The calculations presented here show that we can hope to constrain the initial conditions needed in a MONDian cosmology, which in turn constrain such a cosmology, and these conditions will get more precise with more detailed simulations.\\
\indent
What is more, the formation of massive early-type galaxies is also not understood today, as only $3-4$ percent of the observed galaxies with a baryonic mass larger than about $10^{10}\mathrm{M_\odot}$ are ellipticals \citep{Delgado-Serrano2010}. Amongst others, this problem will be addressed in an upcoming publication (Wittenburg et al. in prep.), where we will show how early-type galaxies might form in the Milgromian framework. Ellipticals could be results of rare mergers, or collisions that strip the outer discs. But we can already conclude that elliptical galaxies are an exception in MOND, since every collapsing proto-galactic gas cloud has angular momentum. The typical outcome of galaxy formation in MOND is therefore a rotationally supported disk galaxy, as is also observed in the real universe \citep{Delgado-Serrano2010}.

\acknowledgments
BF acknowledges funding from the Agence Nationale de la Recherche (ANR project ANR-18-CE31-0006) and from the European Research Council (ERC) under the European Union’s Horizon 2020 research and innovation programme (grant agreement No. 834148).

\bibliographystyle{aasjournal}

\begin{thebibliography}{}
\expandafter\ifx\csname natexlab\endcsname\relax\def\natexlab#1{#1}\fi
\providecommand{\url}[1]{\href{#1}{#1}}
\providecommand{\dodoi}[1]{doi:~\href{http://doi.org/#1}{\nolinkurl{#1}}}
\providecommand{\doeprint}[1]{\href{http://ascl.net/#1}{\nolinkurl{http://ascl.net/#1}}}
\providecommand{\doarXiv}[1]{\href{https://arxiv.org/abs/#1}{\nolinkurl{https://arxiv.org/abs/#1}}}

\bibitem[{{Banik} {et~al.}(2018{\natexlab{a}}){Banik}, {Milgrom}, \&
  {Zhao}}]{Banik2018a}
{Banik}, I., {Milgrom}, M., \& {Zhao}, H. 2018{\natexlab{a}}, ArXiv e-prints.
\newblock \doarXiv{1808.10545}

\bibitem[{{Banik} {et~al.}(2018{\natexlab{b}}){Banik}, {O'Ryan}, \&
  {Zhao}}]{Banik2018}
{Banik}, I., {O'Ryan}, D., \& {Zhao}, H. 2018{\natexlab{b}}, \mnras, 477, 4768,
  \dodoi{10.1093/mnras/sty919}

\bibitem[{{Banik} \& {Zhao}(2018)}]{BanickZhao2018}
{Banik}, I., \& {Zhao}, H. 2018, \mnras, 473, 419,
  \dodoi{10.1093/mnras/stx2350}

\bibitem[{{Barkana} \& {Loeb}(2001)}]{Barkana&Loeb2001}
{Barkana}, R., \& {Loeb}, A. 2001, \physrep, 349, 125,
  \dodoi{10.1016/S0370-1573(01)00019-9}

\bibitem[{{Bekenstein} \& {Milgrom}(1984)}]{Bekenstein&Milgrom1984}
{Bekenstein}, J., \& {Milgrom}, M. 1984, \apj, 286, 7, \dodoi{10.1086/162570}

\bibitem[{{Bertone} \& {Tait}(2018)}]{Bertone2018}
{Bertone}, G., \& {Tait}, T.~M.~P. 2018, \nat, 562, 51,
  \dodoi{10.1038/s41586-018-0542-z}

\bibitem[{{Bigiel} {et~al.}(2008){Bigiel}, {Leroy}, {Walter}, {Brinks}, {de
  Blok}, {Madore}, \& {Thornley}}]{Bigiel2008}
{Bigiel}, F., {Leroy}, A., {Walter}, F., {et~al.} 2008, \aj, 136, 2846,
  \dodoi{10.1088/0004-6256/136/6/2846}

\bibitem[{{B{\'{\i}}lek} {et~al.}(2018){B{\'{\i}}lek}, {Thies}, {Kroupa}, \&
  {Famaey}}]{Bilek2018}
{B{\'{\i}}lek}, M., {Thies}, I., {Kroupa}, P., \& {Famaey}, B. 2018, \aap, 614,
  A59, \dodoi{10.1051/0004-6361/201731939}

\bibitem[{{Bleuler} \& {Teyssier}(2014)}]{Teyssier2014sink}
{Bleuler}, A., \& {Teyssier}, R. 2014, \mnras, 445, 4015,
  \dodoi{10.1093/mnras/stu2005}

\bibitem[{{Brada} \& {Milgrom}(1995)}]{Brada1995}
{Brada}, R., \& {Milgrom}, M. 1995, \mnras, 276, 453,
  \dodoi{10.1093/mnras/276.2.453}

\bibitem[{{Brada} \& {Milgrom}(1999)}]{Brada1999}
---. 1999, \apj, 519, 590, \dodoi{10.1086/307402}

\bibitem[{{Candlish} {et~al.}(2015){Candlish}, {Smith}, \&
  {Fellhauer}}]{Candlish2015}
{Candlish}, G.~N., {Smith}, R., \& {Fellhauer}, M. 2015, \mnras, 446, 1060,
  \dodoi{10.1093/mnras/stu2158}

\bibitem[{{Catelan} \& {Theuns}(1996)}]{Catelan1996}
{Catelan}, P., \& {Theuns}, T. 1996, \mnras, 282, 436,
  \dodoi{10.1093/mnras/282.2.436}

\bibitem[{{Combes}(2014)}]{Combes2014}
{Combes}, F. 2014, \aap, 571, A82, \dodoi{10.1051/0004-6361/201424990}

\bibitem[{{Courty} \& {Alimi}(2004)}]{Courty+Alimi2004}
{Courty}, S., \& {Alimi}, J.~M. 2004, \aap, 416, 875,
  \dodoi{10.1051/0004-6361:20031736}

\bibitem[{{Dalcanton} {et~al.}(1997){Dalcanton}, {Spergel}, \&
  {Summers}}]{DalcatonSpergel&Summers1997}
{Dalcanton}, J.~J., {Spergel}, D.~N., \& {Summers}, F.~J. 1997, \apj, 482, 659,
  \dodoi{10.1086/304182}

\bibitem[{{de Blok} \& {McGaugh}(1997)}]{McGaugh&deBlock1997}
{de Blok}, W.~J.~G., \& {McGaugh}, S.~S. 1997, \mnras, 290, 533,
  \dodoi{10.1093/mnras/290.3.533}

\bibitem[{{de la Rosa} {et~al.}(2016){de la Rosa}, {La Barbera}, {Ferreras},
  {S{\'a}nchez Almeida}, {Dalla Vecchia}, {Mart{\'\i}nez-Valpuesta}, \&
  {Stringer}}]{delaRosa2016}
{de la Rosa}, I.~G., {La Barbera}, F., {Ferreras}, I., {et~al.} 2016, \mnras,
  457, 1916, \dodoi{10.1093/mnras/stw130}

\bibitem[{{Delgado-Serrano} {et~al.}(2010){Delgado-Serrano}, {Hammer}, {Yang},
  {Puech}, {Flores}, \& {Rodrigues}}]{Delgado-Serrano2010}
{Delgado-Serrano}, R., {Hammer}, F., {Yang}, Y.~B., {et~al.} 2010, \aap, 509,
  A78, \dodoi{10.1051/0004-6361/200912704}

\bibitem[{{Desmond}(2017{\natexlab{a}})}]{Desmond2017a}
{Desmond}, H. 2017{\natexlab{a}}, \mnras, 464, 4160,
  \dodoi{10.1093/mnras/stw2571}

\bibitem[{{Desmond}(2017{\natexlab{b}})}]{Desmond2017b}
---. 2017{\natexlab{b}}, \mnras, 472, L35, \dodoi{10.1093/mnrasl/slx134}

\bibitem[{{Di Cintio} \& {Lelli}(2016)}]{Dicintio2016}
{Di Cintio}, A., \& {Lelli}, F. 2016, \mnras, 456, L127,
  \dodoi{10.1093/mnrasl/slv185}

\bibitem[{{Disney} {et~al.}(2008){Disney}, {Romano}, {Garcia-Appadoo}, {West},
  {Dalcanton}, \& {Cortese}}]{Disney2008}
{Disney}, M.~J., {Romano}, J.~D., {Garcia-Appadoo}, D.~A., {et~al.} 2008, \nat,
  455, 1082, \dodoi{10.1038/nature07366}

\bibitem[{{Dubois} \& {Teyssier}(2008)}]{Dubois+Teyssier2008}
{Dubois}, Y., \& {Teyssier}, R. 2008, \aap, 477, 79,
  \dodoi{10.1051/0004-6361:20078326}

\bibitem[{{Dutton}(2009)}]{Dutton2009}
{Dutton}, A.~A. 2009, \mnras, 396, 121,
  \dodoi{10.1111/j.1365-2966.2009.14741.x}

\bibitem[{{Efstathiou} \& {Jones}(1979)}]{Efstathiou1979}
{Efstathiou}, G., \& {Jones}, B.~J.~T. 1979, \mnras, 186, 133,
  \dodoi{10.1093/mnras/186.2.133}

\bibitem[{{Einstein}(1916)}]{Einstein1916}
{Einstein}, A. 1916, Annalen der Physik, 354, 769,
  \dodoi{10.1002/andp.19163540702}

\bibitem[{{Faber} \& {Jackson}(1976)}]{Faber&Jackson1976}
{Faber}, S.~M., \& {Jackson}, R.~E. 1976, \apj, 204, 668,
  \dodoi{10.1086/154215}

\bibitem[{{Fall} \& {Efstathiou}(1980)}]{Fall&Efstathiou1980}
{Fall}, S.~M., \& {Efstathiou}, G. 1980, \mnras, 193, 189,
  \dodoi{10.1093/mnras/193.2.189}

\bibitem[{{Famaey} \& {McGaugh}(2012)}]{Famaey&McGaugh2012}
{Famaey}, B., \& {McGaugh}, S.~S. 2012, Living Reviews in Relativity, 15, 10,
  \dodoi{10.12942/lrr-2012-10}

\bibitem[{{Freeman}(1970)}]{Freeman1970}
{Freeman}, K.~C. 1970, \apj, 160, 811, \dodoi{10.1086/150474}

\bibitem[{{Herpich} {et~al.}(2017){Herpich}, {Tremaine}, \&
  {Rix}}]{Herpich2017}
{Herpich}, J., {Tremaine}, S., \& {Rix}, H.-W. 2017, \mnras, 467, 5022,
  \dodoi{10.1093/mnras/stx352}

\bibitem[{{Hubble}(1926)}]{Hubble1926}
{Hubble}, E.~P. 1926, \apj, 64, 321, \dodoi{10.1086/143018}

\bibitem[{{Ibata} {et~al.}(2013){Ibata}, {Lewis}, {Conn}, {Irwin},
  {McConnachie}, {Chapman}, {Collins}, {Fardal}, {Ferguson}, {Ibata}, {Mackey},
  {Martin}, {Navarro}, {Rich}, {Valls-Gabaud}, \& {Widrow}}]{Ibata2013}
{Ibata}, R.~A., {Lewis}, G.~F., {Conn}, A.~R., {et~al.} 2013, \nat, 493, 62,
  \dodoi{10.1038/nature11717}

\bibitem[{{Javanmardi} \& {Kroupa}(2020)}]{Javanmardi2020}
{Javanmardi}, B., \& {Kroupa}, P. 2020, \mnras, 493, L44,
  \dodoi{10.1093/mnrasl/slaa001}

\bibitem[{{Je{\v{r}}{\'a}bkov{\'a}} {et~al.}(2018){Je{\v{r}}{\'a}bkov{\'a}},
  {Hasani Zonoozi}, {Kroupa}, {Beccari}, {Yan}, {Vazdekis}, \&
  {Zhang}}]{Jerabkova2018}
{Je{\v{r}}{\'a}bkov{\'a}}, T., {Hasani Zonoozi}, A., {Kroupa}, P., {et~al.}
  2018, \aap, 620, A39, \dodoi{10.1051/0004-6361/201833055}

\bibitem[{{Keller} \& {Wadsley}(2017)}]{Keller2017}
{Keller}, B.~W., \& {Wadsley}, J.~W. 2017, \apjl, 835, L17,
  \dodoi{10.3847/2041-8213/835/1/L17}

\bibitem[{{Kennicutt}(1998)}]{Kennicutt1998}
{Kennicutt}, Robert~C., J. 1998, \apj, 498, 541, \dodoi{10.1086/305588}

\bibitem[{{Kroupa}(2012)}]{Kroupa2012}
{Kroupa}, P. 2012, Publications of the Astronomical Society of Australia, 29,
  395, \dodoi{10.1071/AS12005}

\bibitem[{{Kroupa}(2015)}]{Kroupa2015}
---. 2015, Canadian Journal of Physics, 93, 169, \dodoi{10.1139/cjp-2014-0179}

\bibitem[{{Kroupa} {et~al.}(2005){Kroupa}, {Theis}, \&
  {Boily}}]{Kroupaetal2005}
{Kroupa}, P., {Theis}, C., \& {Boily}, C.~M. 2005, \aap, 431, 517,
  \dodoi{10.1051/0004-6361:20041122}

\bibitem[{{Kroupa} {et~al.}(2013){Kroupa}, {Weidner}, {Pflamm-Altenburg},
  {Thies}, {Dabringhausen}, {Marks}, \& {Maschberger}}]{Kroupa2013}
{Kroupa}, P., {Weidner}, C., {Pflamm-Altenburg}, J., {et~al.} 2013, {The
  Stellar and Sub-Stellar Initial Mass Function of Simple and Composite
  Populations}, ed. T.~D. {Oswalt} \& G.~{Gilmore}, Vol.~5, 115,
  \dodoi{10.1007/978-94-007-5612-0_4}

\bibitem[{{Kroupa} {et~al.}(2010){Kroupa}, {Famaey}, {de Boer},
  {Dabringhausen}, {Pawlowski}, {Boily}, {Jerjen}, {Forbes}, {Hensler}, \&
  {Metz}}]{Kroupa2010}
{Kroupa}, P., {Famaey}, B., {de Boer}, K.~S., {et~al.} 2010, \aap, 523, A32,
  \dodoi{10.1051/0004-6361/201014892}

\bibitem[{{Lange} {et~al.}(2015){Lange}, {Driver}, {Robotham}, {Kelvin},
  {Graham}, {Alpaslan}, {Andrews}, {Baldry}, {Bamford}, {Bland-Hawthorn},
  {Brough}, {Cluver}, {Conselice}, {Davies}, {Haeussler}, {Konstantopoulos},
  {Loveday}, {Moffett}, {Norberg}, {Phillipps}, {Taylor},
  {L{\'o}pez-S{\'a}nchez}, \& {Wilkins}}]{Lange2015}
{Lange}, R., {Driver}, S.~P., {Robotham}, A. S.~G., {et~al.} 2015, \mnras, 447,
  2603, \dodoi{10.1093/mnras/stu2467}

\bibitem[{{Larson}(1981)}]{Larson1981}
{Larson}, R.~B. 1981, \mnras, 194, 809, \dodoi{10.1093/mnras/194.4.809}

\bibitem[{{Lelli} {et~al.}(2013){Lelli}, {Fraternali}, \&
  {Verheijen}}]{Lelli2013}
{Lelli}, F., {Fraternali}, F., \& {Verheijen}, M. 2013, \mnras, 433, L30,
  \dodoi{10.1093/mnrasl/slt053}

\bibitem[{{Lelli} {et~al.}(2016{\natexlab{a}}){Lelli}, {McGaugh}, \&
  {Schombert}}]{Lelli2016a}
{Lelli}, F., {McGaugh}, S.~S., \& {Schombert}, J.~M. 2016{\natexlab{a}}, \apjl,
  816, L14, \dodoi{10.3847/2041-8205/816/1/L14}

\bibitem[{{Lelli} {et~al.}(2016{\natexlab{b}}){Lelli}, {McGaugh}, {Schombert},
  \& {Pawlowski}}]{Lelli2016c}
{Lelli}, F., {McGaugh}, S.~S., {Schombert}, J.~M., \& {Pawlowski}, M.~S.
  2016{\natexlab{b}}, \apjl, 827, L19, \dodoi{10.3847/2041-8205/827/1/L19}

\bibitem[{{Lelli} {et~al.}(2017){Lelli}, {McGaugh}, {Schombert}, \&
  {Pawlowski}}]{Lelli2017}
---. 2017, \apj, 836, 152, \dodoi{10.3847/1538-4357/836/2/152}

\bibitem[{{Llinares} {et~al.}(2008){Llinares}, {Knebe}, \&
  {Zhao}}]{Llinares2008}
{Llinares}, C., {Knebe}, A., \& {Zhao}, H. 2008, \mnras, 391, 1778,
  \dodoi{10.1111/j.1365-2966.2008.13961.x}

\bibitem[{{Ludlow} {et~al.}(2017){Ludlow}, {Ben{\'{\i}}tez-Llambay},
  {Schaller}, {Theuns}, {Frenk}, {Bower}, {Schaye}, {Crain}, {Navarro},
  {Fattahi}, \& {Oman}}]{Ludlow2017}
{Ludlow}, A.~D., {Ben{\'{\i}}tez-Llambay}, A., {Schaller}, M., {et~al.} 2017,
  Physical Review Letters, 118, 161103, \dodoi{10.1103/PhysRevLett.118.161103}

\bibitem[{{L{\"u}ghausen} {et~al.}(2015){L{\"u}ghausen}, {Famaey}, \&
  {Kroupa}}]{Lueghausen2015}
{L{\"u}ghausen}, F., {Famaey}, B., \& {Kroupa}, P. 2015, Canadian Journal of
  Physics, 93, 232, \dodoi{10.1139/cjp-2014-0168}

\bibitem[{{Mancini} {et~al.}(2019){Mancini}, {Daddi}, {Juneau}, {Renzini},
  {Rodighiero}, {Cappellari}, {Rodr{\'\i}guez-Mu{\~n}oz}, {Liu}, {Pannella},
  {Baronchelli}, {Franceschini}, {Bergamini}, {D'Eugenio}, \&
  {Puglisi}}]{Mancini2019}
{Mancini}, C., {Daddi}, E., {Juneau}, S., {et~al.} 2019, \mnras, 489, 1265,
  \dodoi{10.1093/mnras/stz2130}

\bibitem[{{McGaugh}(2011)}]{StacyM2011reply}
{McGaugh}, S. 2011, ArXiv e-prints, arXiv:1109.1599.
\newblock \doarXiv{1109.1599}

\bibitem[{{McGaugh}(2014)}]{McGaugh2014}
---. 2014, Galaxies, 2, 601, \dodoi{10.3390/galaxies2040601}

\bibitem[{{McGaugh}(2004)}]{McGaugh2004}
{McGaugh}, S.~S. 2004, \apj, 609, 652, \dodoi{10.1086/421338}

\bibitem[{{McGaugh}(2005)}]{McGaugh2005}
---. 2005, \apj, 632, 859, \dodoi{10.1086/432968}

\bibitem[{{McGaugh}(2012)}]{McGaugh2012}
---. 2012, \aj, 143, 40, \dodoi{10.1088/0004-6256/143/2/40}

\bibitem[{{McGaugh} {et~al.}(2000){McGaugh}, {Schombert}, {Bothun}, \& {de
  Blok}}]{McGaugh2000}
{McGaugh}, S.~S., {Schombert}, J.~M., {Bothun}, G.~D., \& {de Blok}, W.~J.~G.
  2000, \apjl, 533, L99, \dodoi{10.1086/312628}

\bibitem[{{Metz} {et~al.}(2007){Metz}, {Kroupa}, \& {Jerjen}}]{Metz2007}
{Metz}, M., {Kroupa}, P., \& {Jerjen}, H. 2007, \mnras, 374, 1125,
  \dodoi{10.1111/j.1365-2966.2006.11228.x}

\bibitem[{{Milgrom}(1983)}]{Milgrom1983}
{Milgrom}, M. 1983, \apj, 270, 365, \dodoi{10.1086/161130}

\bibitem[{{Milgrom}(1999)}]{Milgrom1999}
---. 1999, Physics Letters A, 253, 273, \dodoi{10.1016/S0375-9601(99)00077-8}

\bibitem[{{Milgrom}(2009)}]{Milgrom2009}
---. 2009, \apj, 698, 1630, \dodoi{10.1088/0004-637X/698/2/1630}

\bibitem[{{Milgrom}(2010)}]{Milgrom2010}
---. 2010, \mnras, 403, 886, \dodoi{10.1111/j.1365-2966.2009.16184.x}

\bibitem[{Milgrom(2014)}]{Scholarpedia2014Moti}
Milgrom, M. 2014, Scholarpedia, 9, 31410, \dodoi{10.4249/scholarpedia.31410}

\bibitem[{{Mor} {et~al.}(2018){Mor}, {Robin}, {Figueras}, \&
  {Antoja}}]{Mor2018}
{Mor}, R., {Robin}, A.~C., {Figueras}, F., \& {Antoja}, T. 2018, \aap, 620,
  A79, \dodoi{10.1051/0004-6361/201833501}

\bibitem[{{M{\"u}ller} {et~al.}(2018){M{\"u}ller}, {Pawlowski}, {Jerjen}, \&
  {Lelli}}]{Muller2018}
{M{\"u}ller}, O., {Pawlowski}, M.~S., {Jerjen}, H., \& {Lelli}, F. 2018,
  Science, 359, 534, \dodoi{10.1126/science.aao1858}

\bibitem[{{Navarro} {et~al.}(2017){Navarro}, {Ben{\'{\i}}tez-Llambay},
  {Fattahi}, {Frenk}, {Ludlow}, {Oman}, {Schaller}, \& {Theuns}}]{Navarro2017}
{Navarro}, J.~F., {Ben{\'{\i}}tez-Llambay}, A., {Fattahi}, A., {et~al.} 2017,
  \mnras, 471, 1841, \dodoi{10.1093/mnras/stx1705}

\bibitem[{{Newton}(1687)}]{Newton1687}
{Newton}, I. 1687, {Philosophiae Naturalis Principia Mathematica. Auctore Js.
  Newton}, \dodoi{10.3931/e-rara-440}

\bibitem[{{Oh} \& {Kroupa}(2016)}]{Oh2016}
{Oh}, S., \& {Kroupa}, P. 2016, \aap, 590, A107,
  \dodoi{10.1051/0004-6361/201628233}

\bibitem[{{Okazaki} \& {Taniguchi}(2000)}]{Okazaki2000}
{Okazaki}, T., \& {Taniguchi}, Y. 2000, \apj, 543, 149, \dodoi{10.1086/317109}

\bibitem[{{Pawlowski}(2018)}]{Pawlowski2018}
{Pawlowski}, M.~S. 2018, Modern Physics Letters A, 33, 1830004,
  \dodoi{10.1142/S0217732318300045}

\bibitem[{{Pawlowski} \& {Kroupa}(2020)}]{Pawlowski2019}
{Pawlowski}, M.~S., \& {Kroupa}, P. 2020, \mnras, 491, 3042,
  \dodoi{10.1093/mnras/stz3163}

\bibitem[{{Pawlowski} {et~al.}(2012){Pawlowski}, {Pflamm-Altenburg}, \&
  {Kroupa}}]{Pawlowski2012}
{Pawlowski}, M.~S., {Pflamm-Altenburg}, J., \& {Kroupa}, P. 2012, \mnras, 423,
  1109, \dodoi{10.1111/j.1365-2966.2012.20937.x}

\bibitem[{{Pflamm-Altenburg} \& {Kroupa}(2009)}]{JPFA2009}
{Pflamm-Altenburg}, J., \& {Kroupa}, P. 2009, \apj, 706, 516,
  \dodoi{10.1088/0004-637X/706/1/516}

\bibitem[{{Posti} {et~al.}(2018){Posti}, {Fraternali}, {Di Teodoro}, \&
  {Pezzulli}}]{Posti2018}
{Posti}, L., {Fraternali}, F., {Di Teodoro}, E.~M., \& {Pezzulli}, G. 2018,
  \aap, 612, L6, \dodoi{10.1051/0004-6361/201833091}

\bibitem[{{Renaud} {et~al.}(2016){Renaud}, {Famaey}, \& {Kroupa}}]{Renaud2016}
{Renaud}, F., {Famaey}, B., \& {Kroupa}, P. 2016, \mnras, 463, 3637,
  \dodoi{10.1093/mnras/stw2331}

\bibitem[{{Rosdahl} {et~al.}(2013){Rosdahl}, {Blaizot}, {Aubert}, {Stranex}, \&
  {Teyssier}}]{Rosdahl2013rt}
{Rosdahl}, J., {Blaizot}, J., {Aubert}, D., {Stranex}, T., \& {Teyssier}, R.
  2013, \mnras, 436, 2188, \dodoi{10.1093/mnras/stt1722}

\bibitem[{{Sancisi}(2004)}]{Sancisi2004}
{Sancisi}, R. 2004, in IAU Symposium, Vol. 220, Dark Matter in Galaxies, ed.
  S.~{Ryder}, D.~{Pisano}, M.~{Walker}, \& K.~{Freeman}, 233.
\newblock \doarXiv{astro-ph/0311348}

\bibitem[{{Sanders}(1990)}]{Sanders1990}
{Sanders}, R.~H. 1990, \aapr, 2, 1, \dodoi{10.1007/BF00873540}

\bibitem[{{Sanders}(1998)}]{Sanders1998}
---. 1998, \mnras, 296, 1009, \dodoi{10.1046/j.1365-8711.1998.01459.x}

\bibitem[{{Serra} {et~al.}(2016){Serra}, {Oosterloo}, {Cappellari}, {den
  Heijer}, \& {J{\'o}zsa}}]{Serra2016}
{Serra}, P., {Oosterloo}, T., {Cappellari}, M., {den Heijer}, M., \&
  {J{\'o}zsa}, G.~I.~G. 2016, \mnras, 460, 1382, \dodoi{10.1093/mnras/stw1010}

\bibitem[{{Smit}(2018)}]{Smit2018}
{Smit}, R. 2018, in American Astronomical Society Meeting Abstracts, Vol. 231,
  American Astronomical Society Meeting Abstracts \#231, \#454.03

\bibitem[{{Smolin}(2017)}]{Smolin2017}
{Smolin}, L. 2017, \prd, 96, 083523, \dodoi{10.1103/PhysRevD.96.083523}

\bibitem[{{Speagle} {et~al.}(2014){Speagle}, {Steinhardt}, {Capak}, \&
  {Silverman}}]{Speagle2014}
{Speagle}, J.~S., {Steinhardt}, C.~L., {Capak}, P.~L., \& {Silverman}, J.~D.
  2014, \apjs, 214, 15, \dodoi{10.1088/0067-0049/214/2/15}

\bibitem[{{Struck} \& {Elmegreen}(2017)}]{Struck&Elmegreen2017}
{Struck}, C., \& {Elmegreen}, B.~G. 2017, \mnras, 469, 1157,
  \dodoi{10.1093/mnras/stx918}

\bibitem[{{Struck} \& {Elmegreen}(2018)}]{Struck&Elmegreen2018}
---. 2018, \apjl, 868, L15, \dodoi{10.3847/2041-8213/aaedb4}

\bibitem[{{Sutherland} \& {Dopita}(1993)}]{Sutherland&Dopita1993}
{Sutherland}, R.~S., \& {Dopita}, M.~A. 1993, \apjs, 88, 253,
  \dodoi{10.1086/191823}

\bibitem[{{Tamburri} {et~al.}(2014){Tamburri}, {Saracco}, {Longhetti},
  {Gargiulo}, {Lonoce}, \& {Ciocca}}]{Tamburri2014}
{Tamburri}, S., {Saracco}, P., {Longhetti}, M., {et~al.} 2014, \aap, 570, A102,
  \dodoi{10.1051/0004-6361/201424040}

\bibitem[{{Teyssier}(2002)}]{Teyssier2002}
{Teyssier}, R. 2002, \aap, 385, 337, \dodoi{10.1051/0004-6361:20011817}

\bibitem[{{Teyssier} {et~al.}(2013){Teyssier}, {Pontzen}, {Dubois}, \&
  {Read}}]{Teyssier2013}
{Teyssier}, R., {Pontzen}, A., {Dubois}, Y., \& {Read}, J.~I. 2013, \mnras,
  429, 3068, \dodoi{10.1093/mnras/sts563}

\bibitem[{{Thomas} {et~al.}(2017){Thomas}, {Famaey}, {Ibata}, {L{\"u}ghausen},
  \& {Kroupa}}]{Guillaume2017}
{Thomas}, G.~F., {Famaey}, B., {Ibata}, R., {L{\"u}ghausen}, F., \& {Kroupa},
  P. 2017, \aap, 603, A65, \dodoi{10.1051/0004-6361/201730531}

\bibitem[{{Thomas} {et~al.}(2018){Thomas}, {Famaey}, {Ibata}, {Renaud},
  {Martin}, \& {Kroupa}}]{Thomas2018}
{Thomas}, G.~F., {Famaey}, B., {Ibata}, R., {et~al.} 2018, \aap, 609, A44,
  \dodoi{10.1051/0004-6361/201731609}

\bibitem[{{Tiret} \& {Combes}(2007)}]{Tiret2007}
{Tiret}, O., \& {Combes}, F. 2007, \aap, 464, 517,
  \dodoi{10.1051/0004-6361:20066446}

\bibitem[{{Tiret} \& {Combes}(2008)}]{Tiret2008}
---. 2008, \aap, 483, 719, \dodoi{10.1051/0004-6361:200809357}

\bibitem[{{Truelove} {et~al.}(1998){Truelove}, {Klein}, {McKee}, {Holliman},
  {Howell}, {Greenough}, \& {Woods}}]{Truelove1998}
{Truelove}, J.~K., {Klein}, R.~I., {McKee}, C.~F., {et~al.} 1998, \apj, 495,
  821, \dodoi{10.1086/305329}

\bibitem[{{Tully} \& {Verheijen}(1997)}]{Tully&Verheijen1997}
{Tully}, R.~B., \& {Verheijen}, M.~A.~W. 1997, \apj, 484, 145,
  \dodoi{10.1086/304318}

\bibitem[{{van Albada} \& {Sancisi}(1986)}]{vanAlba&Sancisi1986}
{van Albada}, T.~S., \& {Sancisi}, R. 1986, Philosophical Transactions of the
  Royal Society of London Series A, 320, 447, \dodoi{10.1098/rsta.1986.0128}

\bibitem[{{Voglis}(1994)}]{Voglis1994}
{Voglis}, N. 1994, in Lecture Notes in Physics, Berlin Springer Verlag, Vol.
  433, Galactic Dynamics and N-Body Simulations, ed. G.~{Contopoulos}, N.~K.
  {Spyrou}, \& L.~{Vlahos}, 365--417, \dodoi{10.1007/3-540-57983-4_24}

\bibitem[{{Wesson}(1985)}]{Wesson1985}
{Wesson}, P.~S. 1985, \aap, 151, 105

\bibitem[{{Wu} \& {Kroupa}(2015)}]{Wu+Kroupa2015}
{Wu}, X., \& {Kroupa}, P. 2015, \mnras, 446, 330, \dodoi{10.1093/mnras/stu2099}

\end{thebibliography}

\appendix

\begin{table*}[h]
\caption{Fit parameters for Eq. \ref{eq:sigma} for All Models. Missing entries correspond to nonexponential behavior.}
\centering
\begin{tabular}[c]{l c c c c}
\hline
Model & $n_\mathrm{stars} (\mathrm{M_\odot pc^{-2}})$ & $n_\mathrm{gas} (\mathrm{M_\odot pc^{-2}})$ & $r_\mathrm{e,stars} (\mathrm{kpc})$ & $r_\mathrm{e,gas} (\mathrm{kpc})$ \\
Name/No. & & & & \\
\hline
M1/1 & $168.61\pm9.43$ & $12.47\pm0.93$ & $1.61\pm0.04$ & $5.37\pm0.10$ \\
M1sn/2 & $119.06\pm7.89$ & $8.32\pm0.40$ & $1.82\pm0.06$ & $6.28\pm0.09$ \\
M1N/3 & - & $3.36\pm0.12$ & - & $9.62\pm0.11$ \\
M1const/4 & $283.35\pm7.53$ & $2.05\pm0.10$ & $1.19\pm0.01$ & $6.44\pm0.07$ \\
M1Zpoor/5 & $169.50\pm12.89$ & $8.11\pm0.64$ & $1.57\pm0.06$ & $5.77\pm0.11$ \\
M1Zpoorsn/6 & $148.43\pm7.87$ & $8.16\pm0.66$ & $1.64\pm0.04$ & $6.32\pm0.14$ \\
M1l11/7 & $225.42\pm18.66$ & $8.95\pm0.81$ & $1.43\pm0.06$ & $5.73\pm0.13$ \\
M1l13/8 & $136.63\pm8.92$ & $6.50\pm0.58$ & $1.81\pm0.06$ & $6.04\pm0.14$ \\
M2/9 & $845.68\pm16.50$ & $3.32\pm0.31$ & $1.35\pm0.01$ & $7.24\pm0.16$ \\
M2sn/10 & $822.86\pm17.59$ & $10.55\pm0.66$ & $1.35\pm0.01$ & $5.80\pm0.07$ \\
M2N/11 & - & $2.30\pm0.12$ & - & $8.55\pm0.13$ \\
M3/12 & $443.41\pm10.04$ & $4.31\pm0.23$ & $1.98\pm0.02$ & $7.79\pm0.10$ \\
M3sn/13 & $323.24\pm16.04$ & $24.02\pm1.34$ & $2.23\pm0.06$ & $5.72\pm0.06$ \\
M4/14 & - & $20.39\pm1.22$ & - & $7.15\pm0.09$ \\
M4sn/15 & - & $32.27\pm1.54$ & - & $6.40\pm0.06$ \\
\hline
\end{tabular}
\label{tab:4}
\end{table*}

\begin{table*}[h]
\caption{Fit parameters for Eq. \ref{SFHfit} for All Models. The index 1 corresponds to the first exponential part (green line) and 2 to the second part (blue line). The second part for model M1N and M1l11 is constant, therefore fit not with Eq. \ref{SFHfit} but with a constant, $c$. $e_2$ represents $c$ in this case.}
\centering
\begin{tabular}[c]{l c c c c}
\hline
Model & $e_1 (\mathrm{M_\odot Myr^{-1}})$ & $f_1 (\mathrm{Gyr})$ & $e_2 (\mathrm{M_\odot Myr^{-1}})$ & $f_2 (\mathrm{Gyr})$ \\
Name/No. & & & & \\
\hline
M1/1 & $14.67\pm0.56$ & $0.554\pm0.018$ & $0.47\pm0.02$ & $2.53\pm0.06$ \\
M1sn/2 & $8.73\pm0.44$ & $0.722\pm0.035$ & $0.46\pm0.17$ & $1.97\pm0.18$ \\
M1N/3 & $33.46\pm4.10$ & $0.566\pm0.023$ & $0.0226\pm0.0004$ & - \\
M1const/4 & $102.69\pm9.12$ & $0.207\pm0.009$ & $1.60\pm0.09$ & $1.03\pm0.02$ \\
M1Zpoor/5 & $14.80\pm0.59$ & $0.550\pm0.019$ & $0.12\pm0.01$ & $3.75\pm0.19$ \\
M1Zpoorsn/6 & $11.16\pm0.43$ & $0.630\pm0.022$ & $1.04\pm0.07$ & $2.14\pm0.05$ \\
M1l11/7 & $16.50\pm1.17$ & $0.49\pm0.03$ & $0.015\pm0.001$ & - \\
M1l13/8 & $13.91\pm0.54$ & $0.60\pm0.02$ & $1.73\pm0.12$ & $1.25\pm0.03$ \\
M2/9 & $2683.14\pm273.4$ & $0.108\pm0.003$ & $0.85\pm0.07$ & $1.56\pm0.05$ \\
M2sn/10 & $1661.98\pm160.30$ & $0.122\pm0.003$ & $3.37\pm0.10$ & $0.97\pm0.01$ \\
M2N/11 & $14522.20\pm3331.00$ & $0.128\pm0.006$ & $8.05\pm0.60$ & $0.85\pm0.02$ \\
M3/12 & $426.77\pm32.97$ & $0.205\pm0.006$ & $2.14\pm0.08$ & $1.52\pm0.03$ \\
M3sn/13 & $337.91\pm29.49$ & $0.222\pm0.008$ & $2.65\pm0.05$ & $1.55\pm0.02$ \\
M4/14 & $8740.40\pm1832.00$ & $0.122\pm0.008$ & $7.13\pm0.20$ & $1.38\pm0.02$ \\
M4sn/15 & $5019.00\pm1828.00$ & $0.146\pm0.020$ & $2.38\pm0.06$ & $2.67\pm0.03$ \\
\hline
\end{tabular}
\label{tab:5}
\end{table*}

\begin{table*}[h]
\caption{Fit parameters for Eq. \ref{gdtfit} for All Models. The index 1 corresponds to the first exponential part (green line) and 2 to the second part (blue line). The second part for model M1N and M1l11 is constant, therefore fit not with Eq. \ref{gdtfit} but with a constant, $c$. $i_2$ represents $c$ in this case.}
\centering
\begin{tabular}[c]{l c c c c}
\hline
Model & $i_1 (\mathrm{Gyr})$ & $j_1 (\mathrm{Gyr})$ & $i_2 (\mathrm{Gyr})$ & $j_2 (\mathrm{Gyr})$ \\
Name/No. & & & & \\
\hline
M1/1 & $0.36\pm0.02$ & $0.85\pm0.02$ & $4.24\pm0.21$ & $3.17\pm0.08$ \\
M1sn/2 & $0.52\pm0.02$ & $0.90\pm0.01$ & $11.09\pm4.28$ & $2.60\pm0.31$ \\
M1N/3 & $0.32\pm0.04$ & $0.84\pm0.02$ & $117.92\pm2.27$ & - \\
M1const/4 & $0.13\pm0.02$ & $0.58\pm0.03$ & $0.85\pm0.06$ & $1.42\pm0.04$ \\
M1Zpoor/5 & $0.38\pm0.02$ & $0.89\pm0.02$ & $11.83\pm1.17$ & $3.92\pm0.20$ \\
M1Zpoorsn/6 & $0.46\pm0.02$ & $0.87\pm0.02$ & $1.72\pm0.14$ & $2.18\pm0.05$ \\
M1l11/7 & $0.46\pm0.03$ & $0.94\pm0.02$ & $87.34\pm6.90$ & - \\
M1l13/8 & $0.38\pm0.17$ & $0.99\pm0.03$ & $1.01\pm0.05$ & $1.45\pm0.02$ \\
M2/9 & $0.051\pm0.011$ & $0.38\pm0.03$ & $2.87\pm0.28$ & $2.09\pm0.07$ \\
M2sn/10 & $0.076\pm0.013$ & $0.42\pm0.02$ & $0.92\pm0.11$ & $1.07\pm0.03$ \\
M2N/11 & $0.014\pm0.003$ & $0.31\pm0.02$ & - & - \\
M3/12 & $0.054\pm0.009$ & $0.34\pm0.02$ & $2.12\pm0.10$ & $1.96\pm0.03$ \\
M3sn/13 & $0.054\pm0.013$ & $0.32\pm0.02$ & $1.62\pm0.11$ & $1.66\pm0.03$ \\
M4/14 & $0.027\pm0.008$ & $0.30\pm0.27$ & $1.15\pm0.03$ & $1.75\pm0.02$ \\
M4sn/15 & $0.028\pm0.011$ & $0.31\pm0.03$ & $2.41\pm0.11$ & $2.60\pm0.04$ \\
\hline
\end{tabular}
\label{tab:6}
\end{table*}

\begin{table*}[h]
\caption{Parameters for the $\Sigma_\mathrm{z}$ profile fitted with Eq. \ref{eq:sigmaz} for All Models. The Newtonian models do not show an exponential profile after the thin disk; therefore, these regions are fitted with Eq. \ref{eq:powerlaw}. Also, some MONDian models do not show a thick disk, because either their profile can be fitted with a single, simple exponential function (M1sn and M1l11) or the profile has a more complicated form in that region than a simple exponential function or power law (M4 and M4sn).}
\centering
\begin{tabular}[c]{l c c c c}
\hline
Model & $n_\mathrm{z,thin} (2\times10^7\mathrm{M_\odot pc^{-1}})$ & $n_\mathrm{z,thick} (2\times10^7\mathrm{M_\odot pc^{-1}})$ & $z_\mathrm{e,thin} (\mathrm{kpc})$ & $z_\mathrm{e,thick} (\mathrm{kpc})$ \\
Name/No. & & $N:$ $a(2\times10^7\mathrm{M_\odot pc^{-1}})$ & & $N:$ $b$ \\
\hline
M1/1 & $0.316\pm0.009$ & $0.305\pm0.008$ & $0.096\pm0.004$ & $0.606\pm0.011$ \\
M1sn/2 & $0.190\pm0.002$ & $0.509\pm0.004$ & - & - \\
M1N/3 & $0.147\pm0.004$ & $0.491\pm0.023$ & $0.025\pm0.001$ & $2.255\pm0.030$ \\
M1const/4 & $0.336\pm0.010$ & $0.228\pm0.007$ & $0.142\pm0.007$ & $0.350\pm0.007$ \\
M1Zpoor/5 & $0.353\pm0.009$ & $0.253\pm0.008$ & $0.078\pm0.004$ & $0.576\pm0.009$ \\
M1Zpoorsn/6 & $0.271\pm0.003$ & $0.319\pm0.006$ & $0.126\pm0.004$ & $0.540\pm0.007$ \\
M1l11/7 & $0.315\pm0.008$ & $0.323\pm0.006$ & - & - \\
M1l13/8 & $0.267\pm0.008$ & $0.328\pm0.013$ & $0.124\pm0.001$ & $0.659\pm0.004$ \\
M2/9 & $1.714\pm0.030$ & $0.186\pm0.003$ & $0.184\pm0.017$ & $0.459\pm0.016$ \\
M2sn/10 & $1.385\pm0.025$ & $0.227\pm0.003$ & $0.158\pm0.010$ & $0.512\pm0.011$ \\
M2N/11 & $1.509\pm0.030$ & $0.265\pm0.006$ & $0.063\pm0.001$ & $2.796\pm0.032$ \\
M3/12 & $1.730\pm0.047$ & $0.240\pm0.006$ & $0.308\pm0.017$ & $0.607\pm0.017$ \\
M3sn/13 & $1.259\pm0.016$ & $0.315\pm0.006$ & $0.290\pm0.013$ & $0.661\pm0.013$ \\
M4/14 & $5.530\pm0.172$ & $0.350\pm0.011$ & - & - \\
M4sn/15 & $5.747\pm0.165$ & $0.353\pm0.010$ & - & - \\
\hline
\end{tabular}
\label{tab:7}
\end{table*}

\newpage

\begin{figure*}[h]
\includegraphics[width=1.0\linewidth]{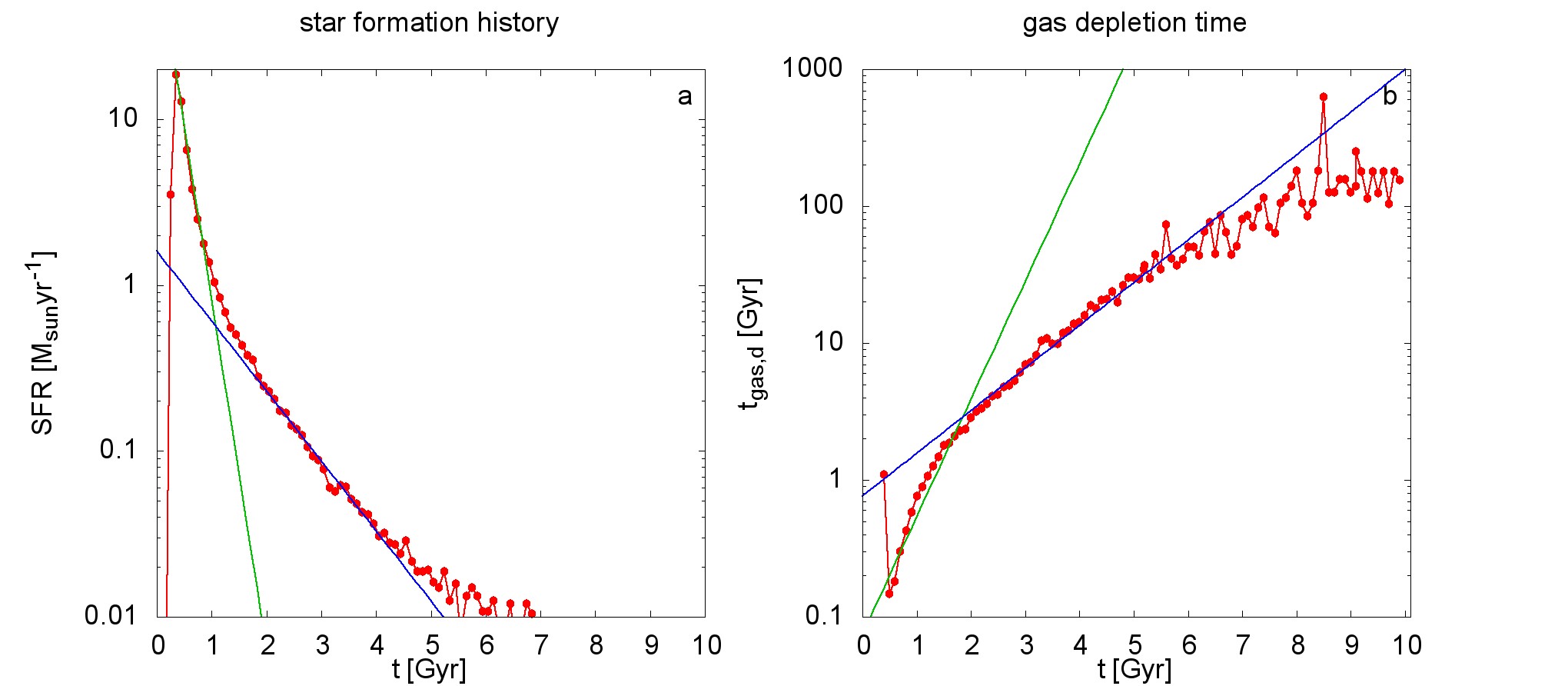}
\caption{The SFH and the evolution of the gas-depletion time for M1const. The red dots show the data, the green line shows the exponential fit for the first part of the decaying star formation rate (SFR) / increasing gas-depletion time directly after the initial collapse and the blue line shows the exponential fit for the shallower part after most of the gas was converted into stellar particles during and shortly after the collapse.}
\label{fig:SFHM1const}
\end{figure*}
\begin{figure*}[h]
\includegraphics[width=1.0\linewidth]{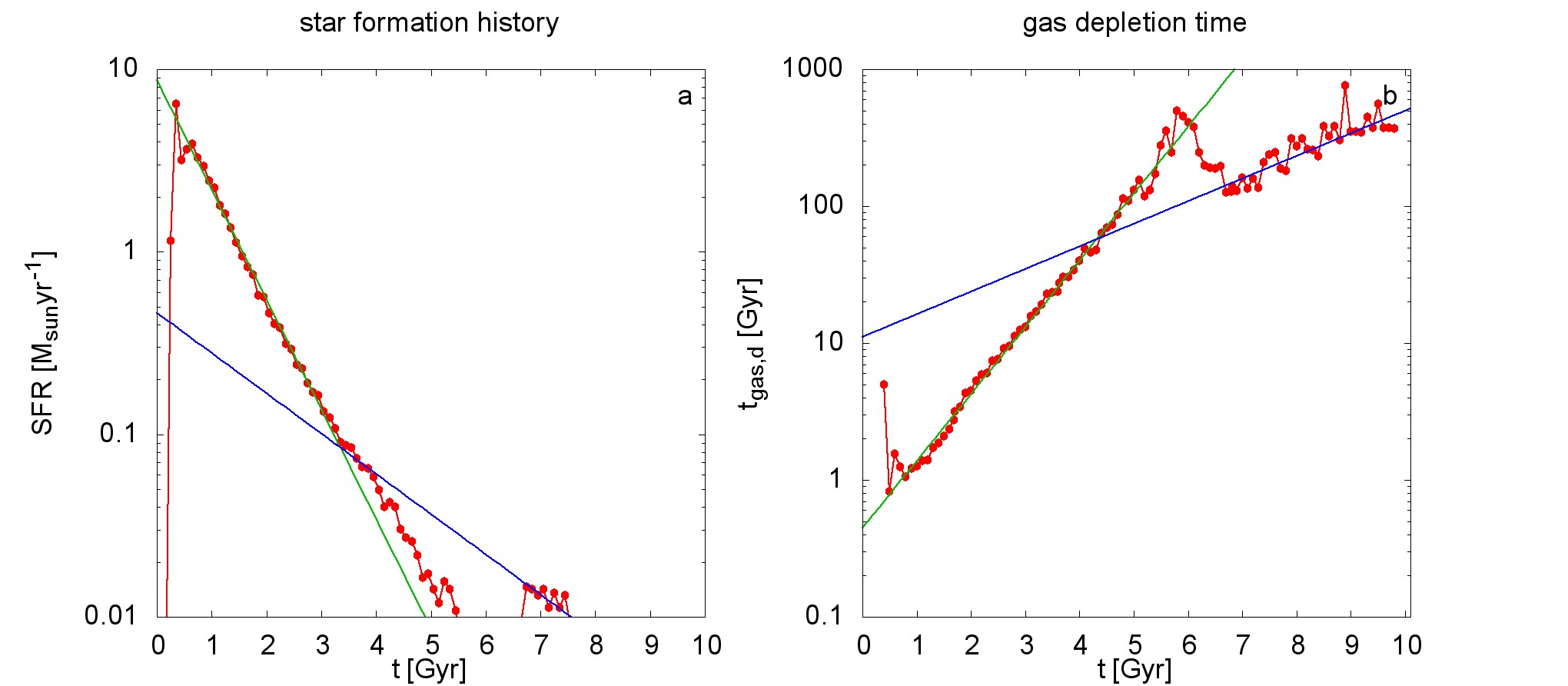}
\caption{As Fig. \ref{fig:SFHM1const} but for M1sn.}
\label{fig:SFHM1sn}
\end{figure*}
\newpage
\begin{figure*}[h]
\includegraphics[width=1.0\linewidth]{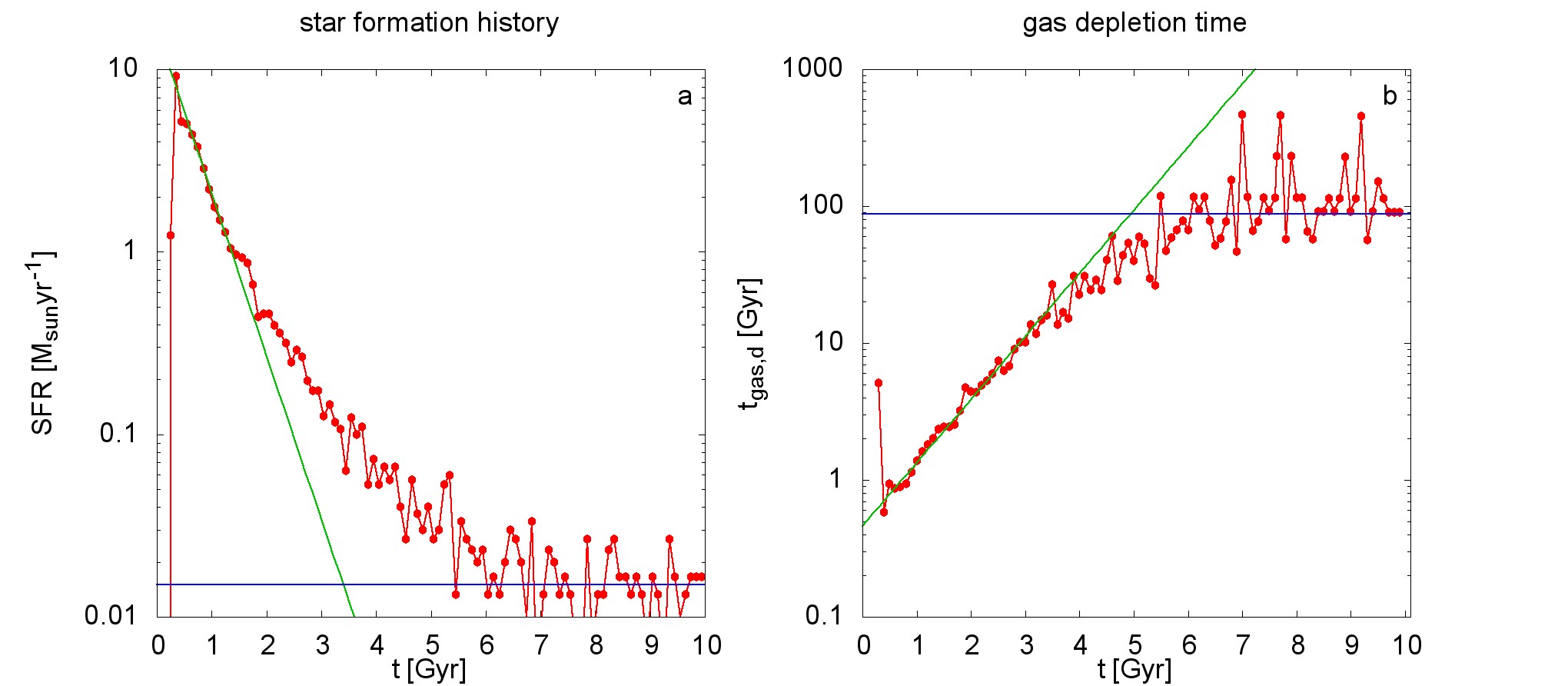}
\caption{As Fig. \ref{fig:SFHM1const} but for M1l11.}
\label{fig:SFHM1l11}
\end{figure*}
\begin{figure*}[h]
\includegraphics[width=1.0\linewidth]{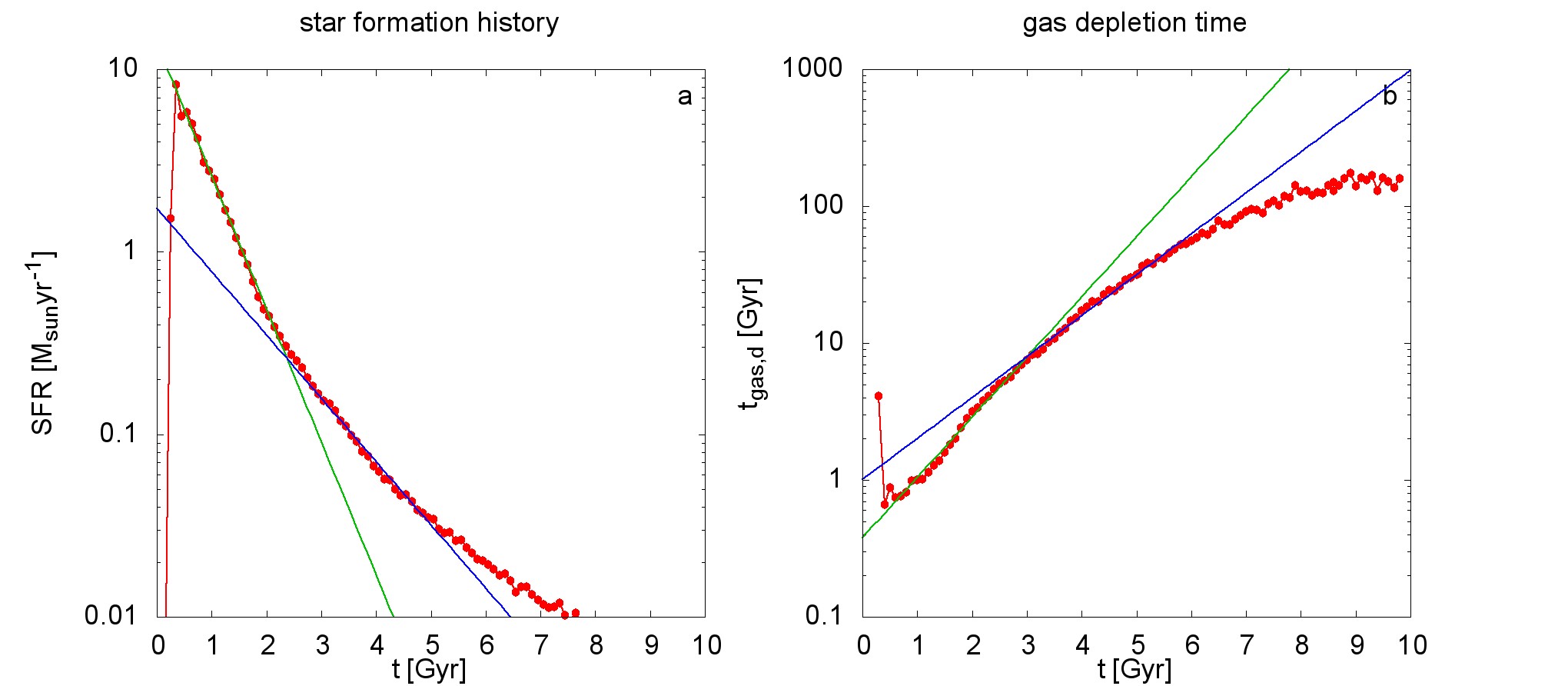}
\caption{As Fig. \ref{fig:SFHM1const} but for M1l13.}
\label{fig:SFHM1l13}
\end{figure*}
\newpage
\begin{figure*}[h]
\includegraphics[width=1.0\linewidth]{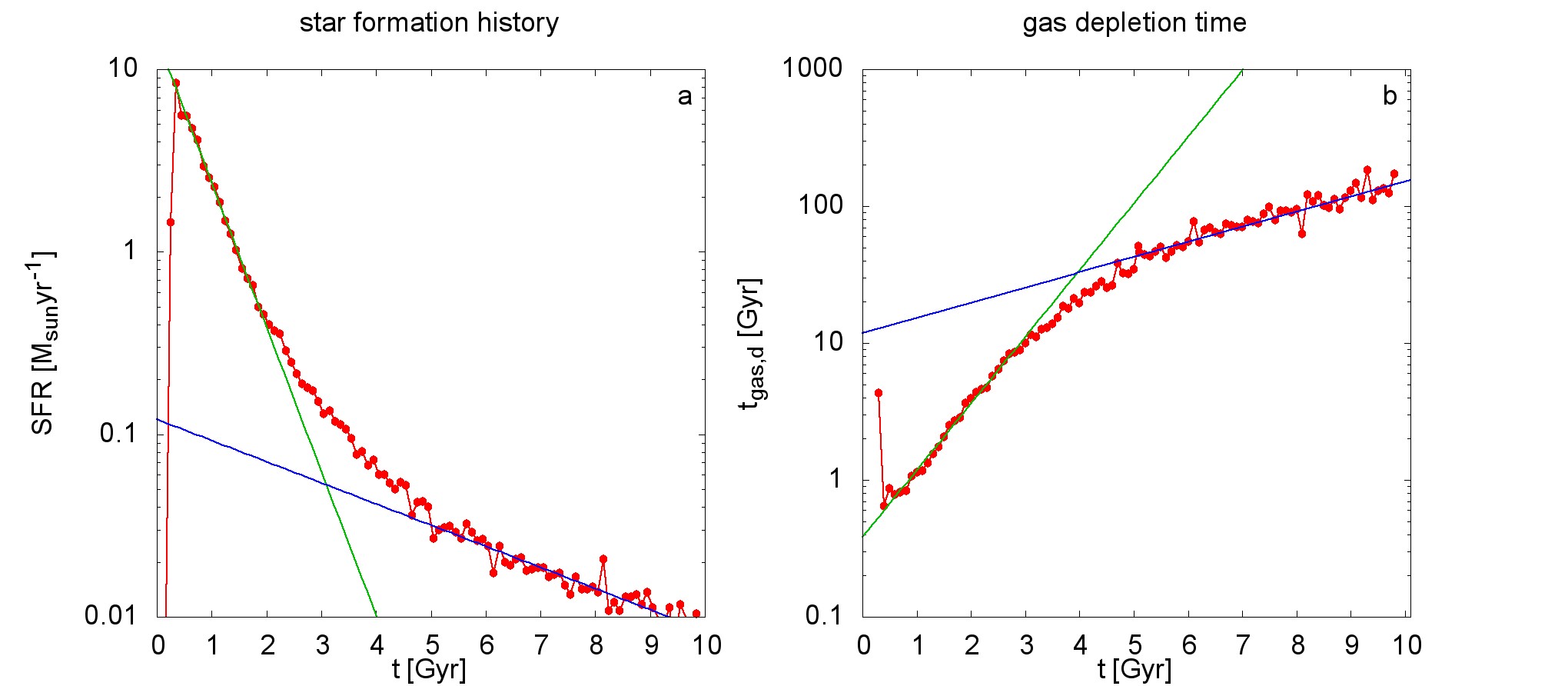}
\caption{As Fig. \ref{fig:SFHM1const} but for M1Zpoor.}
\label{fig:SFHM1Zpoor}
\end{figure*}
\begin{figure*}[h]
\includegraphics[width=1.0\linewidth]{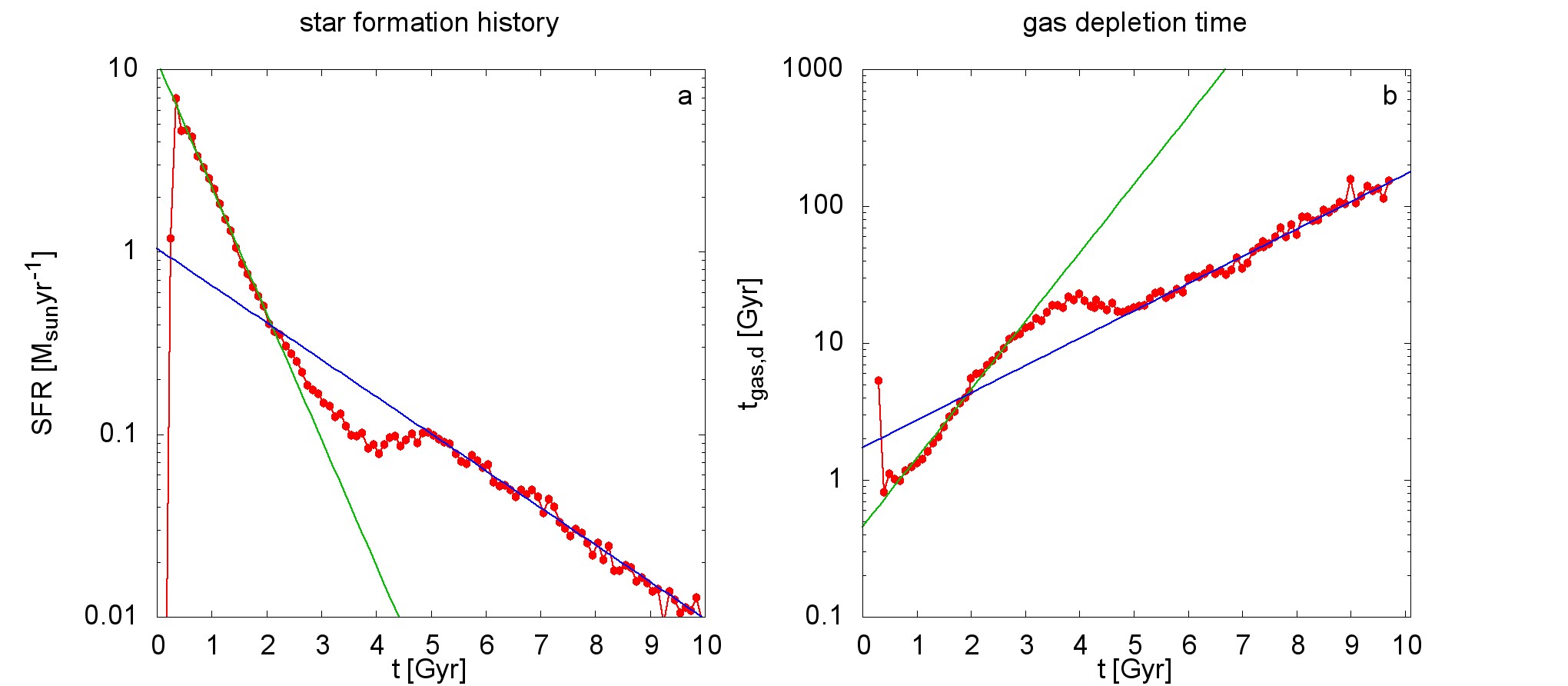}
\caption{As Fig. \ref{fig:SFHM1const} but for M1Zpoorsn.}
\label{fig:SFHM1Zpoorsn}
\end{figure*}
\newpage
\begin{figure*}[h]
\includegraphics[width=1.0\linewidth]{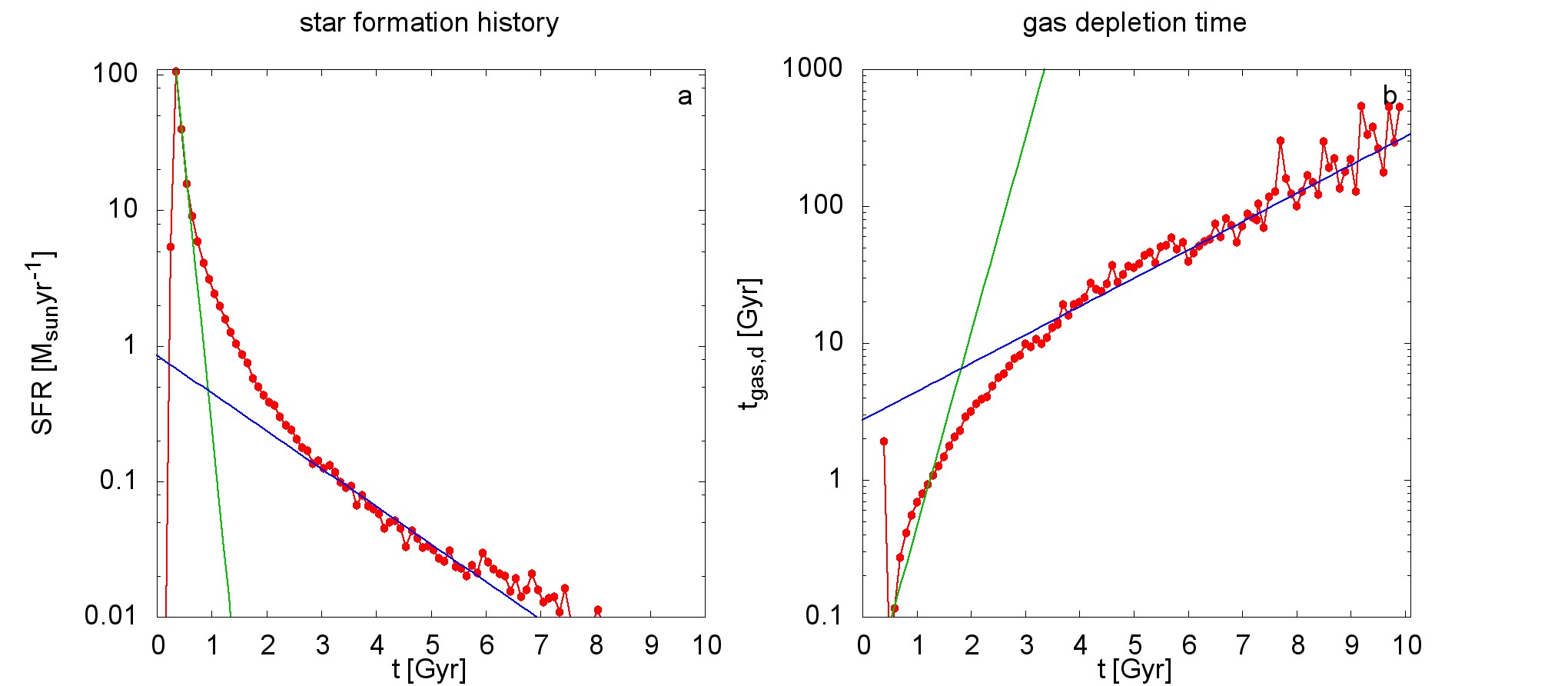}
\caption{As Fig. \ref{fig:SFHM1const} but for M2.}
\label{fig:SFHM2}
\end{figure*}
\begin{figure*}[h]
\includegraphics[width=1.0\linewidth]{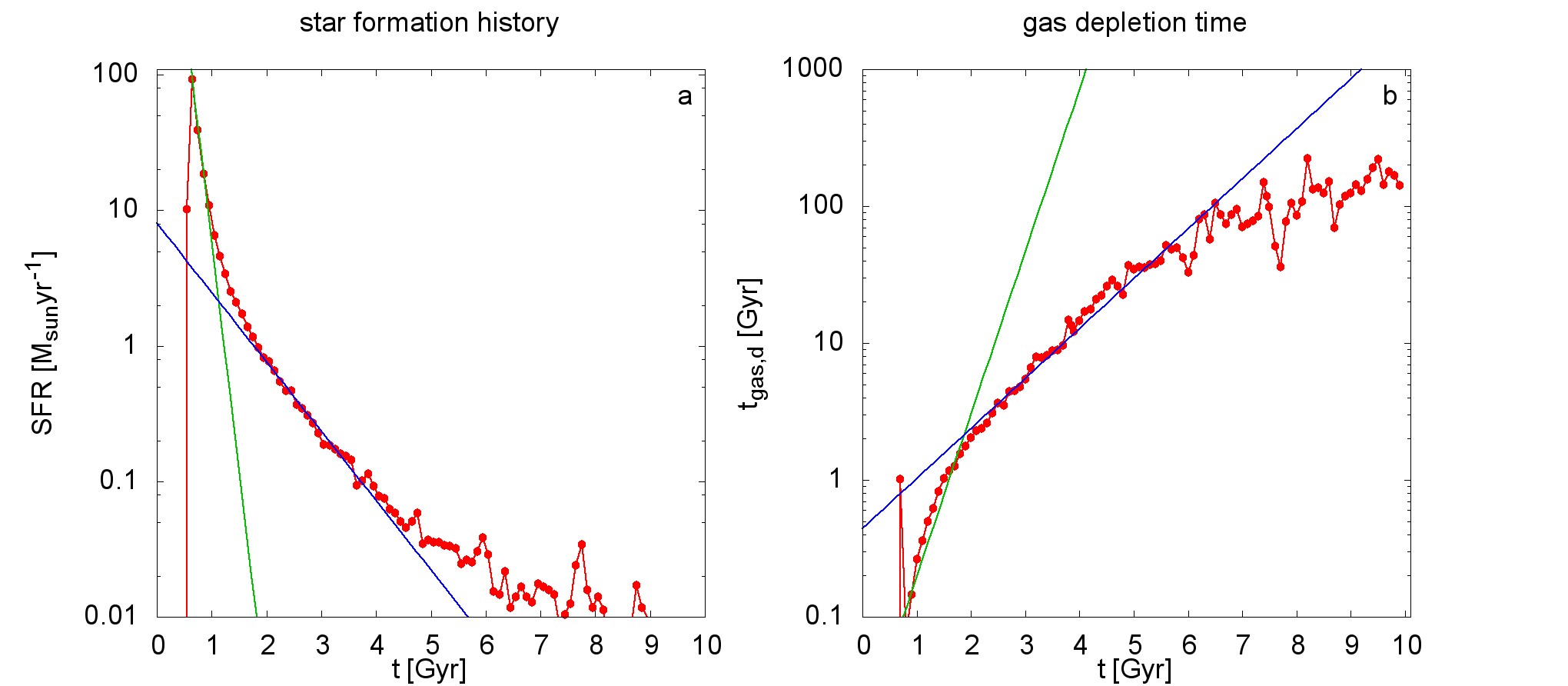}
\caption{As Fig. \ref{fig:SFHM1const} but for M2N.}
\label{fig:SFHM2N}
\end{figure*}
\newpage
\begin{figure*}[h]
\includegraphics[width=1.0\linewidth]{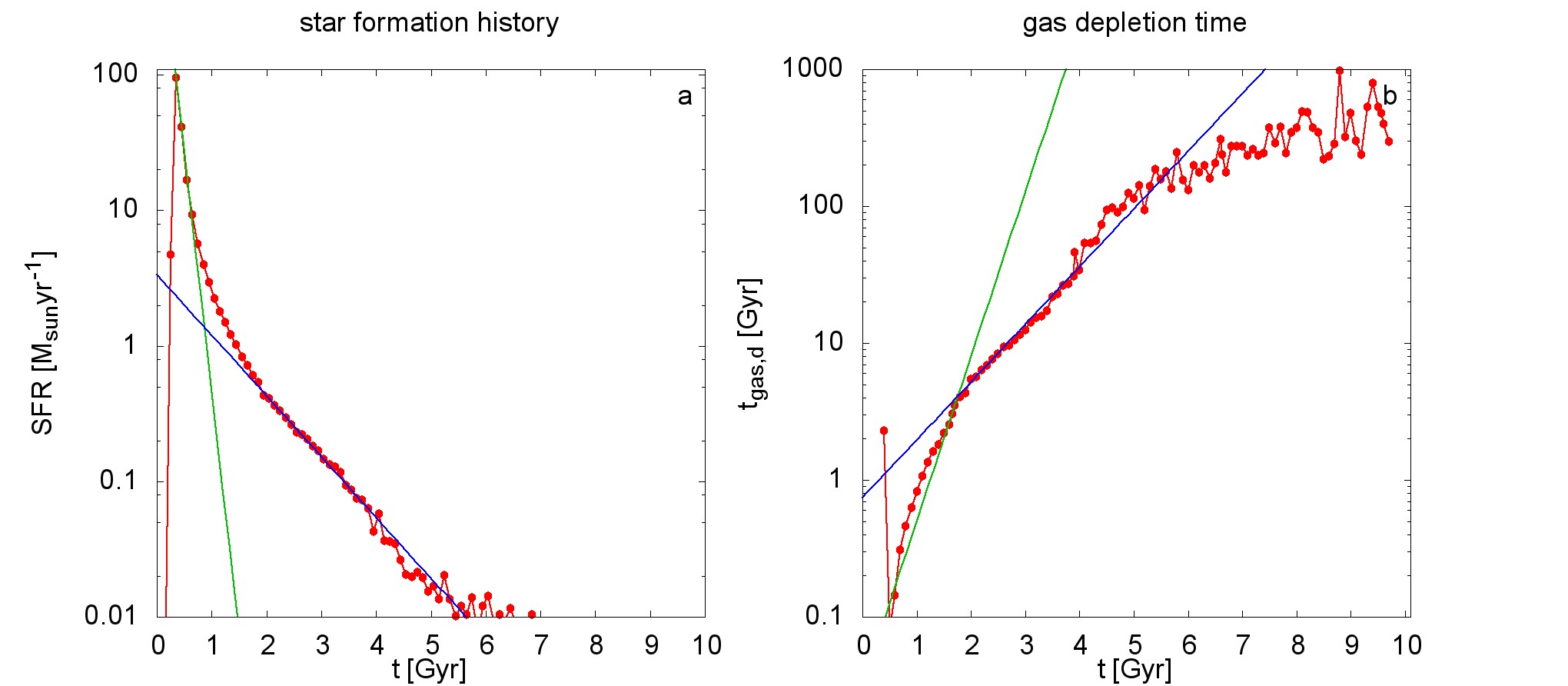}
\caption{As Fig. \ref{fig:SFHM1const} but for M2sn.}
\label{fig:SFHM2sn}
\end{figure*}
\begin{figure*}[h]
\includegraphics[width=1.0\linewidth]{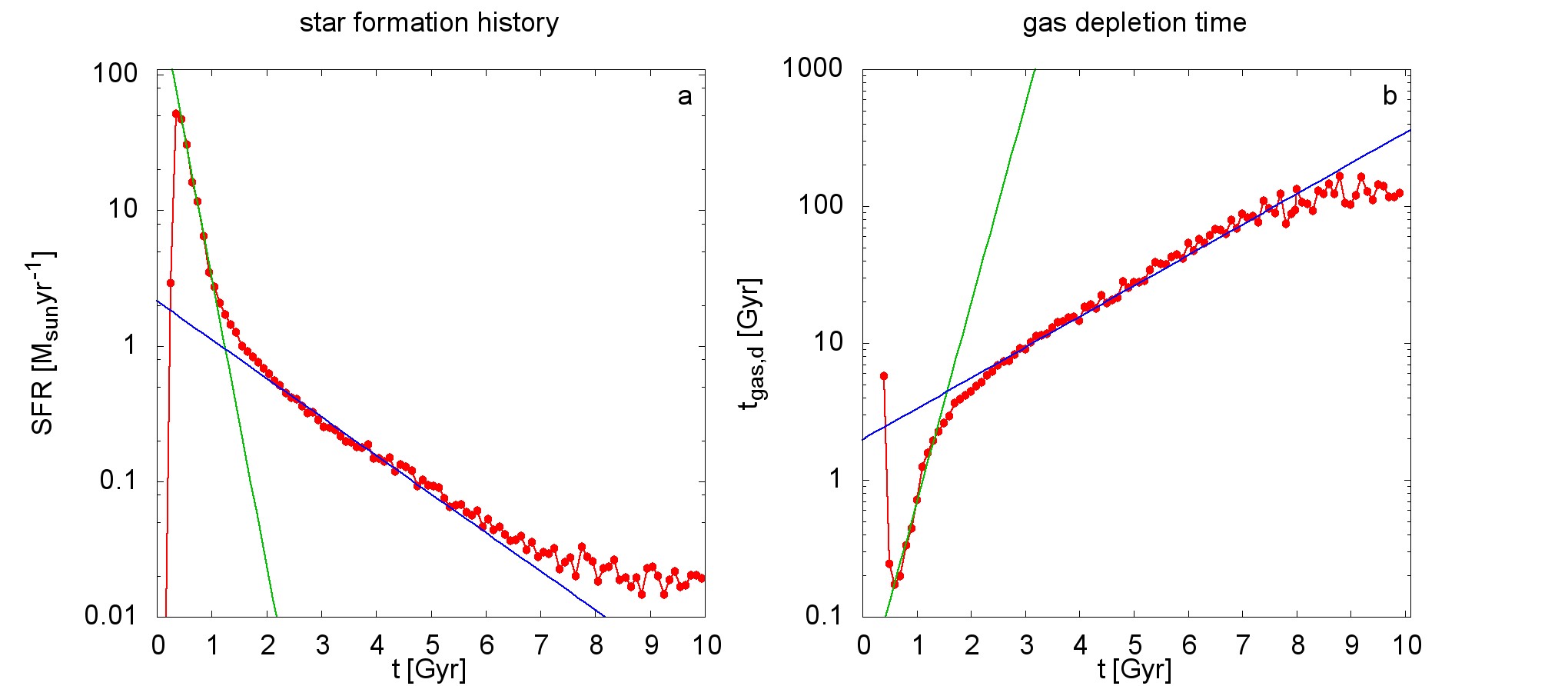}
\caption{As Fig. \ref{fig:SFHM1const} but for M3.}
\label{fig:SFHM3}
\end{figure*}
\newpage
\begin{figure*}[h]
\includegraphics[width=1.0\linewidth]{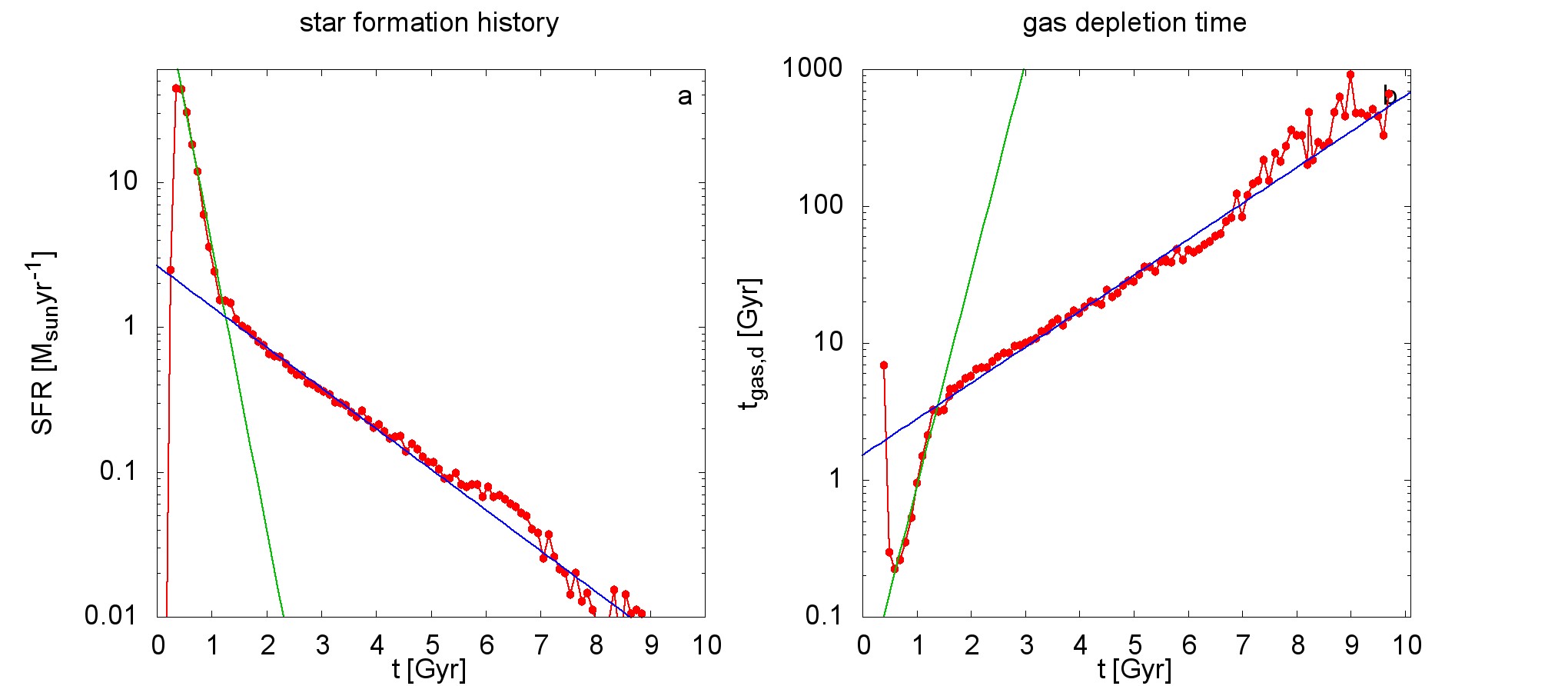}
\caption{As Fig. \ref{fig:SFHM1const} but for M3sn.}
\label{fig:SFHM3sn}
\end{figure*}
\begin{figure*}[h]
\includegraphics[width=1.0\linewidth]{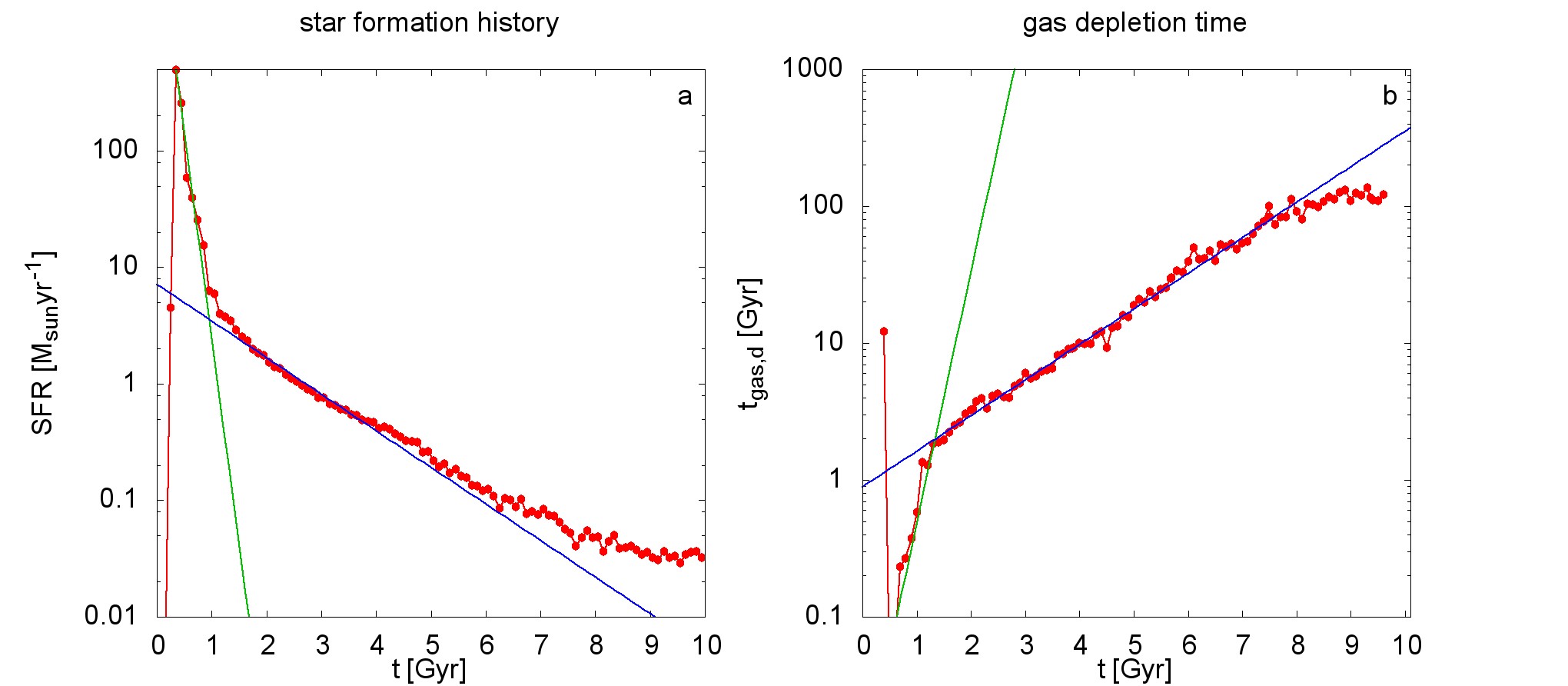}
\caption{As Fig. \ref{fig:SFHM1const} but for M4.}
\label{fig:SFHM4}
\end{figure*}
\newpage
\begin{figure*}[h]
\includegraphics[width=1.0\linewidth]{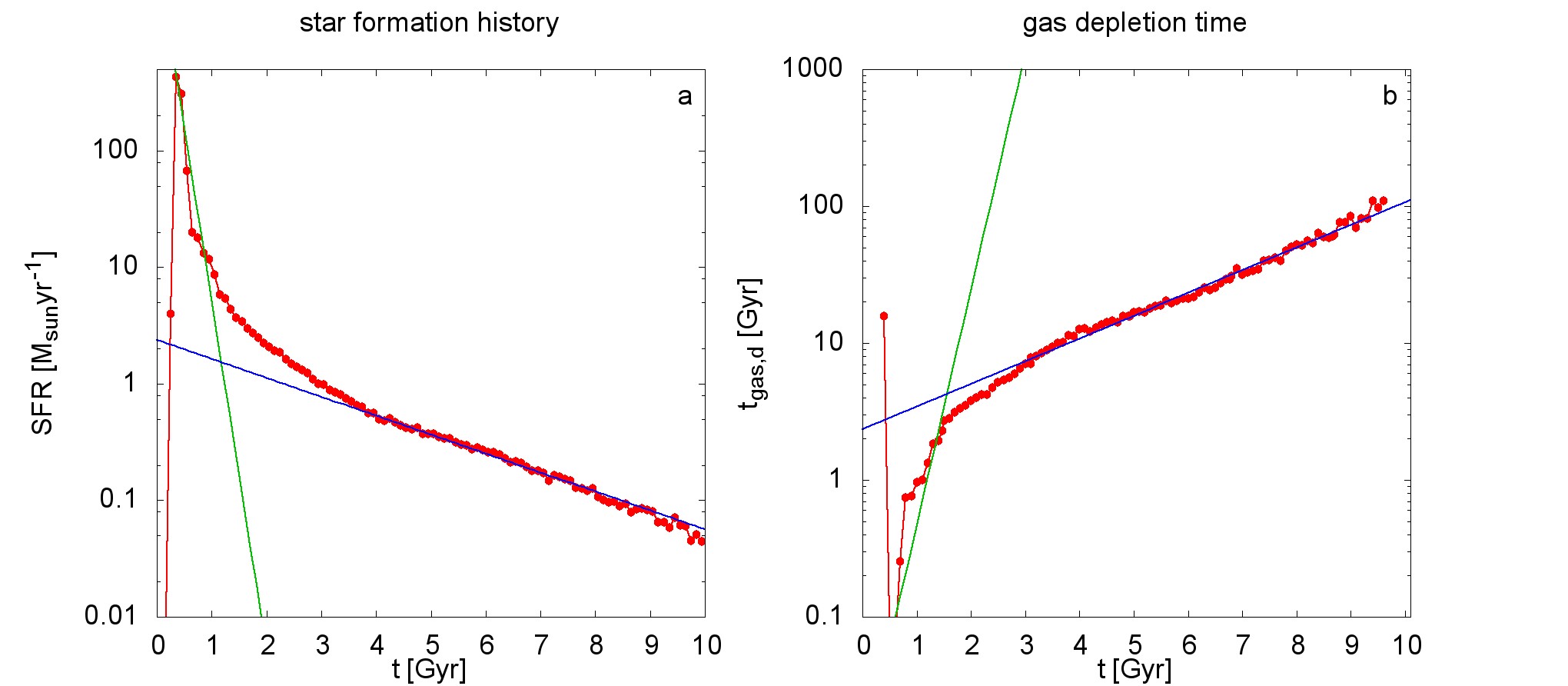}
\caption{As Fig. \ref{fig:SFHM1const} but for M4sn.}
\label{fig:SFHM4sn}
\end{figure*}
\begin{figure*}[h]
\begin{minipage}[t]{0.49\linewidth}
\includegraphics[width=1.0\linewidth]{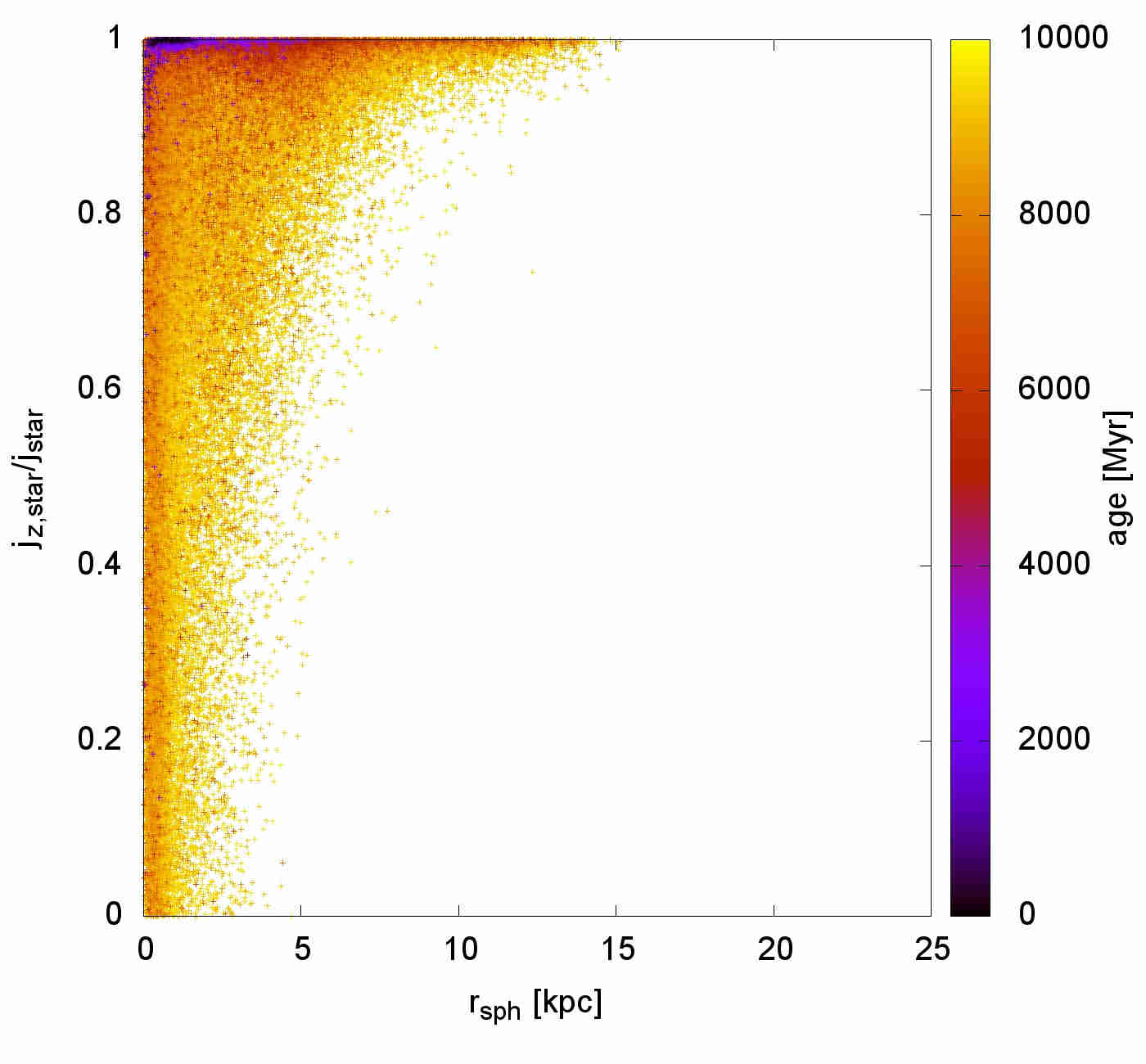}
\end{minipage}
\begin{minipage}[t]{0.49\linewidth}
\includegraphics[width=1.0\linewidth]{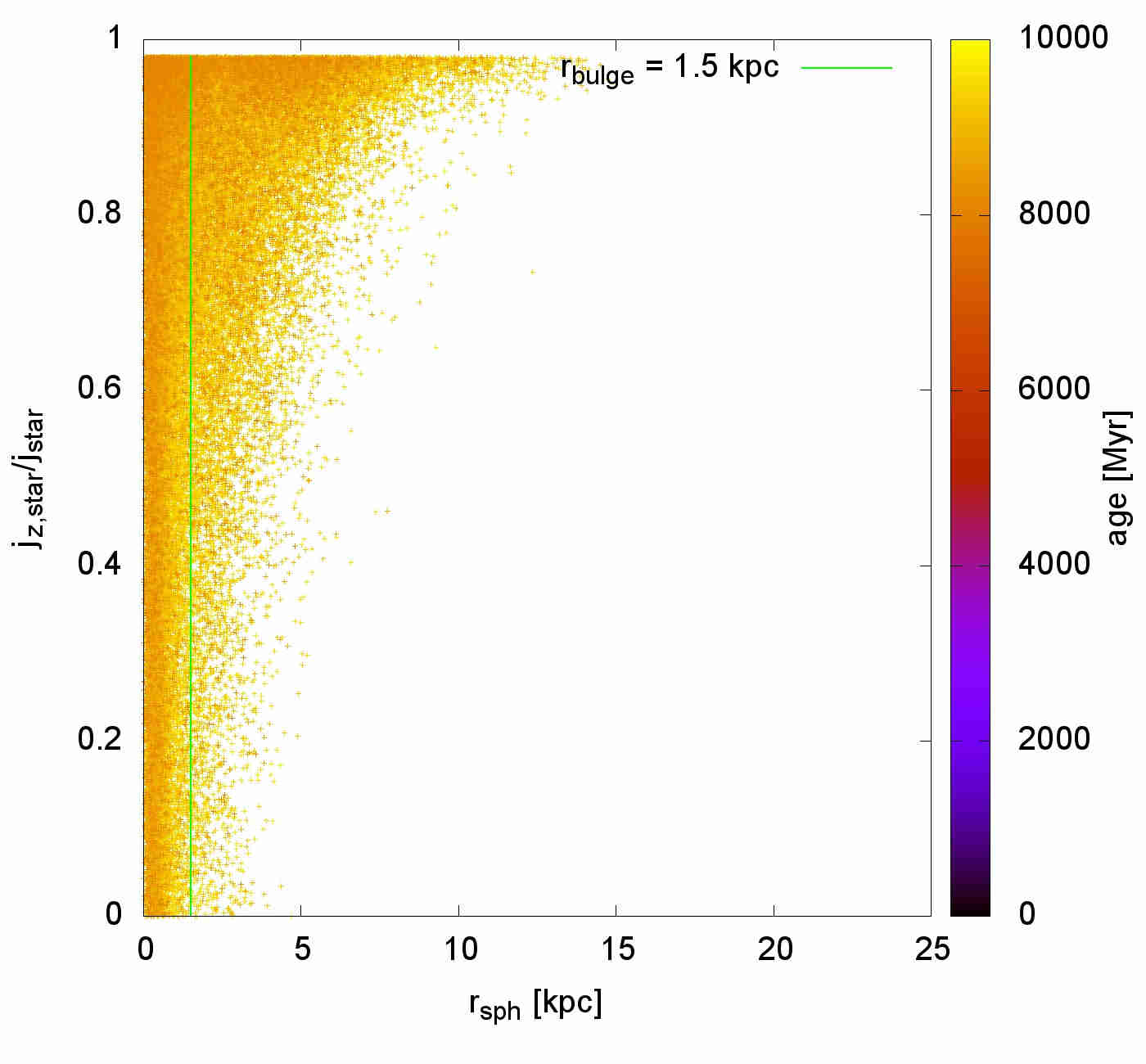}
\end{minipage}
\caption{As Fig. \ref{fig:jz/jM1} but for M1sn.} 
\label{fig:jz/jM1sn}
\end{figure*}
\newpage
\begin{figure*}[h]
\begin{minipage}[t]{0.49\linewidth}
\includegraphics[width=1.0\linewidth]{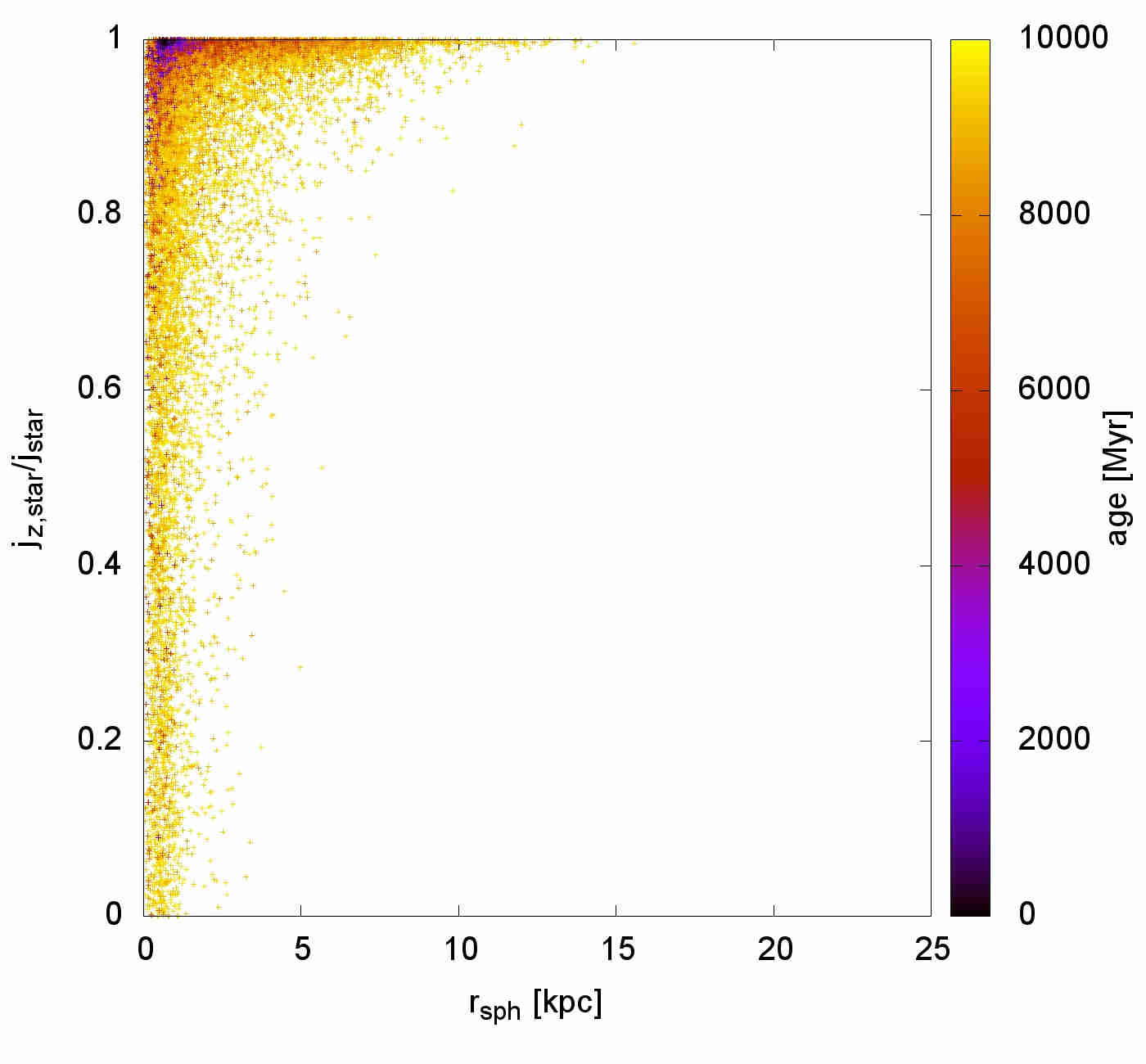}
\end{minipage}
\begin{minipage}[t]{0.49\linewidth}
\includegraphics[width=1.0\linewidth]{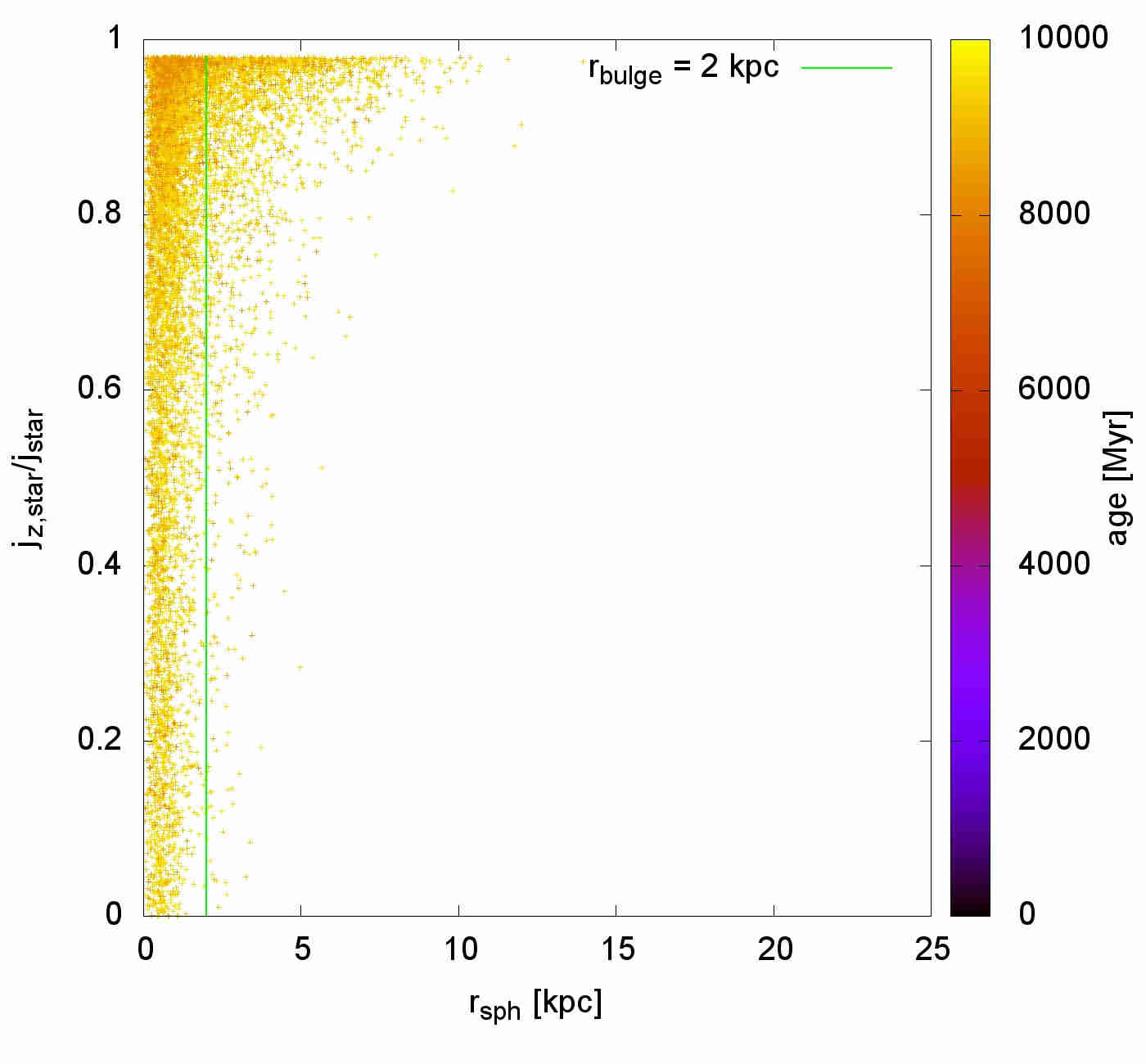}
\end{minipage}
\caption{As Fig. \ref{fig:jz/jM1} but for M1l11.} 
\label{fig:jz/jM1l11}
\end{figure*}
\begin{figure*}[h]
\begin{minipage}[t]{0.49\linewidth}
\includegraphics[width=1.0\linewidth]{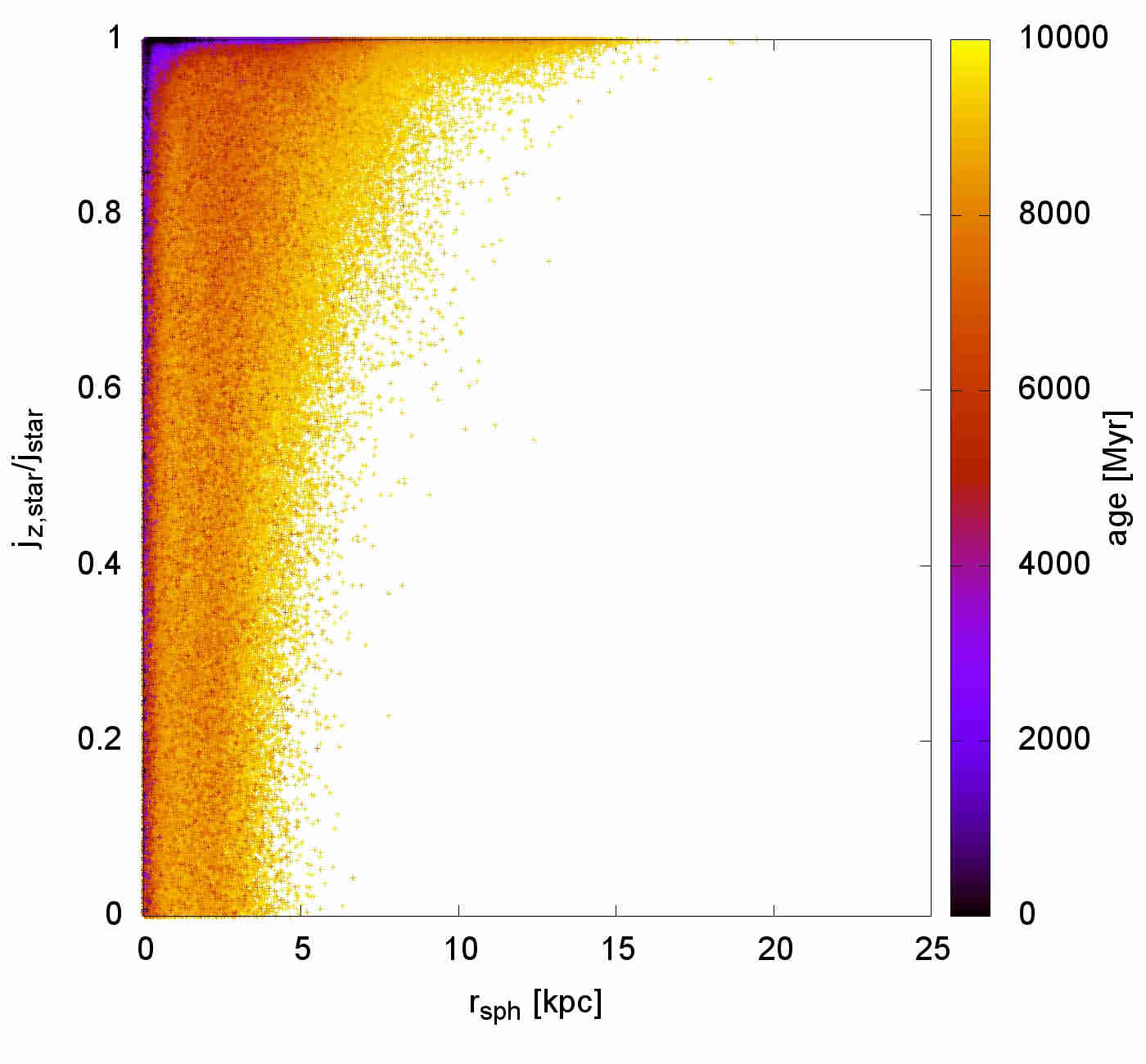}
\end{minipage}
\begin{minipage}[t]{0.49\linewidth}
\includegraphics[width=1.0\linewidth]{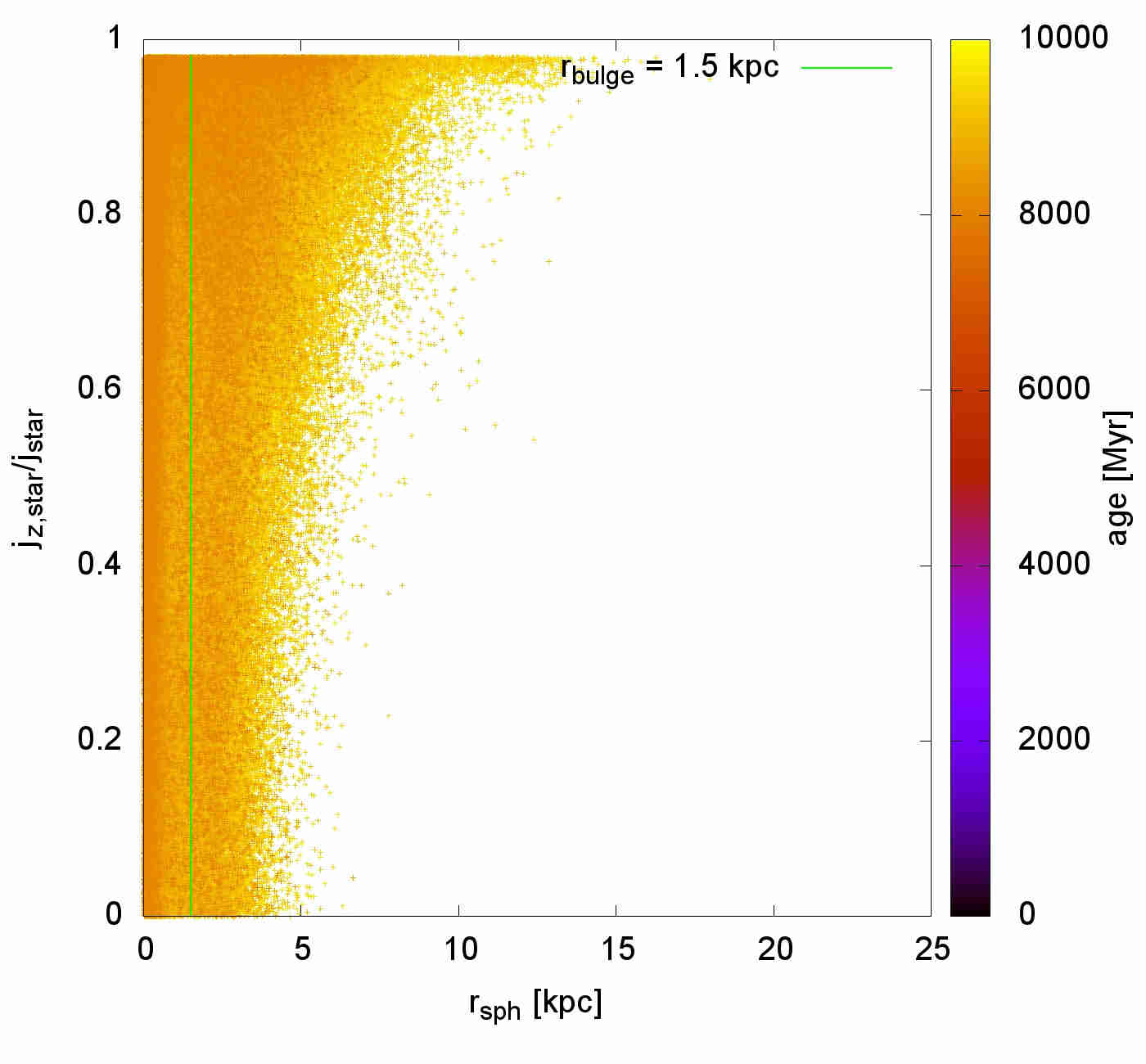}
\end{minipage}
\caption{As Fig. \ref{fig:jz/jM1} but for M1l13.} 
\label{fig:jz/jM1l13}
\end{figure*}
\newpage
\begin{figure*}[h]
\begin{minipage}[t]{0.49\linewidth}
\includegraphics[width=1.0\linewidth]{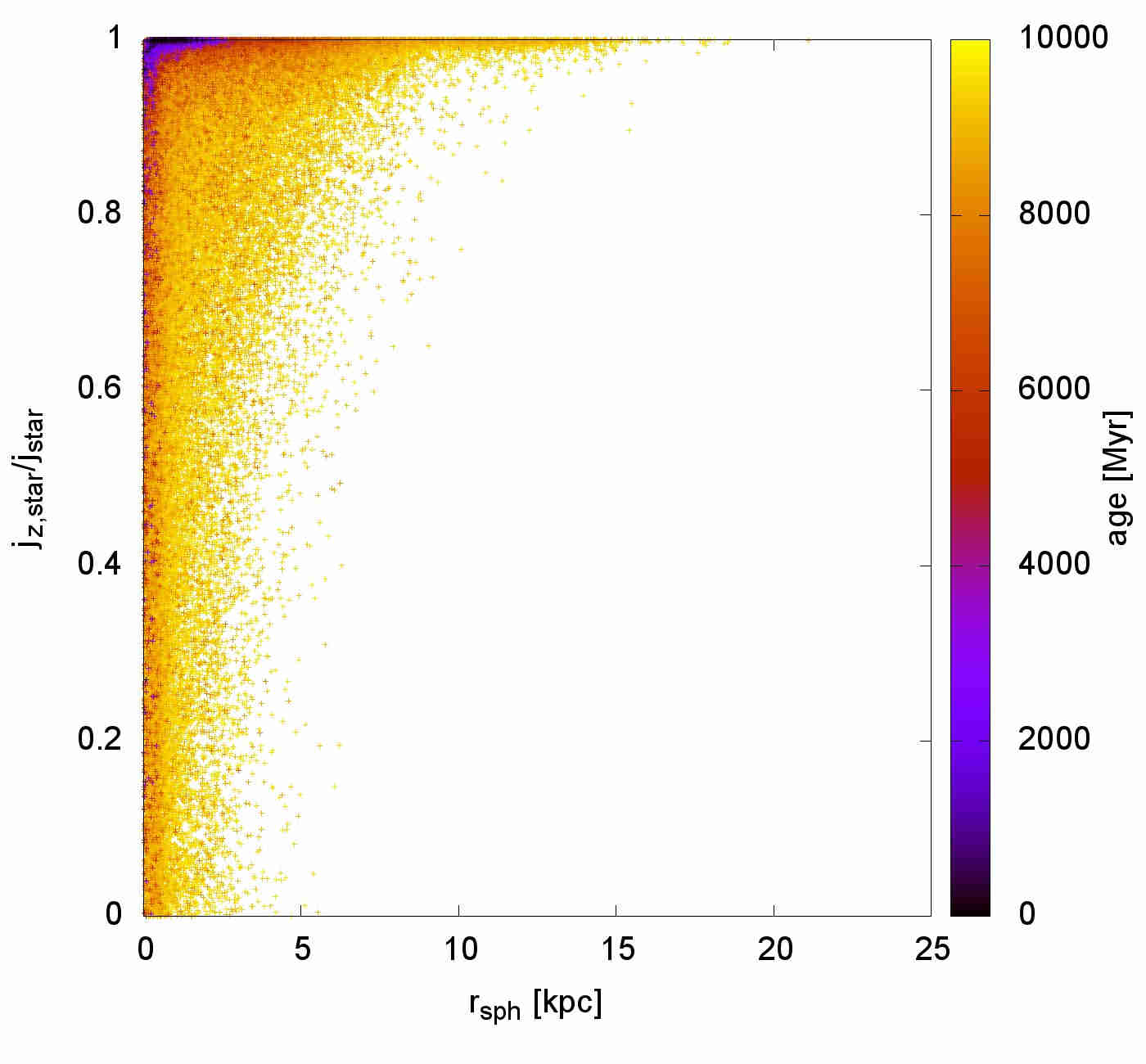}
\end{minipage}
\begin{minipage}[t]{0.49\linewidth}
\includegraphics[width=1.0\linewidth]{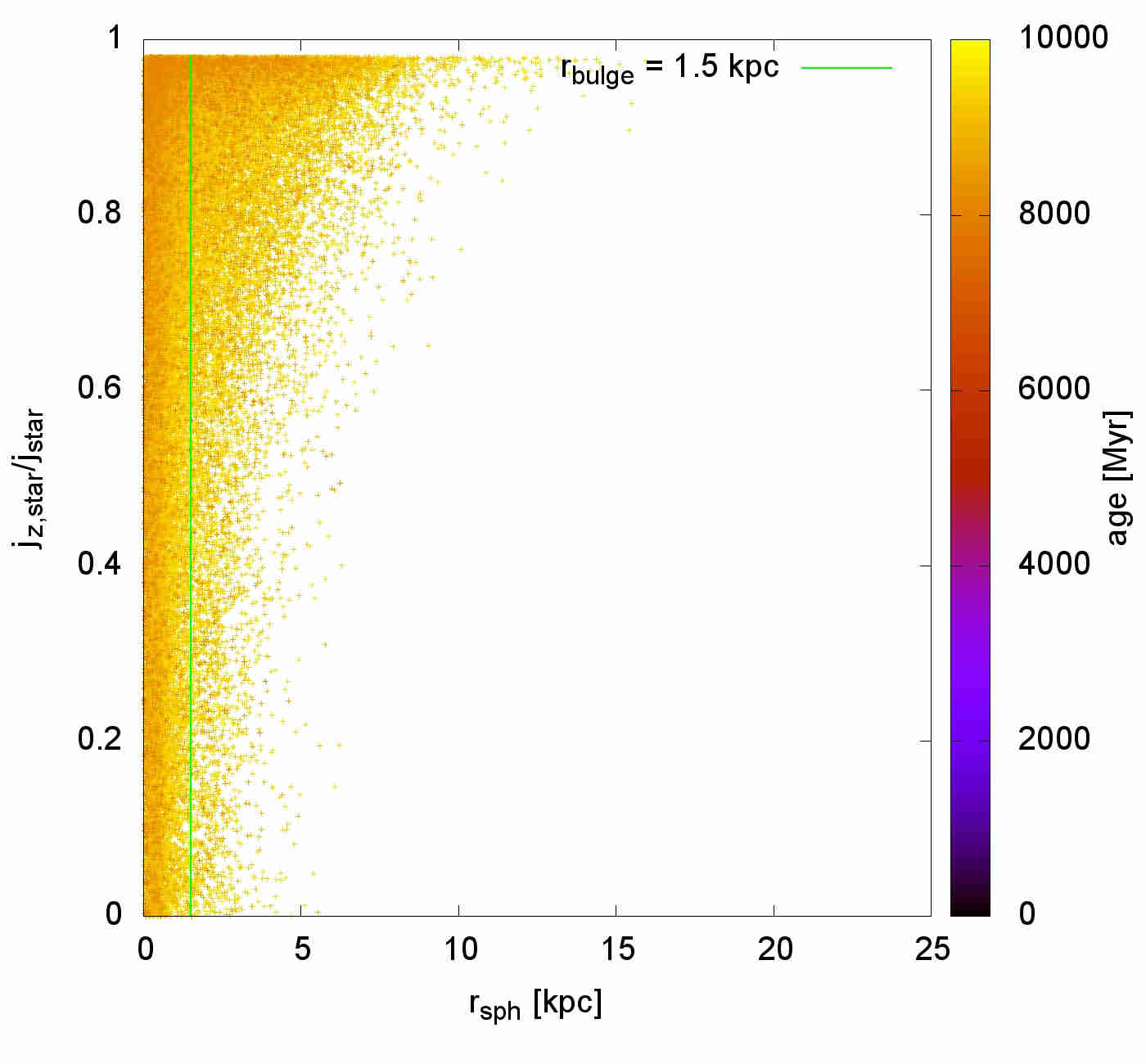}
\end{minipage}
\caption{As Fig. \ref{fig:jz/jM1} but for M1Zpoor.} 
\label{fig:jz/jM1Zpoor}
\end{figure*}
\begin{figure*}[h]
\begin{minipage}[t]{0.49\linewidth}
\includegraphics[width=1.0\linewidth]{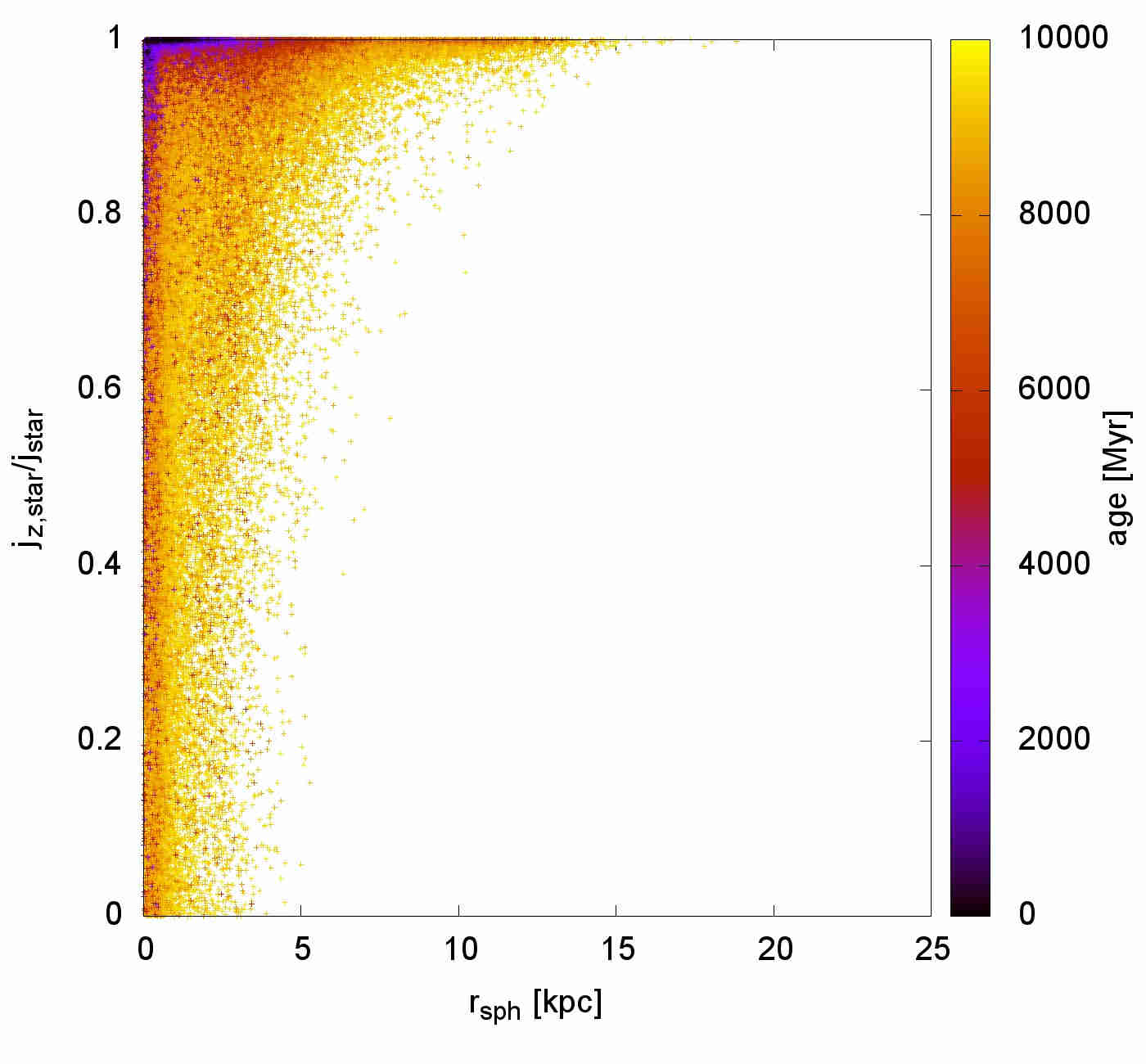}
\end{minipage}
\begin{minipage}[t]{0.49\linewidth}
\includegraphics[width=1.0\linewidth]{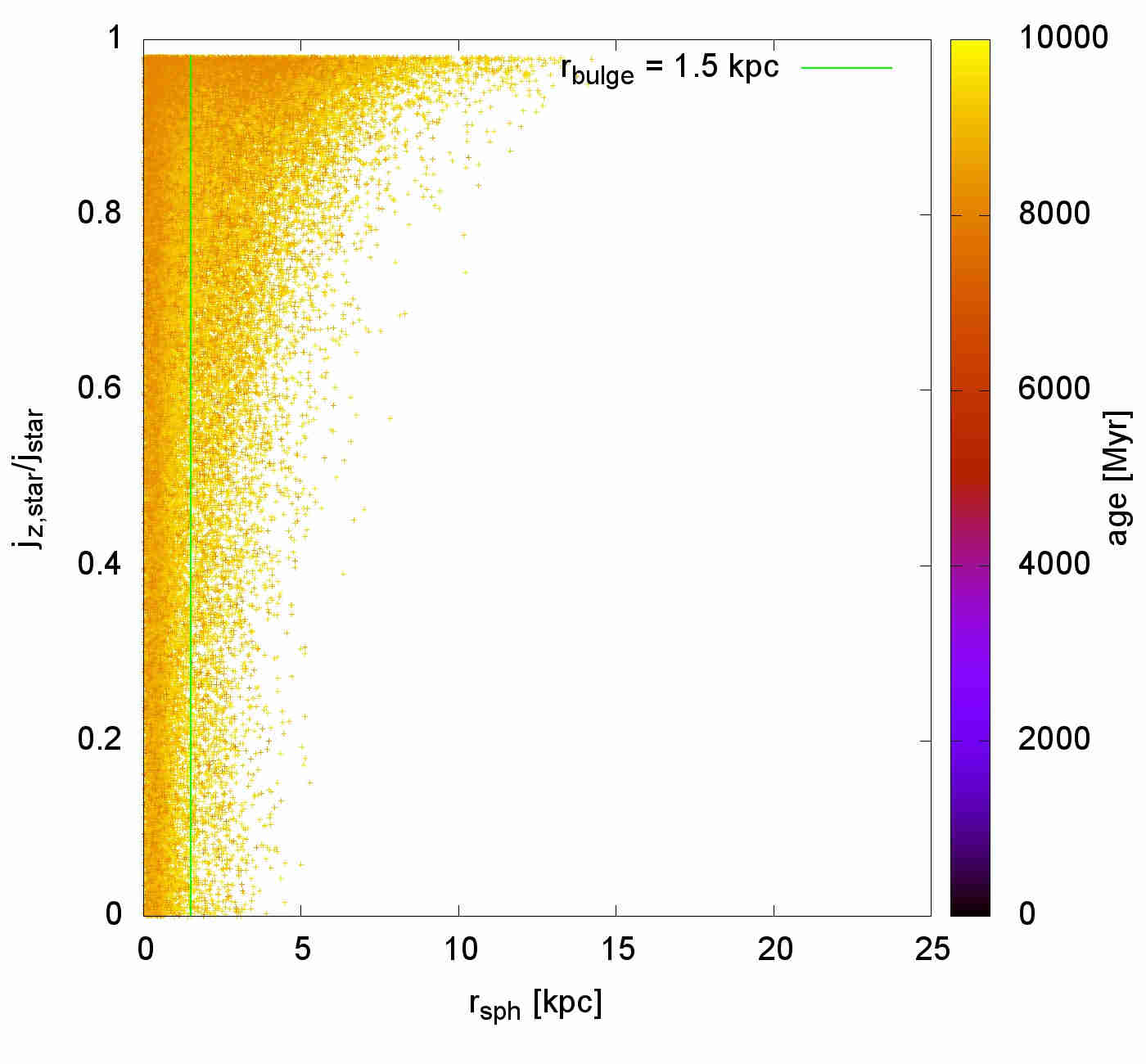}
\end{minipage}
\caption{As Fig. \ref{fig:jz/jM1} but for M1Zpoorsn.} 
\label{fig:jz/jM1Zpoorsn}
\end{figure*}
\newpage
\begin{figure*}[h]
\begin{minipage}[t]{0.49\linewidth}
\includegraphics[width=1.0\linewidth]{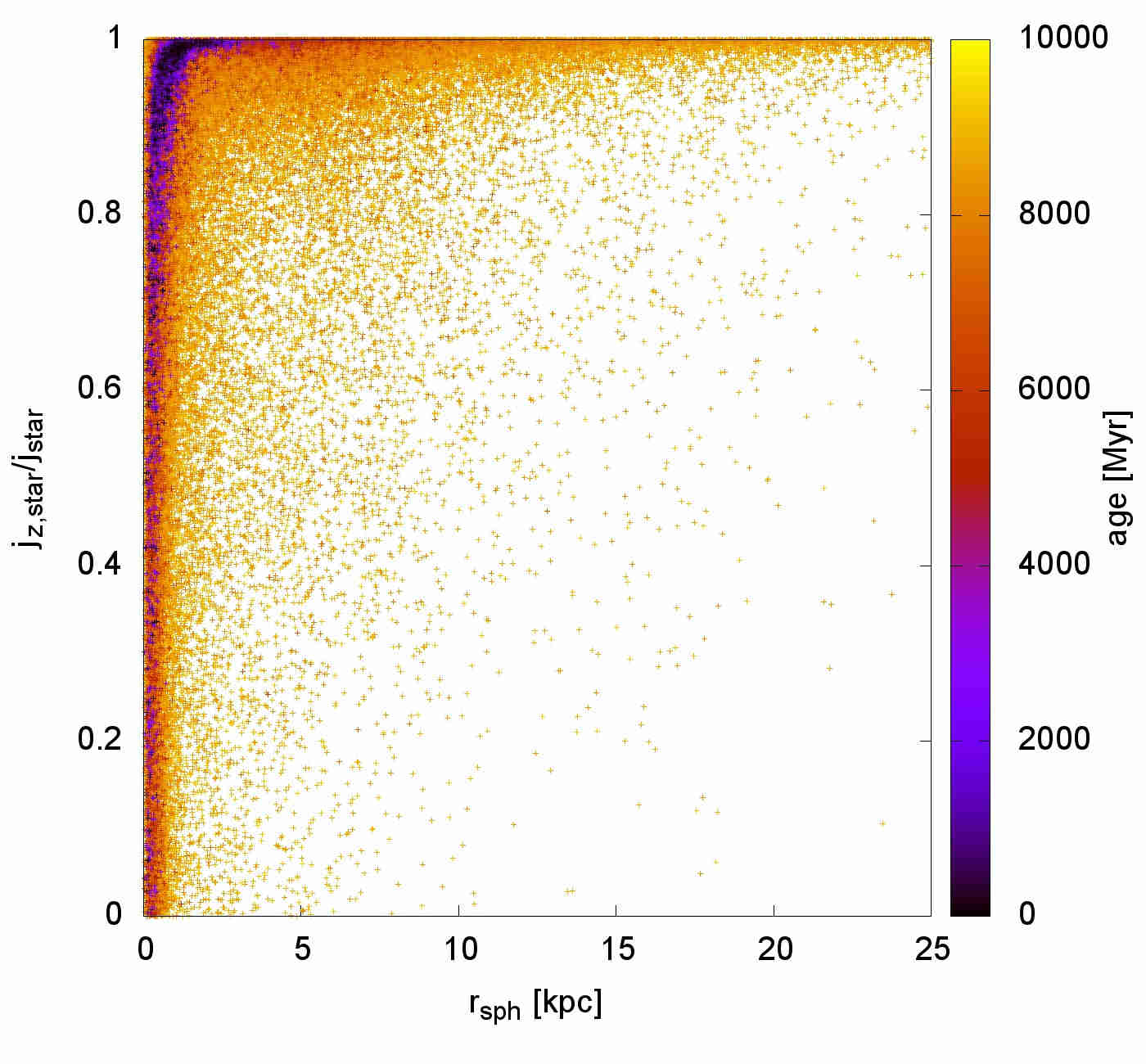}
\end{minipage}
\begin{minipage}[t]{0.49\linewidth}
\includegraphics[width=1.0\linewidth]{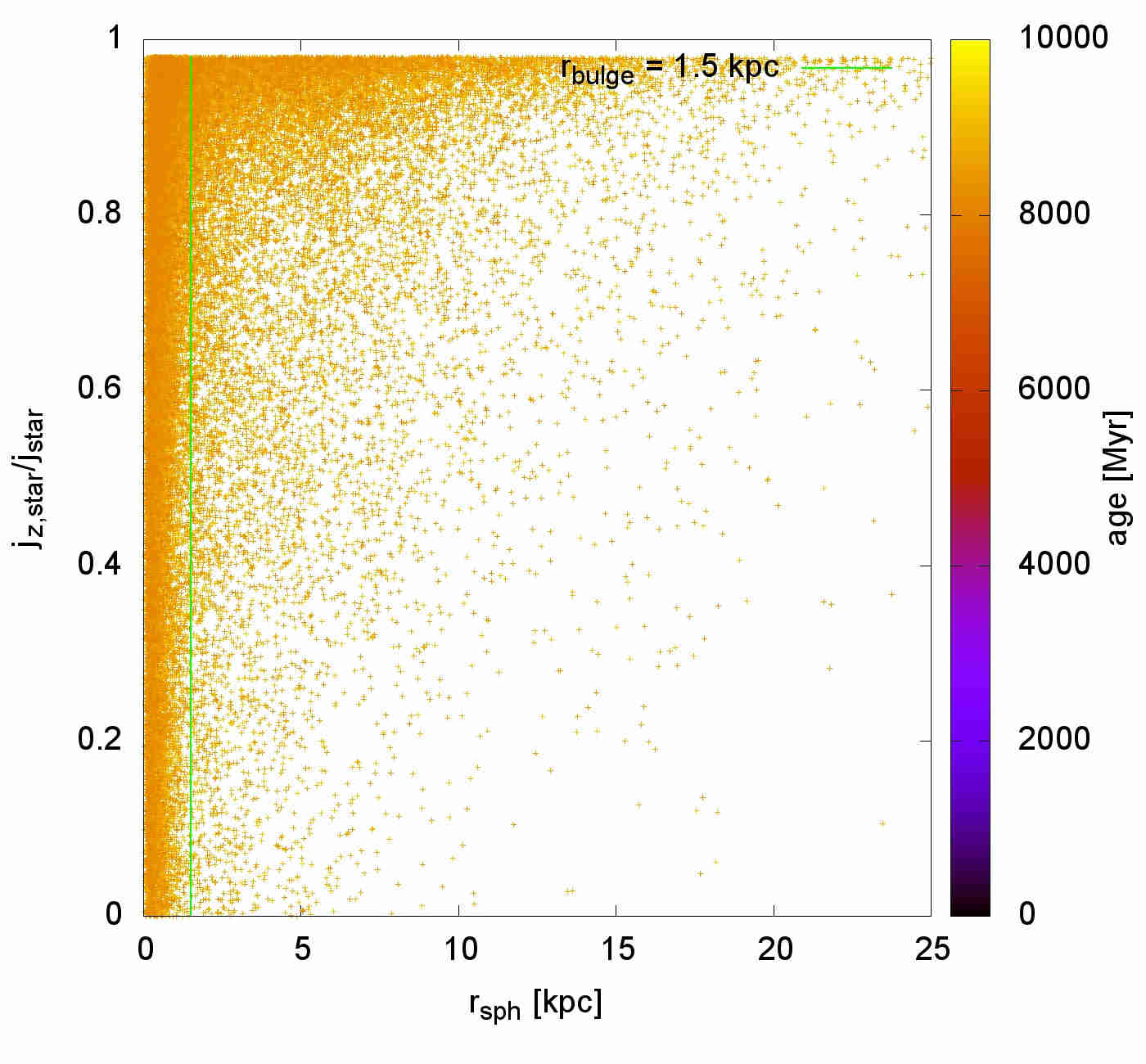}
\end{minipage}
\caption{As Fig. \ref{fig:jz/jM1} but for M1N.} 
\label{fig:jz/jM1N}
\end{figure*}
\begin{figure*}[h]
\begin{minipage}[t]{0.49\linewidth}
\includegraphics[width=1.0\linewidth]{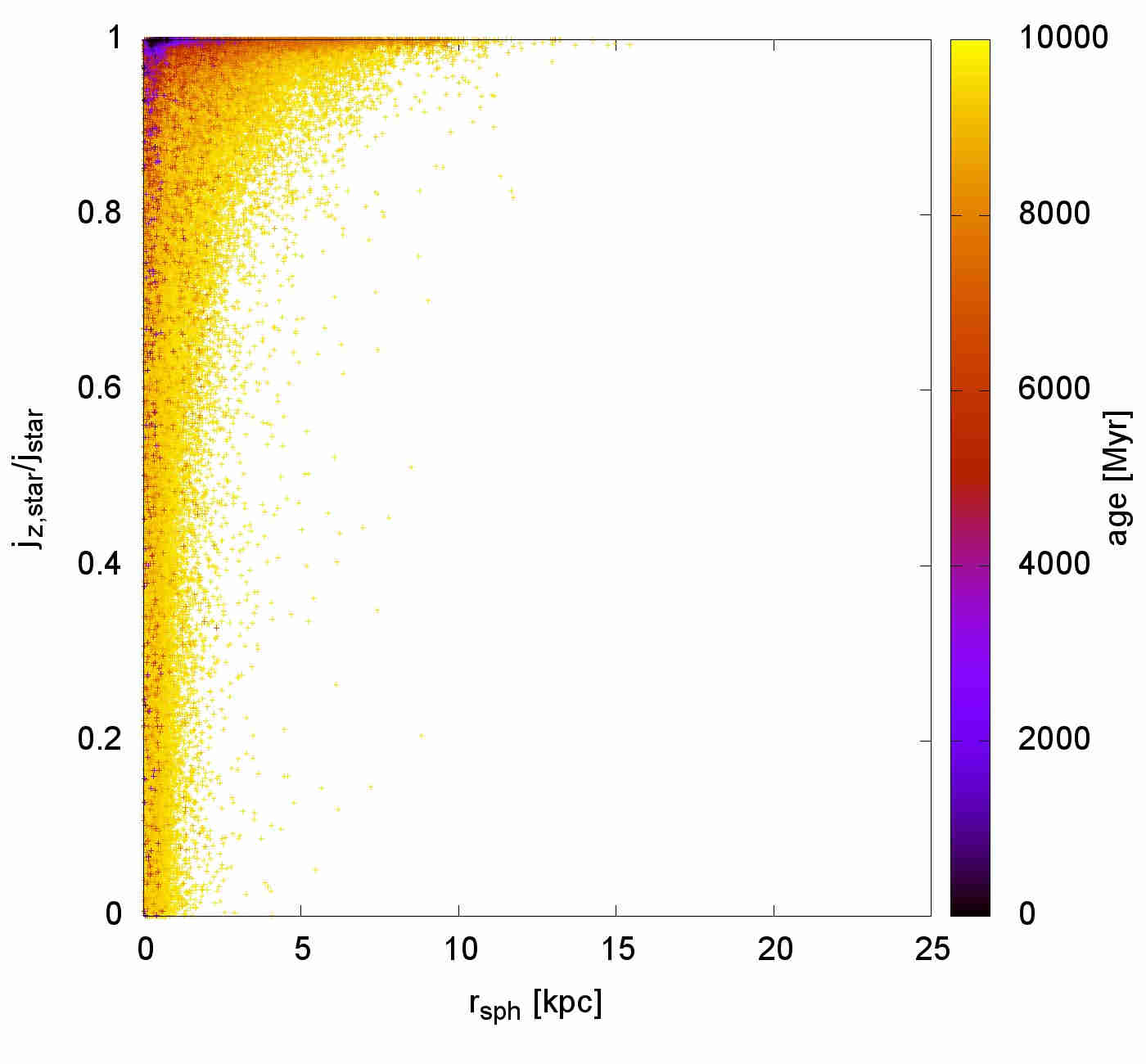}
\end{minipage}
\begin{minipage}[t]{0.49\linewidth}
\includegraphics[width=1.0\linewidth]{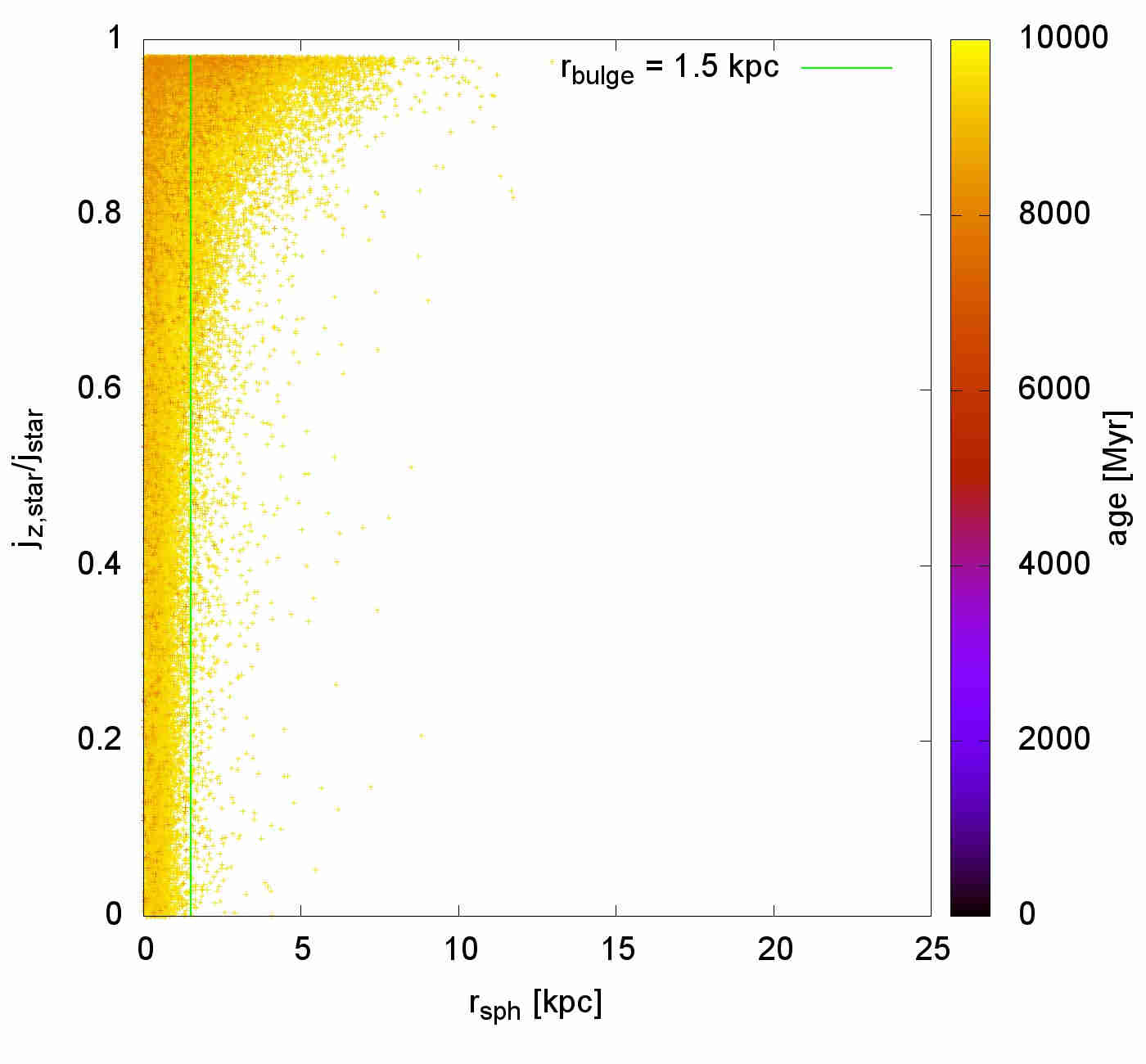}
\end{minipage}
\caption{As Fig. \ref{fig:jz/jM1} but for M1const.} 
\label{fig:jz/jM1const}
\end{figure*}
\newpage
\begin{figure*}[h]
\begin{minipage}[t]{0.49\linewidth}
\includegraphics[width=1.0\linewidth]{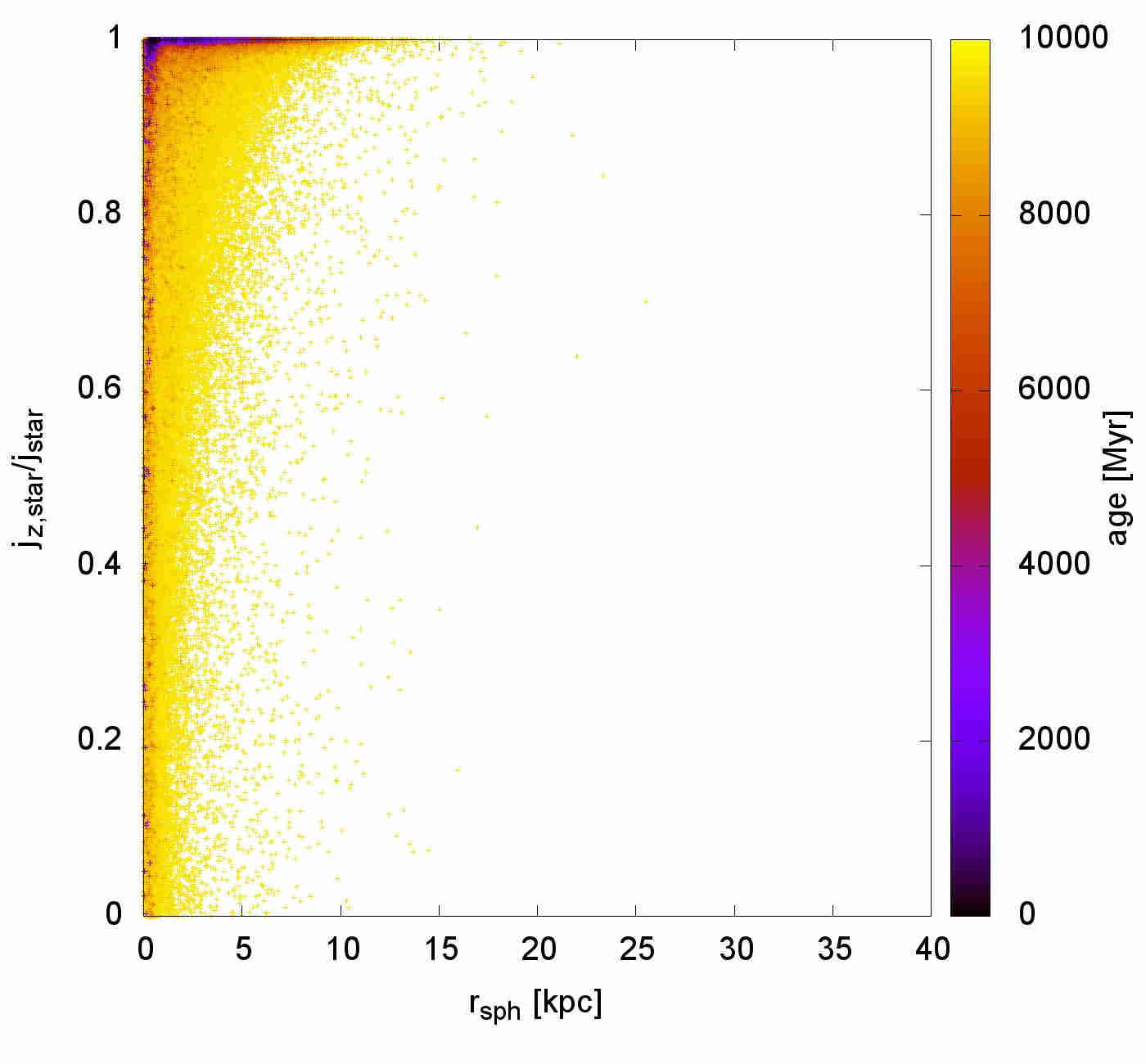}
\end{minipage}
\begin{minipage}[t]{0.49\linewidth}
\includegraphics[width=1.0\linewidth]{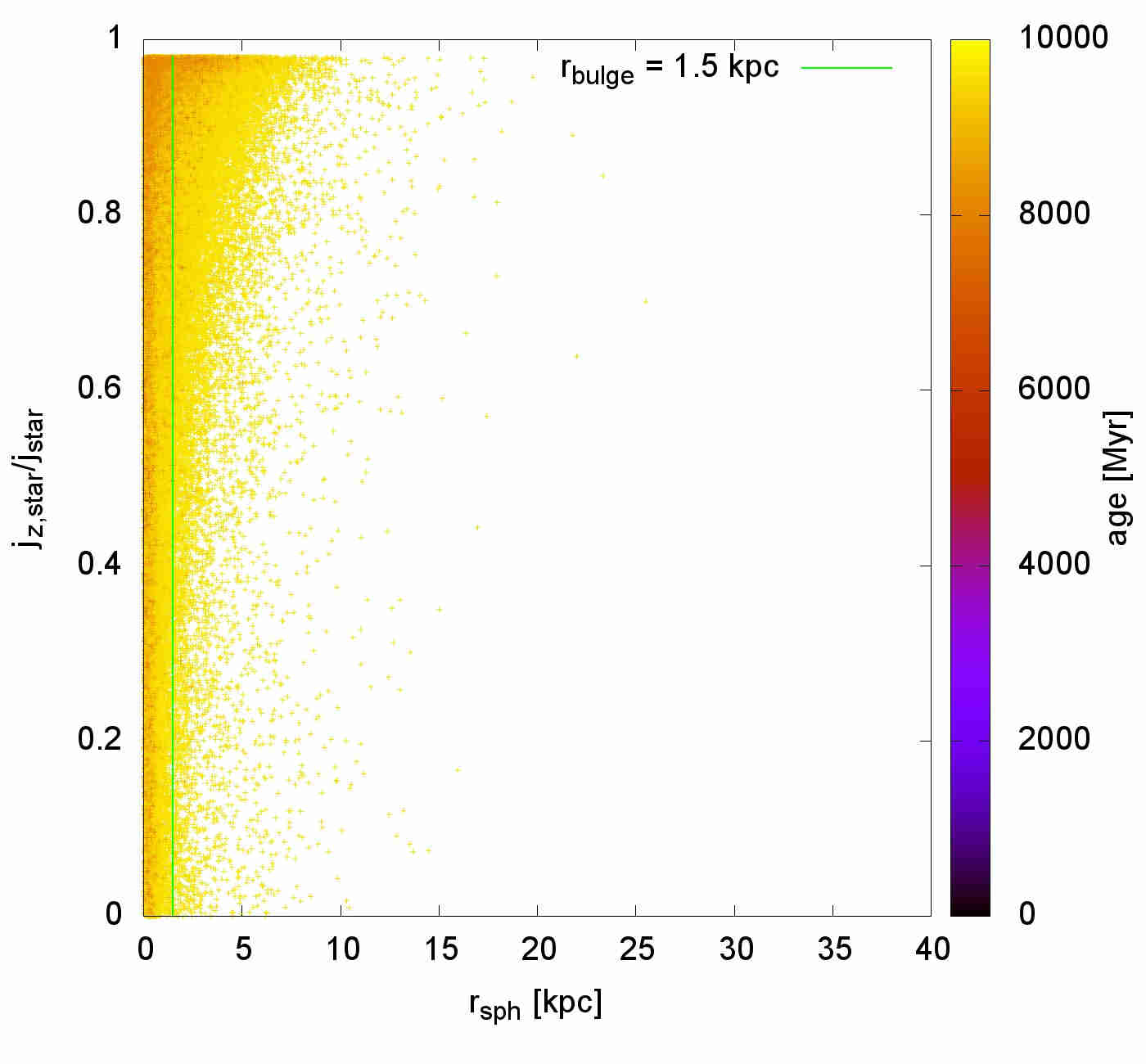}
\end{minipage}
\caption{As Fig. \ref{fig:jz/jM1} but for M2.} 
\label{fig:jz/jM2}
\end{figure*}
\begin{figure*}[h]
\begin{minipage}[t]{0.49\linewidth}
\includegraphics[width=1.0\linewidth]{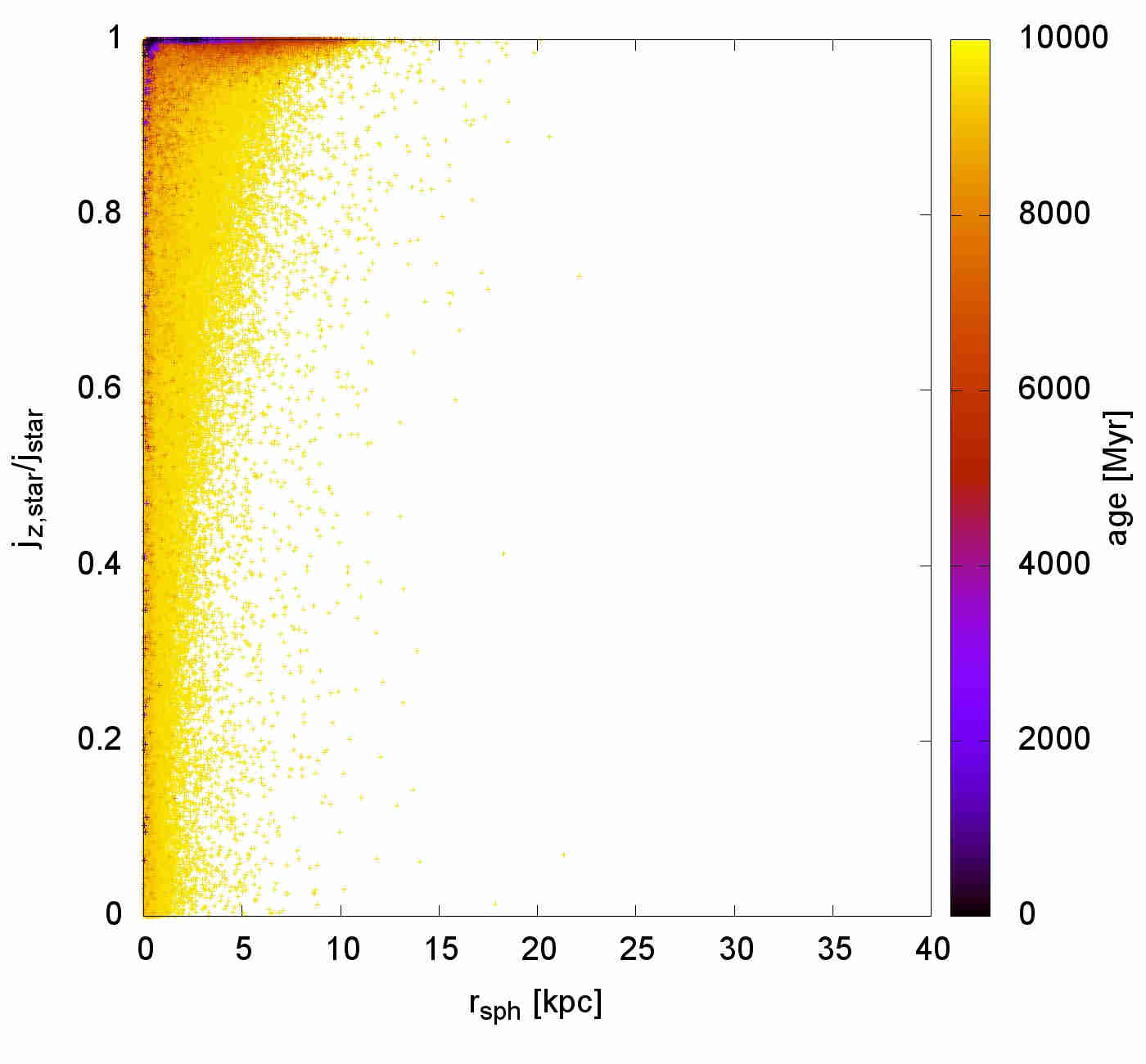}
\end{minipage}
\begin{minipage}[t]{0.49\linewidth}
\includegraphics[width=1.0\linewidth]{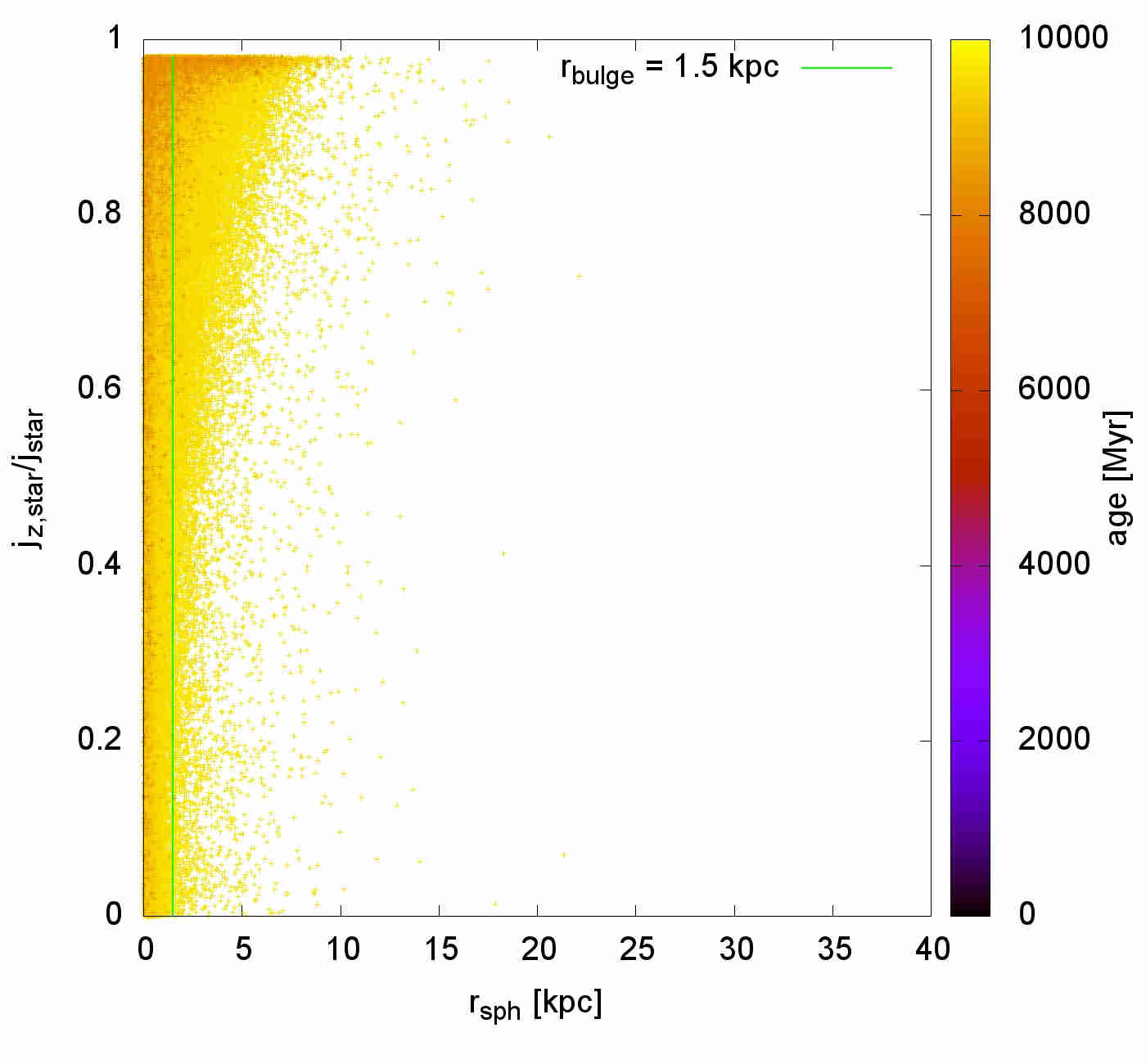}
\end{minipage}
\caption{As Fig. \ref{fig:jz/jM1} but for M2sn.} 
\label{fig:jz/jM2sn}
\end{figure*}
\newpage
\begin{figure*}[h]
\begin{minipage}[t]{0.49\linewidth}
\includegraphics[width=1.0\linewidth]{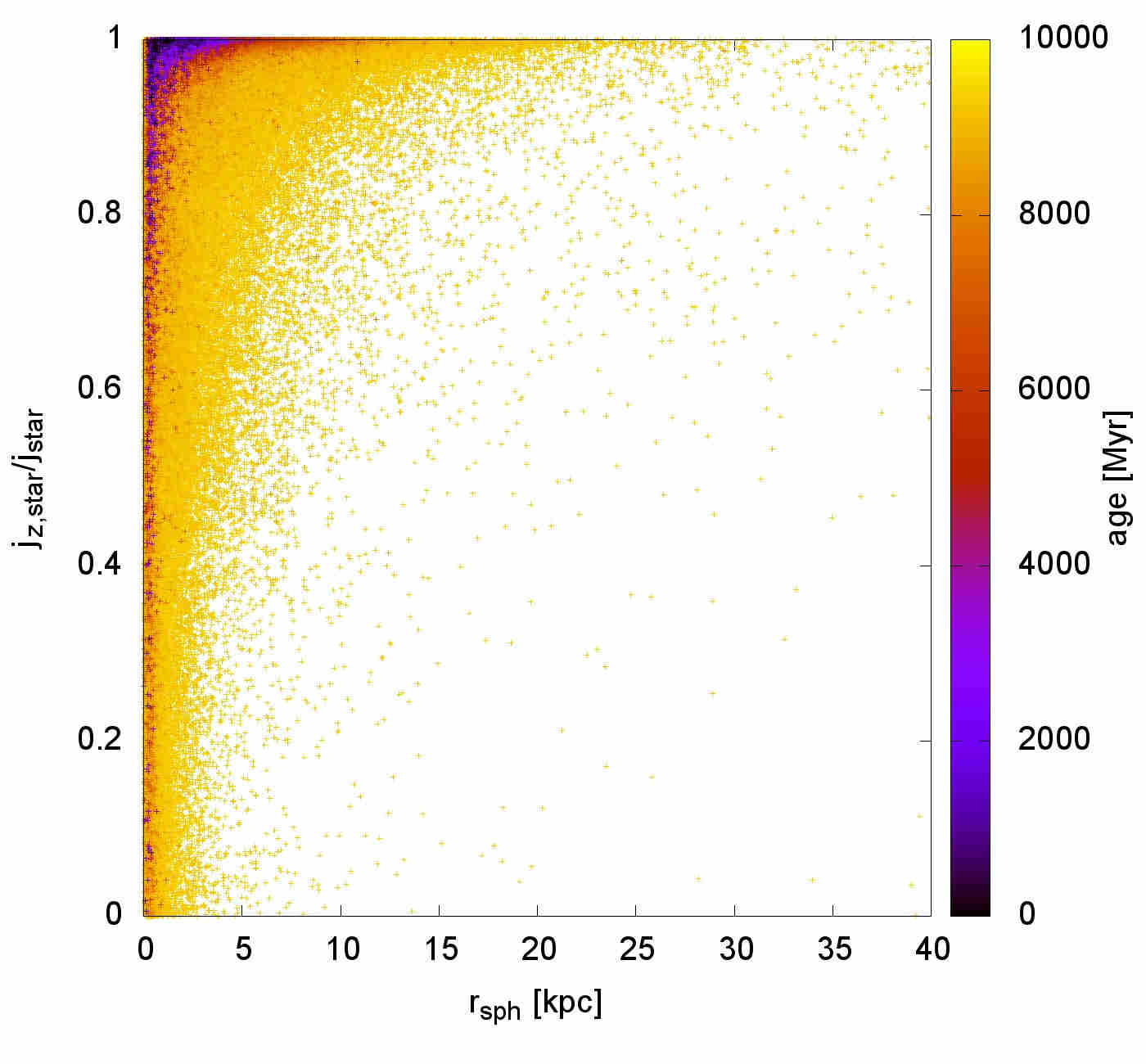}
\end{minipage}
\begin{minipage}[t]{0.49\linewidth}
\includegraphics[width=1.0\linewidth]{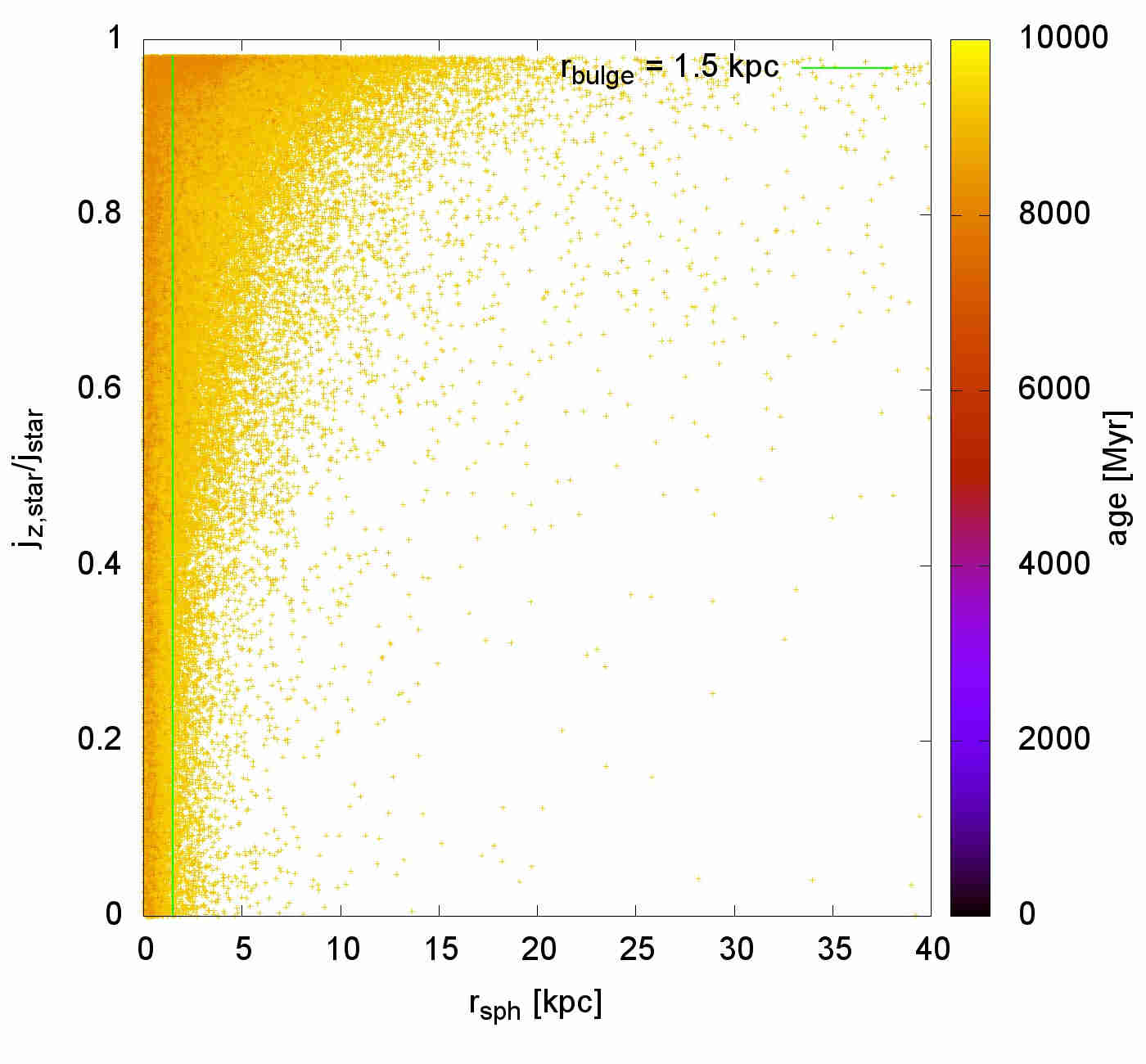}
\end{minipage}
\caption{As Fig. \ref{fig:jz/jM1} but for M2N.} 
\label{fig:jz/jM2N}
\end{figure*}
\begin{figure*}[h]
\begin{minipage}[t]{0.49\linewidth}
\includegraphics[width=1.0\linewidth]{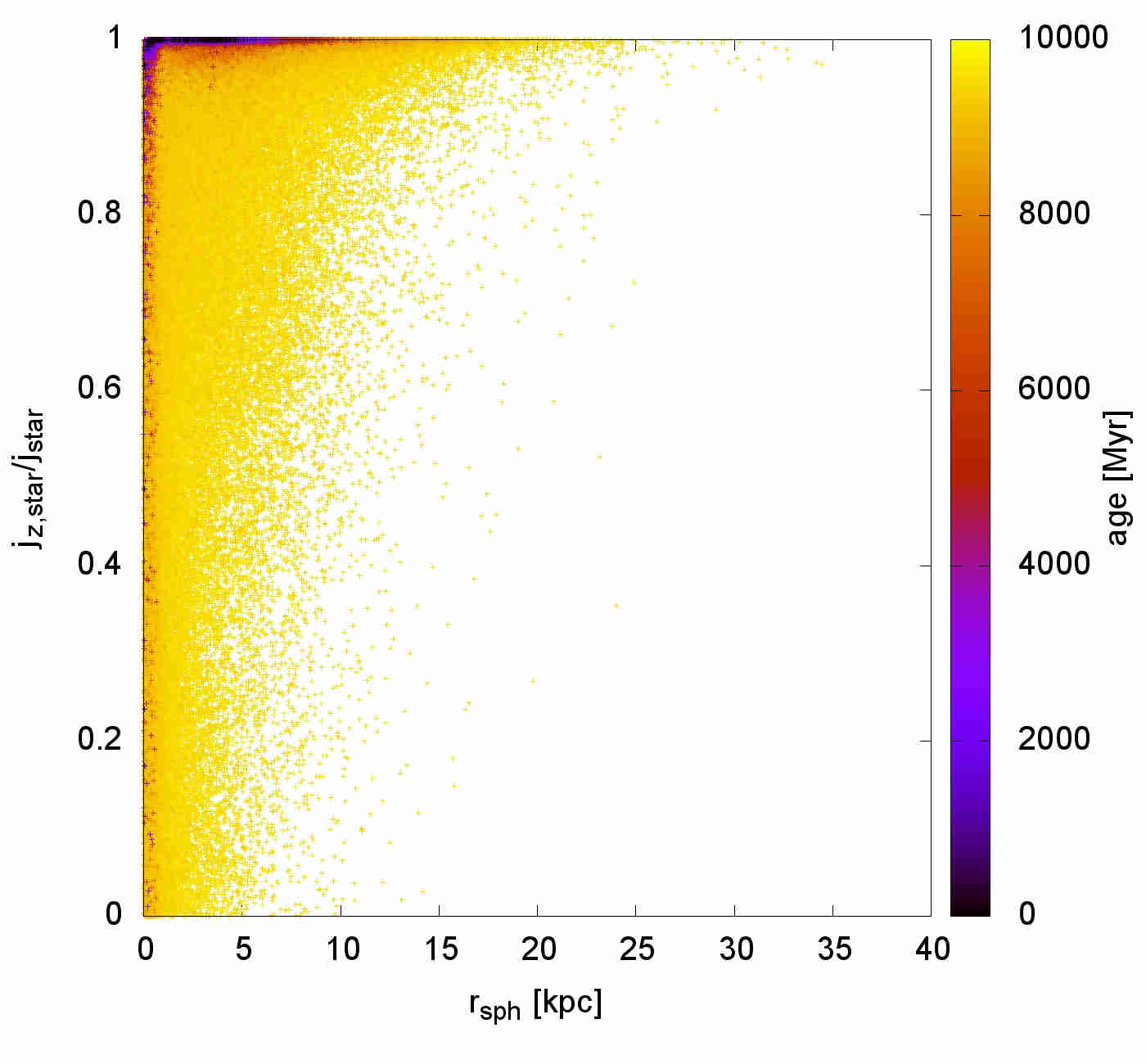}
\end{minipage}
\begin{minipage}[t]{0.49\linewidth}
\includegraphics[width=1.0\linewidth]{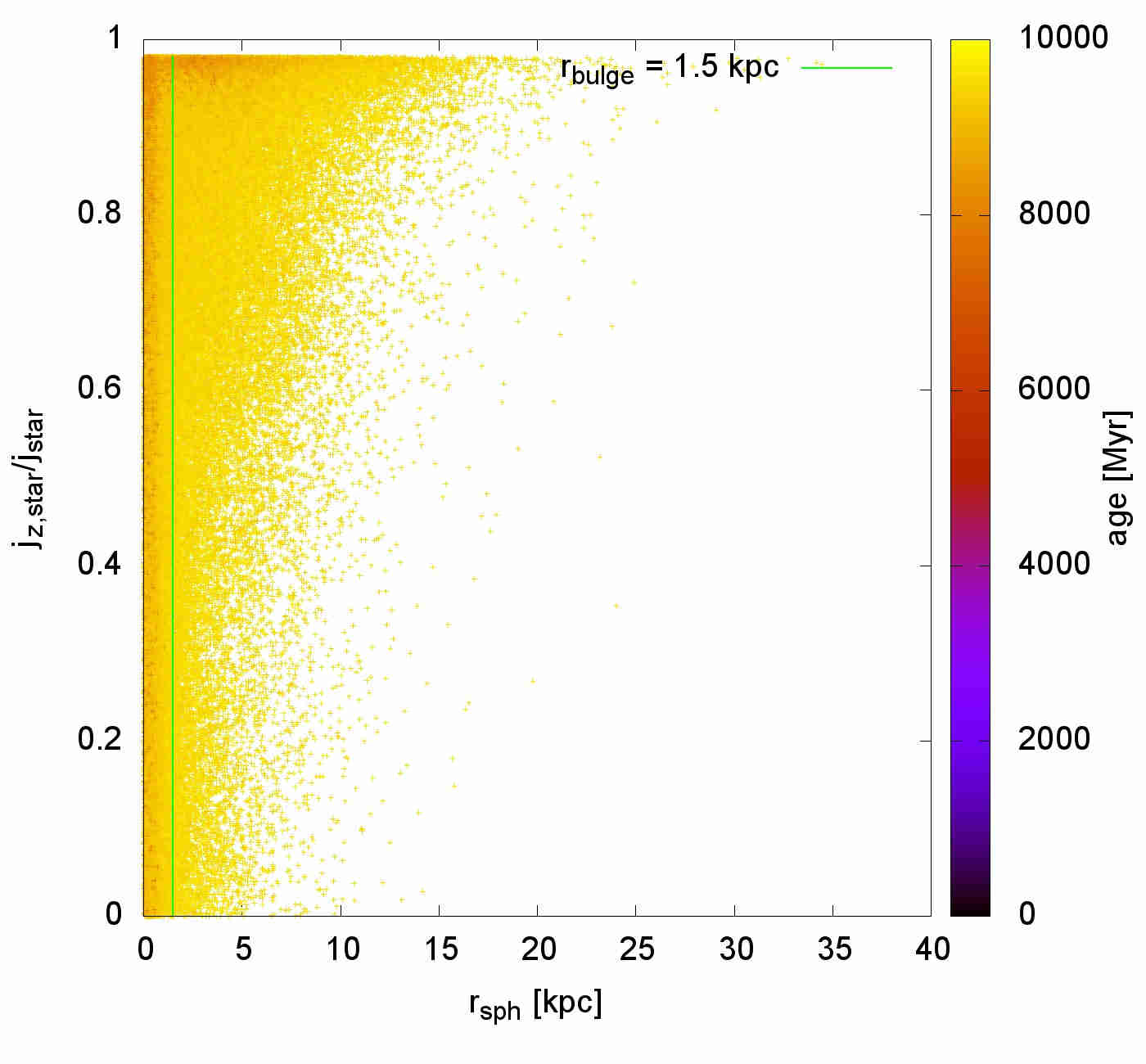}
\end{minipage}
\caption{As Fig. \ref{fig:jz/jM1} but for M3.} 
\label{fig:jz/jM3}
\end{figure*}
\newpage
\begin{figure*}[h]
\begin{minipage}[t]{0.49\linewidth}
\includegraphics[width=1.0\linewidth]{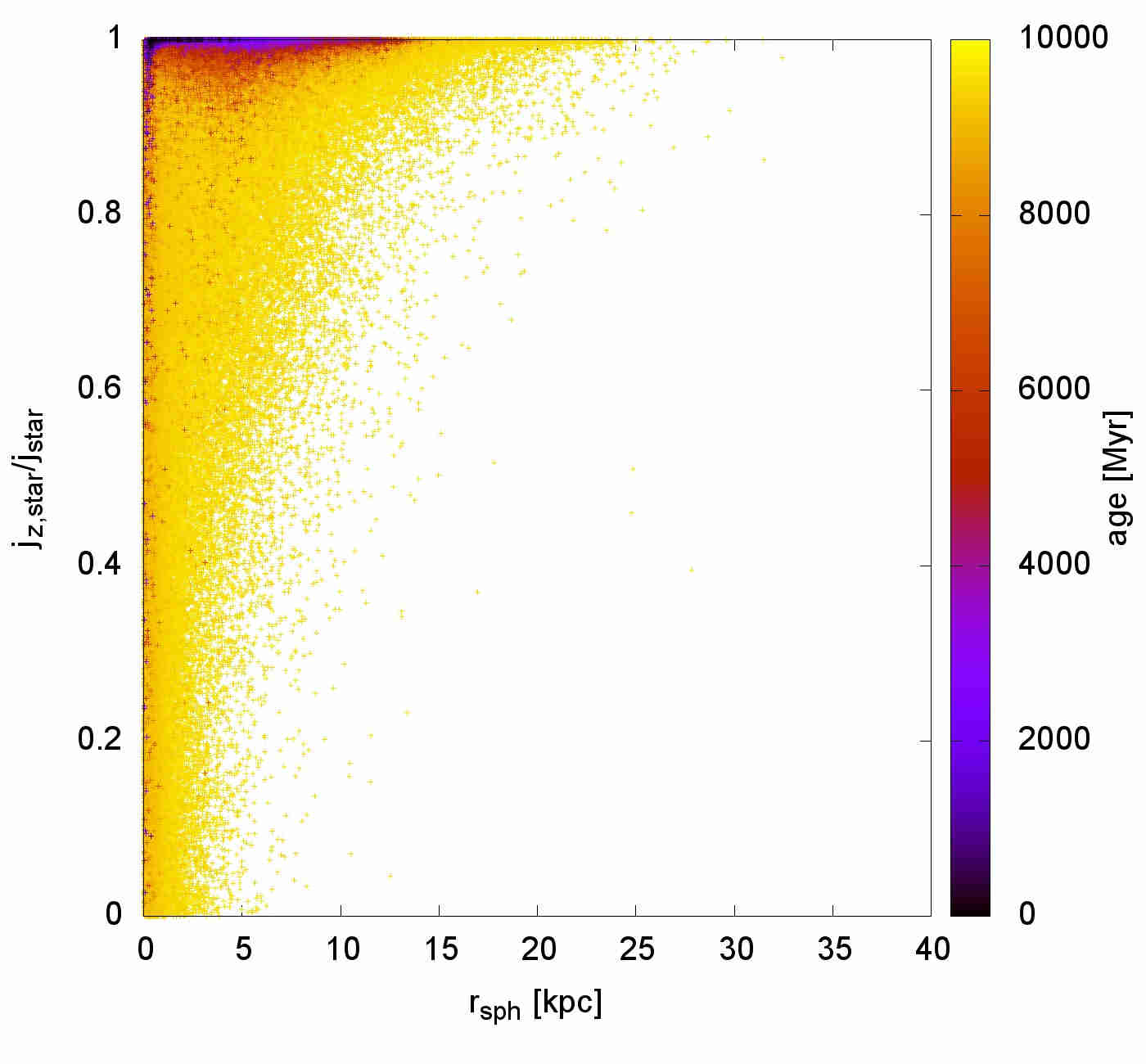}
\end{minipage}
\begin{minipage}[t]{0.49\linewidth}
\includegraphics[width=1.0\linewidth]{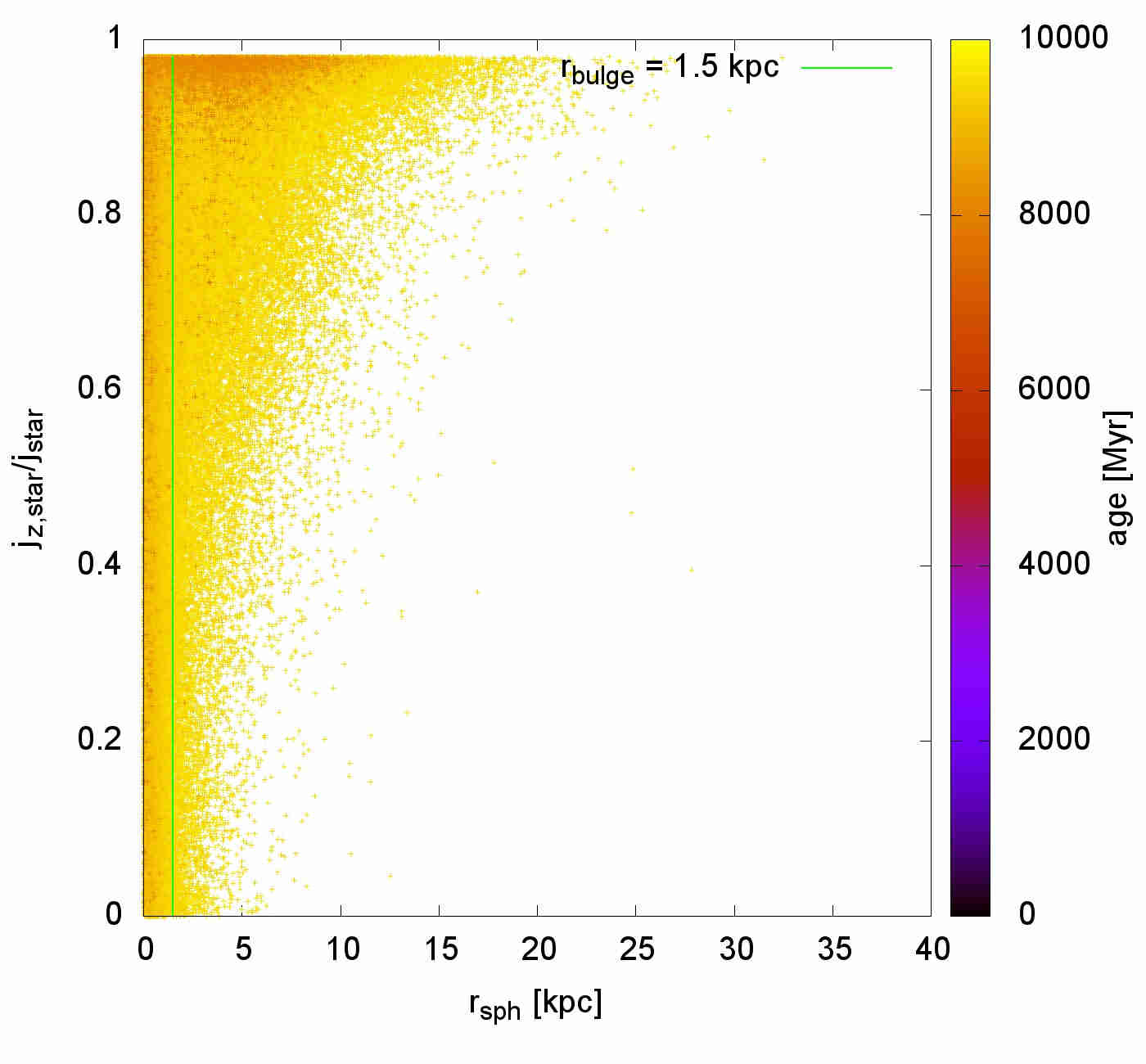}
\end{minipage}
\caption{As Fig. \ref{fig:jz/jM1} but for M3sn.} 
\label{fig:jz/jM3sn}
\end{figure*}
\begin{figure*}[h]
\begin{minipage}[t]{0.49\linewidth}
\includegraphics[width=1.0\linewidth]{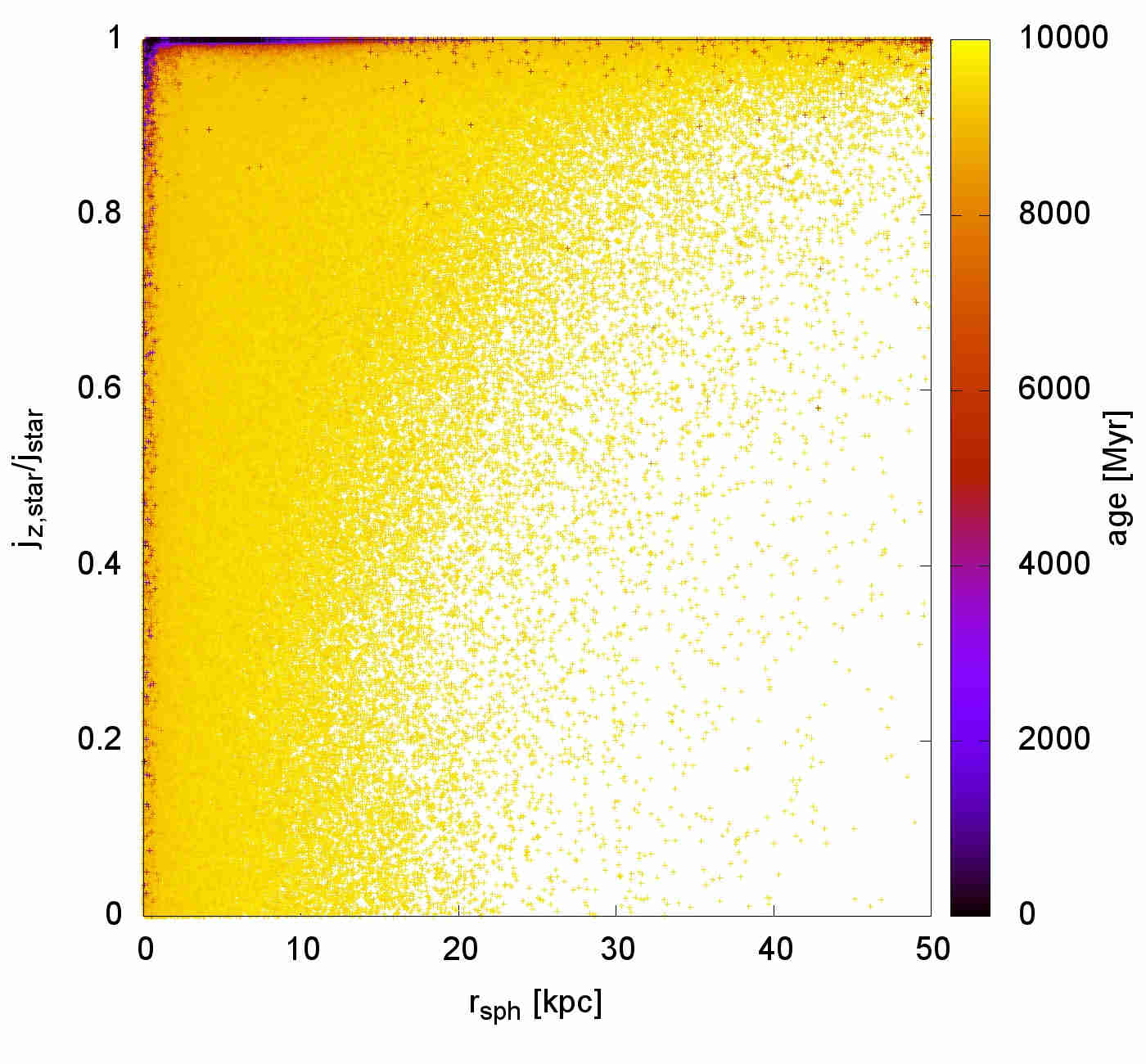}
\end{minipage}
\begin{minipage}[t]{0.49\linewidth}
\includegraphics[width=1.0\linewidth]{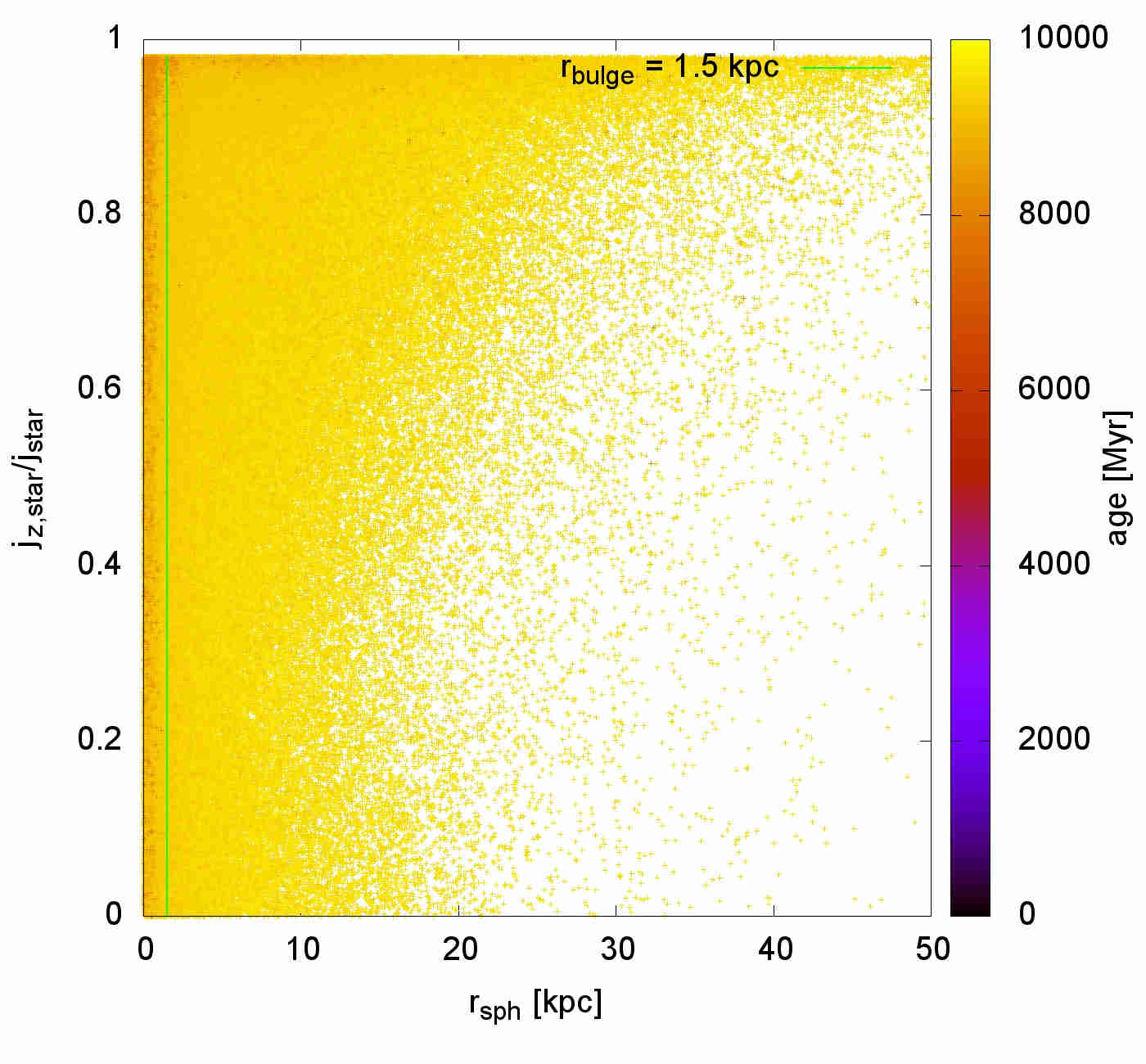}
\end{minipage}
\caption{As Fig. \ref{fig:jz/jM1} but for M4.} 
\label{fig:jz/jM4}
\end{figure*}
\newpage
\begin{figure*}[h]
\begin{minipage}[t]{0.49\linewidth}
\includegraphics[width=1.0\linewidth]{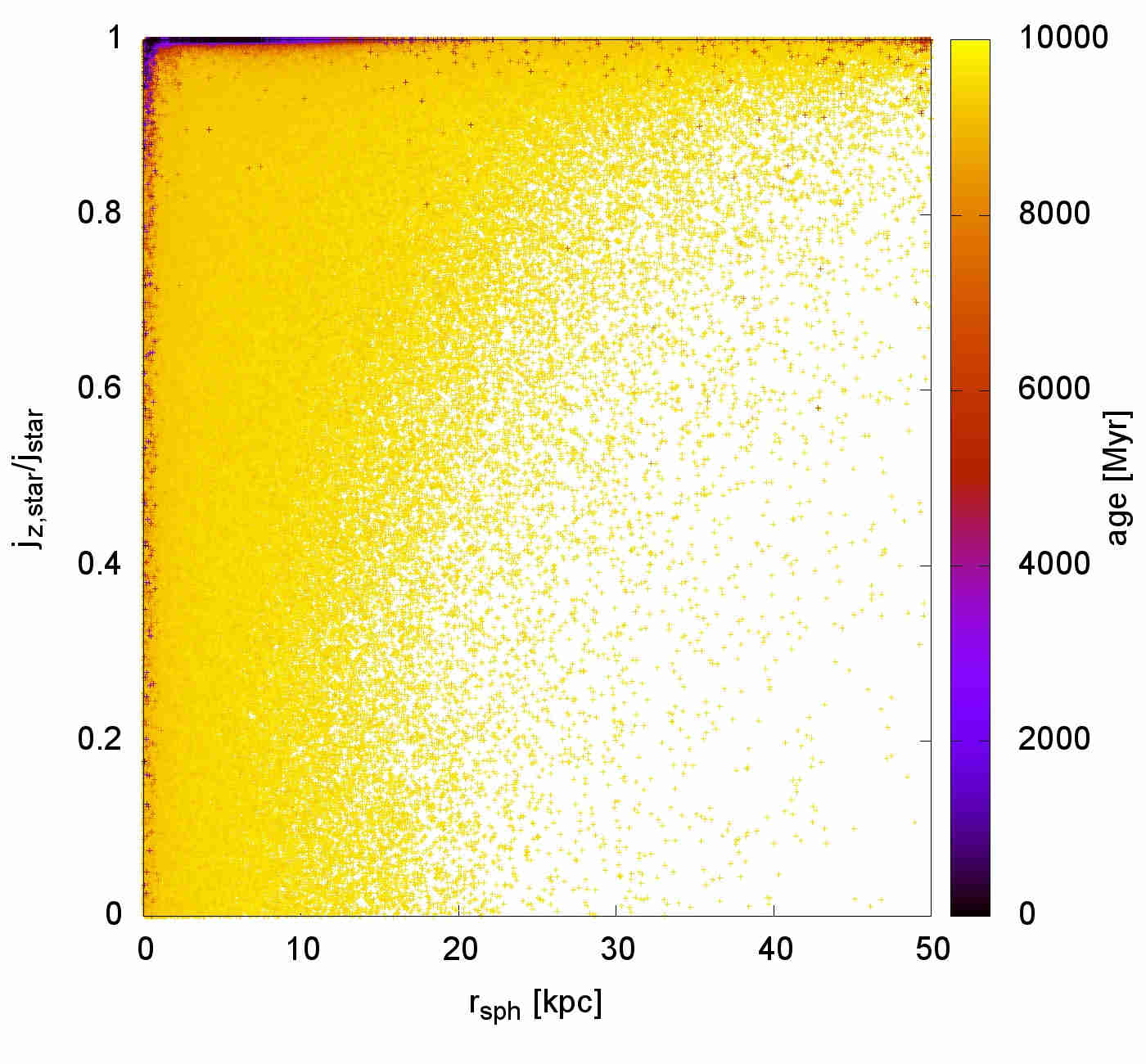}
\end{minipage}
\begin{minipage}[t]{0.49\linewidth}
\includegraphics[width=1.0\linewidth]{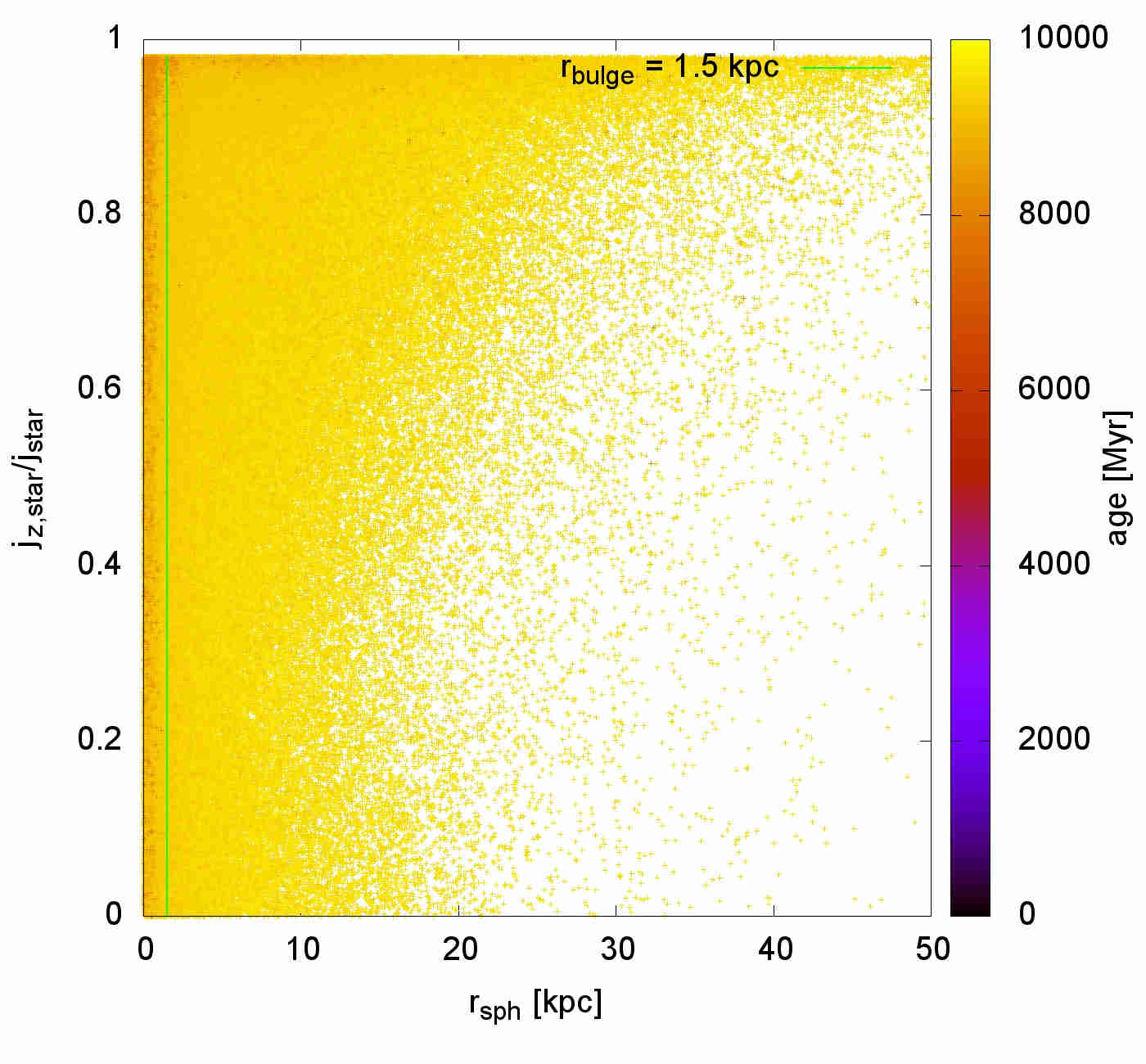}
\end{minipage}
\caption{As Fig. \ref{fig:jz/jM1} but for M4sn.} 
\label{fig:jz/jM4sn}
\end{figure*}
\begin{figure*}[h]
\begin{minipage}[t]{0.49\linewidth}
\includegraphics[width=1.0\linewidth]{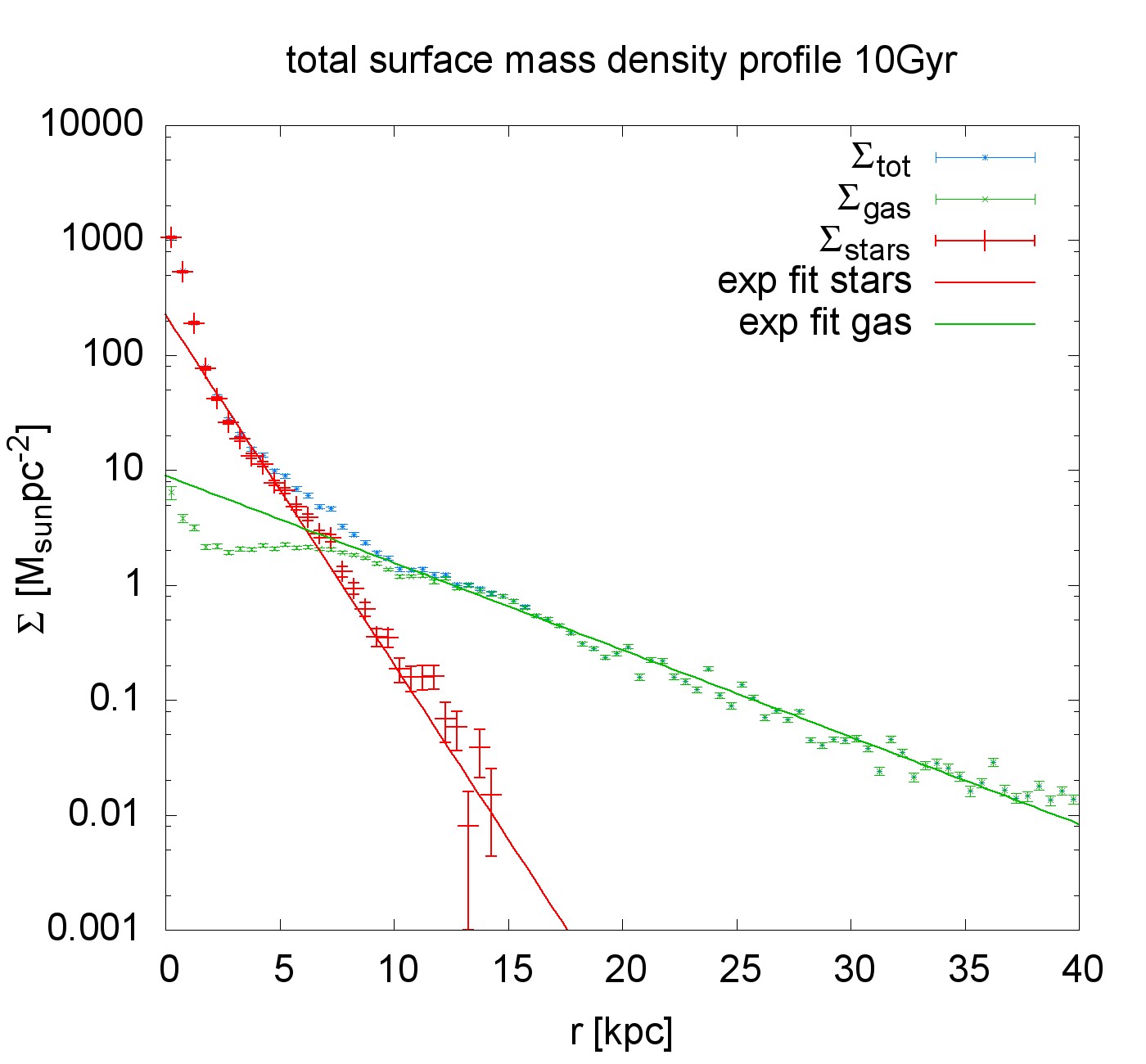}
\end{minipage}
\begin{minipage}[t]{0.49\linewidth}
\includegraphics[width=1.0\linewidth]{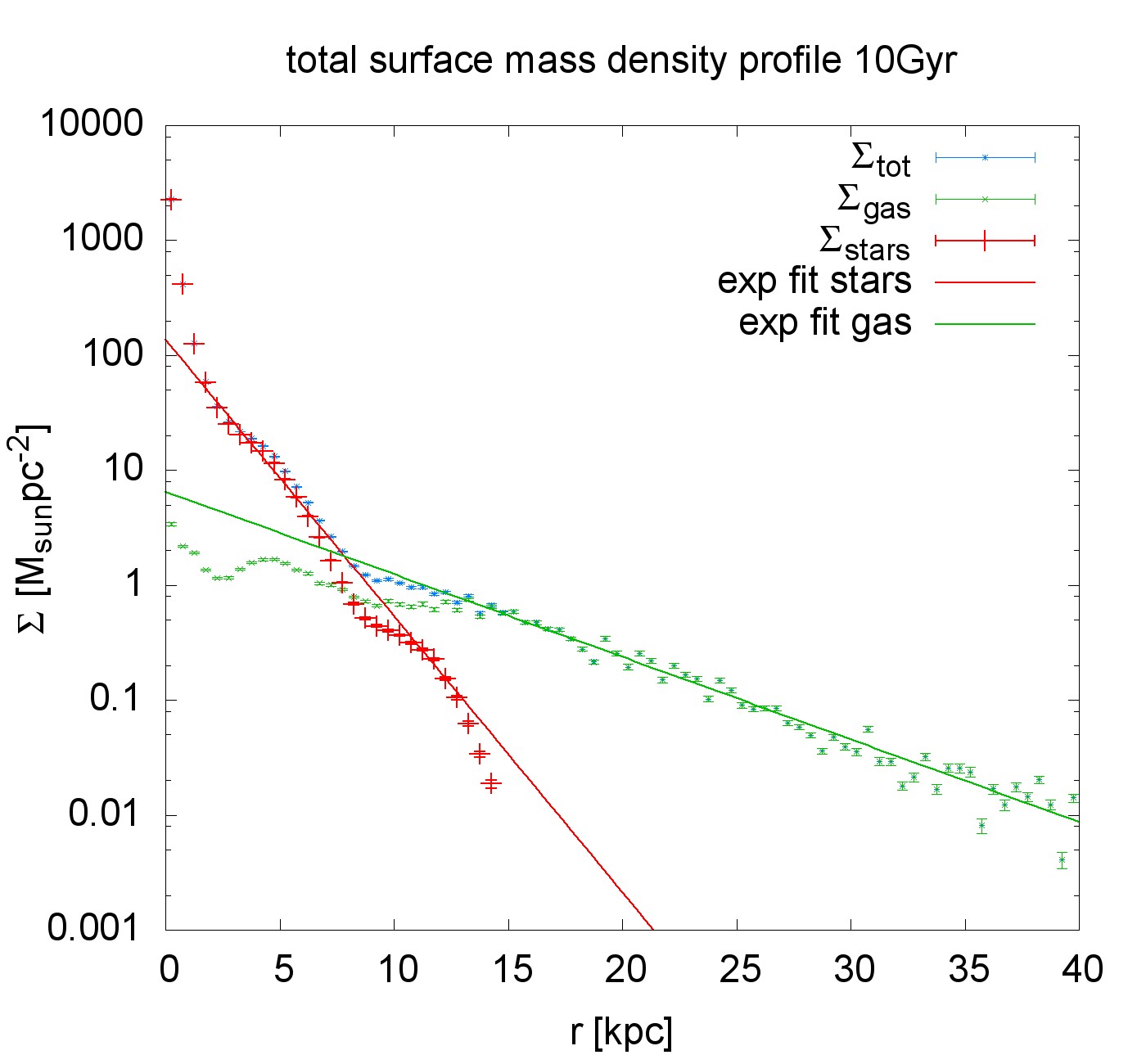}
\end{minipage}
\caption{as Fig. \ref{fig:ddcomparisonr2010Gyr} but for M1l11 and M1l13.}
\label{fig:ddcomparisonl11l13}
\end{figure*}
\newpage
\begin{figure*}[h]
\begin{minipage}[t]{0.49\linewidth}
\includegraphics[width=1.0\linewidth]{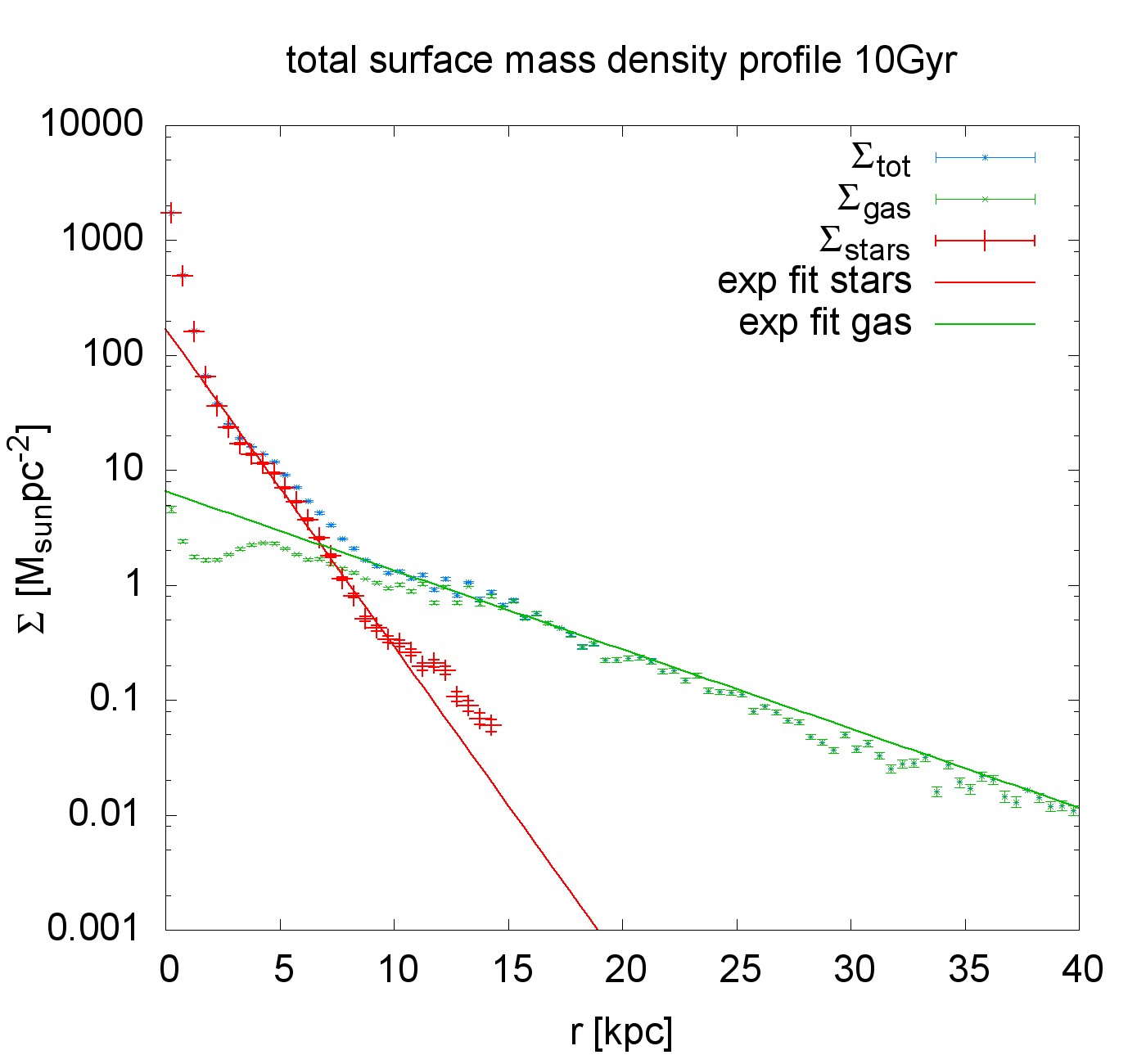}
\end{minipage}
\begin{minipage}[t]{0.49\linewidth}
\includegraphics[width=1.0\linewidth]{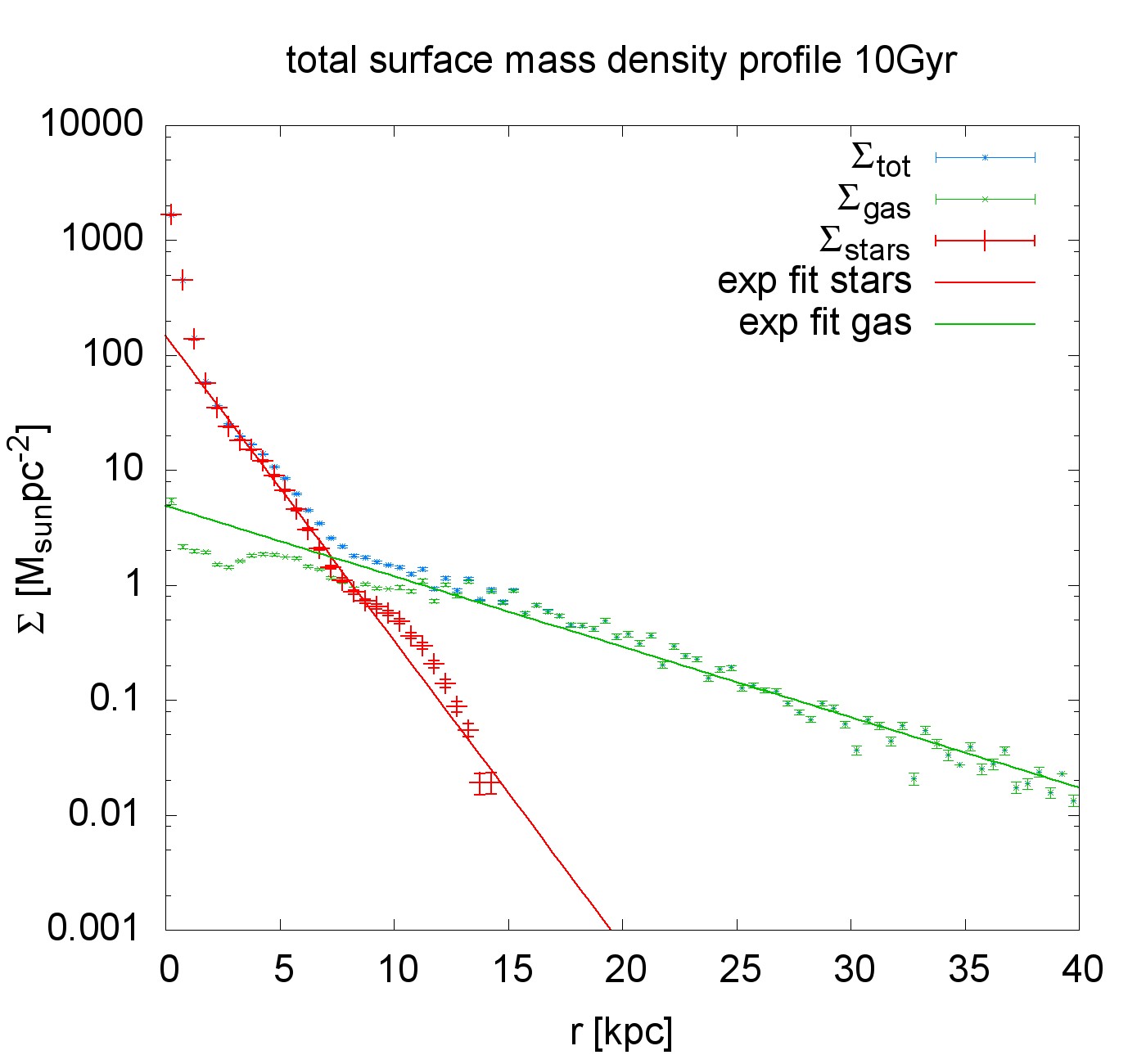}
\end{minipage}
\caption{as Fig. \ref{fig:ddcomparisonr2010Gyr} but for M1Zpoor and M1Zpoorsn.}
\label{fig:ddcomparisonZpoorZpoorsn}
\end{figure*}
\begin{figure*}[h]
\begin{minipage}[t]{0.49\linewidth}
\includegraphics[width=1.0\linewidth]{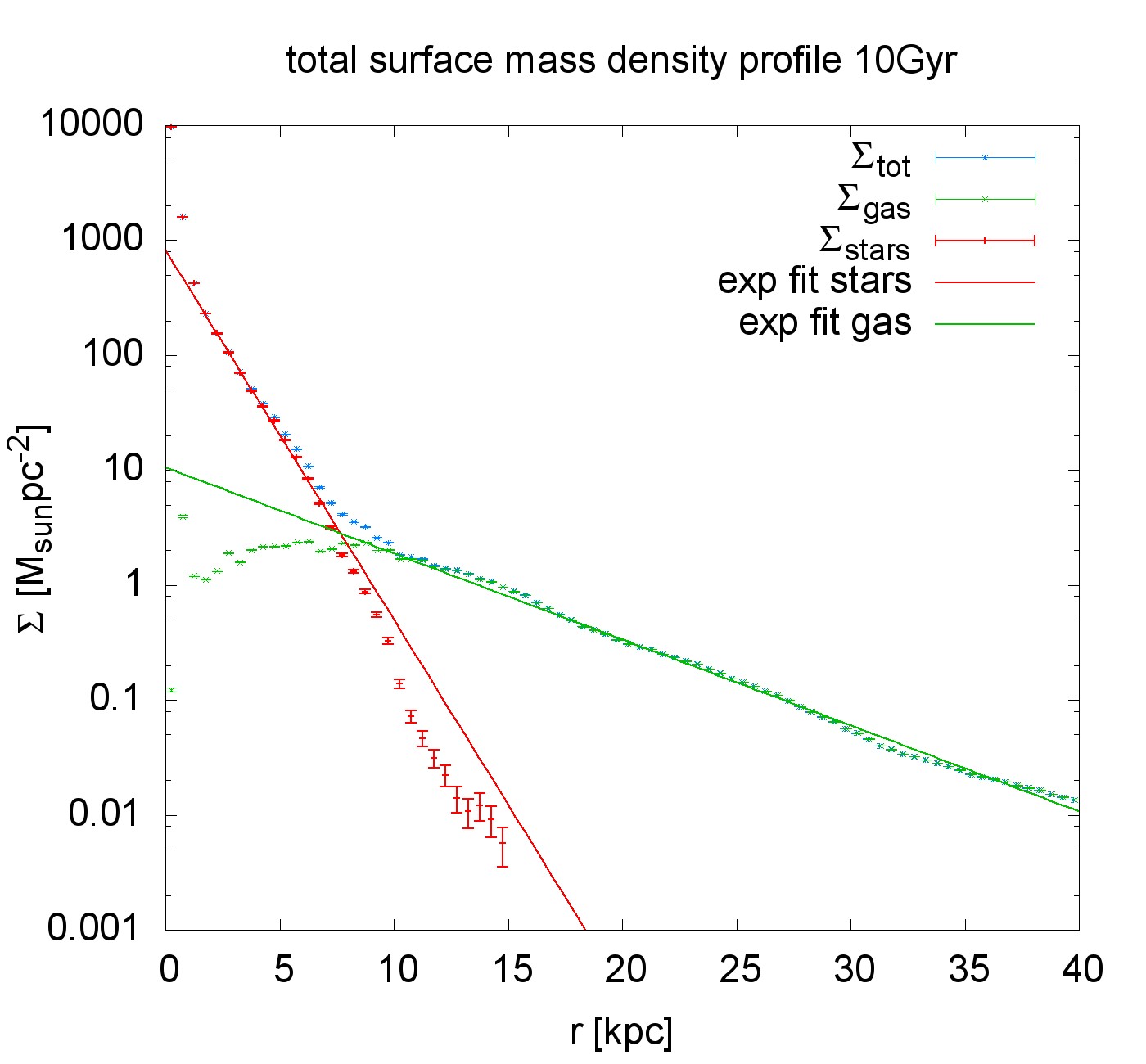}
\end{minipage}
\begin{minipage}[t]{0.49\linewidth}
\includegraphics[width=1.0\linewidth]{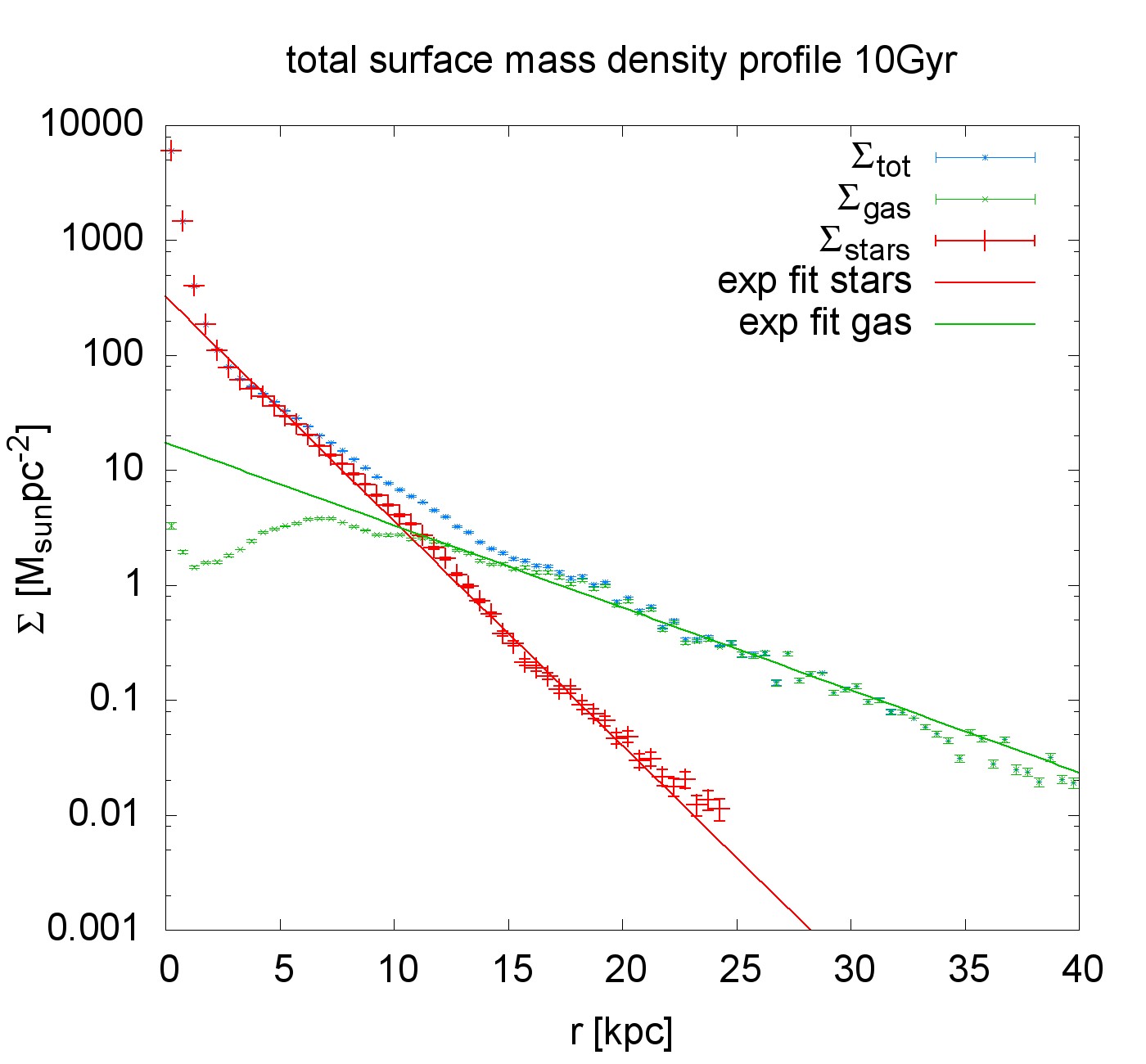}
\end{minipage}
\caption{as Fig. \ref{fig:ddcomparisonr2010Gyr} but for M2sn and M3sn.}
\label{fig:ddcomparisonM2snM3sn}
\end{figure*}
\newpage
\begin{figure*}[h]
\begin{minipage}[t]{0.49\linewidth}
\includegraphics[width=1.0\linewidth]{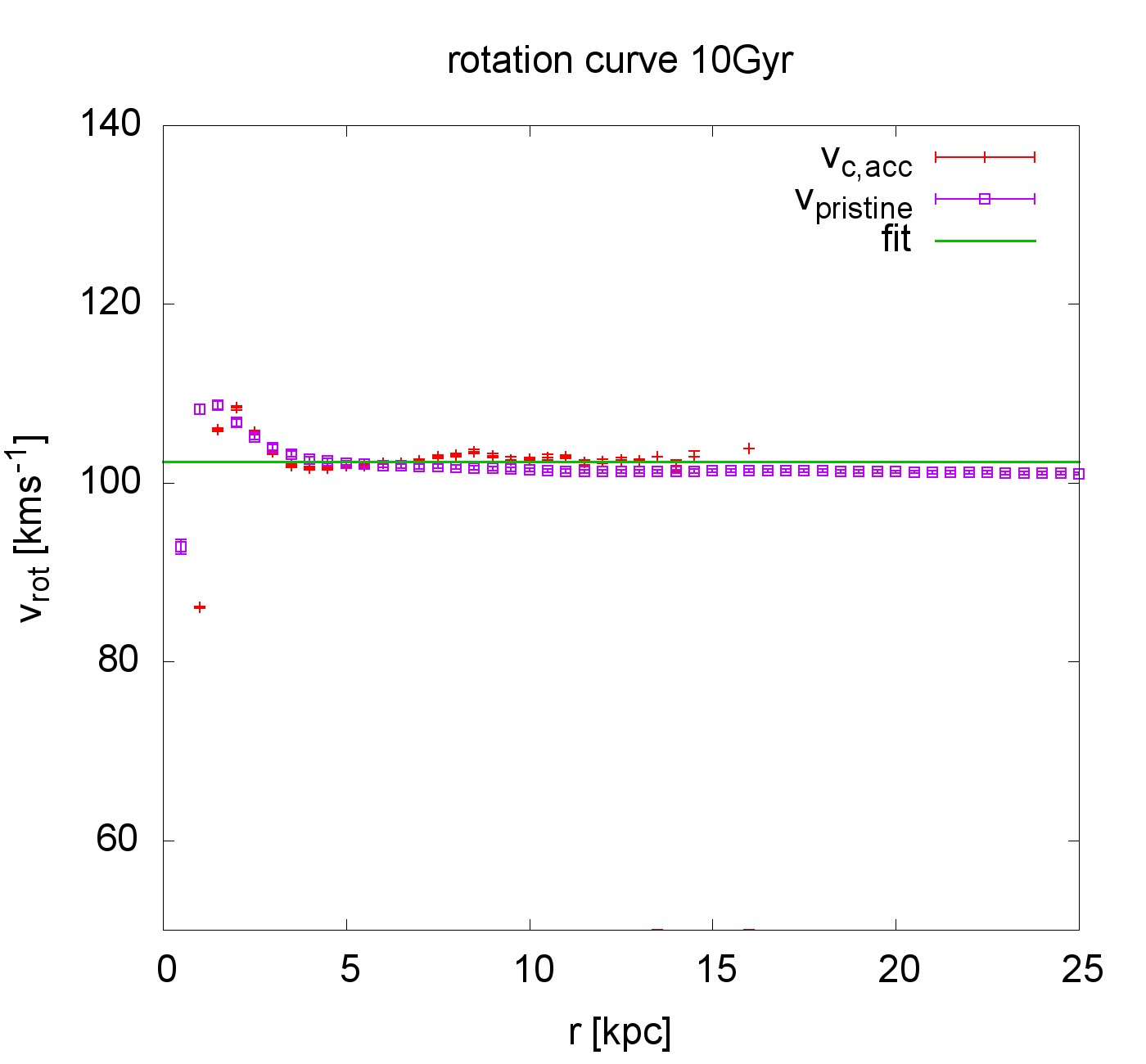}
\end{minipage}
\begin{minipage}[t]{0.49\linewidth}
\includegraphics[width=1.0\linewidth]{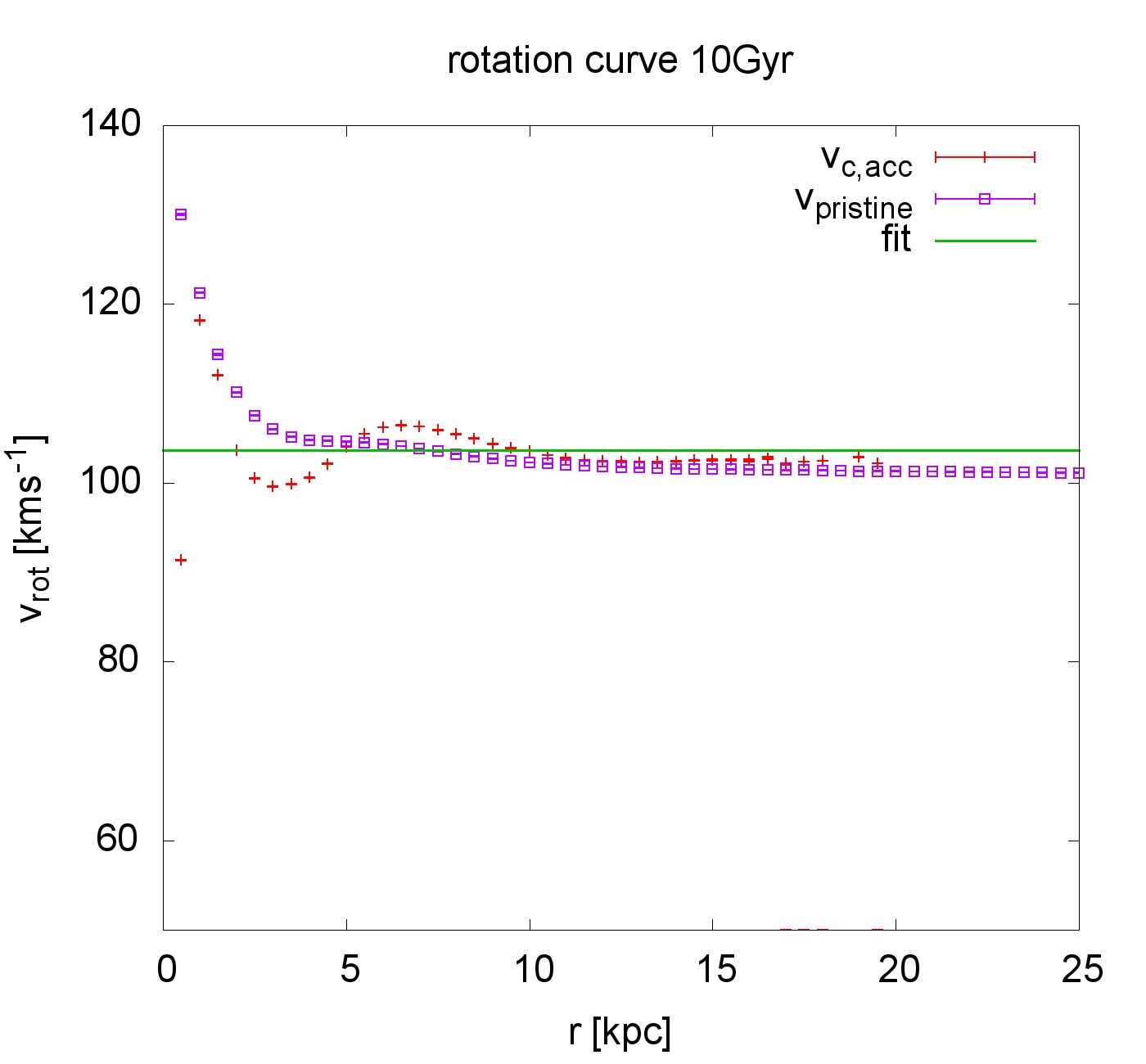}
\end{minipage}
\caption{as Fig. \ref{fig:rccomparisonr2010Gyr} but for M1l11 and M1l13.}
\label{fig:rccomparisonl11l13}
\end{figure*}
\begin{figure*}[h]
\begin{minipage}[t]{0.49\linewidth}
\includegraphics[width=1.0\linewidth]{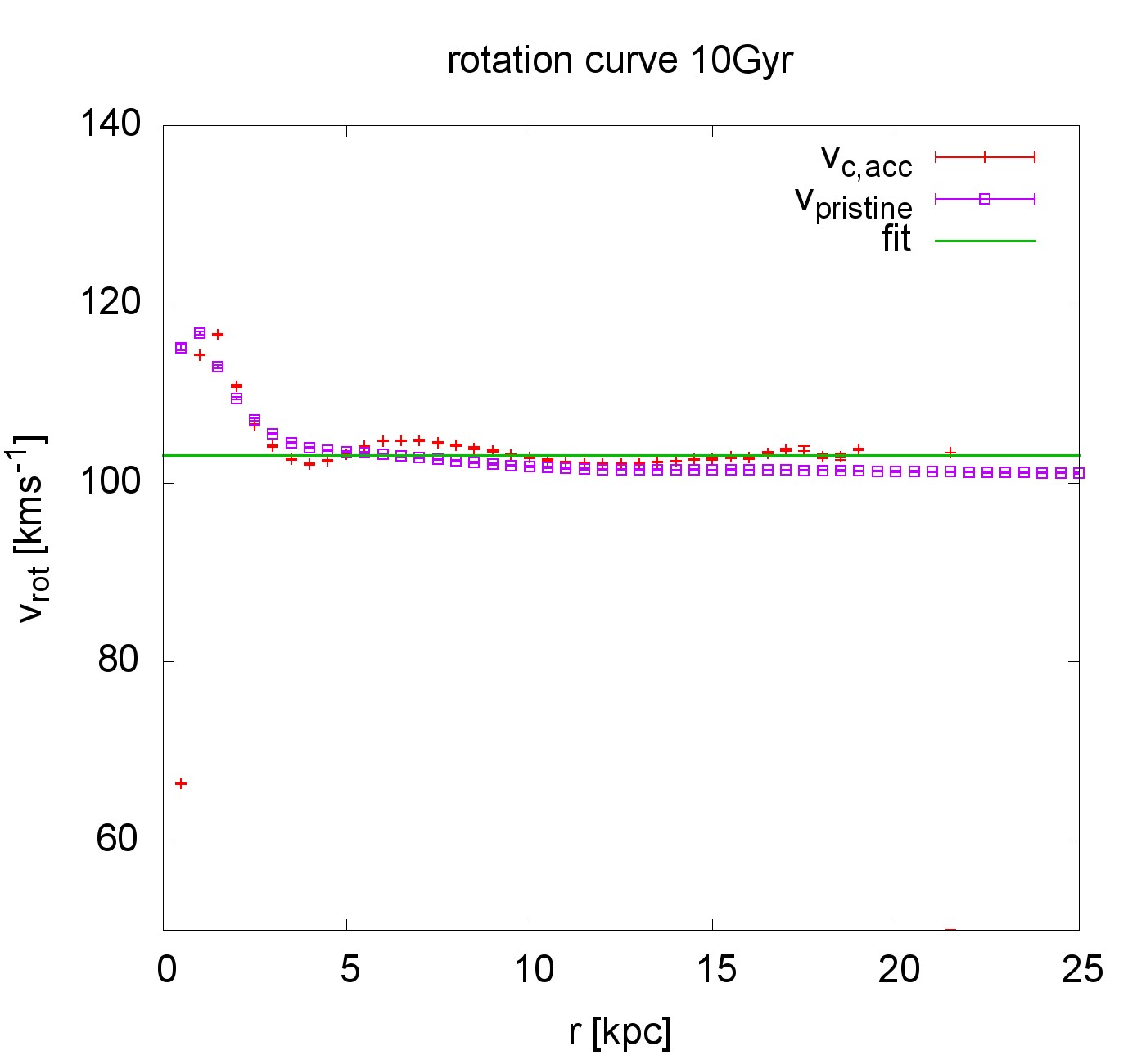}
\end{minipage}
\begin{minipage}[t]{0.49\linewidth}
\includegraphics[width=1.0\linewidth]{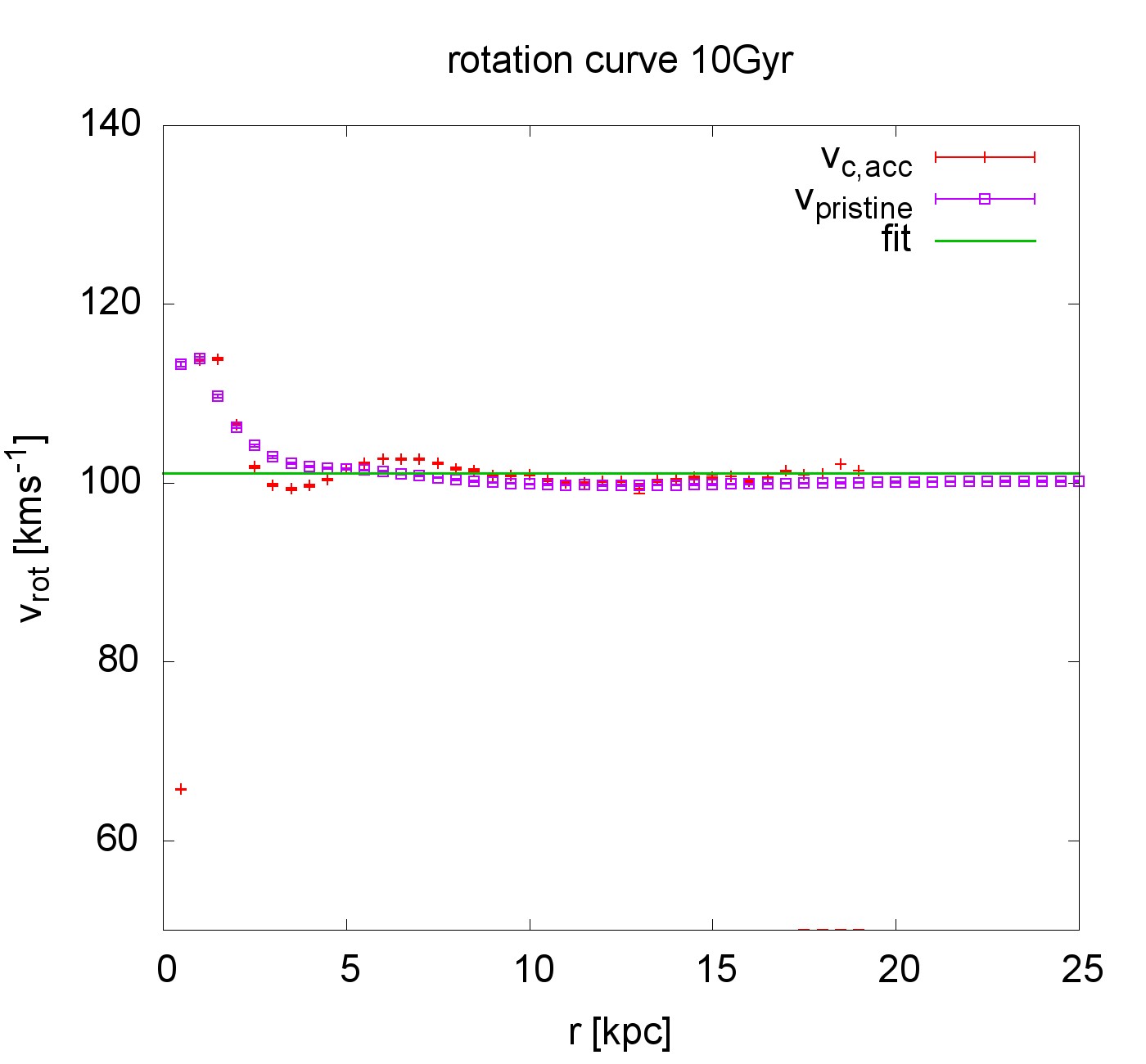}
\end{minipage}
\caption{as Fig. \ref{fig:rccomparisonr2010Gyr} but for M1Zpoor and M1Zpoorsn.}
\label{fig:rccomparisonZpoorZpoorsn}
\end{figure*}
\newpage
\begin{figure*}[h]
\begin{minipage}[t]{0.49\linewidth}
\includegraphics[width=1.0\linewidth]{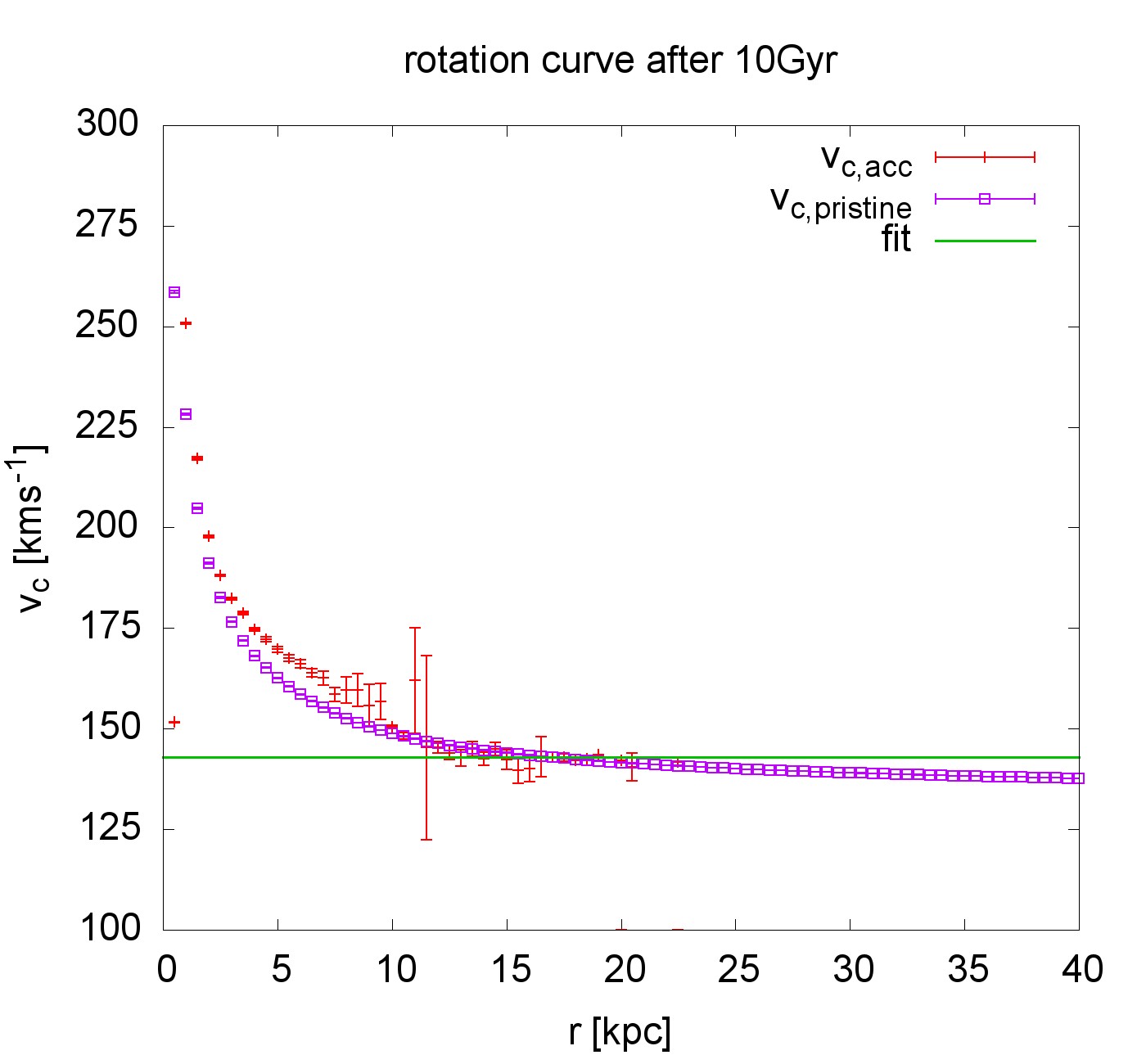}
\end{minipage}
\begin{minipage}[t]{0.49\linewidth}
\includegraphics[width=1.0\linewidth]{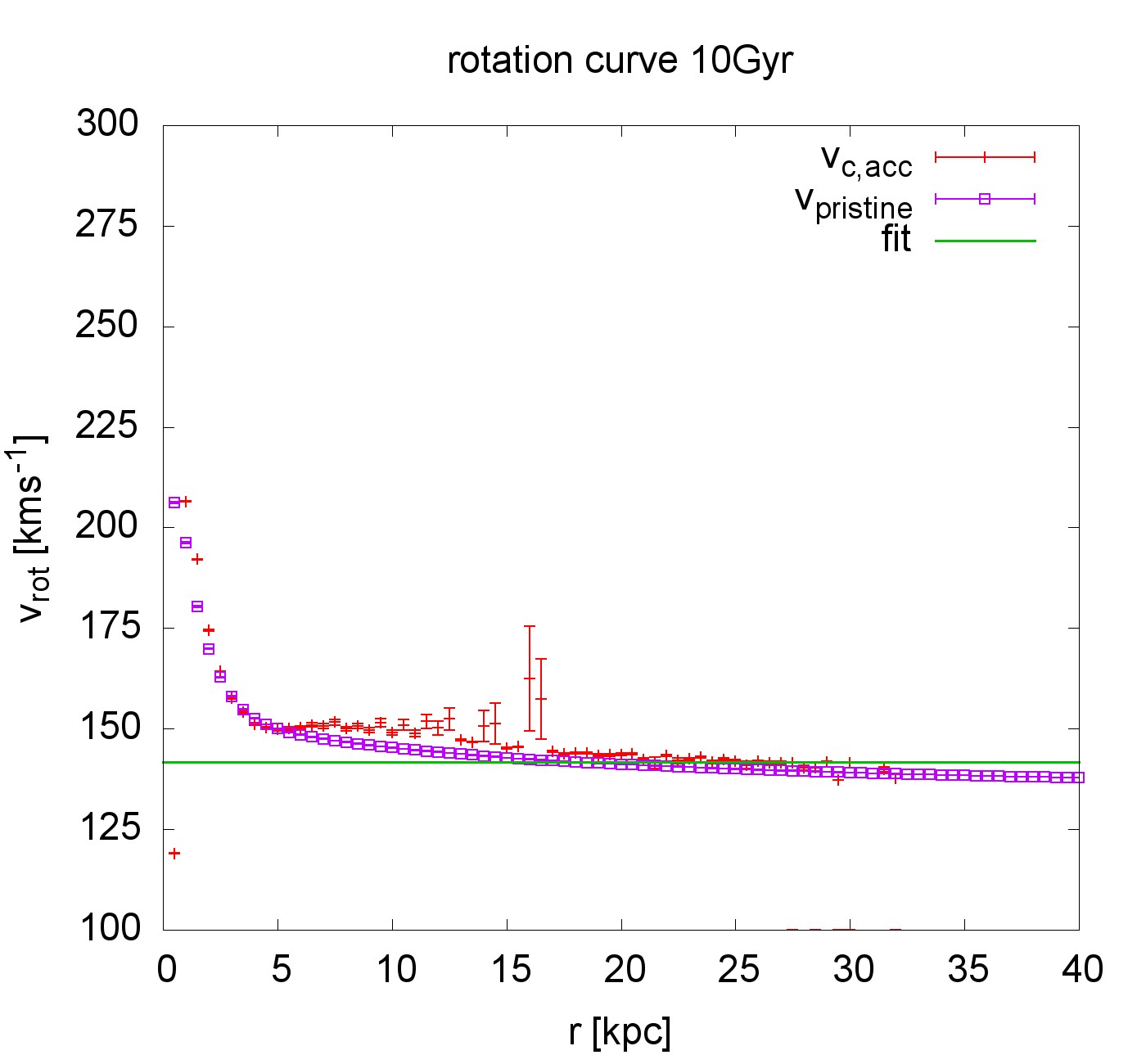}
\end{minipage}
\caption{as Fig. \ref{fig:rccomparisonr2010Gyr} but for M2sn and M3sn.}
\label{fig:rccomparisonM2snM3sn}
\end{figure*}
\begin{figure*}[h]
\begin{minipage}[t]{0.49\linewidth}
\includegraphics[width=1.0\linewidth]{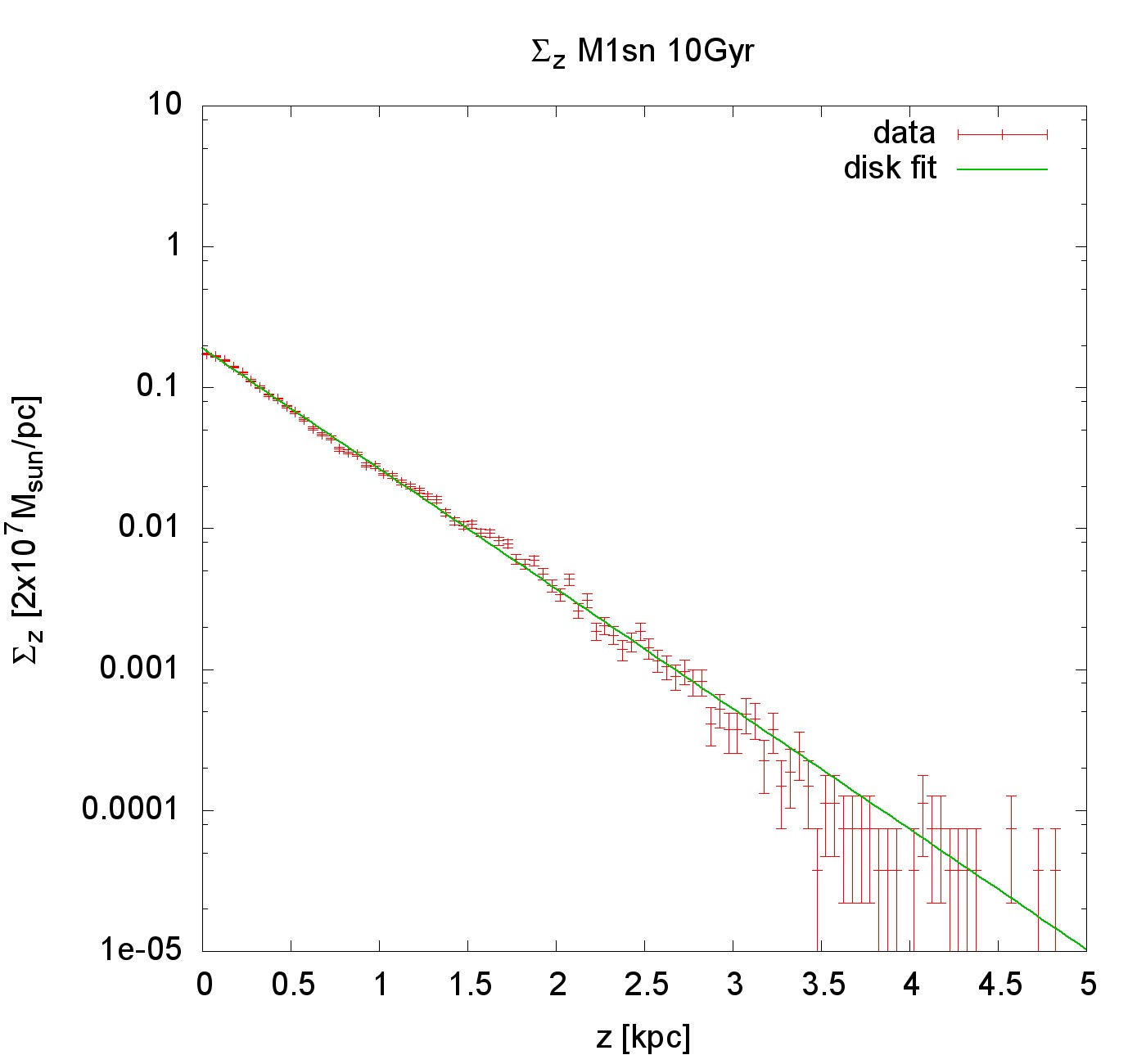}
\end{minipage}
\begin{minipage}[t]{0.49\linewidth}
\includegraphics[width=1.0\linewidth]{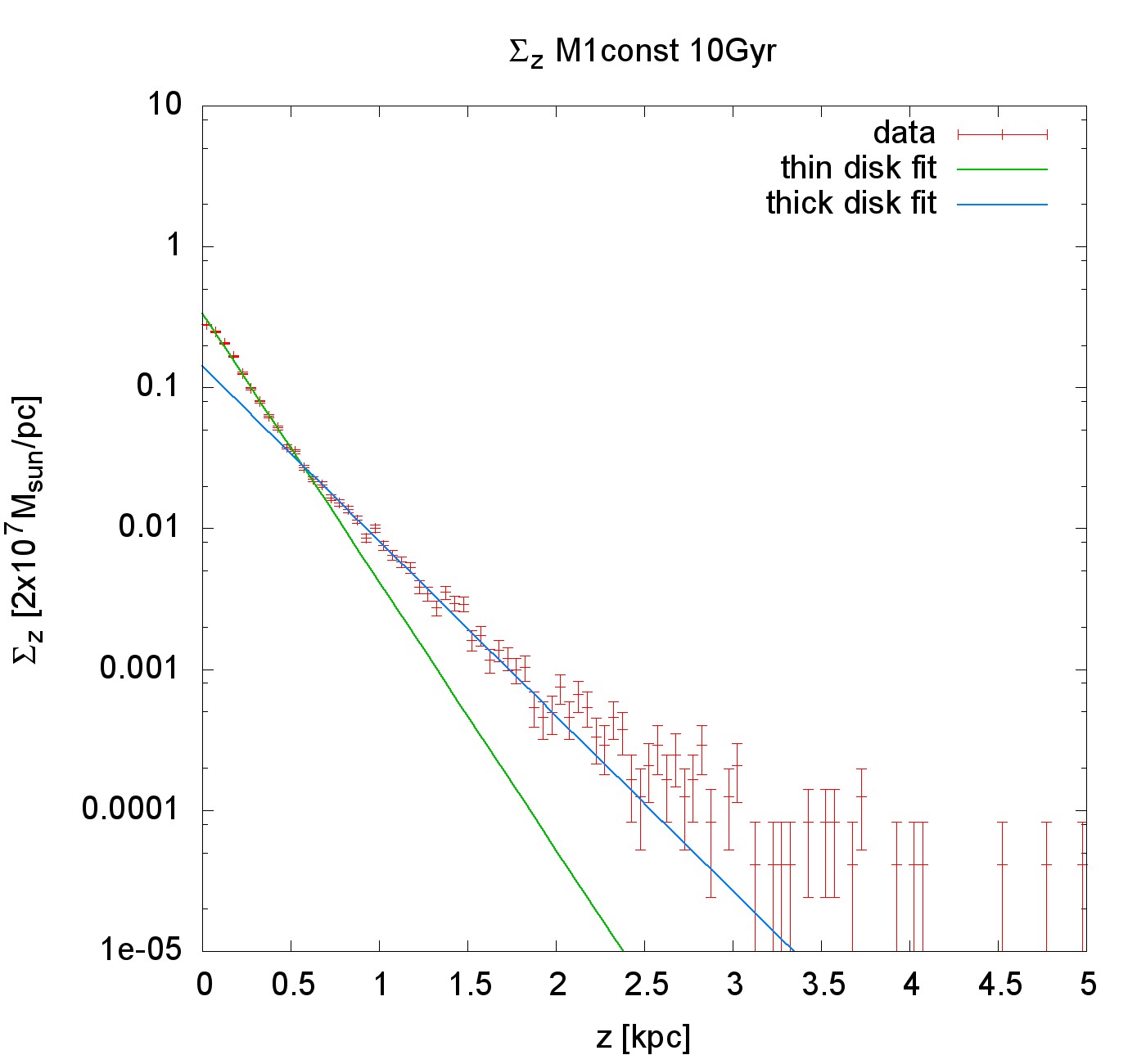}
\end{minipage}
\caption{As Fig. \ref{fig:sigmazM1} \textit{Left panel:}M1sn, \textit{Right panel:} M1const.} 
\label{fig:sigmazM1snM1const}
\end{figure*}
\newpage
\begin{figure*}[h]
\begin{minipage}[t]{0.49\linewidth}
\includegraphics[width=1.0\linewidth]{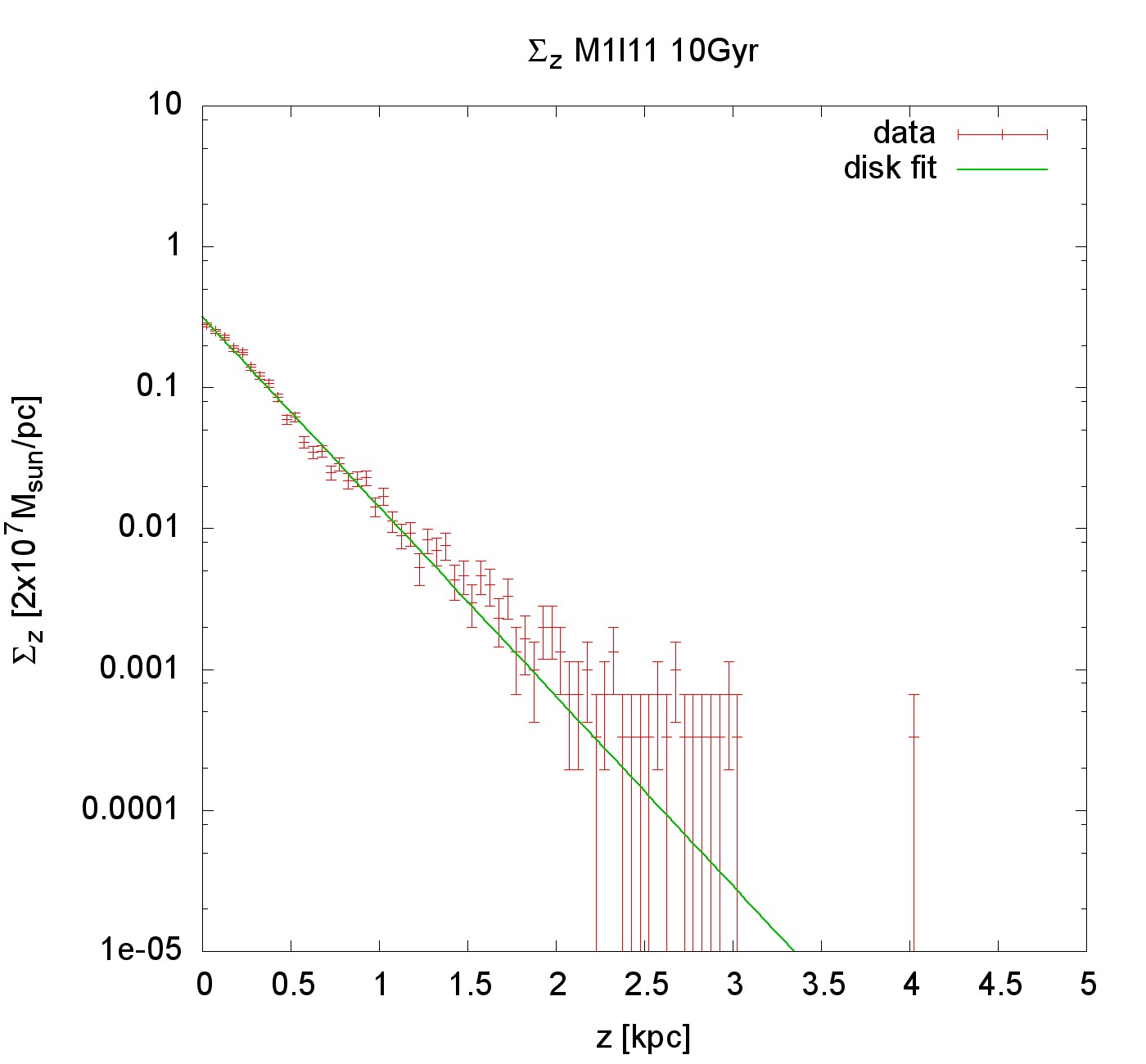}
\end{minipage}
\begin{minipage}[t]{0.49\linewidth}
\includegraphics[width=1.0\linewidth]{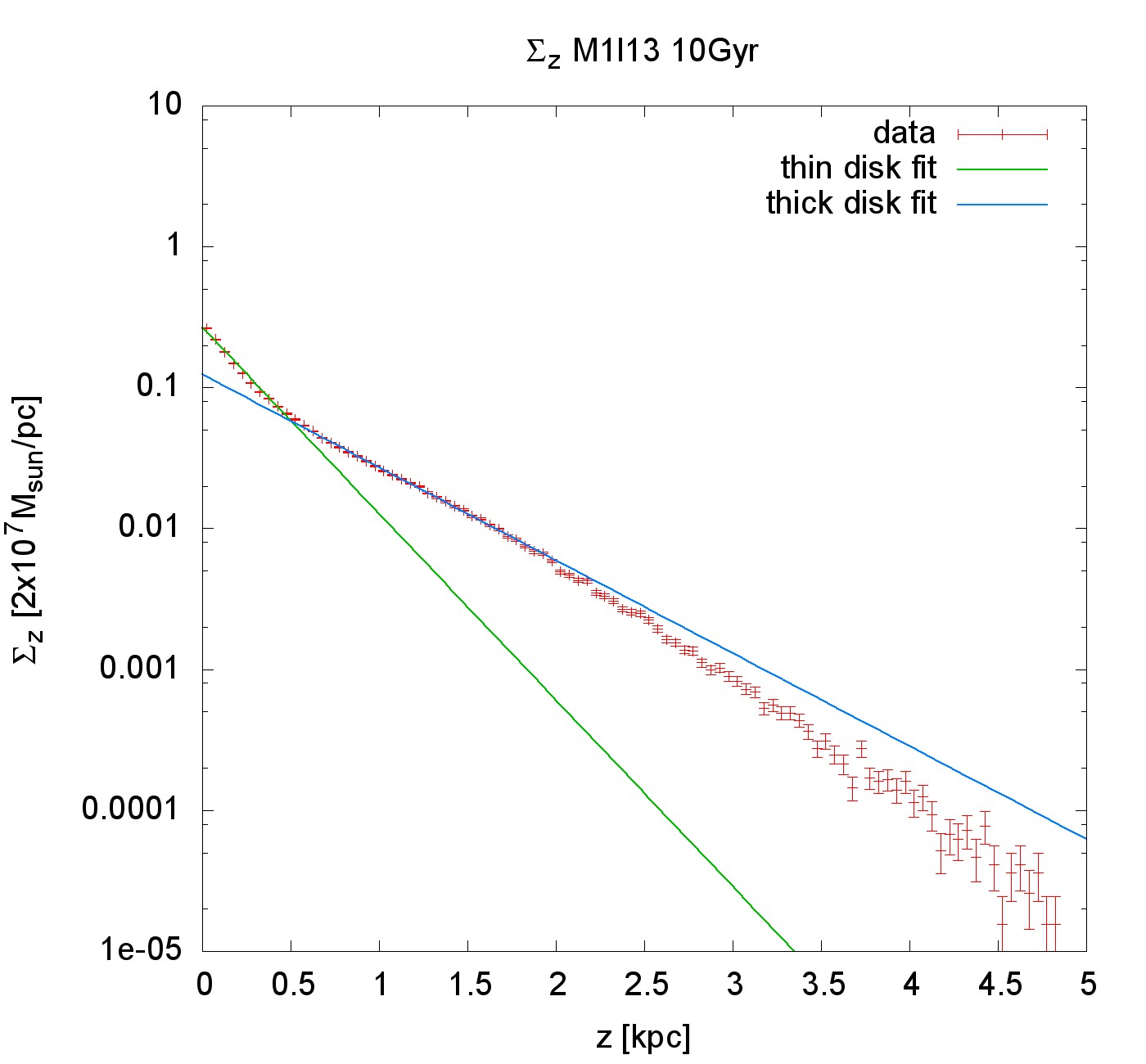}
\end{minipage}
\caption{As Fig. \ref{fig:sigmazM1} \textit{Left panel:}M1l11, \textit{Right panel:} M1l13.} 
\label{fig:sigmazM1l11M1l13}
\end{figure*}
\begin{figure*}[h]
\begin{minipage}[t]{0.49\linewidth}
\includegraphics[width=1.0\linewidth]{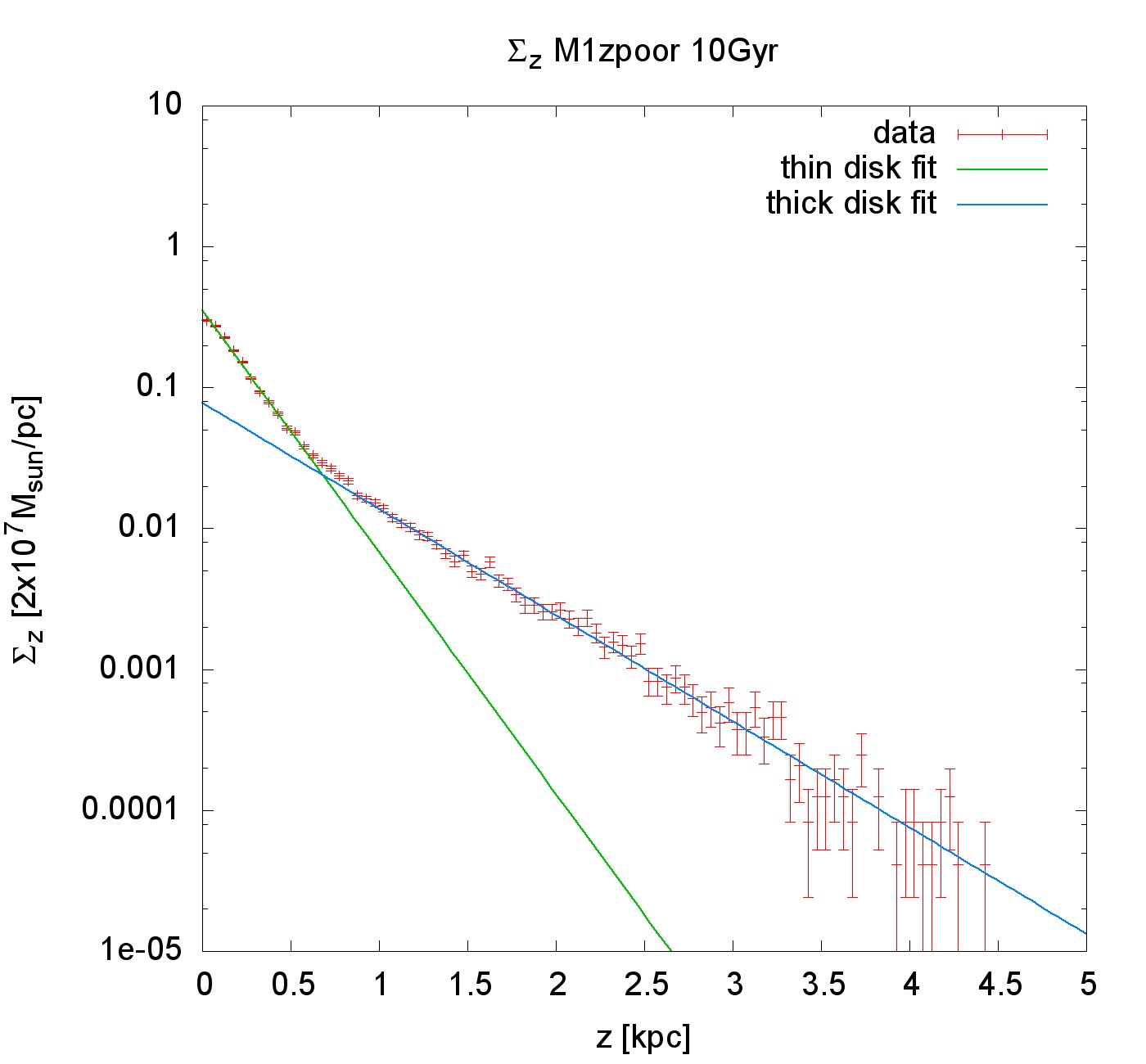}
\end{minipage}
\begin{minipage}[t]{0.49\linewidth}
\includegraphics[width=1.0\linewidth]{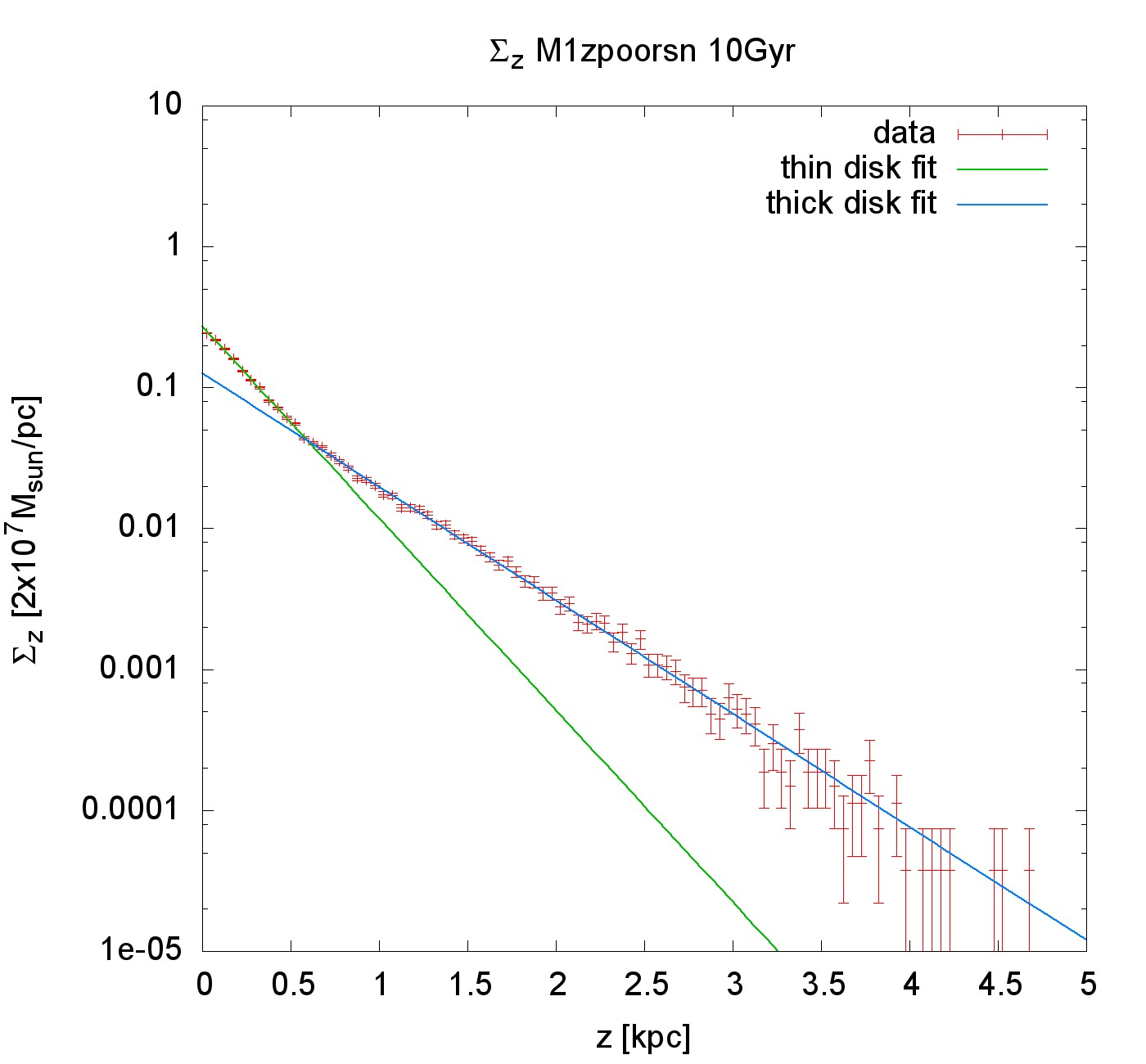}
\end{minipage}
\caption{As Fig. \ref{fig:sigmazM1} \textit{Left panel:}M1Zpoor, \textit{Right panel:} M1Zpoorsn.} 
\label{fig:sigmazM1ZpoorM1Zpoorsn}
\end{figure*}
\newpage
\begin{figure*}[h]
\begin{minipage}[t]{0.49\linewidth}
\includegraphics[width=1.0\linewidth]{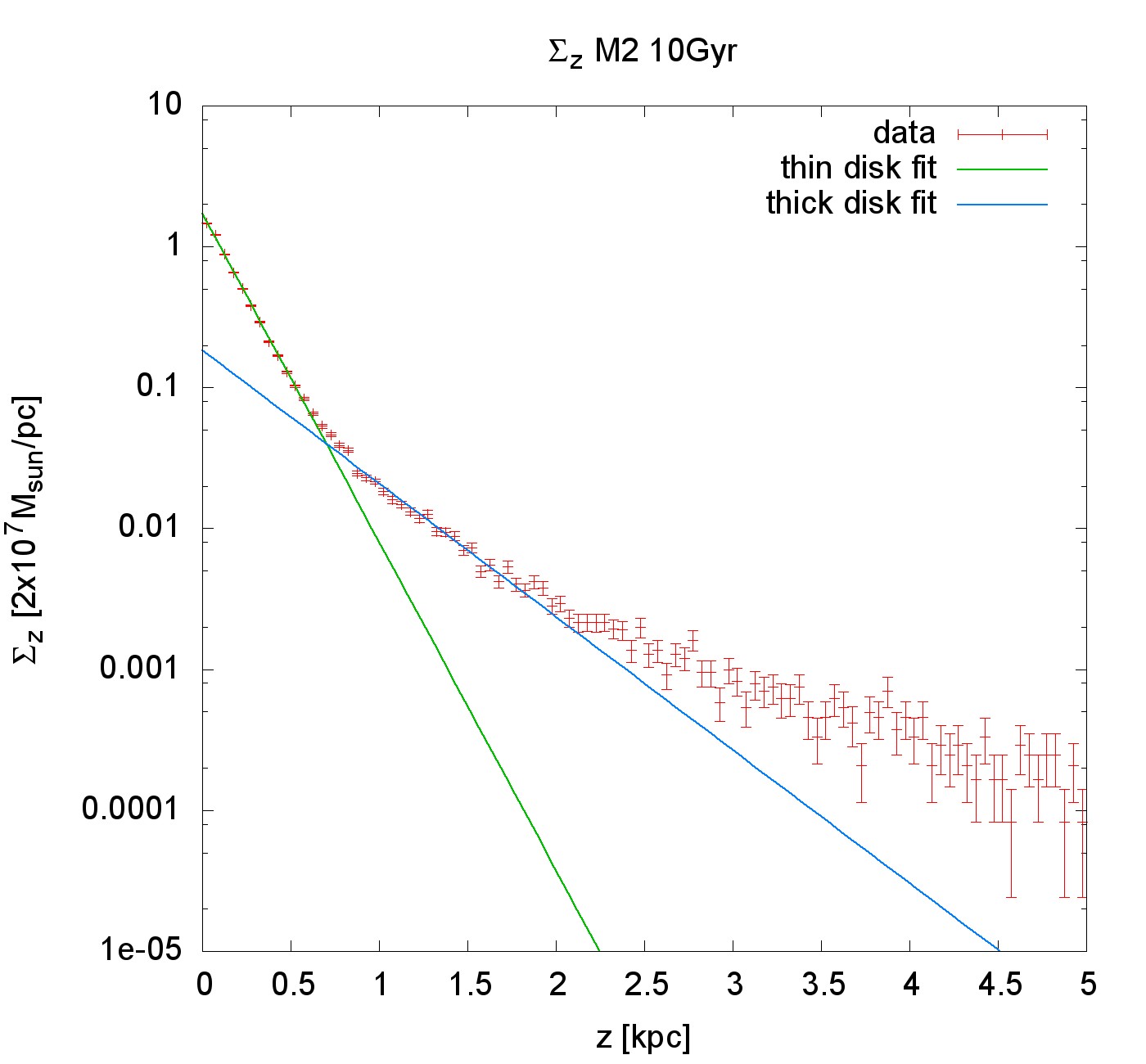}
\end{minipage}
\begin{minipage}[t]{0.49\linewidth}
\includegraphics[width=1.0\linewidth]{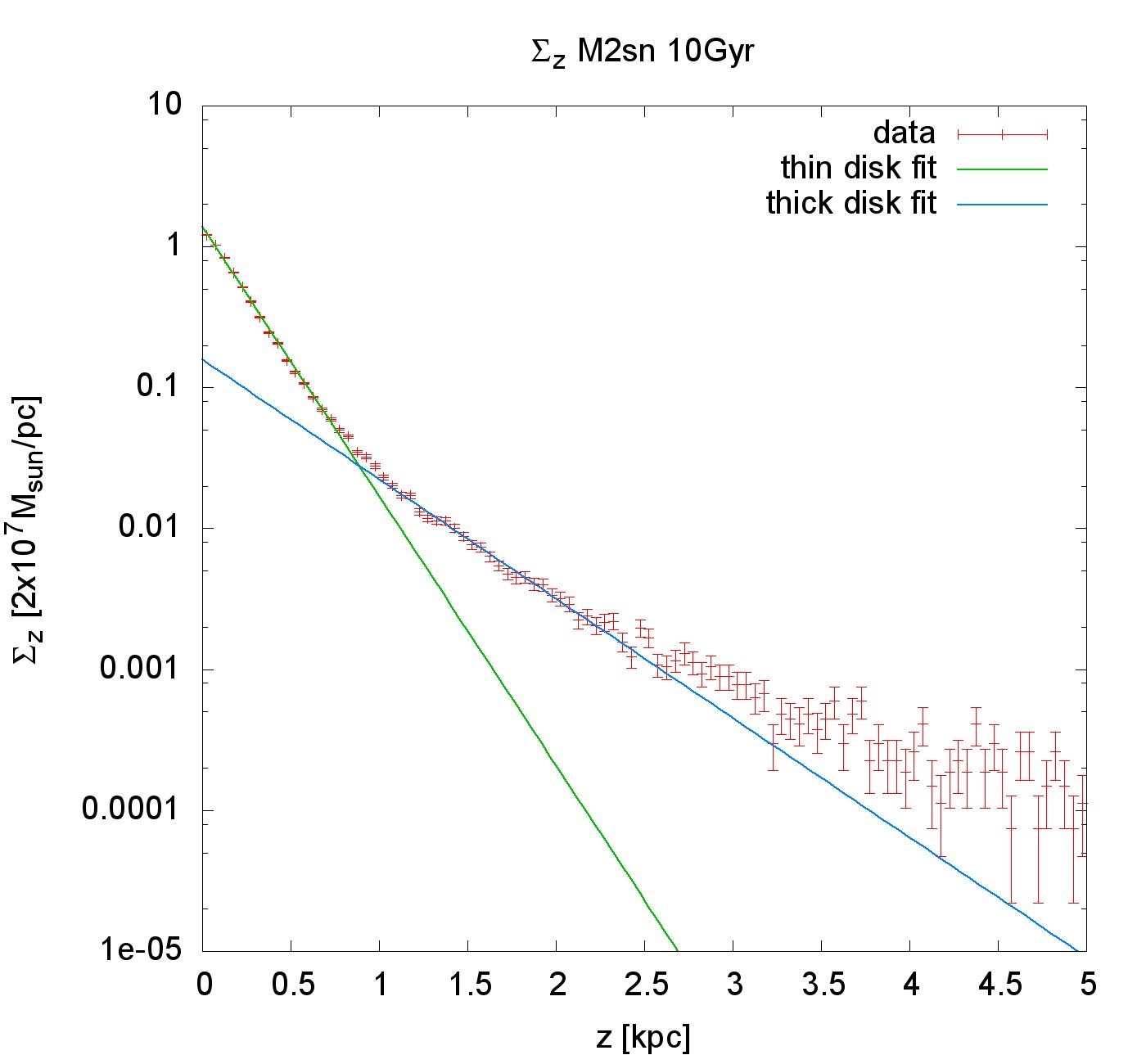}
\end{minipage}
\caption{As Fig. \ref{fig:sigmazM1} \textit{Left panel:}M2, \textit{Right panel:} M2sn.} 
\label{fig:sigmazM2M2sn}
\end{figure*}
\begin{figure*}[h]
\begin{minipage}[t]{0.49\linewidth}
\includegraphics[width=1.0\linewidth]{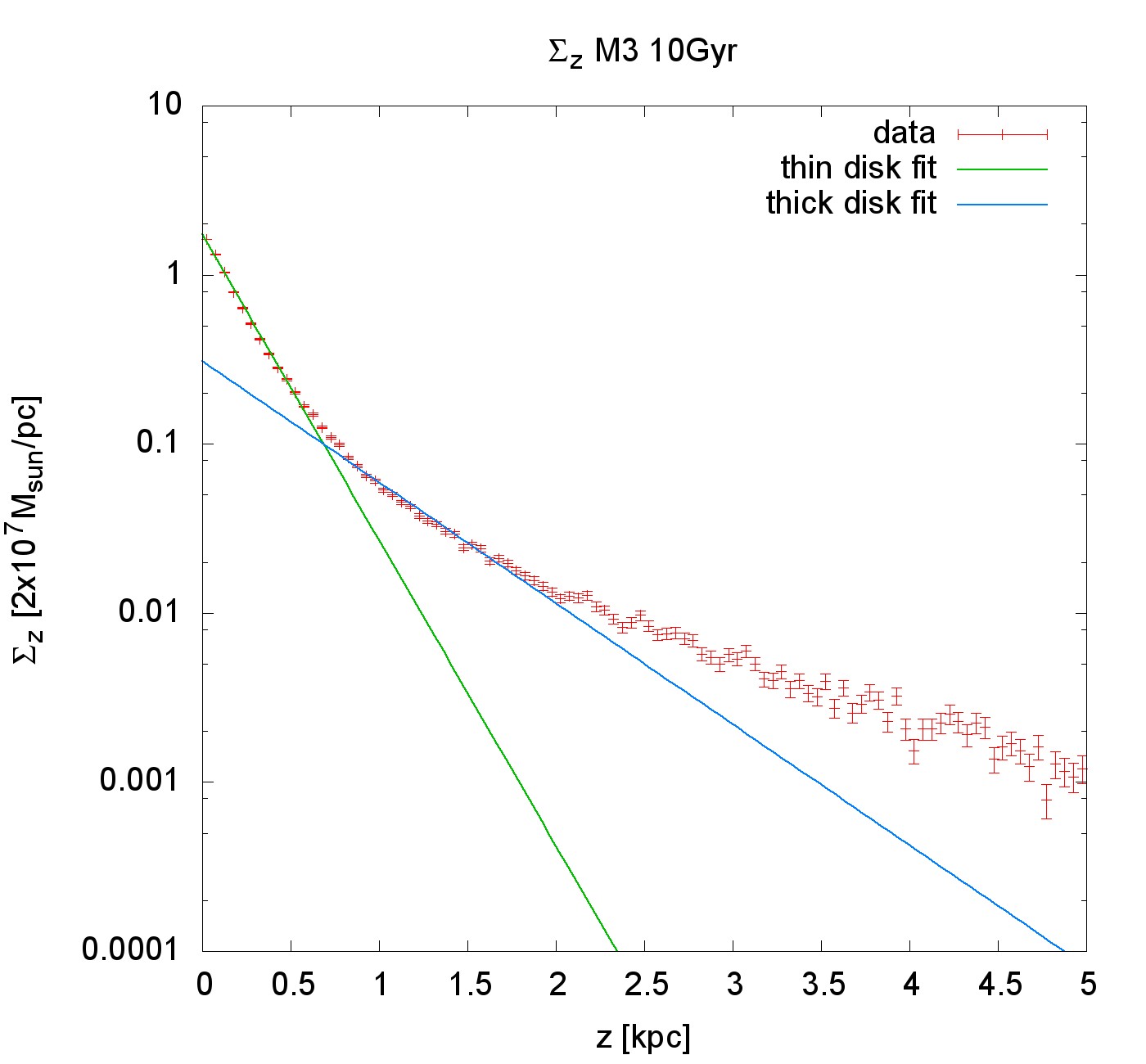}
\end{minipage}
\begin{minipage}[t]{0.49\linewidth}
\includegraphics[width=1.0\linewidth]{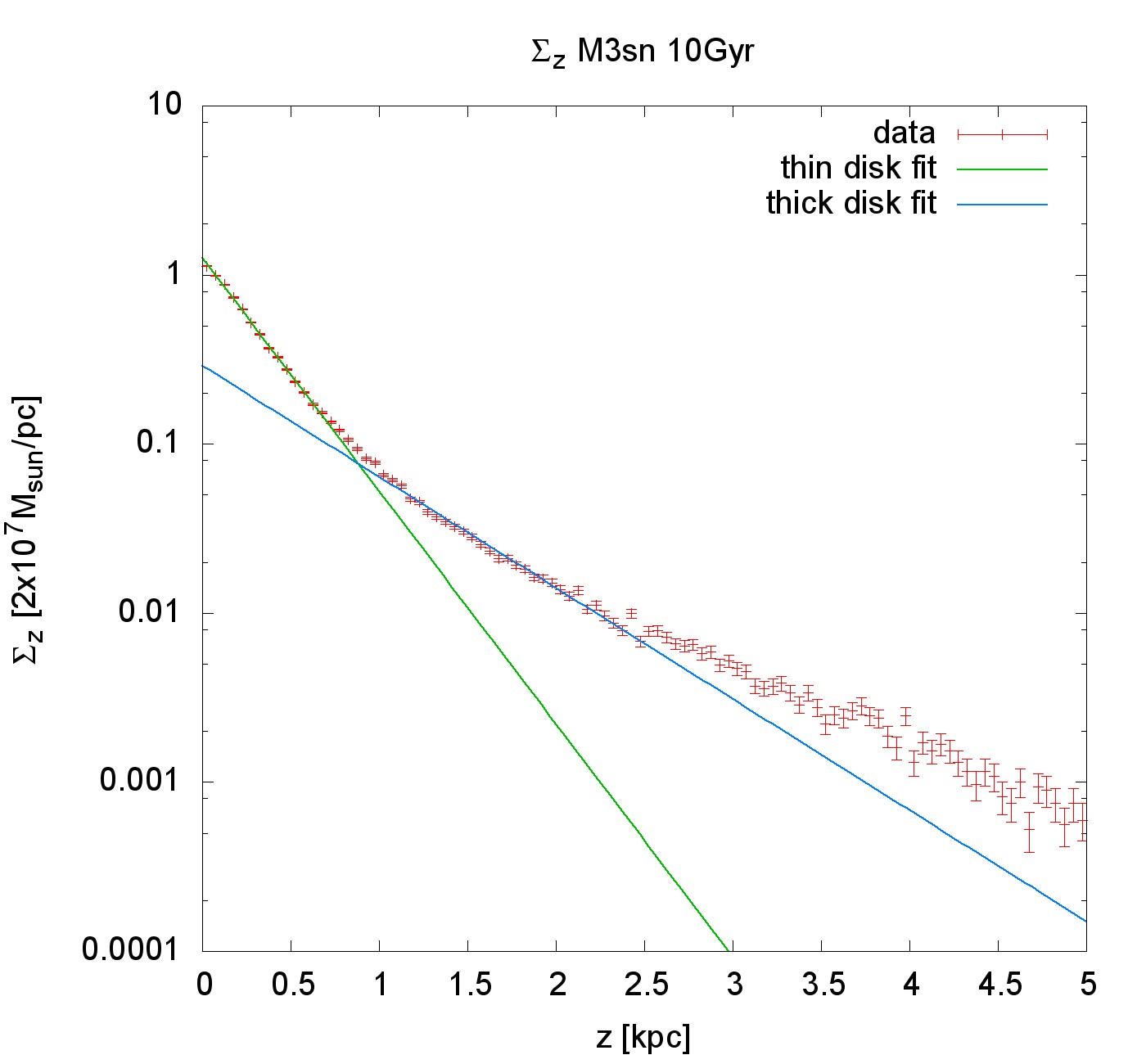}
\end{minipage}
\caption{As Fig. \ref{fig:sigmazM1} \textit{Left panel:}M3, \textit{Right panel:} M3sn.} 
\label{fig:sigmazM3M3sn}
\end{figure*}
\newpage
\begin{figure*}[h]
\begin{minipage}[t]{0.49\linewidth}
\includegraphics[width=1.0\linewidth]{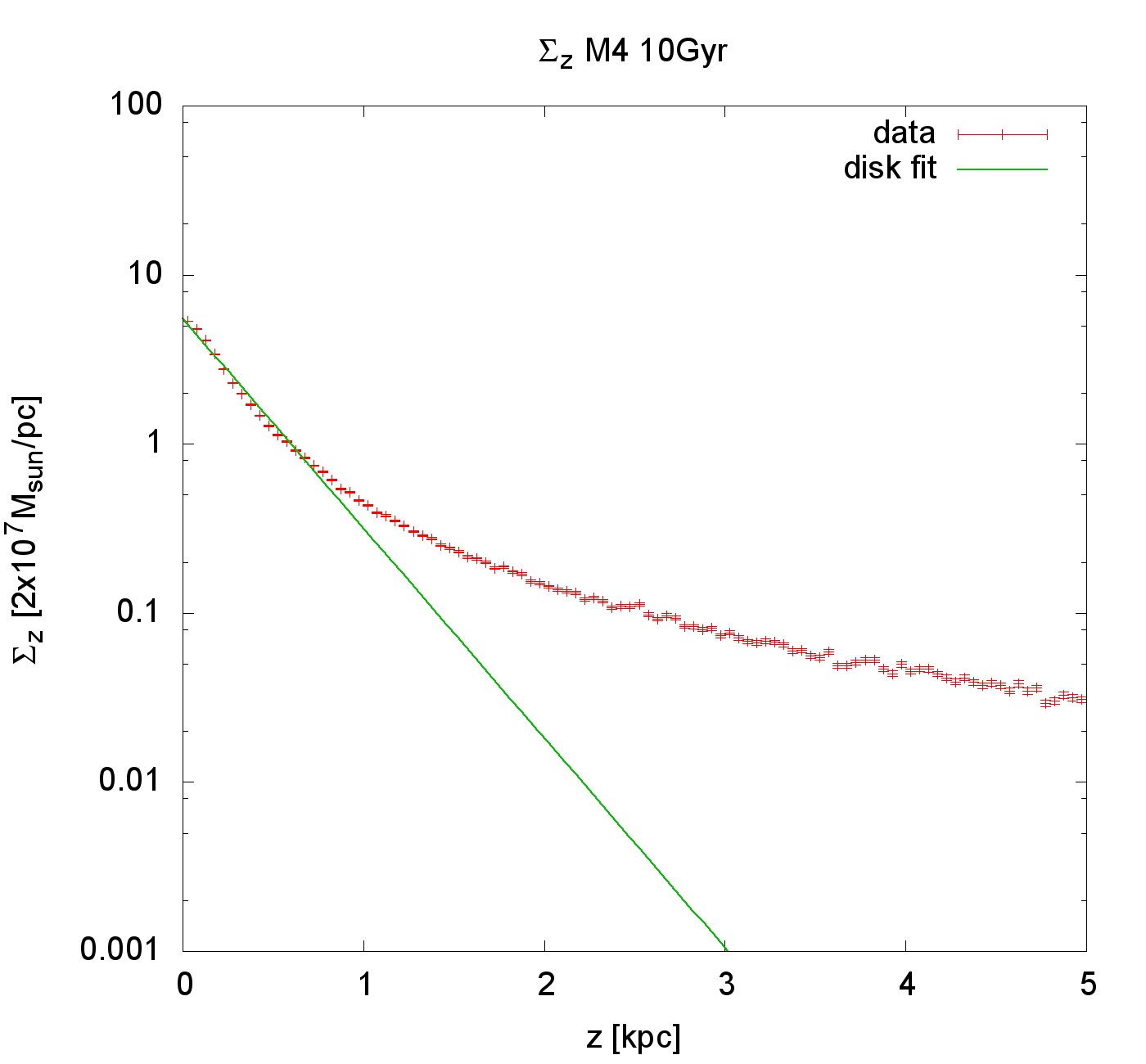}
\end{minipage}
\begin{minipage}[t]{0.49\linewidth}
\includegraphics[width=1.0\linewidth]{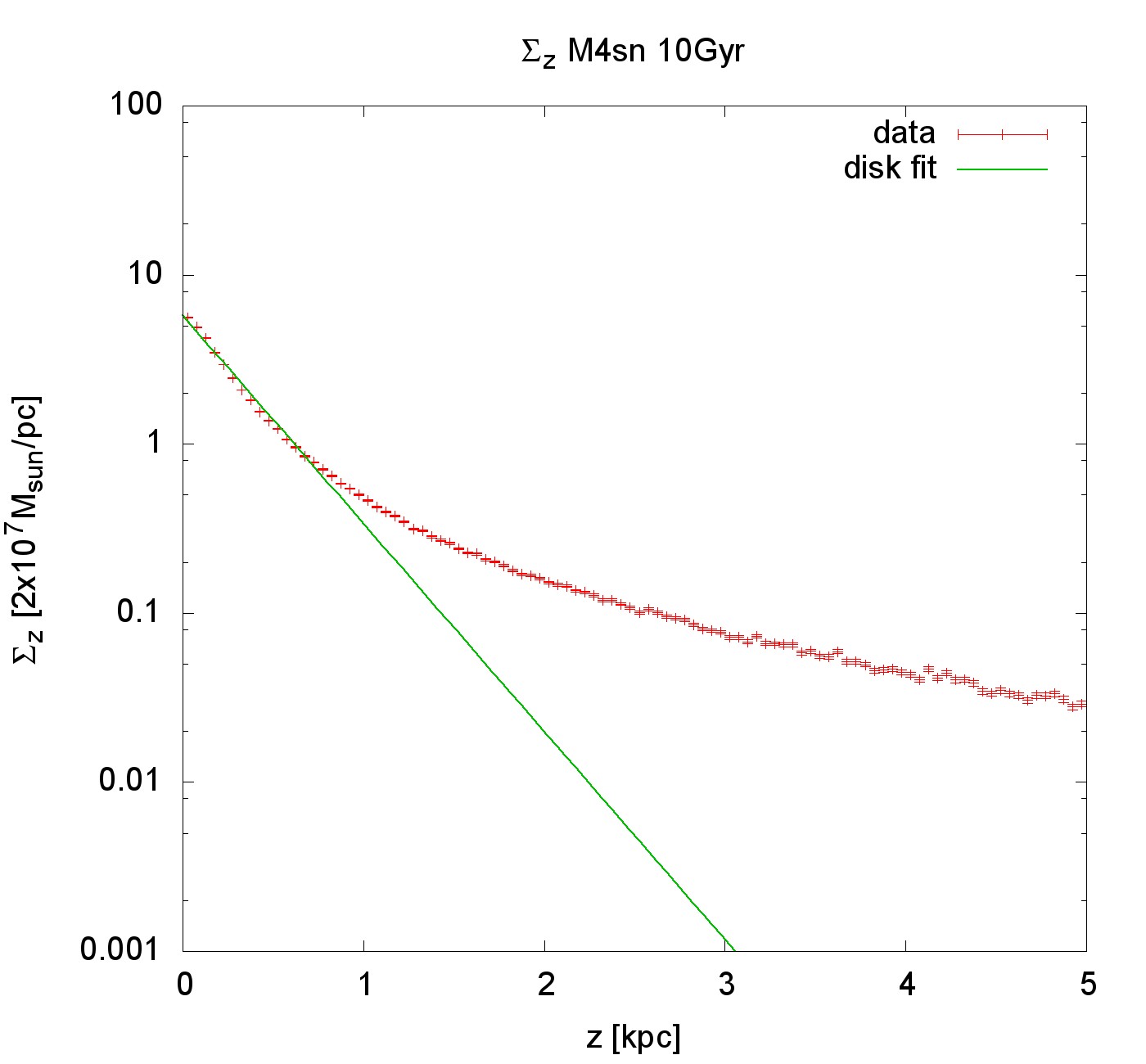}
\end{minipage}
\caption{As Fig. \ref{fig:sigmazM1} \textit{Left panel:}M4, \textit{Right panel:} M4sn.} 
\label{fig:sigmazM4M4sn}
\end{figure*}
\begin{figure*}[h]
\begin{minipage}[t]{0.49\linewidth}
\includegraphics[width=1.0\linewidth]{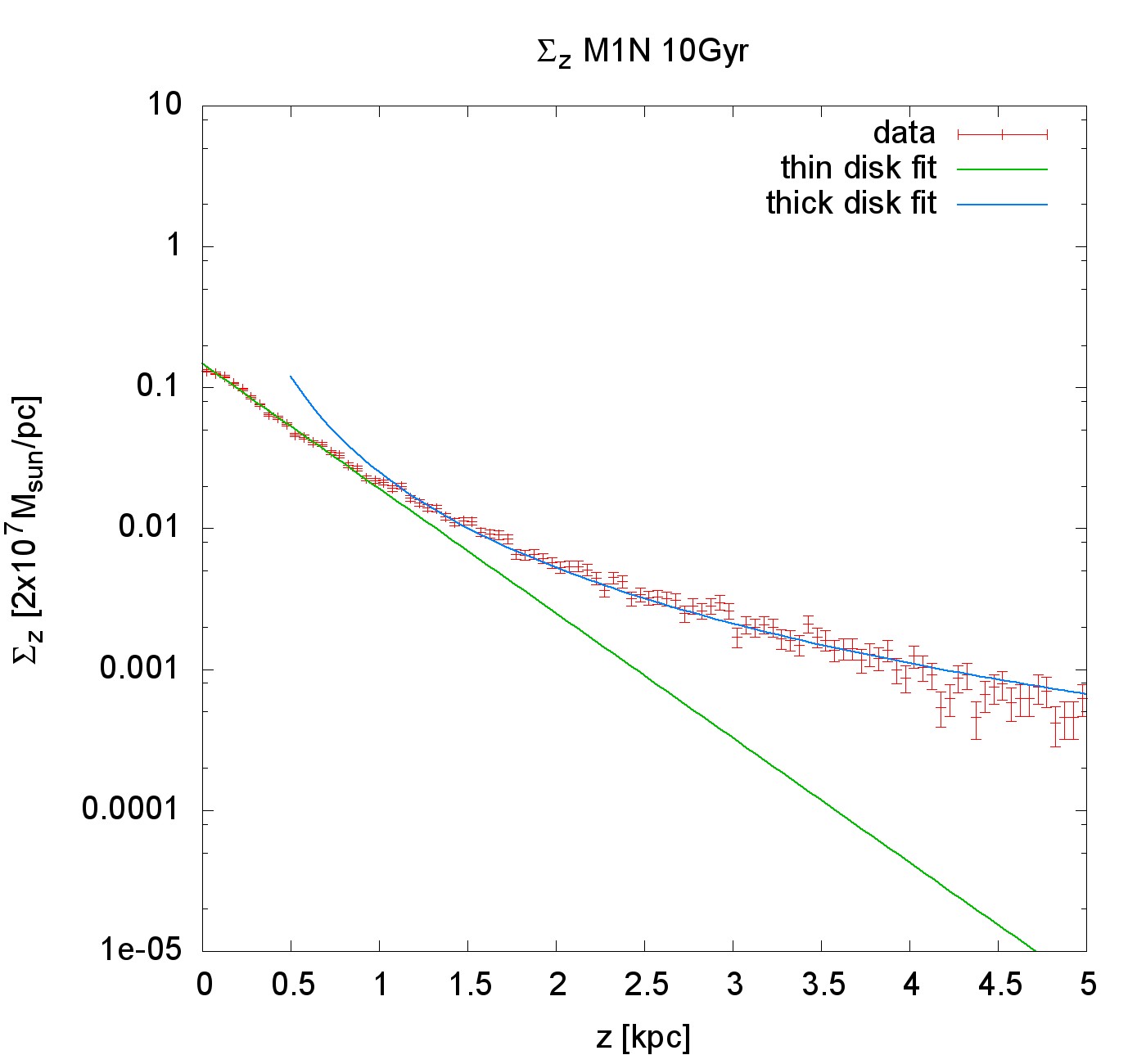}
\end{minipage}
\begin{minipage}[t]{0.49\linewidth}
\includegraphics[width=1.0\linewidth]{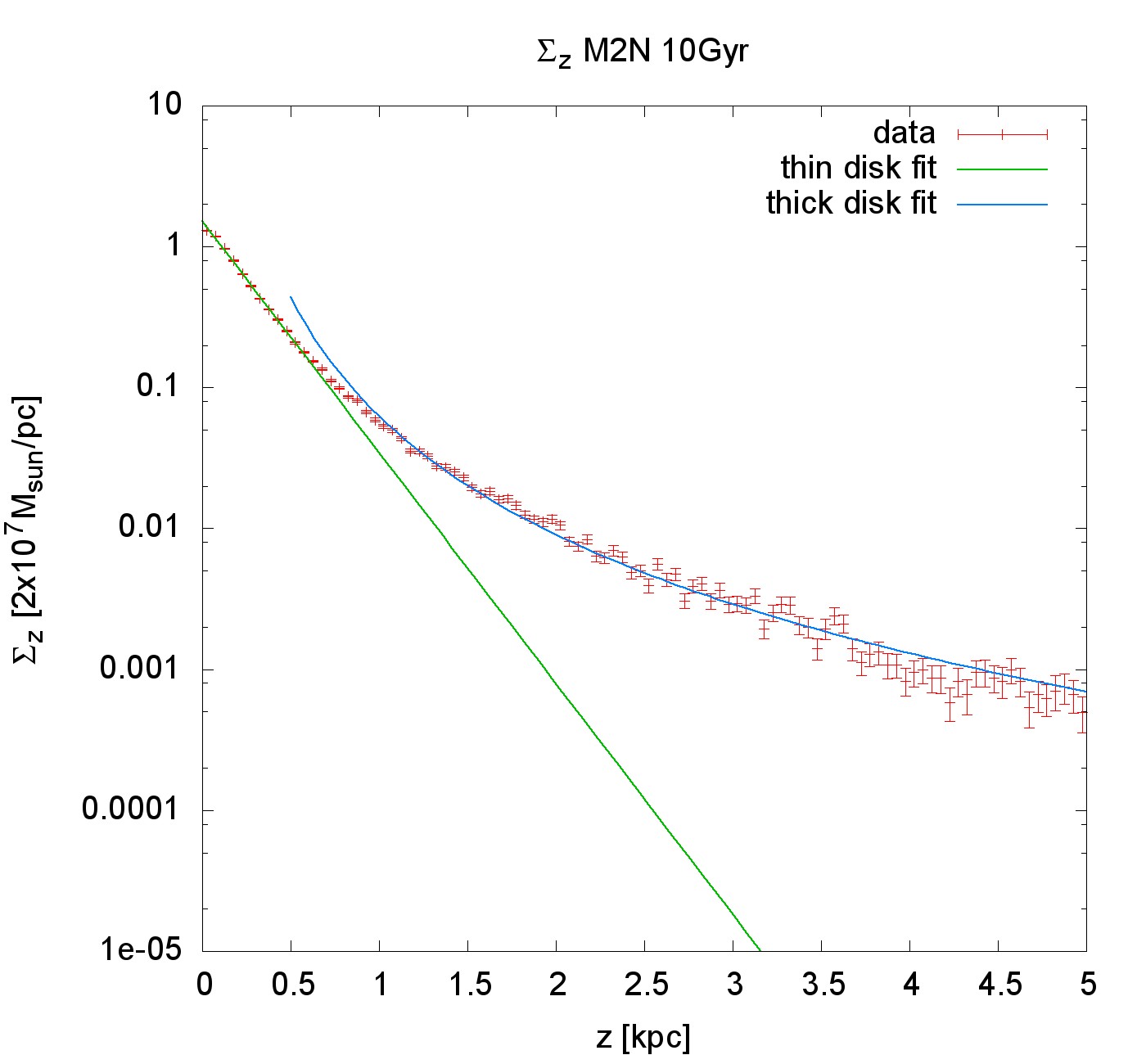}
\end{minipage}
\caption{As Fig. \ref{fig:sigmazM1}, but the thick disk fit is done with Eq. \ref{eq:powerlaw} as the profile is rather curved than straight. \textit{Left panel:} M1N, \textit{Right panel:} M2N.} 
\label{fig:sigmazM1NM2N}
\end{figure*}
\end{document}